\documentclass[12pt, openright, a4paper]{article}
 \usepackage{subcaption}
\usepackage{subfiles}
\usepackage{times}
\usepackage{amsmath}
\usepackage{array}
\usepackage{delarray}
\usepackage{epsfig}
\usepackage{natbib}
\usepackage[subfigure]{tocloft}
\usepackage{float}
\usepackage{amsfonts}
\RequirePackage{verbatim}
\usepackage{lscape}
\usepackage{ragged2e}
\usepackage{soul}
\usepackage[section]{placeins}
\usepackage{bbm}
\usepackage{amssymb}
\usepackage{caption}
\usepackage{graphicx}
\usepackage{booktabs}
\usepackage{tabularx}

\let\oldsubsection\subsection
\renewcommand{\subsection}{\FloatBarrier\oldsubsection}

\usepackage{atbegshi}

\usepackage{tikz}
\usetikzlibrary{arrows}

\newcolumntype{C}[1]{>{\centering\arraybackslash}p{#1}}

\usepackage[hyphens]{url}

\makeatletter
\def\url@leostyle{%
  \@ifundefined{selectfont}{\def\UrlFont{\sf}}{\def\UrlFont{\small\ttfamily}}}
\makeatother
\makeatletter
\newcommand\URLhyphenOn{\def\do@url@hyp{\do\-}}
\newcommand\URLhyphenOff{\def\do@url@hyp{}}
\makeatother

\usepackage{paralist}
\usepackage{enumitem}
\usepackage[doublespacing]{setspace}

\usepackage{geometry}
\newgeometry{
    top=1in,
    bottom=1in,
    outer=1in,
    inner=1in,
}
\usepackage{romannum}

\usepackage[colorlinks=true,linkcolor=blue,allcolors=blue, citecolor=blue]{hyperref}

\begin{document}

\pagenumbering{arabic}
\clearpage

\bf
\Large
\begin{center}
Gender Differences in Healthcare Utilisation: Causal Evidence from Unexpected Adverse Health Shocks

\vspace{1cm}
\normalsize \normalfont

\noindent

Nadja van 't Hoff\footnote{Assistant Professor, Quantitative Economics Section, University of Amsterdam}, Giovanni Mellace\footnote{Professor, Department of Economics, University of Southern Denmark}, Seetha Menon\footnote{Associate Professor, Department of Economics, University of Southern Denmark} \\

\normalsize \normalfont
\vspace{1cm}

\begin{abstract}
\noindent
\textit{Women live longer than men yet report worse health. One common reading of this male-female health-survival paradox is that women also engage more with healthcare. We challenge that reading with causal evidence from the plausibly random timing of first-time non-fatal heart attacks and strokes in Danish administrative data. After such a shock, men increase their statin use and their general-practitioner visits substantially more than women. We find no evidence that the gap reflects a lower willingness among women to seek or take up care: men and women fill the same number of prescriptions, but women receive lower doses per fill, and the gap widens across drug classes where physicians have more discretion over what to prescribe. This points to provider behaviour rather than patient demand. Despite the additional treatment, men fare no better than women on mortality or morbidity, or on the labour-market outcomes we can measure. Given women's general survival advantage, this suggests women might have fared even better had they been treated as intensively as men. The gap arises even within a universal healthcare system and a country with comparatively low gender inequality, so removing barriers to access is not by itself sufficient to close it.}

\end{abstract}
\vspace{1cm}
\noindent

\end{center}

{\par\normalfont\normalsize

  \par

  \vspace{1em}

  \begingroup
  \itshape\small
  \noindent
We are grateful to Esm\'{e}e Zwiers, Melissa Thomasson, and Therese Nilsson for helpful comments, and to participants at EEA 2026, the Essen Health Conference 2026, COMPIE 2026, IAAE 2025, IHEA 2025, the Royal Economic Society Conference 2025, the EUI Alumni Conference in Economics, and seminars at LMU Munich and the University of Siena. Financial support from the Danish Independent Research Foundation (Grant ID: 2033-00076B) is gratefully acknowledged.
  \par
  \endgroup
}

\normalsize \normalfont

\newpage


\section{Introduction}

In high-income countries women outlive men by roughly four to six years, yet over the life course they report poorer self-assessed health, more chronic conditions and higher disability rates \citep{case2005sex,oksuzyan2008men,blaakilde2012dode,pedersen2014sundhed}. This male-female health-survival paradox has persisted across decades and institutional settings, and it remains only partly understood \citep{cutler2005explains}. The behavioural and clinical channels that sustain it are hard to isolate, because the gender differences visible in cross-sectional data confound biology, prior morbidity, healthcare-seeking propensity and provider response.

This paper provides causal evidence on one component of the paradox: how men and women differ in their engagement with healthcare after a major, unexpected adverse health event. In the language of the health-capital model \citep{grossman,grossman2000human}, such an event is a sudden depreciation of health capital that prompts reinvestment in health. The optimal response depends on the remaining horizon, wages and the return to that investment, so the model does not require men and women to respond equally. If anything, women's longer life expectancy implies a larger optimal reinvestment for women, so demand-side theory does not predict that men should raise their care by more. This matters because secondary prevention, productivity recovery and long-run mortality are largely determined in the years following an acute event \citep{Heidenreich2011,Roth2018}, so a gender gap carries first-order welfare implications. Whatever drives such a gap (patient demand, provider supply, or both) bears on the design of healthcare systems, on the literature linking physician-patient gender concordance to clinical outcomes \citep{cabral2024gender,greenwood2018patient}, and on debates about statistical discrimination in medicine \citep{schulman1999effect}. We study this in Denmark, a country with universal, free healthcare and relatively low gender inequality, which narrows the set of admissible explanations because access constraints cannot play the explanatory role they might play elsewhere.

We find a large gender gap in post-shock cardiovascular care. Men increase their secondary-prevention treatment by more than women do in the years after a first heart attack or stroke. Aggregated over the five post-shock years, men's statin volume and primary-care contacts both rise more than women's, on the order of an additional general-practitioner visit per year and around twenty additional defined daily doses (DDD) of statins per year. The design identifies the causal effect of the shock separately for men and for women. The difference between these effects is not itself causally identified, so we treat it as a comparison of two causal effects and examine it directly with a propensity-score-matched sample and a triple-difference specification. The estimates are precise, and pre-trends are small or negligible. The gap holds under alternative estimators, alternative comparison groups, severity strata and a balanced-panel restriction. The triple-difference specification yields the same result, and matching leaves the general-practitioner gap unchanged while narrowing the statin gap at longer horizons.

The gap is not explained by the leading alternative explanations. It does not reflect differential severity, since it survives within severity strata. It does not reflect survival selection, since we detect no gender difference in the shock's mortality effect and the results hold in a balanced panel conditioning on five-year survival. It does not reflect women filling fewer prescriptions or discontinuing more often: the number of separate statin pickups rises by about two per year for both genders, with only a small male excess in the earliest post-shock years, so the volume gap predominantly reflects lower-dose prescriptions per pickup rather than fewer prescriptions. It does not reflect a mechanical ceiling on women: their higher baseline primary-care use, which we examine directly, does not drive the gap.

 Across the cardiovascular secondary-prevention drugs, the male-female gap is small and statistically insignificant for antiplatelets, the most protocolised post-infarction therapy and the one with the most sex-neutral guidelines, moderate for statins, larger for beta blockers and largest for ACE inhibitors. The gap thus grows along the dimension in which clinical guidelines and provider judgement permit more variation. Generic patient demand, adherence or differential health needs would each predict a similar gap across drug classes, so this ordering is hard to reconcile with those explanations, although class-specific tolerability and eligibility differences, which we discuss in Section~\ref{sec:gradient}, could contribute to it. The pattern points away from patient behaviour and towards how care is delivered.

The welfare significance of the gap depends on whether the additional care men receive improves meaningful outcomes. On the margins we observe, it does not: the shock's effect on mortality is statistically indistinguishable across genders at every horizon up to twenty years, the labour-market outcome we measure show no gender difference, and men accumulate at least as many new comorbidities as women. The larger male treatment response is therefore not associated with any measurable advantage over women. Because women also hold a well-documented survival advantage, the pattern is consistent with a higher marginal return to secondary-prevention care for women than for men: had their treatment matched men's, women might have fared even better. 

The paper builds on four bodies of work. A growing literature uses cardiovascular and other acute events as exogenous variation in the marginal value of healthcare, estimating effects on labour supply, family behaviour and care productivity \citep{baji2018adaptation,chandra2007productivity,doyle2011returns,fadlon2021family,bockerman2025family,chandra2020identifying,nielsen}; we are the first to apply this design to gender differences in the patient response itself, using a large administrative panel and modern staggered difference-in-differences methods. An extensive medical literature documents lower prescribing of secondary-prevention therapies to women, particularly high-intensity statins \citep{galdas2005men,juel2007men,nobili2011drug,oksuzyan2011sex,smith2009gender,legato2016consideration,Peters2018sex,nanna2019sex,kiss2024sex,sibbritt2024demographic,Vaccarino2011}; our finding of near-identical pickup counts but lower dose per pickup recovers the same pattern in Danish administrative data, and provides, to our knowledge, the first quasi-experimental estimates of the gender-specific prescribing responses to a health shock, obtained from variation in event timing rather than from cross-sectional comparisons conditional on an event. A complementary economics literature shows that provider characteristics shape healthcare utilisation \citep{currie2016provider,finkelstein2016sources} and that physician-patient gender concordance affects clinical outcomes \citep{greenwood2018patient,cabral2024gender}; while we do not observe the treating physician, near-identical pickups with different doses are consistent with a supply-side channel and motivate a direct concordance test in follow-up work. Our results also speak to evidence on gender differences in preferences and risk-taking \citep{croson2009gender,powell1997gender} and on adverse shocks that themselves shift risk preferences \citep{hanaoka2018risk}, which could in principle change men's and women's demand for preventive care differently. The way the gap varies across drug classes, together with the near-equality of statin-pickup counts (Sections~\ref{sec:demandsupply}--\ref{sec:gradient}), points away from such demand-side channels and towards provider behaviour.

The paper proceeds as follows. Section~\ref{sec:data} describes the registers, the analytical sample and the gender comparison on observables. Section~\ref{sec:identification} sets out the identification strategy and the identifying assumptions. Section~\ref{sec:results} establishes the gender gap in post-shock care. Section~\ref{sec:ruleout} rules out the leading explanations, including the cross-drug pattern in provider discretion. Section~\ref{sec:outcomes_men} asks whether the additional care men receive improves their outcomes. Section~\ref{sec:conclusion} concludes.

\section{Data} \label{sec:data}
Measuring underlying health status presents significant methodological challenges, particularly in survey data where measurement error is common. Inaccuracies in measuring healthcare utilisation or the timing of health shocks can introduce substantial bias into our estimates. To address these limitations, we use reliable and objective administrative measures of health events and healthcare utilisation, characterised by high completeness and verified accuracy through multiple validation studies \citep{schmidt2015danish,ozcan2016danish}.
Our analysis draws on three comprehensive Danish administrative datasets.  The Danish National Patient Register (LPR) records every inpatient and outpatient hospital contact, with ICD-10 diagnoses and procedure codes. The Danish Health Insurance database (Sygesikringsregistret) records every primary-care contact and its associated insurance fee. The Danish Prescription Drug database (L\ae{}gemiddelstatistikregistret) records every prescription filled at a Danish pharmacy, with anatomical-therapeutic-chemical (ATC) codes, defined daily doses (DDD) and patient co-payments. 

\subsection{Measures of health shock and analytical sample}\label{subsec:analyticalsamp}
We restrict the sample to Danish residents aged 50 to 70 who experienced a first-time non-fatal acute myocardial infarction (ICD-10 I21) or cerebral infarction (ICD-10 I63) between 1995 and 2018. We choose this age window for two reasons. At older ages cardiovascular risk equalises between men and women \citep{Vaccarino2011}, and the upper bound also limits selection from competing mortality.

\label{sec:healthshock}
Following \citet{nielsen}, we retain only individuals with a low pre-event CHA\textsubscript{2}DS\textsubscript{2}-VASc score, so that the timing of the event is plausibly orthogonal to underlying cardiovascular risk in our sample.  A CHA\textsubscript{2}DS\textsubscript{2}-VASc score assigns points for congestive heart failure,
hypertension, age $> 75$ (double), diabetes, previous stroke/transient
ischaemic attack/thromboembolism (double), vascular disease, age:
65 to 74, sex (female). Specifically, we exclude men with a risk score of 2 or higher and women with a risk score of 3 or higher, as defined by the European Society of Cardiology and commonly applied by medical professionals. Since female sex contributes one point to the CHA\textsubscript{2}DS\textsubscript{2}-VASc score by construction, applying a threshold of $\geq$3 for women and $\geq$2 for men yields comparable cut-offs in terms of non-sex-related cardiovascular risk factors.

\begin{figure}[htbp]
    \centering
    \includegraphics[width=0.85\linewidth]{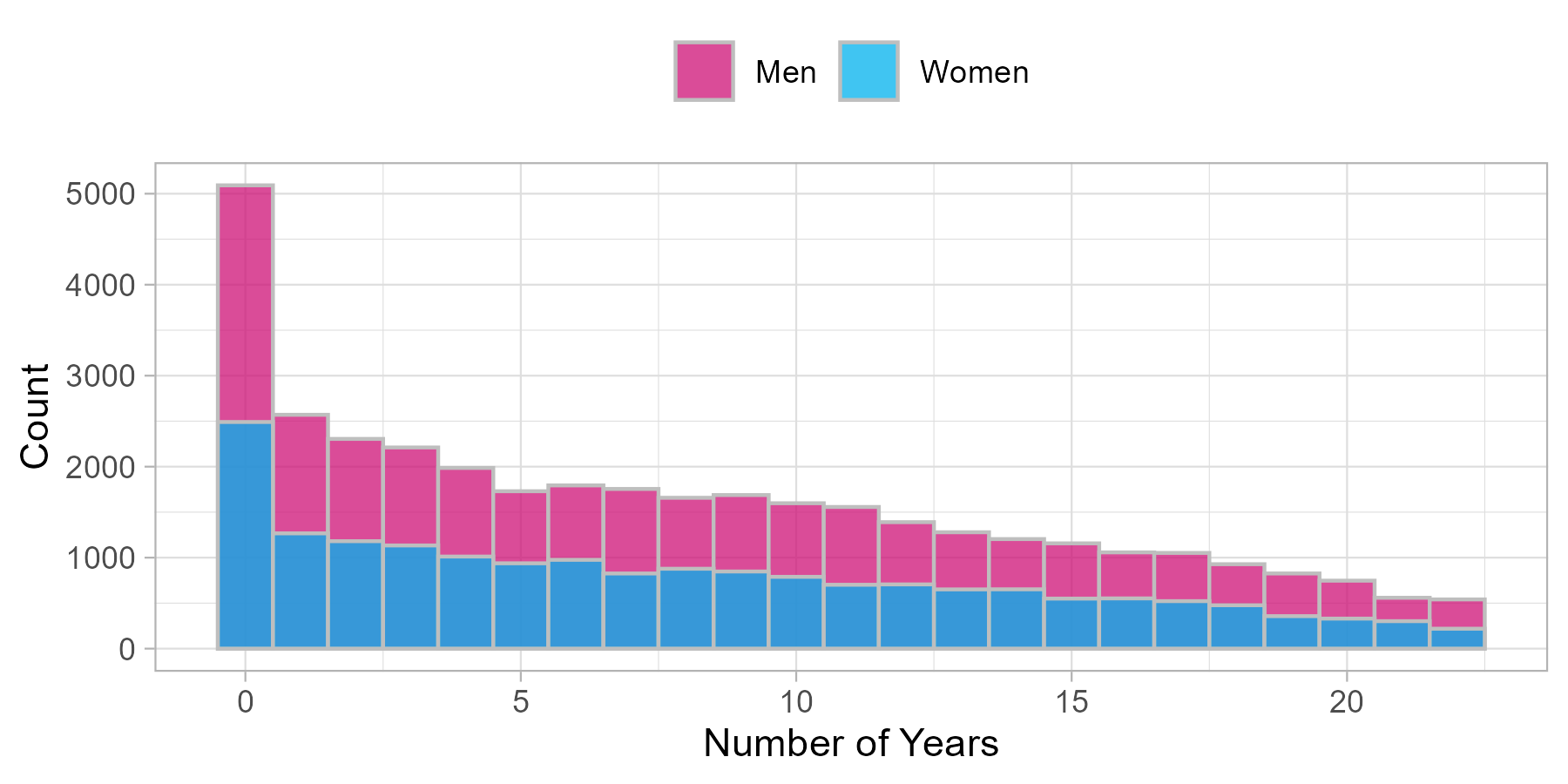}
    \caption{Time until death after the health shock, by gender.}
    \label{fig:time_til_death}
    \begin{minipage}{0.9\textwidth}
    \footnotesize \textit{Notes:} Cumulative distribution of years between the health shock and death, conditional on dying within the observation window. Sample as in Table~\ref{tab:distributionofshock}.
    \end{minipage}
\end{figure}

We define a shock as \emph{non-fatal} if the individual survives at least 12 months after the event, and we exclude individuals who die earlier because their post-shock utilisation is not observed. The analytical sample is therefore a sample of survivors. Conditional on surviving 12 months, subsequent survival profiles are very similar across genders (Figure~\ref{fig:time_til_death}), and Section~\ref{sec:balanced} shows that the results are unchanged in a balanced panel conditioning on five-year survival.

\begin{table}[htbp]
\centering
\caption{Distribution of health shocks by gender.}
\label{tab:distributionofshock}
\begin{tabular}{lccc}
  \hline\hline
 & Men & Women & $p$-value \\
  \hline
Acute myocardial infarction & 58.98\% & 50.05\% & $<0.01$ \\
Cerebral infarction         & 41.02\% & 49.95\% &  \\
\midrule
Total                       & 100\%   & 100\%   &  \\
   \hline\hline
No.\ unique individuals     & 30{,}103 & 15{,}140 & \\
   \hline\hline
\end{tabular}
\\[1em]
\begin{minipage}{0.9\textwidth}
\centering
\footnotesize \textit{Notes:} Sample of Danish residents aged 50--70 experiencing their first non-fatal acute myocardial infarction or cerebral infarction during 1995--2018, conditional on a low pre-event CHA\textsubscript{2}DS\textsubscript{2}-VASc score and 12-month survival. The $p$-value, reported once for the table, is from a single $\chi^{2}$ test of equal distribution of shock types across genders.
\end{minipage}
\end{table}

Table~\ref{tab:distributionofshock} reports the distribution of shock types by gender. Men are disproportionately likely to experience heart attacks, at around 59 per cent compared with 50 per cent for women, while women are about equally likely to experience either condition. This compositional difference, together with documented sex differences in the clinical presentation and management of cardiovascular events \citep{Canto2000,Schulte2023,Joseph2021,Cardeillac2022}, motivates our decision to estimate gender-specific causal effects rather than pool the sample.

\begin{figure}[htbp]
    \centering
    \includegraphics[width=0.85\linewidth]{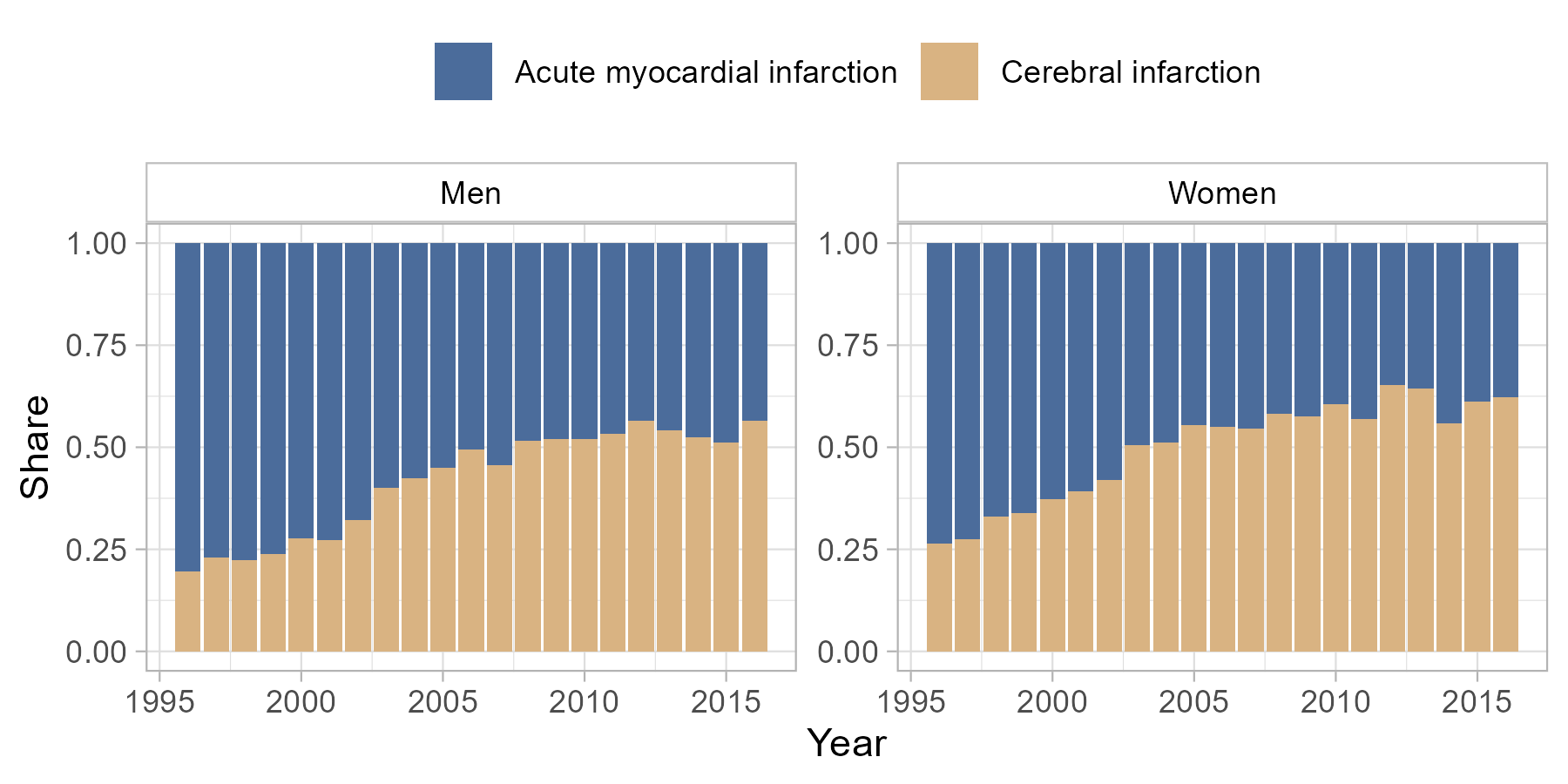}
    \caption{Share of observations by gender and shock type over time.}
    \label{fig:health_shock}
    \begin{minipage}{0.9\textwidth}
    \footnotesize \textit{Notes:} Annual share of first-time non-fatal cardiovascular events by type and gender, 1995--2018. The rising share of cerebral infarctions reflects standard high-income-country trends in cardiovascular disease epidemiology \citep{Li2020,Shah2021}. Sample as in Table~\ref{tab:distributionofshock}.
    \end{minipage}
\end{figure}

Figure~\ref{fig:health_shock} shows that the time profile of incidence is broadly similar across genders. The increasing share of cerebral infarctions relative to acute myocardial infarctions over time for both men and women is consistent with broader trends in high-income countries, where improved prevention and treatment of coronary heart disease, population ageing, and advances in stroke diagnostics have shifted the composition of major cardiovascular events (see, for example, \cite{Li2020} and \cite{Shah2021} for evidence on diverging long-run trends in myocardial infarction and stroke incidence).  Figure~\ref{fig:time_til_death} shows survival profiles by gender, which are very similar. This is reassuring, and we return to it in Section~\ref{sec:balanced} to assess differential attrition as a driver of our results.

\subsection{Outcomes}\label{sec:outcomes}

We study three sets of outcomes designed to capture distinct aspects of post-shock care. The first set comprises three utilisation outcomes. We measure defined daily doses of statins per year, which captures the intensity of lipid-lowering therapy. We measure the number of primary-care contacts per year, which captures ambulatory follow-up. We measure the number of inpatient days per year, which captures the severe margin of utilisation. Together these three measures capture important channels through which post-event care is delivered in Denmark.

The second set extends the analysis to the full set of cardiovascular secondary-prevention drugs. We measure annual doses of antiplatelets (ATC code B01AC), beta blockers (ATC code C07) and ACE inhibitors (ATC codes C09A to C09B). We count the number of distinct cardiovascular drug classes filled in a year, an extensive-margin measure of how comprehensive the regimen is. We decompose statin use by molecule, separating atorvastatin, simvastatin, rosuvastatin and the remaining molecules. Together with statins, these cover the four drug classes recommended by United States and European post-infarction guidelines, and they form the basis of the cross-drug analysis in Section~\ref{sec:gradient}.

The third set comprises three expenditure outcomes. We measure patient out-of-pocket payments for statins, total annual public-insurance expenditure on cardiovascular medications, and the gross fees paid by the public health insurance system to primary-care providers under its fee schedule. We deflate all three to 2015 Danish kroner (DKK) using the Statistics Denmark Consumer Price Index,\footnote{Obtained from \url{https://www.statbank.dk/20072}.} and we report kroner figures alongside their euro equivalents at the average 2015 reference rate of about 7.46 kroner to the euro.

We additionally examine annual mortality, both within the five-year follow-up window used for the utilisation outcomes and within an extended 20-year window for the cohorts where this is feasible (Section~\ref{sec:longrun}).

We construct two cumulative comorbidity indices from administrative diagnosis records to track subsequent morbidity, based on the Elixhauser classification and separated into cardiac and non-cardiac subindices. The \emph{cardiac} subindex comprises five conditions: congestive heart failure, cardiac arrhythmias, valvular disease, pulmonary circulation disorders and peripheral vascular disease. The \emph{non-cardiac} index aggregates the remaining 26 Elixhauser conditions, such as renal failure, depression and metabolic disorders, capturing broader systemic deterioration not specific to the cardiovascular system. Both are cumulative first-occurrence counts: for each patient in each year, the index equals the number of distinct conditions ever recorded up to and including that year, rising only on a genuinely new diagnosis and never decreasing. We estimate the post-shock trajectory of each index separately by gender, using the same staggered difference-in-differences design with not-yet-treated patients as the control group. The cardiac subindex captures the morbidity margin most directly exposed to differences in secondary-prevention care, while the non-cardiac index serves as a placebo for generic gender differences in diagnostic contact.

\begin{figure}[htbp]
    \centering
    \includegraphics[width=0.65\textwidth]{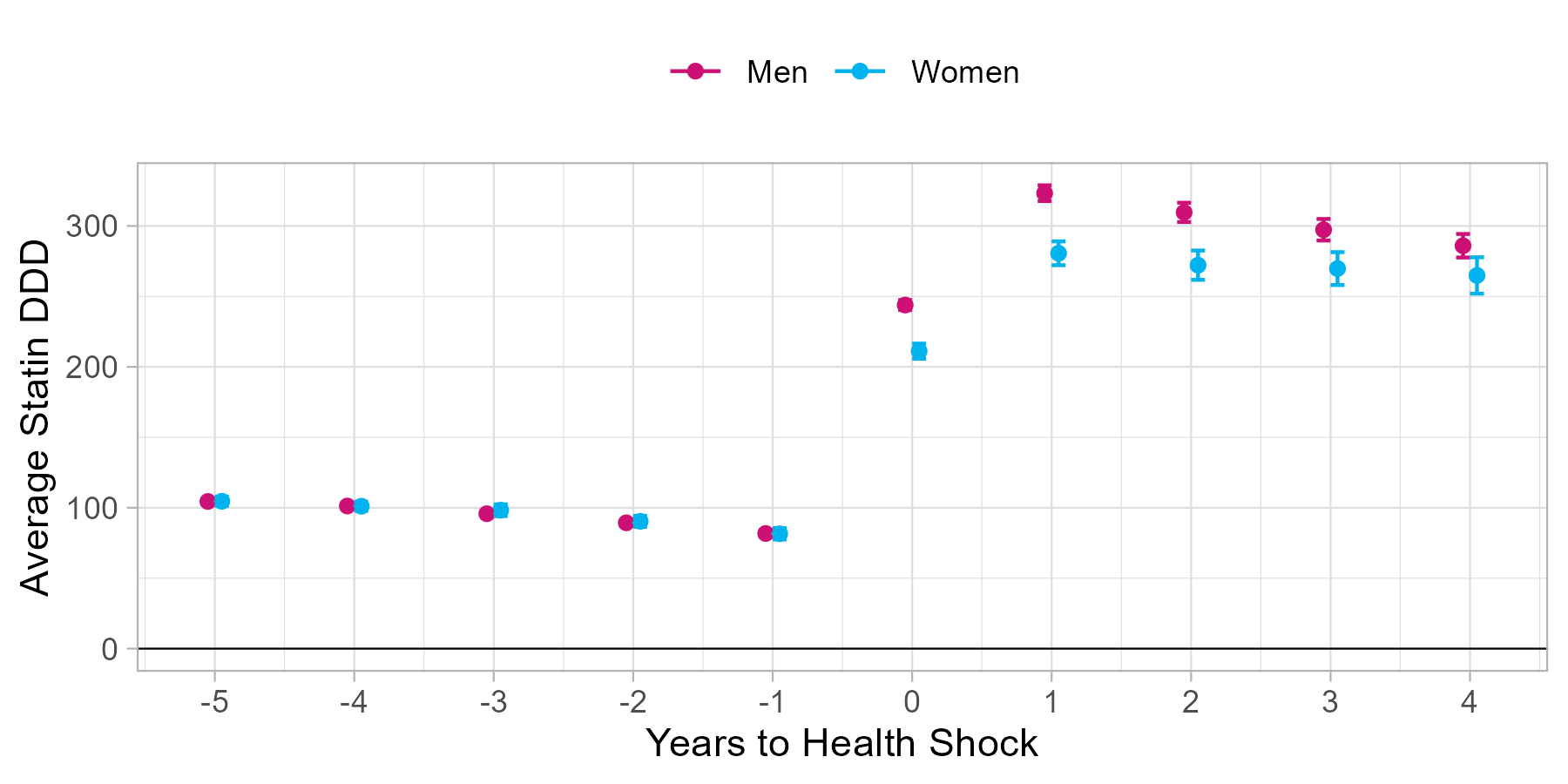}
    \vspace{0.5cm}

    \includegraphics[width=0.65\textwidth]{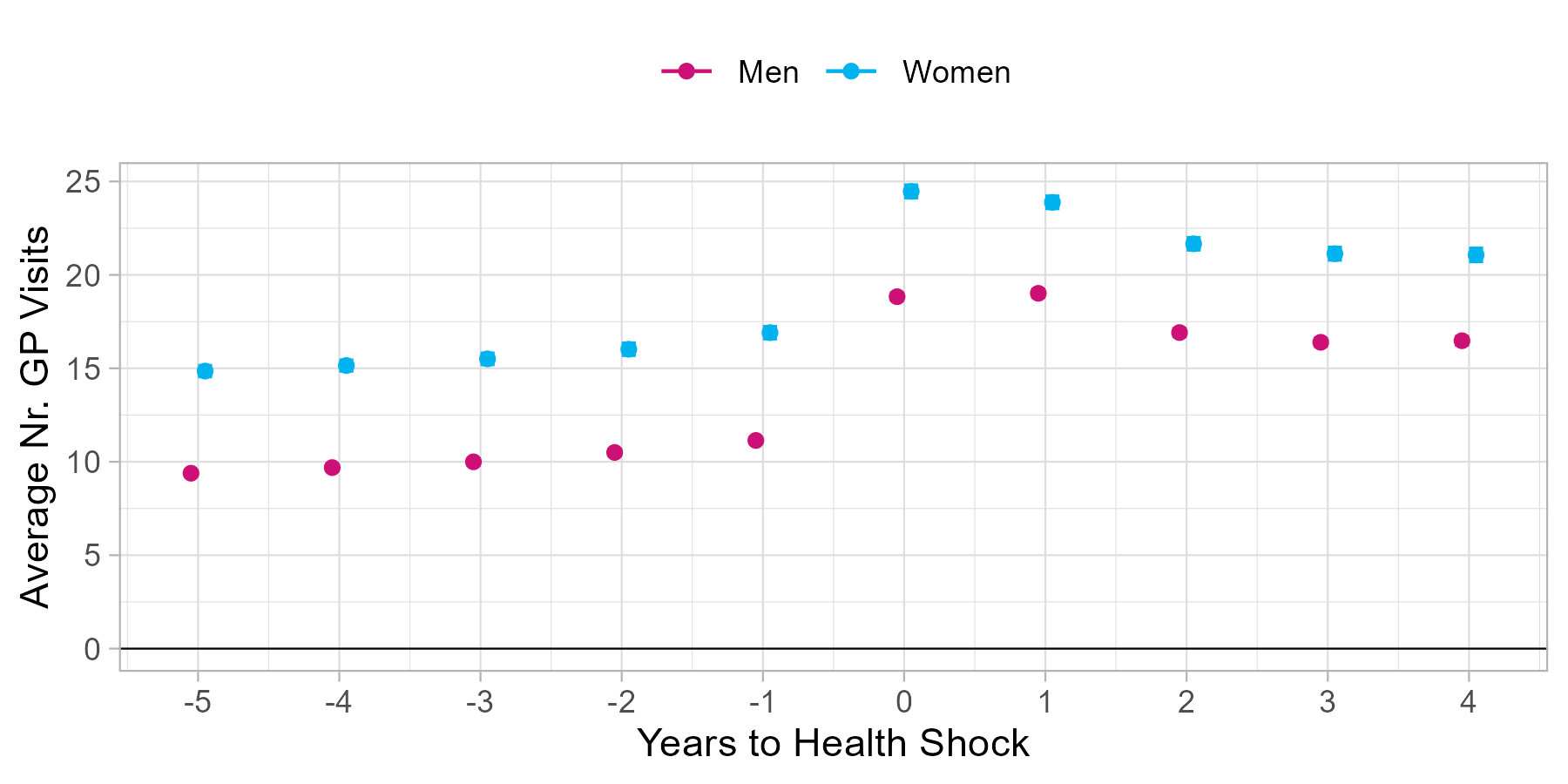}
    \vspace{0.5cm}

    \includegraphics[width=0.65\textwidth]{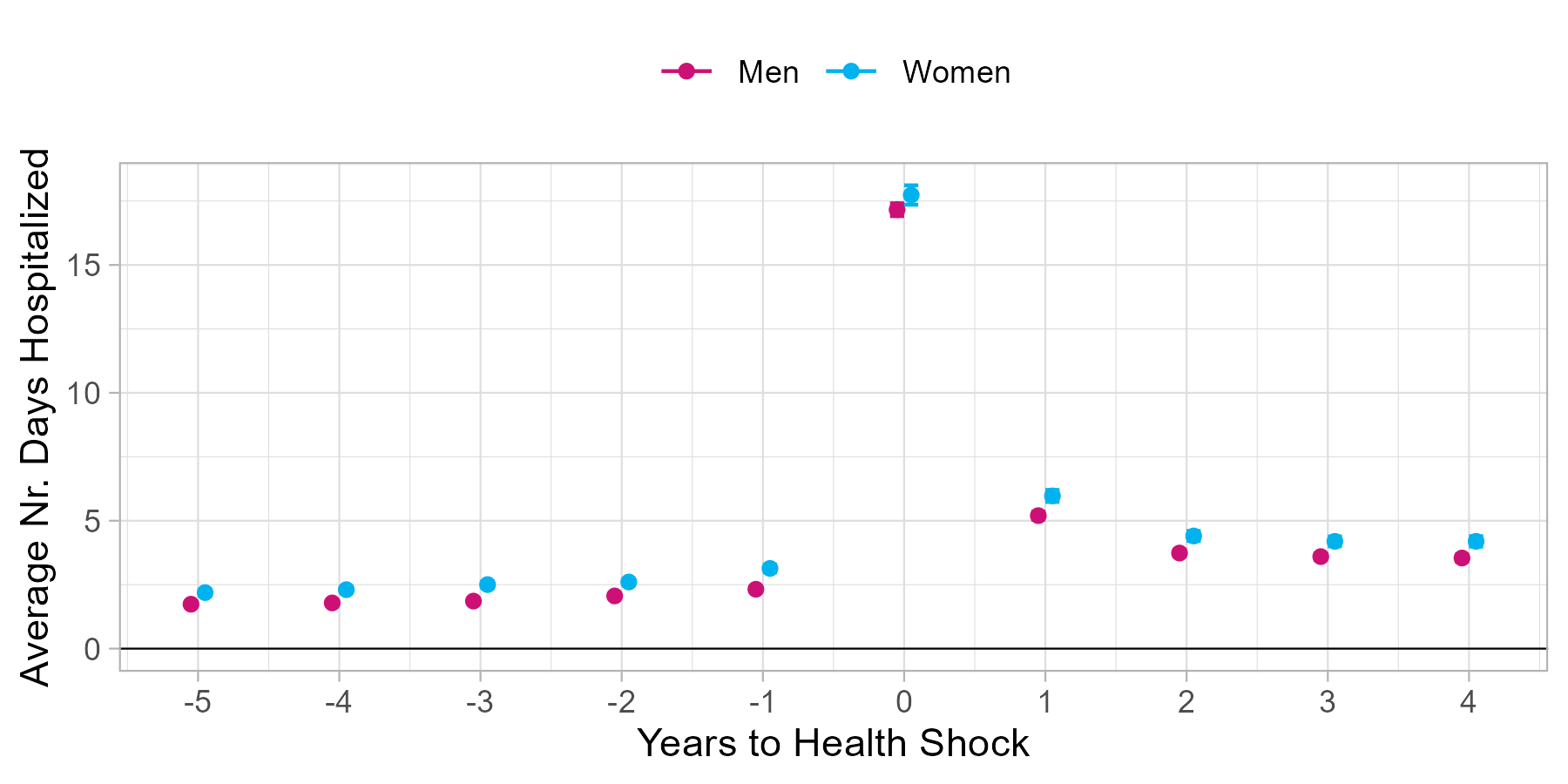}

\caption{Mean healthcare utilisation by event-time and gender.}
\label{fig:outcomes descriptives}
\begin{minipage}{0.9\textwidth}
\footnotesize \textit{Notes:} Cohort-mean outcomes by event time (years relative to the health shock), with year fixed effects partialled out. Top panel: statin DDD; middle panel: number of GP visits; bottom panel: days hospitalised.
\end{minipage}
\end{figure}

\begin{table}[htbp]
\centering
\caption{Mean healthcare utilisation in the five years before and after the shock.}
\label{tab:outcomes pre and post}
\begin{tabular}{lccc}
  \hline\hline
 & 5 yrs pre-shock & 5 yrs post-shock & $p$-value \\
  \hline
  \emph{Men} & & & \\
  \hline
Statins (DDD, sum over 5 yrs) & 180.83 & 1{,}574.25 & $<0.01$ \\
GP visits (sum over 5 yrs)    & 42.54  & 84.45      & $<0.01$ \\
Days hospitalised (sum over 5 yrs) & 8.18 & 32.54 & $<0.01$ \\
  \hline
  \emph{Women} & & & \\
  \hline
Statins (DDD, sum over 5 yrs) & 208.39 & 1{,}467.11 & $<0.01$ \\
GP visits (sum over 5 yrs)    & 68.62  & 107.87     & $<0.01$ \\
Days hospitalised (sum over 5 yrs) & 11.08 & 35.63 & $<0.01$ \\
   \hline\hline
\end{tabular}
\\[1em]
\begin{minipage}{0.9\textwidth}
\centering
\footnotesize \textit{Notes:} Sums over the five pre- and five post-shock years for each individual, averaged across the sample. The $p$-value is from a paired test of equal pre/post means within gender.
\end{minipage}
\end{table}

Figure~\ref{fig:outcomes descriptives} and Table~\ref{tab:outcomes pre and post} report mean utilisation before and after the shock for the three outcomes in the first set described above. The level of utilisation differs by gender: men consume more statins, while women have more primary-care contacts in the pre-shock period. Statin dosing in Danish guidelines is set by cardiovascular risk and cholesterol targets rather than body weight, so the gender difference in doses is not mandated by anthropometric scaling \citep{mangione2022statin}. The raw pre/post differences in Table~\ref{tab:outcomes pre and post} are descriptive only: they confound the shock with calendar-time trends, ageing and the emergence of other comorbidities, and we do not interpret them as causal effects. We report the causal counterparts in Section~\ref{sec:results} and benchmark each estimate against the cohort-weighted annual pre-shock mean.

\section{Identification, estimation and inference} \label{sec:identification}

Our object of interest is the gender-specific causal effect of a first-time, unanticipated cardiovascular event on healthcare utilisation. The core idea of the design is to compare individuals who have experienced a health shock with \emph{not-yet-treated} individuals who will face the same shock within the following years.

\subsection{Gender-specific group-time average treatment effects}

Two-way fixed-effects models perform poorly in settings with staggered treatment adoption and heterogeneous treatment effects \citep{borusyak2024revisiting,goodman2021difference,de2020two}, and our setting has both features. We therefore implement the doubly robust estimator proposed by \cite{callaway2021difference}, which adapts the \cite{sant2020doubly} methodology to multiple time periods, estimates a treatment effect for each cohort in each time period, and delivers uniform inference across event times.\footnote{\cite{callaway2021difference} is not the only method that addresses the issues with the traditional two-way fixed-effects model, but the alternatives are less suitable for our setting: \cite{de2020two} does not capture the evolution of treatment effects over time, \cite{borusyak2024revisiting} requires more pre-treatment periods to predict counterfactual trends, and \cite{sun2021estimating} imposes an unconditional parallel-trends assumption.} 

Formally, let $Y_{i,t}$ denote utilisation in calendar year $t$ for individual $i$, and let $G_i \in \{1995,\ldots, \allowbreak 2018\}$ denote the calendar year of $i$'s health shock. Define the binary indicator $D_{i,t} = \mathbf{1}\{t \geq G_i\}$.
We identify the group-time average treatment effect on the treated for cohort $g$ at calendar year $t$, separately within gender $s\in\{F,M\}$:
\begin{equation}
\label{eq:attgt_def}
ATT_{s}(g,t) \;=\; \mathbb{E}\!\left[\, Y_{i,t}(g) - Y_{i,t}(0) \,\Big|\, G_i = g,\, S_i = s\,\right],
\end{equation}
where $Y_{i,t}(g)$ is the potential outcome in year $t$ for individuals first treated in year $g$ and $Y_{i,t}(0)$ is the never-treated potential outcome.

Identification of $ATT_{s}(g,t)$ rests on (i) no anticipation, (ii) random treatment timing within a five-year window of cohort $g$, and (iii) conditional parallel trends with respect to a vector of covariates $X_i$. Under these assumptions, $ATT_{s}(g,t)$ is identified by the doubly robust estimand
\begin{equation}
\label{eq:attgt_dr}
ATT_{s}(g,t) \;=\; \mathbb{E}\!\left[\,\left( \frac{G_g}{\mathbb{E}(G_g)} - \frac{\frac{p_{g,t}(X)\,(1-D_t)\,(1-G_g)}{1-p_{g,t}(X)}}{\mathbb{E}\!\left[\frac{p_{g,t}(X)\,(1-D_t)\,(1-G_g)}{1-p_{g,t}(X)}\right]} \right) \left(Y_t - Y_{g-1} - m_{g,t}(X)\right) \,\Big|\, S_i = s\,\right],
\end{equation}
where $p_{g,t}(X)=\Pr(G_g=1\,|\,X,\,G_g+(1-D_t)(1-G_g)=1)$ is the propensity of being treated in year $g$ versus being not-yet-treated by year $t$, and $m_{g,t}(X)=\mathbb{E}[Y_t-Y_{g-1}\,|\,X,\,D_t=0,\,G_g=0]$ is the not-yet-treated outcome regression. We estimate $p_{g,t}(\cdot)$ by logistic regression and $m_{g,t}(\cdot)$ by ordinary least squares; the second-stage estimator plugs the fitted nuisance functions into the empirical analogue of \eqref{eq:attgt_dr}. The estimator attains the semiparametric efficiency bound when both nuisance models are correctly specified and is consistent if either one is. Inference uses the multiplier bootstrap of \cite{callaway2021difference}, which delivers simultaneous confidence bands across event times by reweighting individual influence-function contributions with Rademacher draws and taking the $(1-\alpha)$ quantile of the maximum studentised statistic.

The covariate vector $X_i$ contains age (in years) at the start of year $g$ and log equivalised disposable household income (from Statistics Denmark's personal income registers), region of residence, household type and educational attainment, all except for age measured at $t=g-1$ and therefore \emph{predetermined} with respect to the shock. For the Callaway and Sant’Anna specification, we include only age and income as covariates due to multicollinearity. The findings are similar in alternative specifications, including two-way fixed-effects and triple-difference models with the full set of covariates. Fixing income at its last pre-shock value ensures it is unaffected by the event itself; it enters only to absorb covariate-specific trends in baseline utilisation that are uncorrelated with shock timing but matter for the conditional parallel trends assumption, not as a post-shock outcome to be conditioned on.\footnote{The unconditional version of the estimator yields nearly identical estimates (Appendix~\ref{app:unconditional}), consistent with quasi-random treatment timing within our restricted sample.} \emph{All standard errors are clustered at the individual level via the multiplier bootstrap}; this is the appropriate level of clustering given that treatment is assigned at the individual level and outcomes are observed in repeated annual snapshots of the same individuals. All dynamic effects are normalised to the final pre-shock year $g-1$ ($\tau=-1$), which is the omitted reference (base) period in every event-study figure.

\begin{figure}[htbp]
    \centering
    \includegraphics[width=0.85\linewidth]{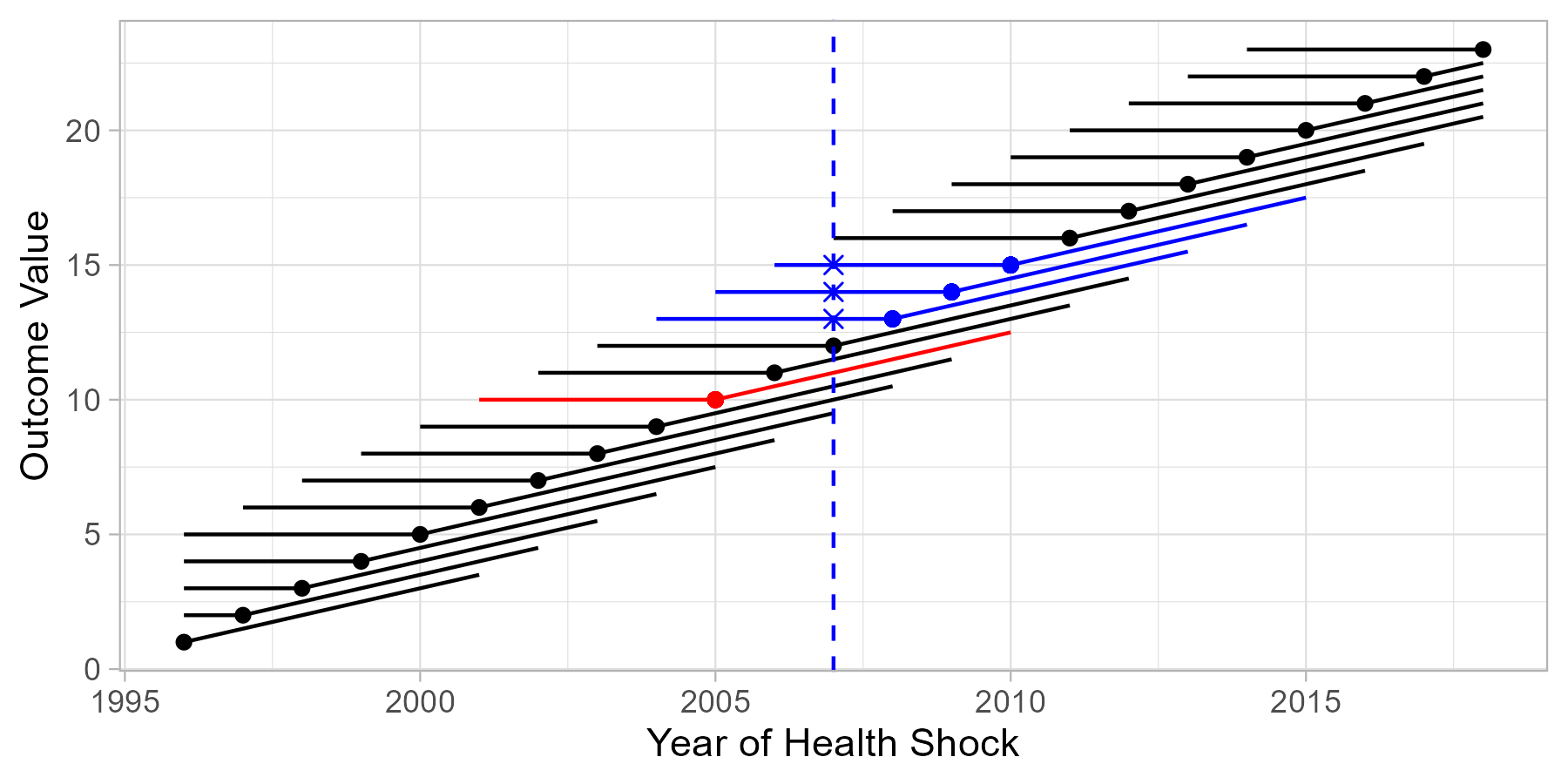}
    \caption{Five-year non-shifting comparison window.}
    \label{fig:five_year_window_explained}
    \begin{minipage}{0.9\textwidth}
    \footnotesize \textit{Notes:} For each treatment cohort $g$, the comparison set comprises individuals first treated in years $g+1$ to $g+5$. This restriction yields up to four pre-treatment and five post-treatment event times per cohort.
    \end{minipage}
\end{figure}

Our panel spans 23 years, and the plausibility of random treatment timing weakens when we compare individuals whose shocks lie far apart in time. Individuals still untreated by 2018 may differ systematically from those treated in 1996, reflecting changes in medical practice, patient demographics and underlying health trends over such a long period. To restrict comparisons to temporally proximate cohorts, we adopt a five-year non-shifting window: cohort $g$ is compared with not-yet-treated cohorts $g' \in \{g+1,\ldots,g+5\}$, so that no comparison ever pairs a 1996 cohort with a 2018 one. Figure~\ref{fig:five_year_window_explained} illustrates the resulting comparison sets. To take a concrete example, individuals treated in 2005 can be compared, when we estimate the effect two years after the shock (in 2007), with individuals who will experience their shock in 2008, 2009 or 2010. The restriction allows us to estimate up to four pre-treatment and five post-treatment effects for each cohort. The window also creates some limitations at the panel boundaries: cohorts treated in 2015 or later have fewer available control cohorts, which reduces precision but is unlikely to matter given our sample size, while cohorts treated before 2000 have fewer periods available for pre-treatment estimation. Section~\ref{sec:robustness} reports robustness to three- and seven-year windows.

\subsection{Matched-sample and Triple DiD}\label{sec:gender_contrasts}

The objects identified by Equation~\eqref{eq:attgt_dr} are gender-specific causal effects. Their difference,
\begin{equation}
\label{eq:gap}
\Delta(g,t) \;=\; ATT_{M}(g,t) - ATT_{F}(g,t),
\end{equation}
is \emph{not} causally identified by the design alone: it conflates the gender difference in the response to the shock with any unobserved gender difference in the counterfactual trajectory $Y_{i,t}(0)$. We therefore report $\widehat{\Delta}(g,t)$ as a descriptive ``gender gap in the causal effect'' and complement it with two strategies that target the gender difference directly.

First, we match each woman to her nearest neighbour among men based on the propensity score estimated at $t=g-1$, and re-estimate Equation~\eqref{eq:attgt_dr} on the matched sample. The estimated gap $\widehat{\Delta}(g,t)$ has a causal interpretation under the standard conditional independence assumption.

The second strategy is a triple difference-in-differences specification in the spirit of \cite{nielsen}. Let $F_i$ be a binary indicator for female. The triple-difference parameter at event time $\tau$ is
\begin{equation}
\label{eq:tripledid}
\beta^{\text{DDD}}_{\tau} \;=\; \mathbb{E}\!\left[\, (Y_{i,g+\tau} - Y_{i,g-1}) \,\big|\, F_i = 1,\, D_{i,g+\tau}=1\,\right] \,-\, \mathbb{E}\!\left[\, (Y_{i,g+\tau} - Y_{i,g-1}) \,\big|\, F_i = 0,\, D_{i,g+\tau}=1\,\right]
\end{equation}
\begin{equation*}
\quad -\, \Big\{\, \mathbb{E}\!\left[\, (Y_{i,g+\tau} - Y_{i,g-1}) \,\big|\, F_i = 1,\, D_{i,g+\tau}=0\,\right] \,-\, \mathbb{E}\!\left[\, (Y_{i,g+\tau} - Y_{i,g-1}) \,\big|\, F_i = 0,\, D_{i,g+\tau}=0\,\right] \Big\}.
\end{equation*}
Under the assumption that gender-specific time trends in the not-yet-treated group accurately proxy the counterfactual gender-specific time trends in the treated group, $\beta^{\text{DDD}}_{\tau}$ identifies the causal gender difference at time $\tau$. The comparison group is a more flexible variant of \cite{nielsen}: rather than requiring control individuals to be treated exactly five years after the focal cohort, we include those treated at any point within the following five-year window, which increases statistical power while preserving temporal comparability. 

The matched-sample and triple-difference designs are complementary: the former relies on selection on observables, the latter on parallel time trends in counterfactuals across genders. Section~\ref{sec:robustness} shows that the triple-difference design delivers the same qualitative conclusions as the gender-specific estimates from Equation~\eqref{eq:attgt_dr}, and that matching leaves the general-practitioner gap unchanged while narrowing the statin gap at longer horizons.

\subsection{Plausibility of identifying assumptions}\label{sec:plausibility}

The key assumptions are no anticipation and random treatment timing within the five-year window. While neither is directly testable, three pieces of evidence are reassuring. \emph{First}, dynamic ATT estimates in the pre-treatment period are small in magnitude for both genders (Figure~\ref{fig:dynamic_att_full}). For men, the only statistically significant pre-period coefficients are at $\tau=-1$ and are negligible in size: about 2.9 DDD on a post-period base of approximately 250 DDD, 0.2 GP visits and 0.2 inpatient days. For women, the significant pre-period coefficients (on GP visits and inpatient days at $\tau=-1$, about 0.6 visits and 0.4 days respectively) are also small and most plausibly reflect a short prodrome: women more often present with non-classical symptoms and are initially mistakenly diagnosed with non-cardiac conditions \citep{Canto2000,Schulte2023,Joseph2021,Cardeillac2022,StuartShor2009,Ali2021}. This attenuates our estimate of the female post-shock response on these two margins and could, if anything, slightly understate the corresponding gender gap; the statin gap, where women's pre-period coefficients are insignificant, is unaffected. \emph{Second}, parallel-trends plots constructed using future-treated cohorts as a placebo control group (Appendix~\ref{sec:trendplots}) yield similar patterns for not just for treated vs not-yet-treated but also for men vs women. Section~\ref{sec:robustness} demonstrates that the main estimates are also insensitive to comparison-window length, covariate inclusion, estimator choice, and control-group definition.


\section{Results} \label{sec:results}

This section documents the gender gap in healthcare use after a health shock. We proceed in two steps: we first examine the volume of care, measured by statin doses, general-practitioner visits and inpatient days, and then turn to the corresponding spending by patients and by the public insurer. For each outcome, we report dynamic and aggregate average treatment effects on the treated, estimated separately by gender. Unless stated otherwise, all estimates pool heart attacks and strokes; Section~\ref{sec:hetbyshock} reports the decomposition by event type and shows that the gap is concentrated after heart attacks.

\begin{figure}[htbp]
    \centering
    \begin{subfigure}{0.65\textwidth}
        \centering
        \includegraphics[width=\textwidth]{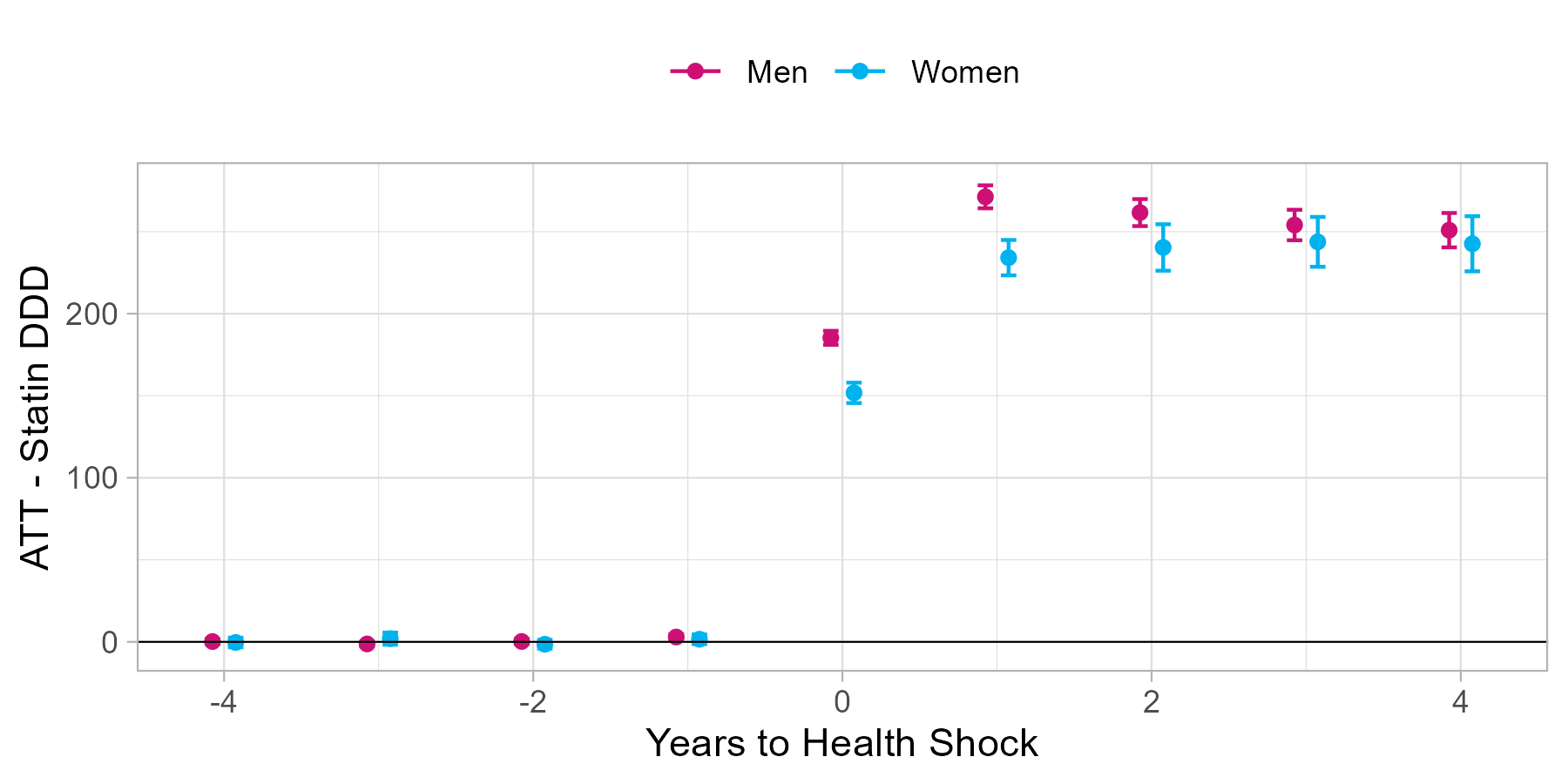}
        \caption{Statin DDD.}
        \label{fig:dynamic_att_statin_ddd}
    \end{subfigure}

    \vspace{0.4cm}

    \begin{subfigure}{0.65\textwidth}
        \centering
        \includegraphics[width=\textwidth]{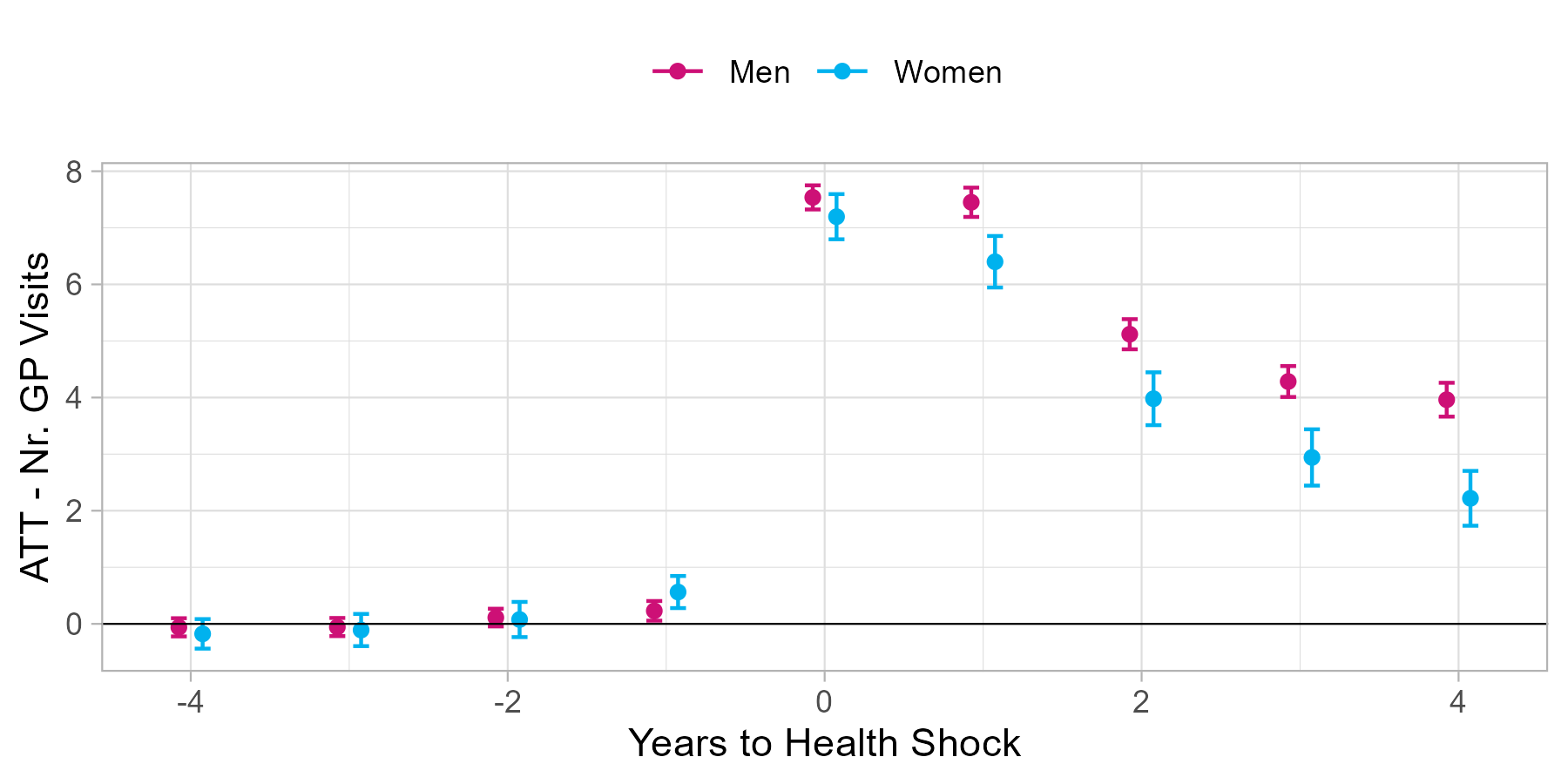}
        \caption{GP visits.}
        \label{fig:dynamic_att_gp_visits}
    \end{subfigure}

    \vspace{0.4cm}

    \begin{subfigure}{0.65\textwidth}
        \centering
        \includegraphics[width=\textwidth]{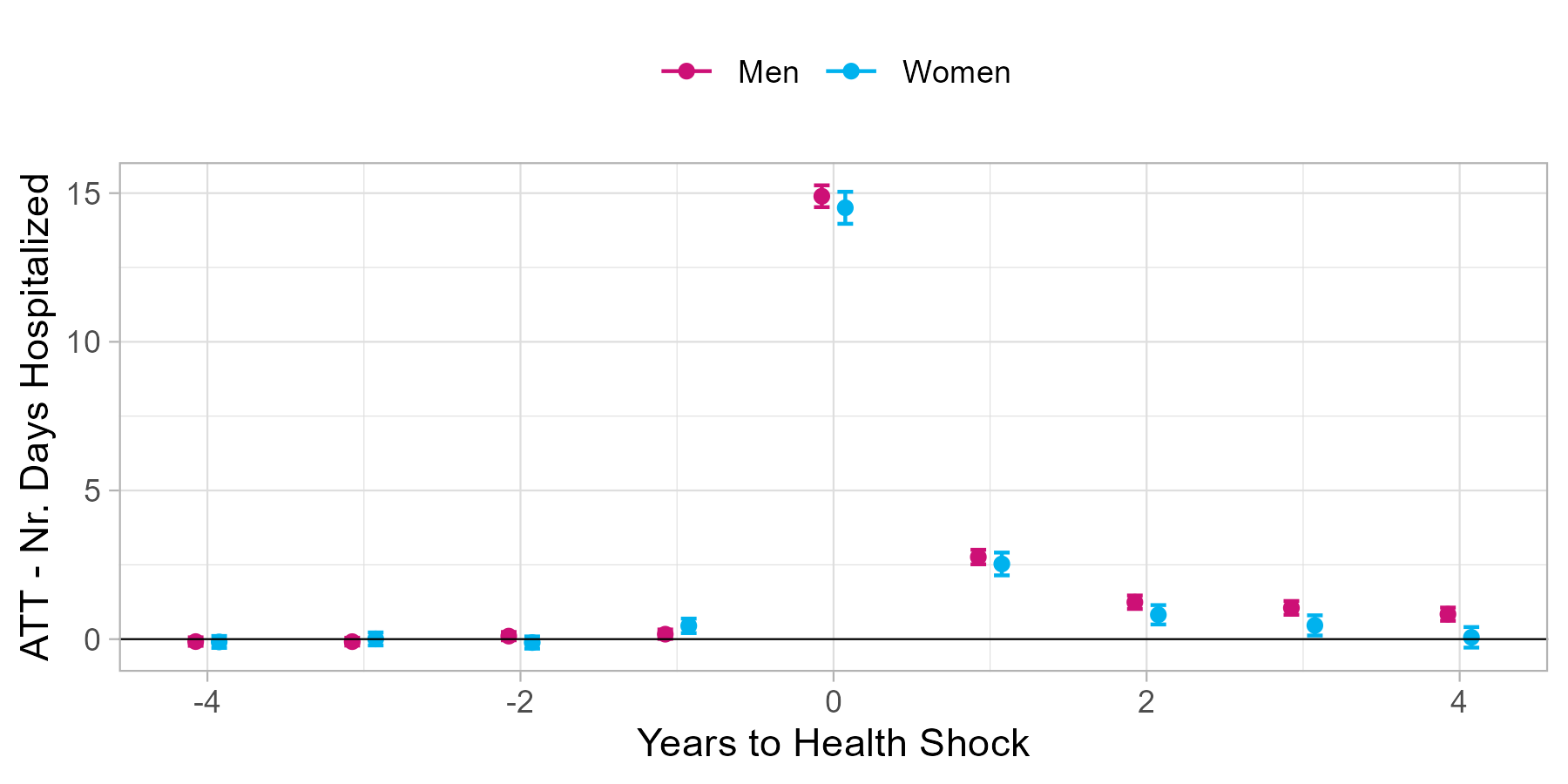}
        \caption{Days hospitalised.}
        \label{fig:dynamic_att_days_treated}
    \end{subfigure}

    \caption{Dynamic ATT estimates by event time and gender.}
    \label{fig:dynamic_att_full}
    \begin{minipage}{0.9\textwidth}
    \footnotesize \textit{Notes:} Each panel reports $\widehat{ATT}_s(g,t)$ aggregated over cohorts within event time, separately by gender, using the \cite{callaway2021difference} doubly robust estimator with a five-year comparison window. Shaded areas denote simultaneous 95\% bootstrap confidence bands (1000 multiplier-bootstrap replications, individual-level clustering). Pre-period estimates are reported as evidence on the no-anticipation/parallel-trends assumption. The event-year coefficient is mechanically attenuated because shocks may occur in any month of the year. Tables \ref{tab:Statinmain}, \ref{tab:GPmain} and \ref{tab:hospmain} in Appendix~\ref{sec:tables_main} report the corresponding numerical estimates.
    \end{minipage}
\end{figure}

We begin with the dynamic estimates, which trace the effect of the shock by event time separately for men and women. Figure~\ref{fig:dynamic_att_full} shows that, for statin volume, both genders exhibit a sharp jump in the year of the shock and then settle at a markedly higher level, while men maintain a higher post-shock level than women. The male effect ranges from 251 to 271 defined daily doses per year over the four post-shock years, whereas the female effect ranges from 234 to 244. The resulting gap is widest in the first two post-shock years, at around 33 to 37 doses per year, and narrows thereafter. 

We turn next to primary-care contacts. In the event year and the year following, both genders increase their visits by about 6 to 8 per year, before the effect declines. 

The third outcome is inpatient days. Both genders are hospitalised for roughly 15 days in the year of the shock, and hospitalisation declines sharply thereafter. In years two to four a statistically detectable but quantitatively small gap emerges: men are hospitalised for about half a day more per year, on a baseline of only one to two days. This gap is small in absolute terms and, in relative terms, comparable to the statin-dose gap (both about 11 per cent of the female effect), while the primary-care gap is larger in relative terms (about 22 per cent). Across all three outcomes the estimates are precise: the confidence bands are narrow and the pre-period coefficients are small (Section~\ref{sec:plausibility}), consistent with the no-anticipation and parallel-trends assumptions.

\subsection{Aggregate effects and benchmarking}

\begin{table}[htbp]
\centering
\caption{Aggregate ATT over the five post-shock years.}
\label{tab:aggregate_att}
\footnotesize
\begin{tabular}{lcccccccc}
  \hline\hline
& ATT & CI lower & CI upper & Std.\ err. & No.\ obs. & $p$-value & Baseline & Rel. \\
&  &  &  &  &  &  & mean & \% change \\
  \hline
\multicolumn{9}{l}{\textbf{Statin DDD}} \\ \hline
Women & 219.48 & 212.85 & 226.11 & 3.40 & 15{,}140 & $<0.01$ & 19.13 & 1{,}147\% \\
Men   & 242.90 & 238.47 & 247.34 & 2.26 & 30{,}103 & $<0.01$ & 18.49 & 1{,}314\% \\
Gap ($M-W$) & 23.42 & 15.42 & 31.42 & 4.08 & & $<0.01$ & & \\ \hline
\multicolumn{9}{l}{\textbf{No.\ of GP visits}} \\ \hline
Women & 4.77 & 4.51 & 5.03 & 0.13 & 15{,}140 & $<0.01$ & 13.82 & 35\% \\
Men   & 5.84 & 5.70 & 5.99 & 0.07 & 30{,}103 & $<0.01$ & 8.12  & 72\% \\
Gap ($M-W$) & 1.07 & 0.78 & 1.36 & 0.15 & & $<0.01$ & & \\ \hline
\multicolumn{9}{l}{\textbf{No.\ of days hospitalised}} \\ \hline
Women & 4.17 & 4.00 & 4.35 & 0.09 & 15{,}140 & $<0.01$ & 2.07 & 201\% \\
Men   & 4.63 & 4.52 & 4.74 & 0.06 & 30{,}103 & $<0.01$ & 1.56 & 296\% \\
Gap ($M-W$) & 0.45 & 0.24 & 0.66 & 0.11 & & $<0.01$ & & \\
\hline\hline
\end{tabular}
\\[0.5em]
\begin{minipage}{0.95\textwidth}
\footnotesize \textit{Notes:} ATT averaged over event times $\tau \in \{0,1,2,3,4\}$, cohort-weighted using the within-event-time cohort sample sizes and event-time-equal-weighted across $\tau$. The gender-specific estimates are computed on disjoint samples of men and women, so the standard error of the gap $M-W$ is the square root of the sum of the two gender-specific squared standard errors, and its confidence interval and $p$-value follow accordingly. The baseline column is the cohort-weighted \emph{annual} pre-shock mean of the outcome, computed using the same weights and over the five pre-shock years; it is therefore comparable to the annual ATT. Confidence intervals at the 95\% level via the multiplier bootstrap with 1{,}000 replications and individual-level clustering. Relative \% change is the ATT divided by the baseline mean, rounded to the nearest per cent.
\end{minipage}
\end{table}

We aggregate the post-shock effects over the five post-event years and benchmark them against cohort-weighted annual pre-shock means; Table~\ref{tab:aggregate_att} reports the results. For statins, men show a larger absolute increase than women, 243 against 219 defined daily doses per year, a gap of 23.4 doses (95\% confidence interval 15.4 to 31.4). Because the pre-shock baselines are comparably low for both genders, men's relative increase is larger as well. For primary-care visits, men's absolute increase is again larger, 5.84 against 4.77 visits per year, a gap of 1.07 visits (0.78 to 1.36). For inpatient days, the absolute gap of 0.45 days per year (0.24 to 0.66) is small, and the large relative magnitudes reflect very low pre-shock baselines rather than a meaningful difference. Each gap is precisely estimated and significant at the 1 per cent level (Table~\ref{tab:aggregate_att}).

Throughout the paper we treat the absolute effect as the primary object. The relative-change column is reported only for comparability: because it divides the effect by the pre-shock baseline, it is mechanically sensitive to gender differences in baselines, and it should not be read as the size of the gap.

The pattern across outcomes is a substantial gap in pharmacotherapy and primary care, a small absolute gap in inpatient days, and no detectable gap in the shock's mortality effect (Section~\ref{sec:balanced}). This is consistent with gender differences operating at the margin of follow-up and preventive care rather than at the margin of survival.

\subsection{Healthcare expenditure}

The gap also appears in spending. We examine three financial outcomes: patient out-of-pocket payments for statins, the gross fee that the public health insurance system pays primary-care providers per visit, and total public-insurance expenditure on cardiovascular medications. The first is borne by the patient, while the latter two are borne by the public insurer. All three are deflated to 2015 Danish kroner, with one euro equal to about 7.46 kroner.

\begin{figure}[htbp]
    \centering
        \includegraphics[width=0.65\textwidth]{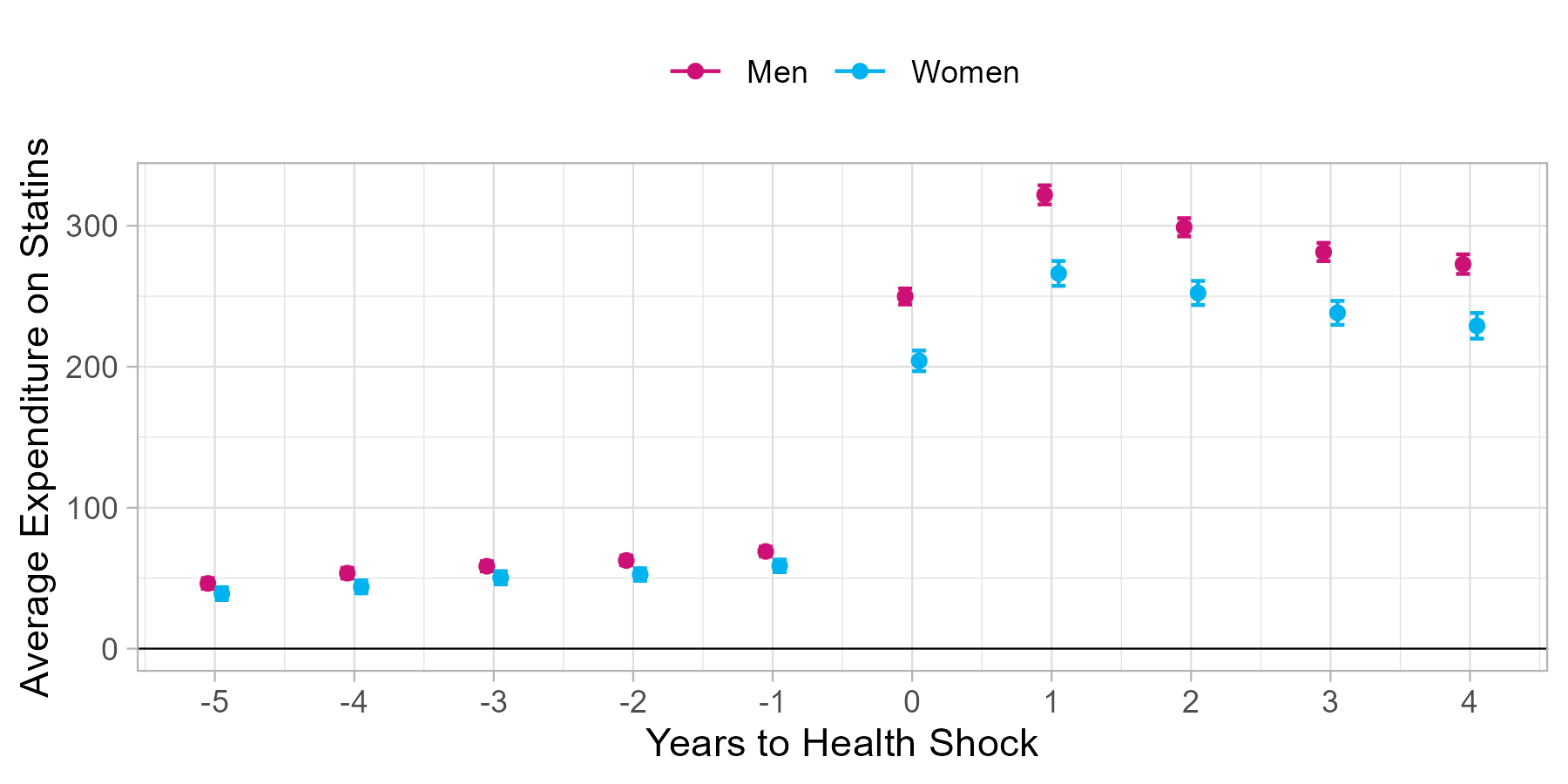}
        \caption{Raw trend: out-of-pocket statin expenditure.}
        \label{fig:trend_expenditure_statins}

        \includegraphics[width=0.65\textwidth]{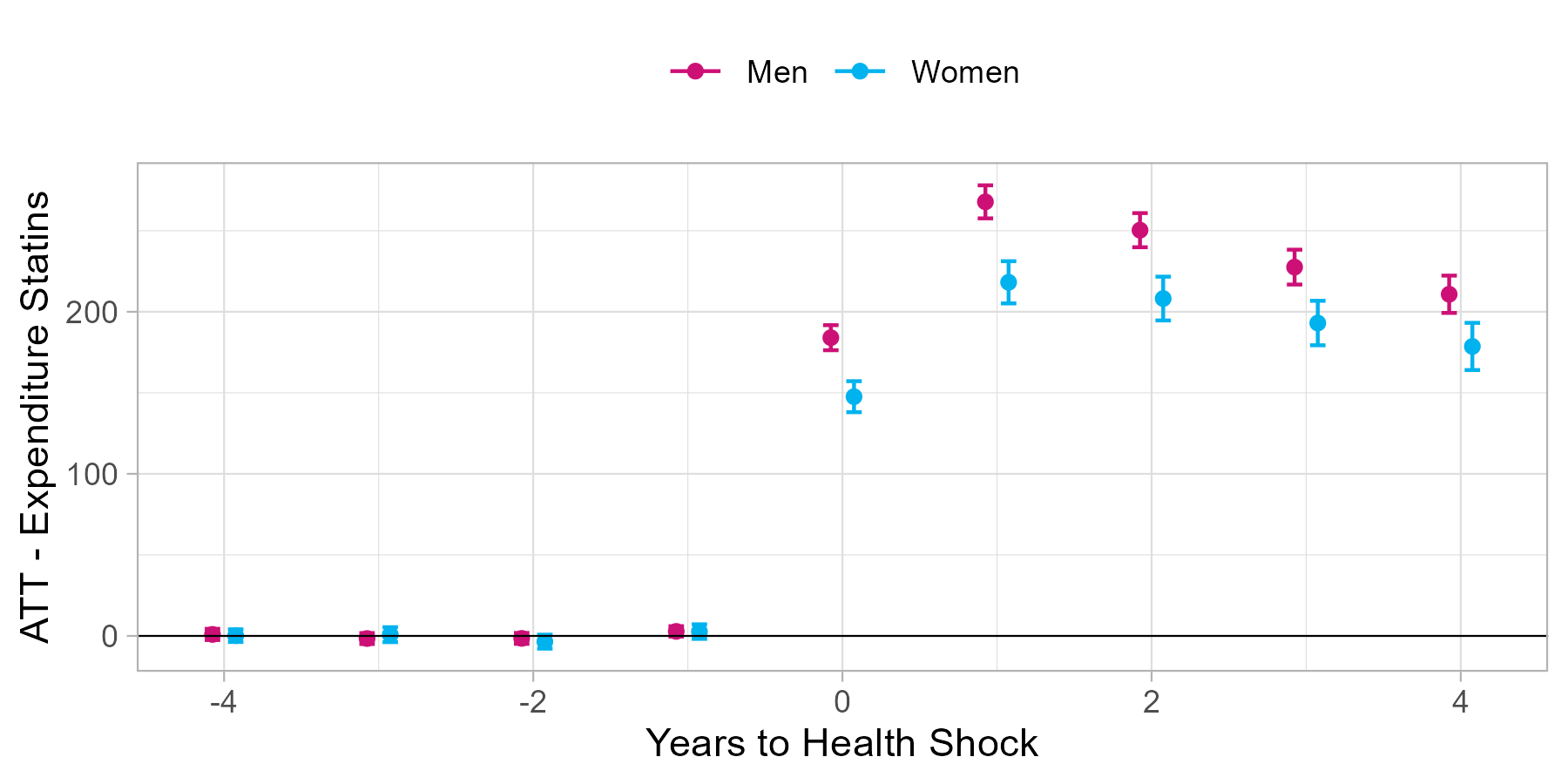}
        \caption{Dynamic ATT: out-of-pocket statin expenditure.}
        \label{fig:expenditure_statins}
\begin{minipage}{0.9\textwidth}
\footnotesize \textit{Notes:} Annual patient co-payments for statins, deflated to 2015 DKK. Estimates use \cite{callaway2021difference}; 95\% simultaneous bootstrap bands. Numerical values in Table~\ref{tab:ATT_expenditure_statins}.
\end{minipage}
\end{figure}

The expenditure results mirror the utilisation results closely. Figure~\ref{fig:trend_expenditure_statins} plots the raw trend in out-of-pocket statin spending, and Figure~\ref{fig:expenditure_statins} shows the corresponding dynamic ATT estimates: spending rises by about 228 kroner per year for men and 189 kroner for women, roughly 31 against 25 euros (Table~\ref{tab:ATT_expenditure_statins}).

\begin{figure}[htbp]
    \centering
        \includegraphics[width=0.65\textwidth]{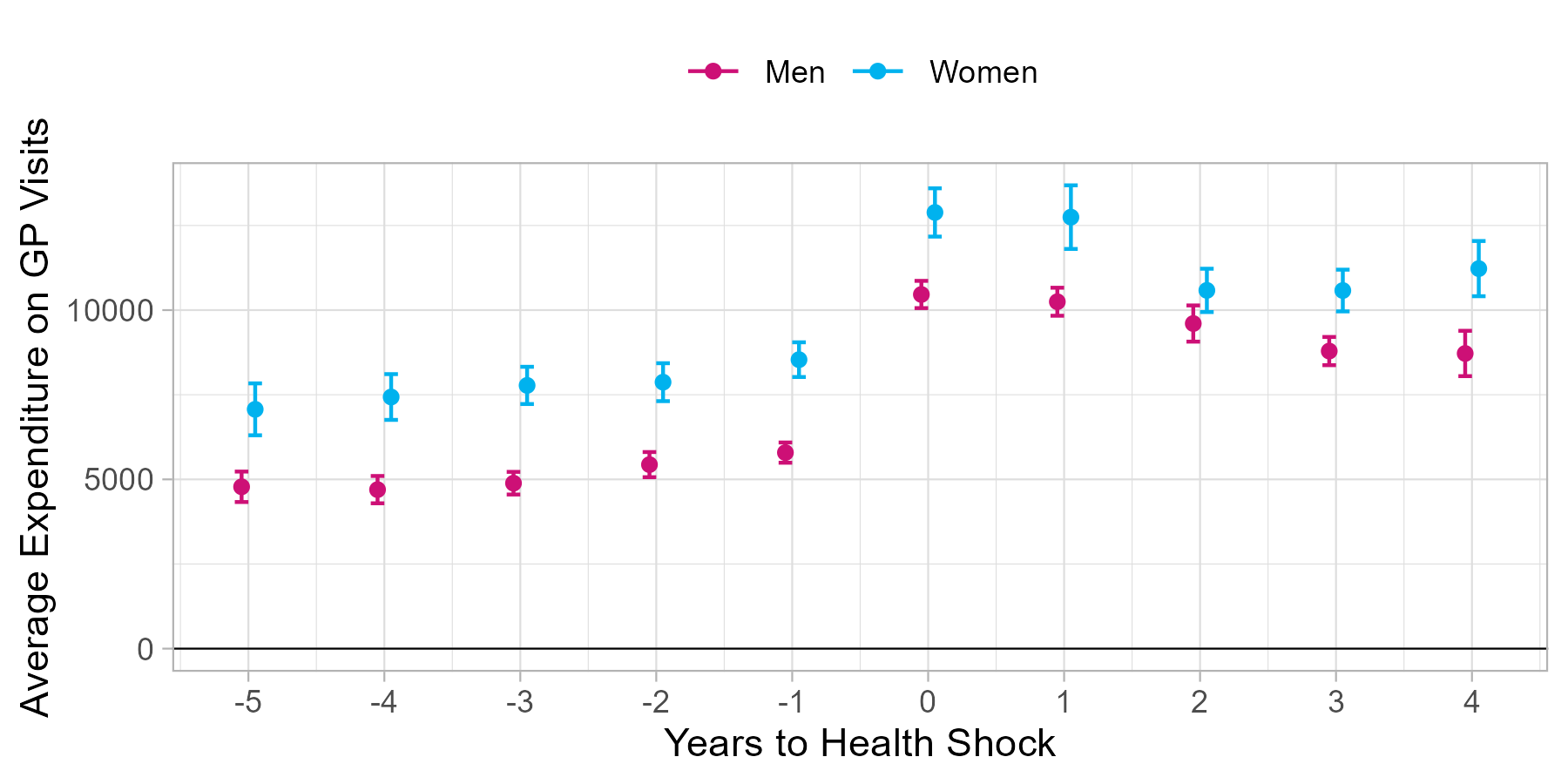}
        \caption{Raw trend: GP visit fees.}
        \label{fig:trend_expenditure_gp_visits}

        \includegraphics[width=0.65\textwidth]{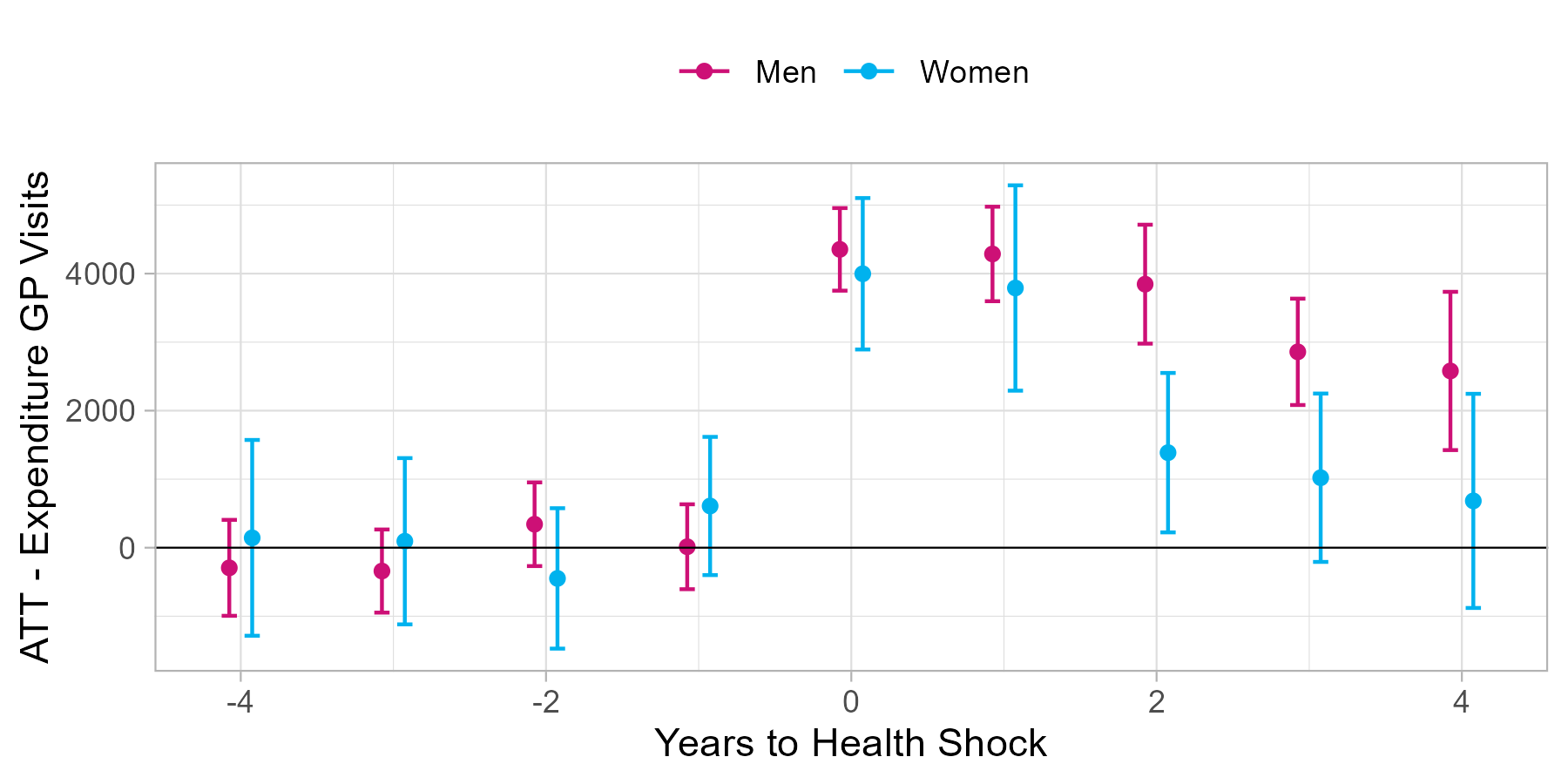}
        \caption{Dynamic ATT: GP visit fees.}
        \label{fig:expenditure_gp_visits}
\begin{minipage}{0.9\textwidth}
\footnotesize \textit{Notes:} Annual gross fees paid by the public health insurance system to primary-care providers, deflated to 2015 DKK. Excludes patient co-payments. Estimates use \cite{callaway2021difference}; 95\% simultaneous bootstrap bands.
\end{minipage}
\end{figure}

Figures~\ref{fig:trend_expenditure_gp_visits} and \ref{fig:expenditure_gp_visits} present the analogous results for primary-care fees. Here we find no consistent gender difference (Table~\ref{tab:ATT_expenditure_gp_visits}). This suggests that men's larger increase in the number of contacts is partly offset by women receiving more intensive services per contact, which is consistent with the prescribing-intensity reading we develop below.

\begin{figure}[htbp]
    \centering
        \includegraphics[width=0.65\textwidth]{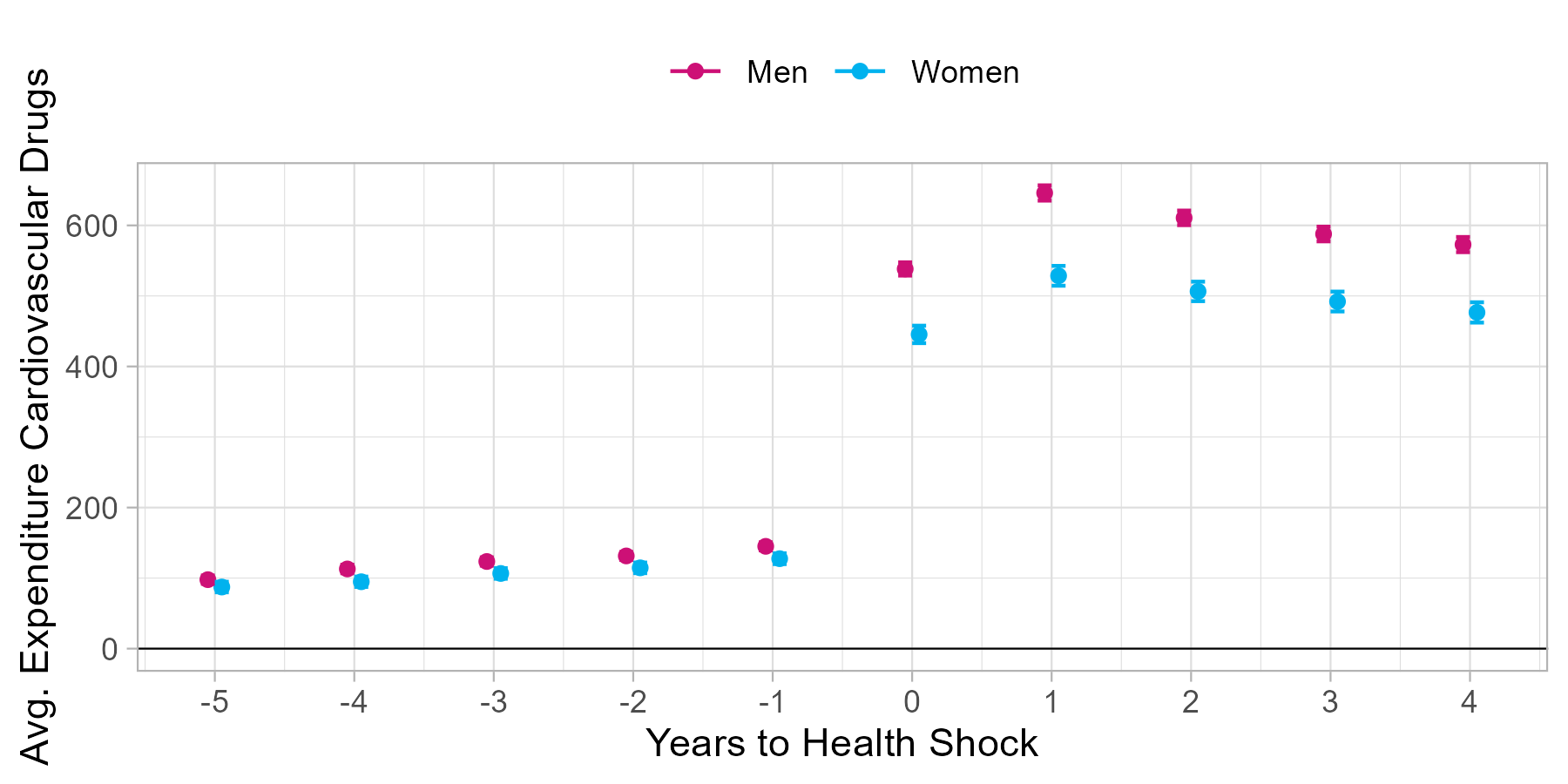}
        \caption{Raw trend: total cardiovascular drug expenditure.}
        \label{fig:trend_expenditure_cardio}

        \includegraphics[width=0.65\textwidth]{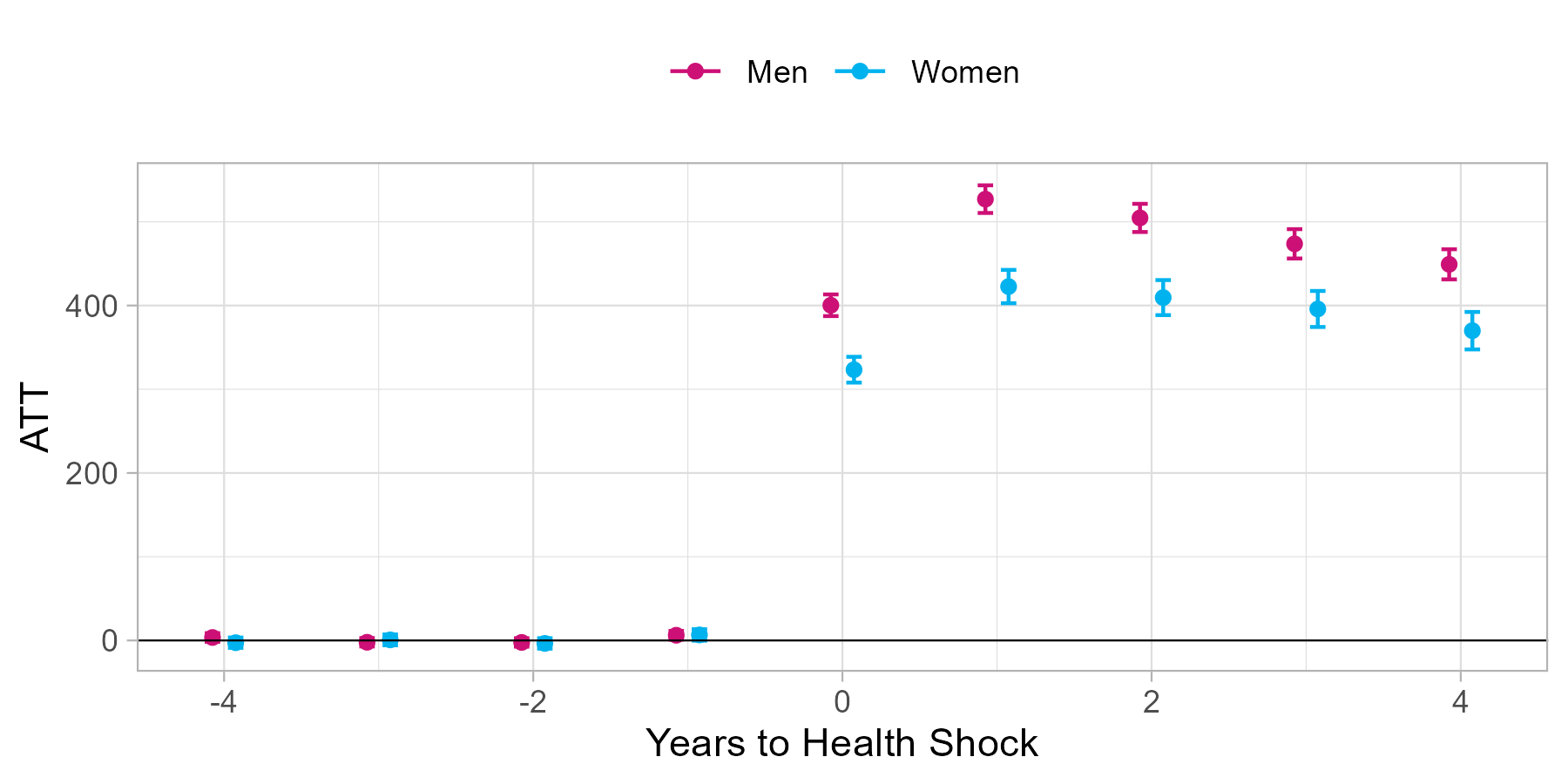}
        \caption{Dynamic ATT: total cardiovascular drug expenditure.}
        \label{fig:expenditure_cardio}
\begin{minipage}{0.9\textwidth}
\footnotesize \textit{Notes:} Total annual public-insurance expenditure on cardiovascular medications (statins, antiplatelets, beta blockers, ACE inhibitors, and other cardiovascular drug classes), deflated to 2015 DKK (1 EUR $\approx$ 7.46 DKK). Estimator and inference as in Figure~\ref{fig:dynamic_att_full}.
\end{minipage}
\end{figure}

Total cardiovascular drug spending shows the same pattern. Figures~\ref{fig:trend_expenditure_cardio} and \ref{fig:expenditure_cardio} present the raw trend and the corresponding dynamic ATT estimates. The aggregate post-shock effect is about 470 kroner per year for men and 383 kroner for women, roughly 63 against 51 euros; the resulting gap of about 87 kroner per year is precisely estimated and significant at the 1 per cent level. The spending gap closely tracks the volume gap, which indicates that the difference in prescribing intensity is a substantively important driver of post-shock pharmacotherapy spending in a universal-coverage system.

\subsection{Robustness}\label{sec:robustness}

The result rests jointly on the gender-specific dynamic effects and on the cross-gender differences, and we assess its robustness along several dimensions; Table~\ref{tab:robustness_summary} summarises the checks and their outcomes. To assess whether systematic pre-shock differences between the genders drive the estimated differences, we re-estimate on a matched sample that pairs women to men by nearest-neighbour propensity-score matching. The matching achieves standardised mean differences of at most 0.03 on every covariate (Table~\ref{tab:match_balance} in Appendix \ref{sec:matched_sample}). The women's estimates are essentially unchanged, as expected since matching retains the full female sample. For general-practitioner visits the matched-sample estimates are also virtually identical to the baseline ones, so the visit gap is robust to composition. For statins, the matched men's effects are somewhat smaller than in the full male sample: the gap remains visible in the first two post-shock years, most clearly in the event year, and is no longer statistically distinguishable at longer horizons (Figure~\ref{fig:matched_results} in Appendix \ref{sec:matched_sample}). Part of the full-sample statin gap at longer horizons therefore reflects the composition of the male sample on matched observables, while the early-year gap, the phase in which post-shock therapy is initiated and titrated, survives matching. If treatment timing is as good as random, adjusting for covariates should not materially affect the estimates. To assess this implication, we additionally estimate the specification without covariates, relying on unconditional parallel trends, and again the estimates change little.

The triple-difference specification, which contrasts the genders within a single estimation framework following \citet{nielsen}, recovers the same gap. So does a two-way fixed-effects specification, though the latter shows some non-zero pre-trends of the kind expected under heterogeneous staggered treatment \citep{borusyak2024revisiting,goodman2021difference,de2020two}. The gap is also present with comparison windows of three, five and seven years, although the seven-year window narrows the statin gap after the second event year and widens the general-practitioner-visit gap. It survives in the balanced panel and within the income and household splits, and cohort-specific estimates show that no single treatment cohort drives the results (Appendix~\ref{sec:group_att}).

\begin{table}[htbp]
\centering
\caption{Summary of robustness checks.}
\label{tab:robustness_summary}
\footnotesize
\begin{tabular}{p{4cm}p{8.5cm}p{2.5cm}}
\toprule
\textbf{Check} & \textbf{What it does} & \textbf{Result} \\
\midrule
Matched sample & Nearest-neighbour PS matching of women to men; standardised mean differences $\leq 0.03$ on every covariate. & GP gap unchanged; statin gap narrows at longer horizons \\
Unconditional DiD & No covariates, relying on unconditional parallel trends. & Unchanged \\
Triple-difference (\cite{nielsen}) & Direct gender contrast within a single estimation framework. & Unchanged \\
Two-way fixed effects & TWFE specification as in \cite{nielsen}; some non-zero pre-trends, as expected under heterogeneous staggered treatment \citep{borusyak2024revisiting,goodman2021difference,de2020two}. & Qualitatively unchanged \\
Comparison window (3 / 5 / 7 yrs) & Alternative non-shifting windows. The 7-yr window narrows the statin gap after $\tau=2$ and widens the GP-visit gap. & Unchanged \\
Balanced panel & Conditioning on 5-year survival; 74\% of men and 77\% of women retained. & Unchanged \\
Income / household splits & Below/above-median equivalised income; single vs.\ non-single households. & Unchanged \\
\bottomrule
\end{tabular}
\\[0.5em]
\begin{minipage}{0.95\textwidth}
\footnotesize \textit{Notes:} ``Unchanged'' means the gender gap in statin DDD and GP visits remains positive, statistically significant, and within the 95\% confidence band of the main estimate; the small inpatient-day gap and the null finding for mortality remain unchanged. For the matched sample, the general-practitioner gap is unchanged while the statin gap remains visible in the first two post-shock years and is no longer statistically distinguishable at longer horizons (Figure~\ref{fig:matched_results}). Underlying estimates: matched sample (Figure~\ref{fig:matched_results}, Appendix~\ref{sec:matched_sample}); unconditional DiD (Figure~\ref{fig:unconditional}, Appendix~\ref{app:unconditional}); triple-difference (Figure~\ref{fig:triple_did}, Appendix~\ref{sec:tdid}); TWFE (Figure~\ref{fig:twfe_estimates_main}, Appendix~\ref{sec:twfe_estimates_main}); comparison-window variants (Figures~\ref{fig:3year_window}--\ref{fig:7year_window}, Appendix~\ref{sec:estimates_7year_window}); balanced panel (Appendix~\ref{app:twfe_balanced}); income/household heterogeneity (Figures~\ref{fig:subgrouping_income}--\ref{fig:subgrouping_household}, Appendices~\ref{sec:income_heterogeneity}--\ref{sec:household_heterogeneity}).
\end{minipage}
\end{table}

\section{Mechanisms} \label{sec:ruleout}

We have established a robust gender gap in post-shock secondary-prevention care: men increase their statin volume and their general-practitioner contact substantially more than women after the same acute cardiovascular event, while the effects on short-run and long-run mortality and on inpatient days show no comparable gap. In this section, we ask what generates the gap. We proceed by elimination, taking each candidate explanation in turn and assessing whether it can account for the patterns in the data. The candidates are differential adherence, differential patient demand, differential severity, survival selection or attrition, a mechanical ceiling on women, and specification choices. None of them accounts for the gap, and what remains points to how secondary-prevention care is delivered rather than to anything on the patient side.

\subsection{Prescribing intensity versus adherence}\label{sec:demandsupply}

\begin{figure}[htbp]
    \centering
        \includegraphics[width=0.65\textwidth]{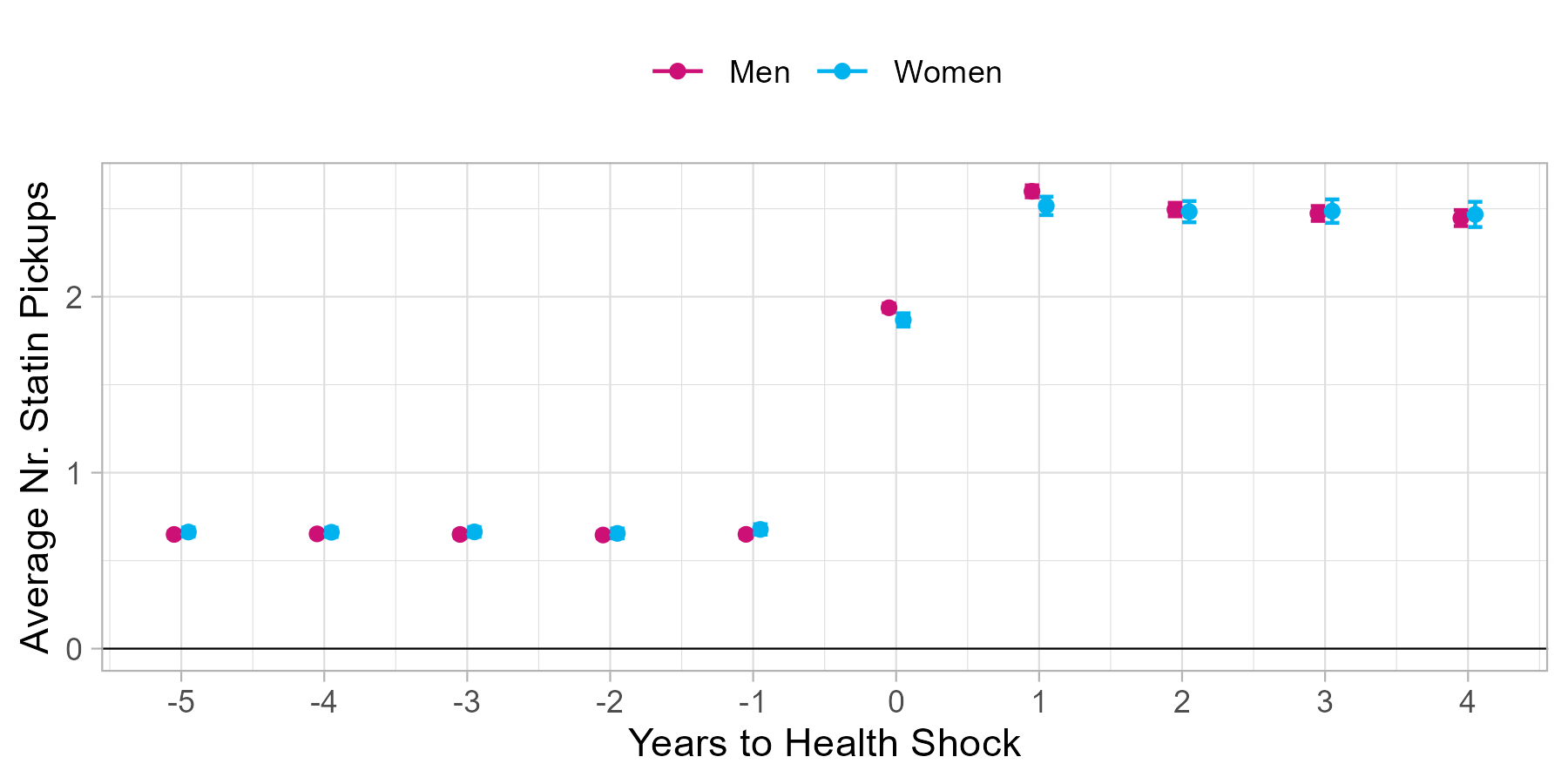}
        \caption{Raw trend: number of statin pickups per year.}
        \label{fig:s_pickupraw}

        \includegraphics[width=0.65\textwidth]{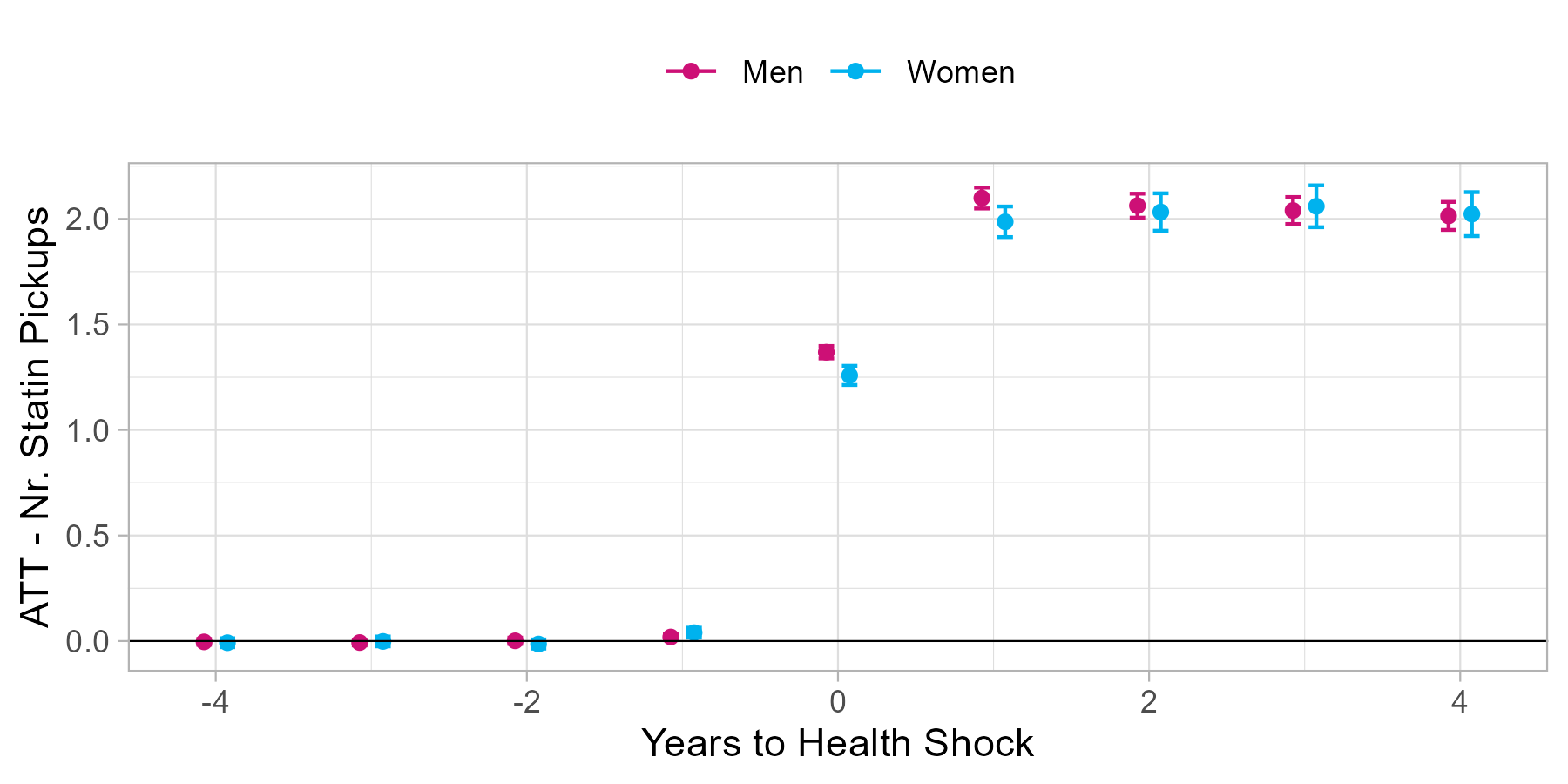}
        \caption{Dynamic ATT on number of statin pickups per year.}
        \label{fig:s_pickupATT}
\begin{minipage}{0.9\textwidth}
\footnotesize \textit{Notes:} Pickups are counted as separate fills at a Danish pharmacy. Top panel: raw cohort means by event-time and gender. Bottom panel: dynamic ATT estimated using \cite{callaway2021difference}, with simultaneous 95\% bootstrap bands. Numerical estimates in Table~\ref{tab:s_pickup_overallATT}.
\end{minipage}
\end{figure}

The statin volume gap could in principle reflect women filling fewer prescriptions rather than receiving lower doses when they do fill. To separate these two aspects, we estimate the dynamic average treatment effect on the annual count of statin pickups, where a pickup is a separate fill dispensed at a Danish pharmacy. Figure~\ref{fig:s_pickupraw} plots the raw trend in the number of pickups, and Figure~\ref{fig:s_pickupATT} reports the corresponding dynamic estimates for men and women. Both genders increase their pickups by around two per year throughout the post-shock period. The effects are not perfectly identical at the start: in the event year men's increase exceeds women's by about 0.11 pickups (1.37 against 1.26; Table~\ref{tab:s_pickup_overallATT}), a similar difference remains in the first post-shock year, and the effects become statistically indistinguishable from the second year onwards. This early gap in fills is, however, far smaller in relative terms than the contemporaneous gap in volume, and it disappears while the volume gap persists. Set against a gap of roughly 23 defined daily doses in annual statin volume, the near-equality of pickup counts implies that women receive lower-dose prescriptions per fill, not fewer fills.

Comparing the two across event time identifies where the gap emerges. In the event year, the male-female gap in volume is about 22 per cent of the female effect, while the gap in fills is about 9 per cent, so roughly two-fifths of the early gap comes from fills and the remainder from the dose written per fill. In the first post-shock year the pattern is similar, with a volume gap of about 16 per cent against a fills gap of about 6 per cent. From the second post-shock year onwards the fills gap disappears, yet a volume gap of 4 to 9 per cent persists. In other words, the early gap combines fills and dose, whereas the later gap comes from dose alone.

This margin is informative because the prescription register records dispensed fills, and a Danish pharmacy dispenses the molecule, pack size and dose written by the prescribing physician. The dose per fill is therefore the dose actually prescribed, and the pickup count already conditions on a prescription being filled. Near-equal pickup counts thus show that women are not failing to fill or to refill their prescriptions. The one margin we do not observe is a prescription that is written but never filled. Non-filling of this kind would, however, lower women's pickup counts, which is the opposite of what we find.

Explanations that rest on women being less likely to fill prescriptions, more likely to discontinue, or less adherent are therefore inconsistent with the near-identical pickup counts. We find the gap in the dose written per fill, which is a feature of how the prescription is set rather than of how the patient follows it. Several supply-side channels are consistent with this pattern: clinical guidelines that under-emphasise women's cardiovascular risk, sex-differential interpretation of risk by physicians, and real or perceived statin tolerability concerns in women. This reading is consistent with the United States evidence in \citet{Peters2018sex} and \citet{nanna2019sex}, who document sex differences in high-intensity statin prescribing, and with the Dutch evidence in \citet{kiss2024sex}, who report that women were less likely to be prescribed high-intensity statins. Our design adds a causal estimate of the post-shock change in prescribing rather than a cross-sectional gap conditional on an event.

We do not observe the prescribing physician in our register linkage, so we cannot test directly whether the gap is concentrated in encounters with male or female providers \citep{greenwood2018patient,cabral2024gender}. We can nevertheless describe where the relevant prescribing decisions are taken. Initial secondary-prevention prescribing in Denmark typically occurs at hospital discharge under the treating cardiologist. The general practitioner then manages maintenance, dose titration and repeat prescriptions over the following years, and this is the margin on which most post-shock fills occur. The dose gap is evident from the event year onwards, implying that prescribing decisions made both at hospital discharge and during subsequent primary care may play a role. Establishing the patient-physician link is the highest-priority extension, and we return to it in the conclusion.

\subsection{The gap across drug classes}\label{sec:gradient}

A second candidate is that women simply demand less care after an acute event, whether through lower health-seeking propensity, lower trust in physicians, or lower willingness to maintain therapy. A demand-side account of this kind makes a clear prediction: it should move every drug class in the same direction, because patient behaviour does not distinguish one secondary-prevention drug from another. To test this prediction, we extend the analysis to the full cardiovascular secondary-prevention regimen and compare the gap across antiplatelet drugs, statins, beta blockers and angiotensin-converting-enzyme inhibitors.

These classes differ in how much latitude the prescribing decision leaves the physician. Antiplatelet therapy after acute myocardial infarction is recommended for essentially every patient irrespective of sex, is started in hospital before discharge, and is maintained at a fixed dose that is not titrated \citep{byrne2023esc,rao2025acc}. Statin therapy is likewise recommended for nearly all patients, but the guidelines set a lipid target rather than a fixed dose, so the intensity of the statin and the molecule chosen remain matters of clinical judgement \citep{mach2020esc,grundy2019aha}. Beta blockers and angiotensin-converting-enzyme inhibitors sit at the other end. Their indications are conditional on comorbidity and their doses are uptitrated to the maximum the patient tolerates, which leaves the widest room for judgement about whom to start and whom to continue \citep{byrne2023esc,rao2025acc}. The same structure holds for strokes \citep{kleindorfer2021guideline,dawson2022european}. \footnote{Over our 1995--2018 window, beta blockers were treated as near-universal therapy after myocardial infarction, so within-sample discretion operated mainly through dose titration and tolerance rather than through the initiation decision. The indication has since narrowed, though only in a specific group. REDUCE-AMI found no benefit in patients with an ejection fraction of at least 50\% \citep{yndigegn2024beta}, and a pooled analysis of individual patient data from five trials in the same population reached the same conclusion \citep{kristensen2026beta}. Outside that group the evidence is not uniform: REBOOT found no effect above 40\% \citep{ibanez2025beta}, whereas BETAMI-DANBLOCK found a lower risk at 40\% and above \citep{munkhaugen2025beta}, and a pooled analysis restricted to mildly reduced ejection fraction found a benefit \citep{rossello2025beta}. The narrowing therefore applies to preserved ejection fraction specifically, and if anything it reinforces the discretion interpretation.}

\begin{table}[htbp]
\centering
\caption{The cross-drug-class gradient: gender gap in post-shock prescribing rises with clinical discretion.}
\label{tab:cardio_gradient}
\footnotesize
\begin{tabular}{lcccc}
\toprule
\textbf{Drug class} & \textbf{Women ATT} & \textbf{Men ATT} & \textbf{Gap (level)} & \textbf{Gap (\% of women's ATT)} \\
\midrule
Antiplatelets (DDD)              & 191.51 & 198.13 & $+6.62$  & $+3.4\%$ \\
Statins (DDD)                    & 219.48 & 242.90 & $+23.42$ & $+10.7\%$ \\
Beta blockers (DDD)              & 46.66  & 57.84  & $+11.18$ & $+24.0\%$ \\
ACE inhibitors (DDD)             & 93.64  & 128.24 & $+34.60$ & $+37.0\%$ \\
Distinct cardio drugs (count)    & 1.38   & 1.50   & $+0.12$  & $+8.7\%$ \\
Total cardio expenditure (2015 DKK) & 383.34 & 470.46 & $+87.12$ & $+22.7\%$ \\
\bottomrule
\end{tabular}
\\[0.5em]
\begin{minipage}{0.95\textwidth}
\footnotesize \textit{Notes:} ATT averaged over event times $\tau \in \{0,1,2,3,4\}$, cohort-weighted using within-event-time sample sizes and event-time-equal-weighted across $\tau$, separately by gender. Drug classes are ordered by their position on the clinical-discretion spectrum: antiplatelets are Class~I, Level~A in U.S.\ and European post-MI guidelines and are mechanically protocolised; statins are Class~I but intensity-discretionary; beta blockers are Class~I/IIa depending on indication; ACE inhibitors are Class~I/IIa with the broadest indication set. The distinct-cardio-drugs count is the number of distinct ATC-5 codes per year. Expenditure is in 2015 DKK (1 EUR $\approx$ 7.46 DKK), measured as the gross fee paid by the public insurance system. Standard errors via the individual-level multiplier bootstrap. The cross-gender gap is statistically significant at the 1\% level for statins, beta blockers, ACE inhibitors, the distinct-drug count and total expenditure, and \emph{not} statistically distinguishable from zero for antiplatelets (two-sided $p \approx 0.27$). Underlying dynamic ATTs in Figure~\ref{fig:cardio_drug_battery} and raw trends in Appendix \ref{app:cardio_drugs}.
\end{minipage}
\end{table}

\begin{figure}[htbp]
    \centering
    \begin{subfigure}{0.48\textwidth}
        \centering
        \includegraphics[width=\textwidth]{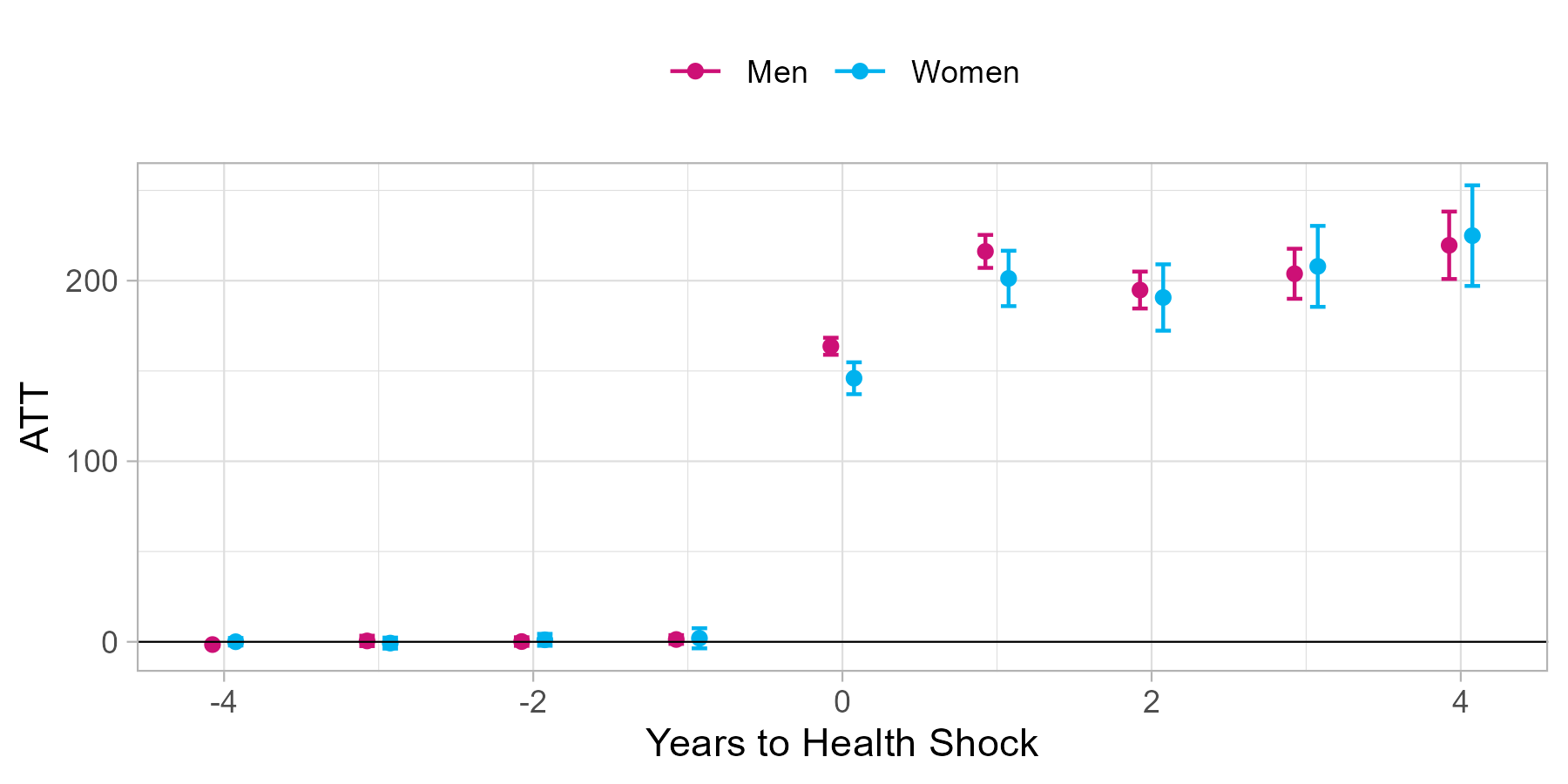}
        \caption{Antiplatelets (DDD).}
        \label{fig:dynamic_att_antiplatelet}
    \end{subfigure}
    \hfill
    \begin{subfigure}{0.48\textwidth}
        \centering
        \includegraphics[width=\textwidth]{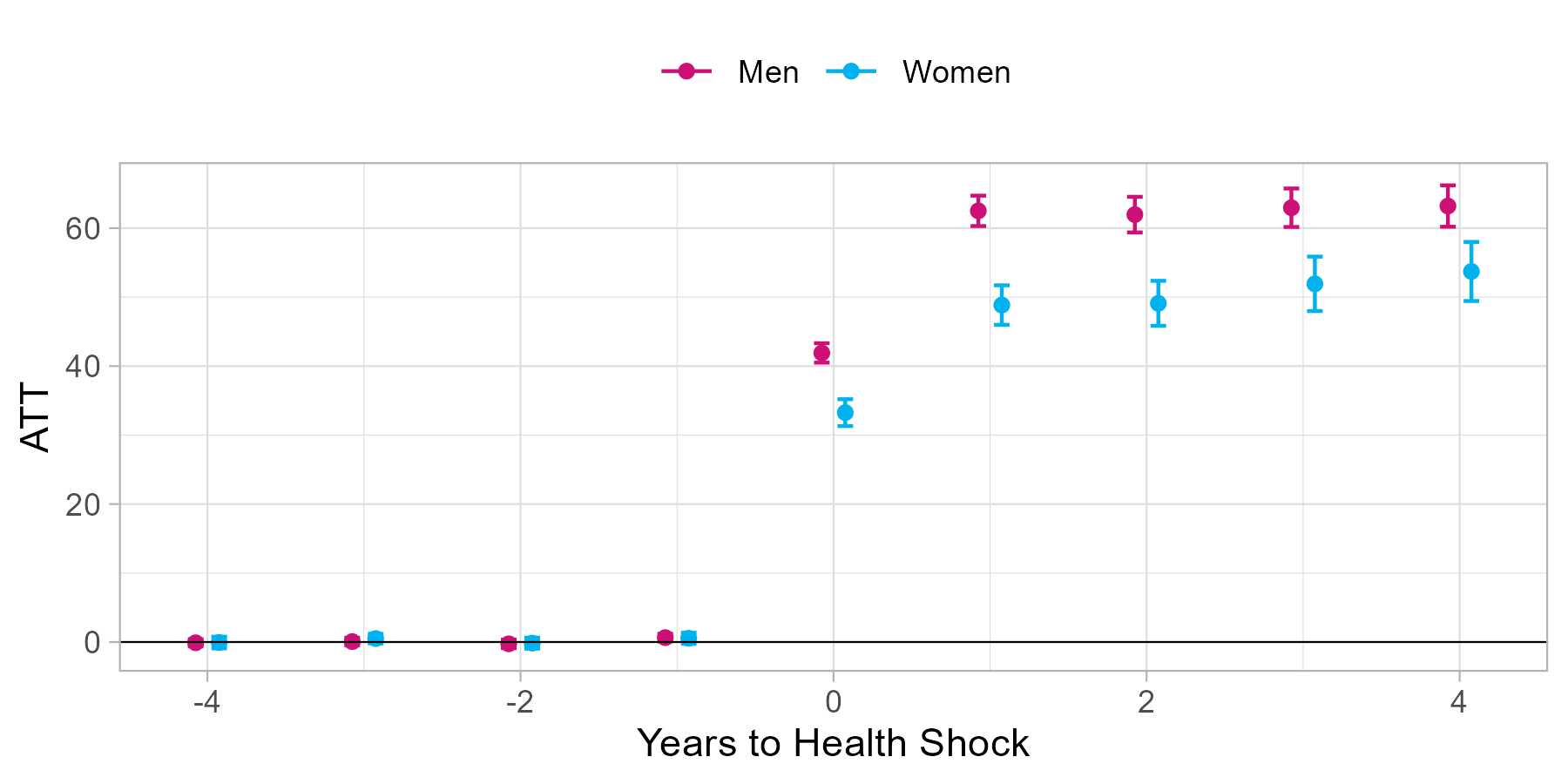}
        \caption{Beta blockers (DDD).}
        \label{fig:dynamic_att_beta_blocker}
    \end{subfigure}

    \vspace{0.3cm}

    \begin{subfigure}{0.48\textwidth}
        \centering
        \includegraphics[width=\textwidth]{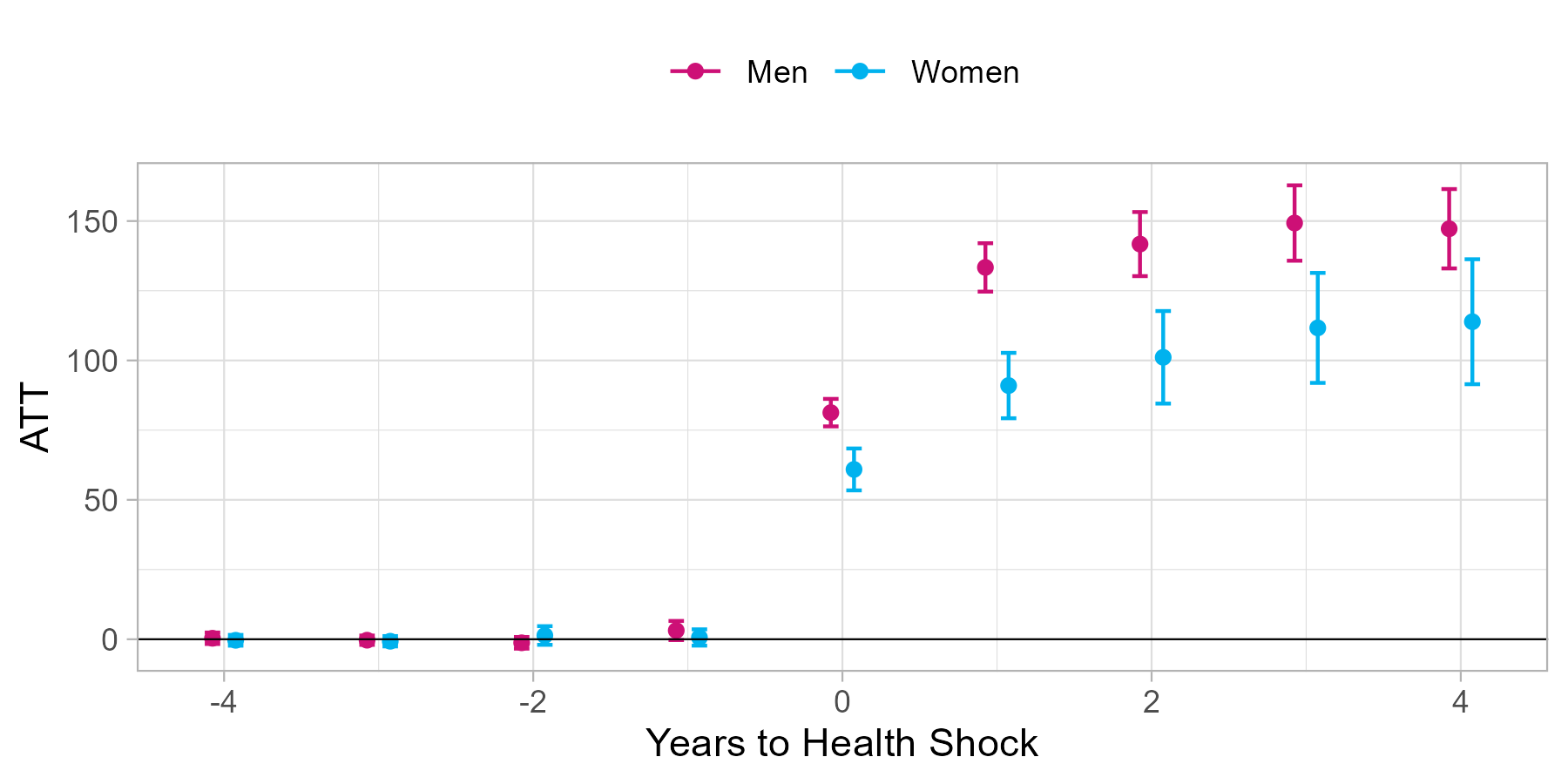}
        \caption{ACE inhibitors (DDD).}
        \label{fig:dynamic_att_ace_inhibitors}
    \end{subfigure}
    \hfill
    \begin{subfigure}{0.48\textwidth}
        \centering
        \includegraphics[width=\textwidth]{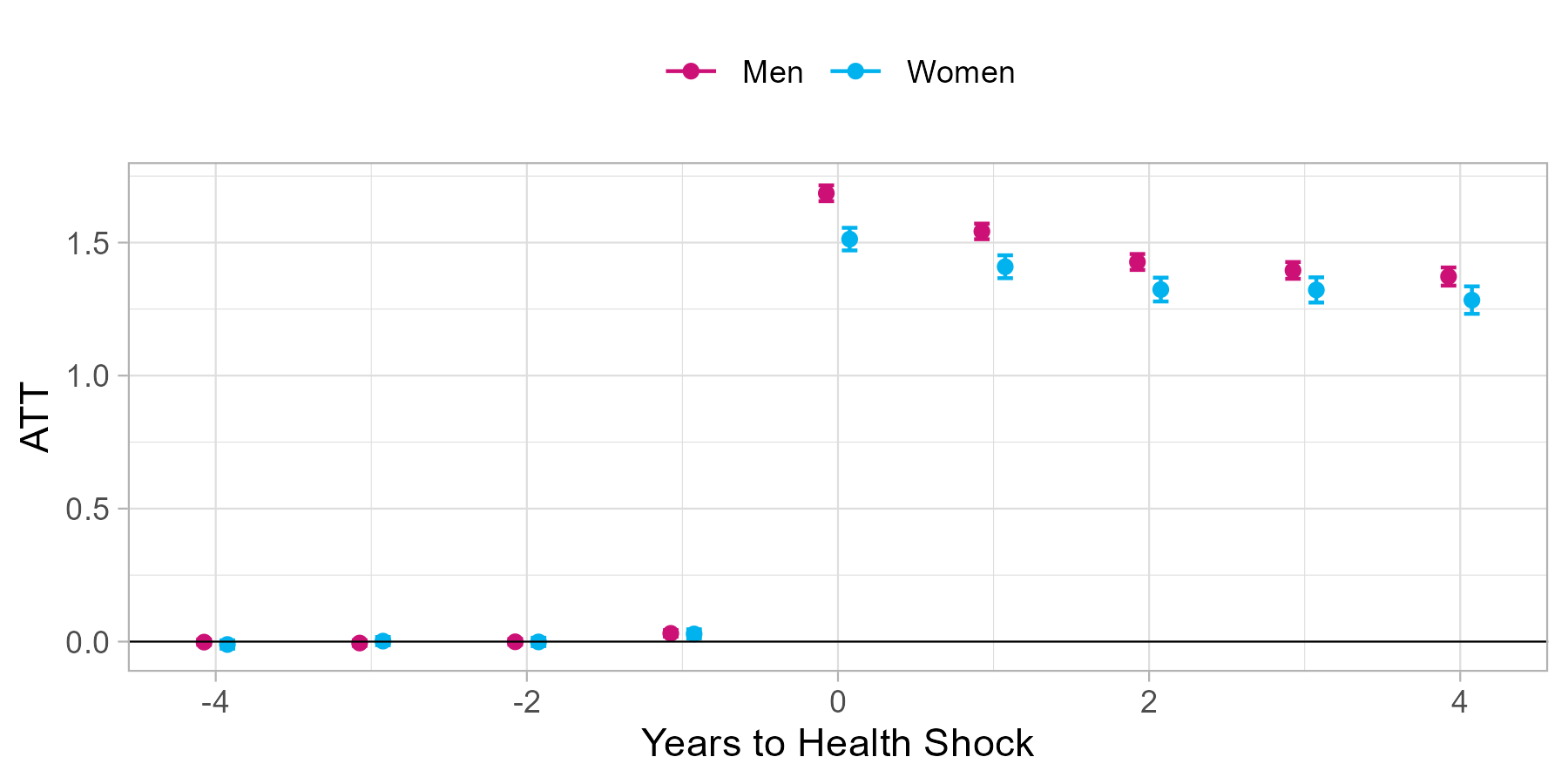}
        \caption{Number of distinct cardiovascular drugs.}
        \label{fig:dynamic_att_nr_cardio}
    \end{subfigure}

    \caption{Dynamic ATT estimates for the cardiovascular drug battery, by gender.}
    \label{fig:cardio_drug_battery}
    \begin{minipage}{0.95\textwidth}
    \footnotesize \textit{Notes:} Each panel reports the dynamic ATT estimated using the \cite{callaway2021difference} doubly robust estimator with a five-year non-shifting comparison window, separately by gender. Vertical bars denote simultaneous 95\% bootstrap confidence intervals. The discretion ordering across panels (antiplatelets $<$ statins $<$ beta blockers $<$ ACE inhibitors) follows U.S.\ and European post-MI clinical guidelines as described in the main text.
    \end{minipage}
\end{figure}

The gap does not appear uniformly across classes. Table~\ref{tab:cardio_gradient} reports the aggregate gap for each drug class, and Figure~\ref{fig:cardio_drug_battery} presents the underlying dynamic estimates. For antiplatelets, the most protocolised therapy, the gap is near zero and statistically indistinguishable from zero, at roughly 3 per cent of women's effect. The gap is moderate for statins, at roughly 11 per cent, larger for beta blockers, at roughly 24 per cent, and largest for angiotensin-converting-enzyme inhibitors, at roughly 37 per cent (raw trends in Appendix~\ref{app:cardio_drugs}). The proportional ordering of the gap across classes tracks the room for clinical judgement rather than anything on the patient side.\footnote{An alternative ordering by Class of Recommendation alone would put statins, antiplatelets and beta blockers at Class~I and ACE inhibitors at Class~I/IIa; the empirical ordering preserves the Class~I/IIa boundary and adds within-Class~I ranking by the degree of intensity-discretion as described in the text.} We compare the gaps in proportional terms because the classes differ more than fourfold in how intensively they are used, with women's post-shock effect ranging from about 47 defined daily doses for beta blockers to about 219 for statins, so a given proportional gap corresponds to very different absolute doses. In absolute doses the ranking is not monotone: the beta-blocker gap of about 11 defined daily doses is smaller than the statin gap of about 23. The number of distinct cardiovascular drugs prescribed per year is also higher for men, consistent with men receiving a fuller regimen rather than a higher dose of a single drug.

The antiplatelet result is the most informative one for assessing differential demand, precisely because antiplatelet therapy is essentially fixed-dose: after a myocardial infarction almost every patient receives the same standard regimen, so the antiplatelet dose leaves little intensive margin on which a prescribing gap could open. What remains is the extensive margin of initiation and persistence, which is exactly the margin a demand-side account operates on. If women had lower health-seeking propensity, lower trust, or higher discontinuation, the gap should therefore appear in antiplatelets, and at a substantial magnitude, because antiplatelet therapy is at least as demanding to maintain as statin therapy. Its near-absence is evidence against a demand-side account. A near-zero gap where the prescription is fixed, set against a widening gap where dose and drug choice are open to judgement, points away from patient demand and towards how care is delivered.

Two caveats apply to this conclusion, since we do not have a clean test that attributes the cross-drug pattern to provider behaviour rather than to unobserved factors that happen to correlate with the room for judgement.\footnote{Because the sample spans ages at which many women transition through menopause, a natural question is whether this transition could account for the patterns we observe. Menopause is a clinically significant transition, with a documented rise in physician consultations and hormone-replacement-therapy use \citep{conti2025menopause}, but clinical guidelines recommend statins for post-menopausal women at elevated cardiovascular risk regardless of HRT use \citep{hirsh2019guidelines,witting2022review}. While menopause may affect baseline health and utilisation, it is therefore unlikely to account for the post-shock gap we document.} The first caveat concerns tolerability, which is not uniform across these classes. Angiotensin-converting-enzyme-inhibitor cough is roughly two to three times more common in women, and statin intolerance is reported more often in women, while antiplatelets have few sex-differential tolerability drivers at standard doses. A purely patient-side persistence channel could therefore reproduce part of the ordering. The second caveat concerns case mix. The beta-blocker and angiotensin-converting-enzyme-inhibitor indications are conditional on comorbidity that our design does not match on, so part of these gaps could reflect differential eligibility rather than differential discretion.

\begin{figure}[htbp]
    \centering
    \includegraphics[width=0.85\textwidth]{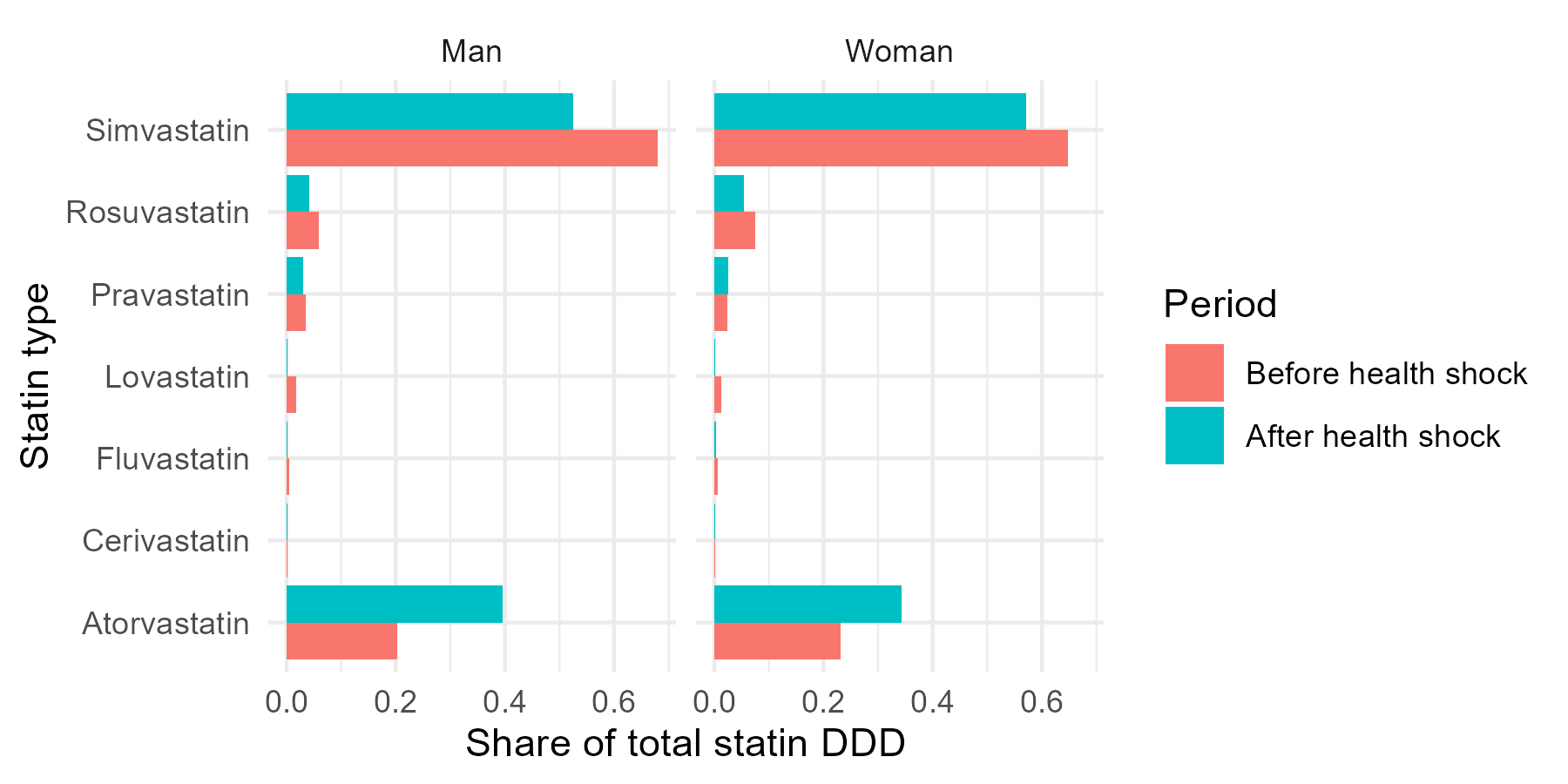}
    \caption{Composition of statin DDD by molecule, before and after the health shock, by gender.}
    \label{fig:statin_types}
    \begin{minipage}{0.9\textwidth}
    \footnotesize \textit{Notes:} Each bar reports the share of total statin DDD accounted for by a given molecule, in the five-year pre-shock window (red) and the five-year post-shock window (teal), separately for men and women. Statin molecules are ordered by intensity: atorvastatin and rosuvastatin are the highest-intensity statins in routine clinical use; simvastatin is moderate-intensity; pravastatin, lovastatin, and fluvastatin are lower-intensity.
    \end{minipage}
\end{figure}

The within-statin evidence reinforces the same reading on the patient-demand margin. To see this, we decompose total statin volume by molecule. Before the shock, both genders' use is concentrated in simvastatin, at roughly 65 to 70 per cent of total volume, reflecting Danish prescribing over the sample period. After the shock, both genders shift towards atorvastatin, the most-prescribed high-intensity statin in routine use, but Figure~\ref{fig:statin_types} shows that the shift is markedly more pronounced among men. The atorvastatin share rises by roughly 19 percentage points for men, from about 21 to 40 per cent, against roughly 10 percentage points for women, from about 23 to 33 per cent. Women are therefore not only receiving lower-dose statin prescriptions but are also less likely to be moved to a higher-intensity statin after the event.

\subsection{Severity of the health shock}\label{sec:severity}

The gap could reflect men having more severe events on average rather than gender as such. To address this concern, we stratify the analysis in two ways. We first split the heart-attack subsample into ST-elevation myocardial infarction, which is generally more severe, and non-ST-elevation infarction.\footnote{ICD-10 codes I21.0--I21.3 for STEMI and I21.4 for NSTEMI.} In our sample, 25.3 per cent of women's heart attacks were ST-elevation, against 34.1 per cent of men's. Within both strata, men's post-shock utilisation exceeds women's, so the gap survives conditioning on this clinical typology.

\begin{figure}[htbp]
    \centering
    \begin{subfigure}{0.85\textwidth}
        \centering
        \includegraphics[width=\textwidth]{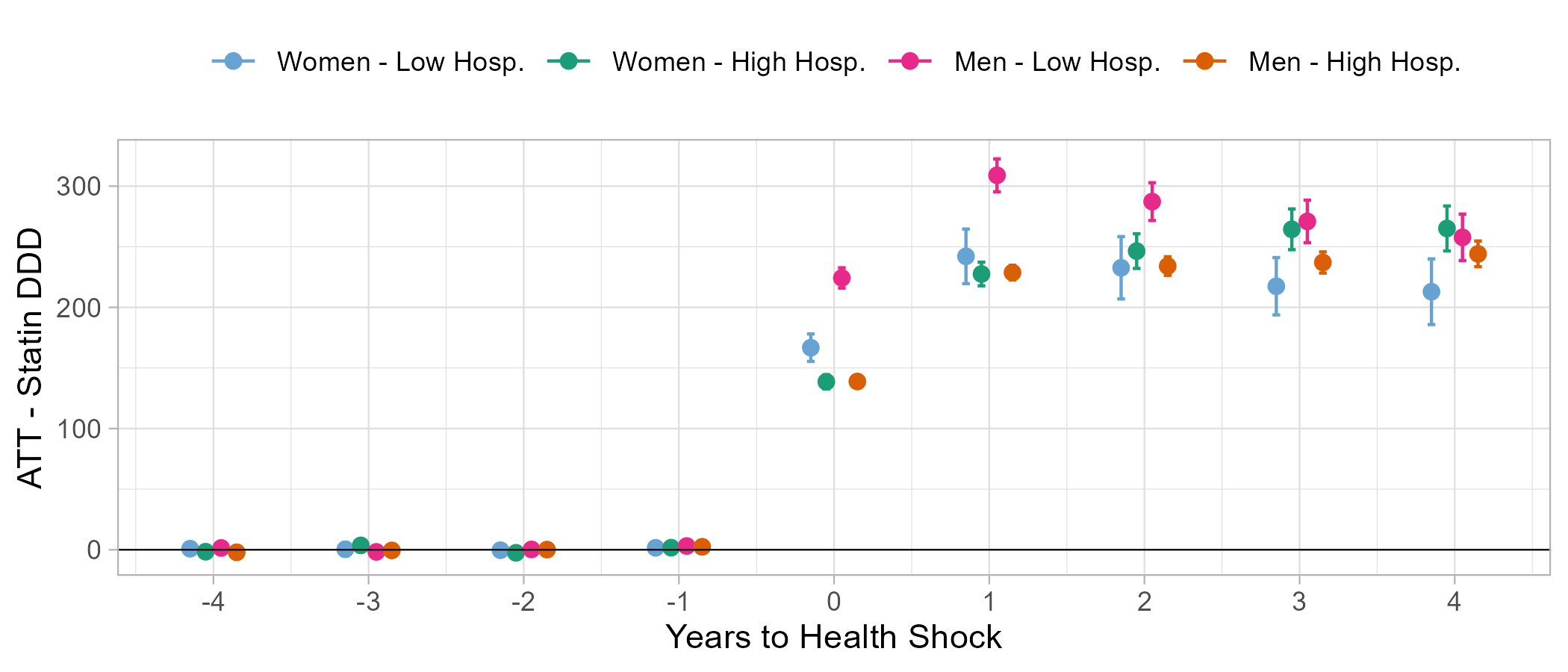}
        \caption{Statin DDD.}
        \label{fig:severity_statin}
    \end{subfigure}

    \vspace{0.3cm}

    \begin{subfigure}{0.85\textwidth}
        \centering
        \includegraphics[width=\textwidth]{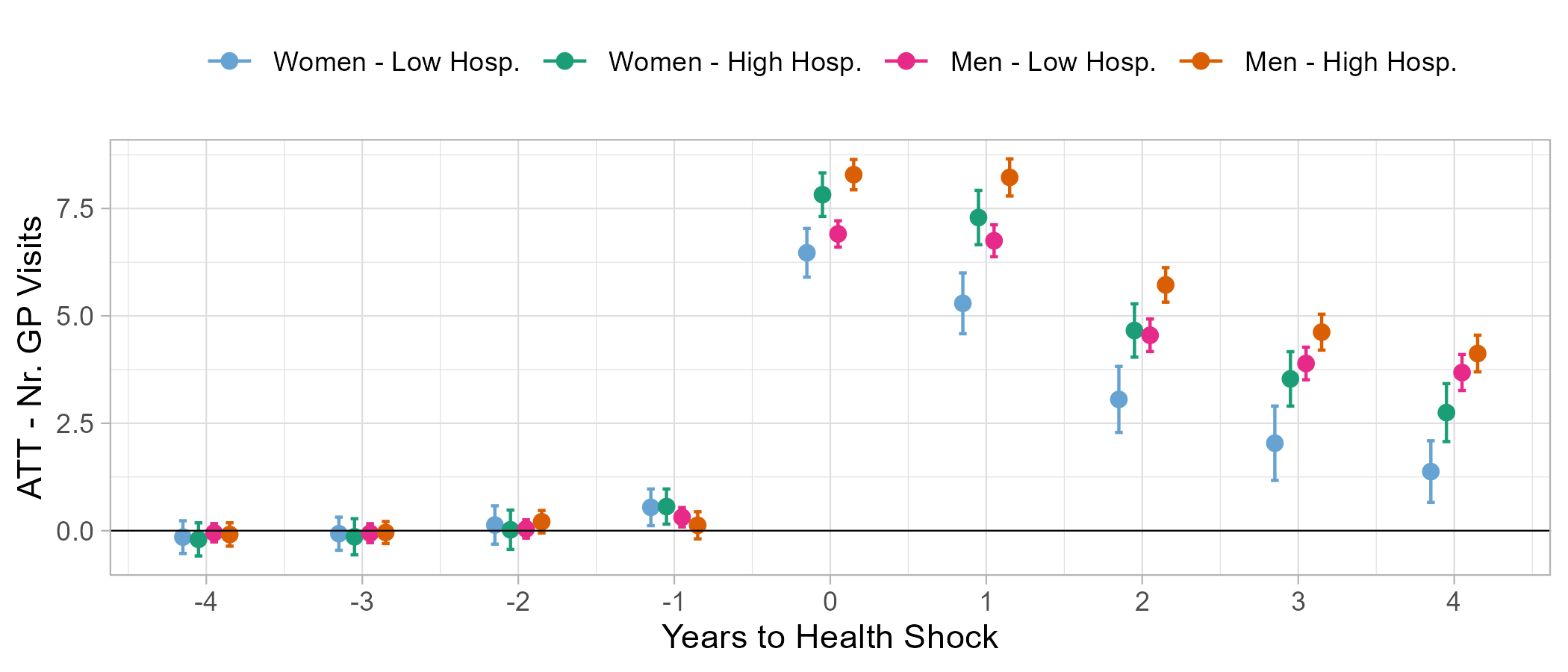}
        \caption{GP visits.}
        \label{fig:severity_gp}
    \end{subfigure}

    \caption{Dynamic ATT by gender and event severity.}
    \label{fig:severity_stratified}
    \begin{minipage}{0.95\textwidth}
    \footnotesize \textit{Notes:} Severity is defined by below- vs.\ above-median number of inpatient days in the event year, computed within shock type. ``Low Hosp.'' denotes the below-median severity group; ``High Hosp.'' denotes the above-median group. Estimator and inference as in Figure~\ref{fig:dynamic_att_full}.
    \end{minipage}
\end{figure}

We then re-estimate the dynamic effect in a four-way split of gender by inpatient-day severity, defined as below or above the median number of days hospitalised in the event year. This measure has the advantage of being observed for every patient regardless of shock type. Figure~\ref{fig:severity_stratified} shows that the gap in statin volume is preserved within both severity strata, as is the gap in general-practitioner visits; Appendix~\ref{sec:intensity} reports the STEMI/NSTEMI split.

\subsection{Gender differences in survival rates}\label{sec:balanced}

\begin{figure}[htbp]
    \centering
        \includegraphics[width=0.65\textwidth]{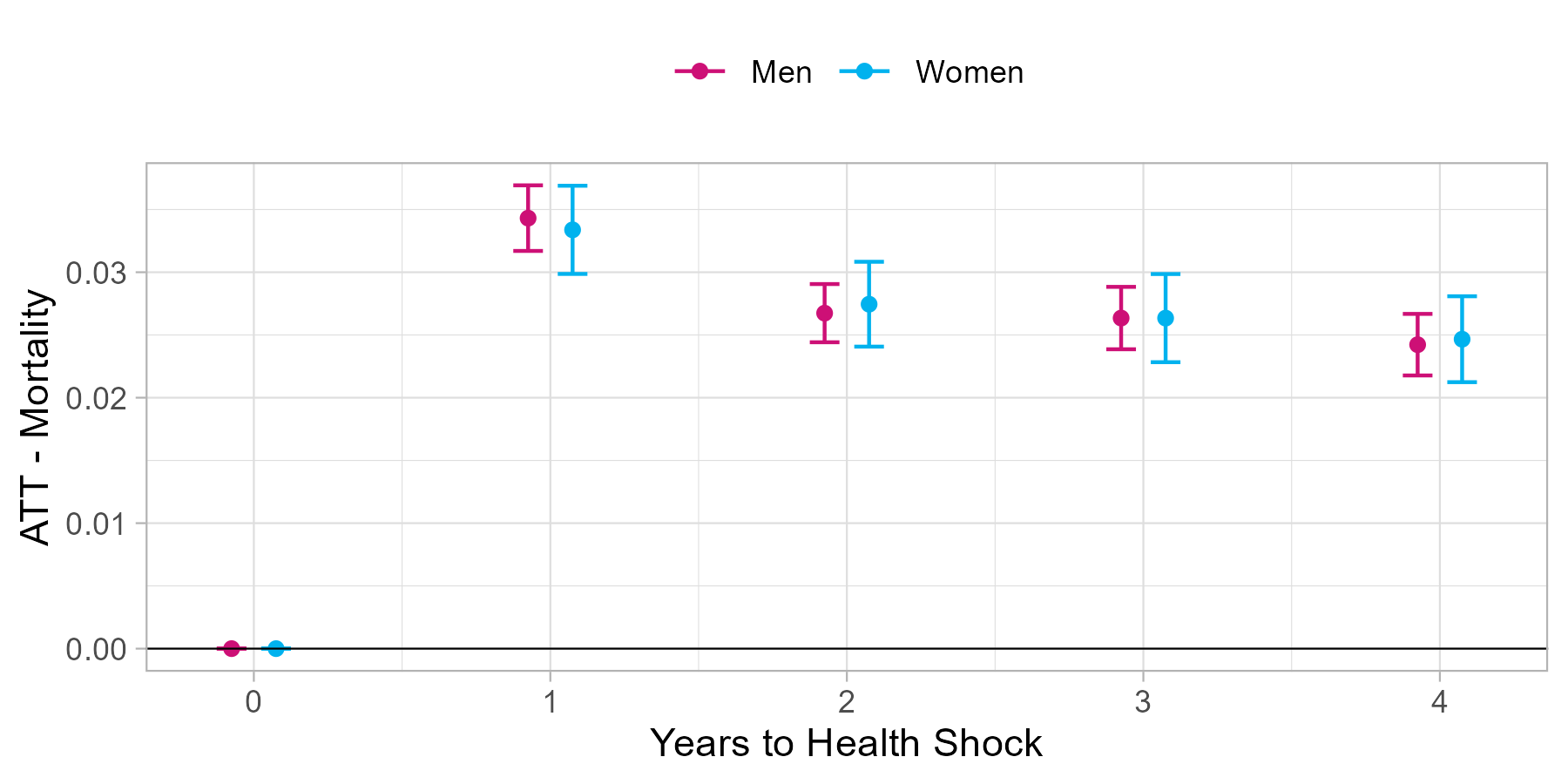}
        \caption{Dynamic ATT for mortality.}
        \label{fig:mortality_ATT}
    \begin{minipage}{0.9\textwidth}
    \footnotesize \textit{Notes:} Annual probability of death conditional on the shock, by event time and gender. Estimated using the \cite{callaway2021difference} doubly robust estimator. Numerical estimates in Table~\ref{tab:mortality}.
    \end{minipage}
\end{figure}

Differential mortality could generate a spurious gap if more severely ill men were selectively removed from the sample. Two pieces of evidence speak against this concern. First, we estimate the effect of the shock on mortality directly. Figure~\ref{fig:mortality_ATT} and Table~\ref{tab:mortality} show that the effect on one- to four-year mortality is similar across genders, at 0.02 to 0.03 in every event year for both groups, with overlapping confidence intervals throughout; the intervals exclude a one-percentage-point gender gap in annual mortality. Second, we re-estimate the main results on a balanced panel that conditions on five-year survival, which retains roughly 74 per cent of men and 77 per cent of women. The balanced-panel estimates reproduce the main estimates closely, which indicates that survivor selection is mild and symmetric across genders.

\begin{table}[htbp]
\centering
\caption{Dynamic ATT for mortality.}
\label{tab:mortality}
\begin{tabular}{m{2em}ccccc}
  \hline\hline
  \multicolumn{1}{p{2cm}}{\textbf{Event Time}} & \textbf{Avg ATT} & \textbf{CI-Lower}  & \multicolumn{1}{p{5em}}{\textbf{CI-Upper}}  & \multicolumn{1}{p{4em}}{\textbf{No.\ unique treated}} & \textbf{$p$-value} \\
  \hline
  \multicolumn{2}{l}{\textbf{Panel A: Women}} & & &  & \\
  \hline
0 & 0.00 & 0.00 & 0.00  & 15{,}140 &  \\
1 & 0.03 & 0.03 & 0.04  & 14{,}363 & $<0.01$ \\
2 & 0.03 & 0.02 & 0.03  & 13{,}046 & $<0.01$ \\
3 & 0.03 & 0.02 & 0.03  & 11{,}934 & $<0.01$ \\
4 & 0.02 & 0.02 & 0.03  & 10{,}918 & $<0.01$ \\
  \hline
  \multicolumn{2}{l}{\textbf{Panel B: Men}} & & & & \\
\hline
0 & 0.00 & 0.00 & 0.00  & 30{,}103 &  \\
1 & 0.03 & 0.03 & 0.04  & 28{,}547 & $<0.01$ \\
2 & 0.03 & 0.02 & 0.03  & 25{,}945 & $<0.01$ \\
3 & 0.03 & 0.02 & 0.03  & 23{,}790 & $<0.01$ \\
4 & 0.02 & 0.02 & 0.03  & 21{,}944 & $<0.01$ \\
\hline\hline
\end{tabular}
\\[0.5em]
\begin{minipage}{0.95\textwidth}
\footnotesize \textit{Notes:} ATT on the annual probability of death, estimated using \cite{callaway2021difference}. Confidence intervals at the 95\% level via the multiplier bootstrap with individual-level clustering.
\end{minipage}
\end{table}

\subsection{Baseline differences in primary-care use}\label{sec:ceiling}

Women use primary care more intensively than men before the shock, at 13.8 visits per year against 8.1. This raises a natural concern: their smaller post-shock general-practitioner response might be mechanically driven by having less room to rise. Three observations suggest otherwise. First, the gap is not confined to general-practitioner visits. It is present, and more cleanly concentrated, in statin volume, where pre-shock baselines are very low and nearly identical across genders, at 19.1 defined daily doses for women and 18.5 for men; a ceiling therefore cannot explain the statin gap. Second, the general-practitioner effect is large for both groups, at 5.8 additional visits per year for men and 4.8 for women: men nearly double their annual primary-care contacts, while women increase theirs by about half in the first event years before tapering. Third, the matched-sample analysis, which compares men and women with similar pre-shock characteristics, leaves the general-practitioner gap intact. We cannot fully exclude that some part of the general-practitioner gap reflects this mechanical ceiling, and we revisit this caveat in the conclusion.

\subsection{Heart attacks versus strokes}\label{sec:hetbyshock}

Finally, we examine whether the gap differs by type of health shock. Heart attacks and strokes differ in how they are managed clinically. Secondary prevention after a heart attack centres on aggressive lipid-lowering and antiplatelet therapy, while post-stroke care emphasises blood-pressure control, anticoagulation in selected patients and structured rehabilitation \citep{Heidenreich2011,Roth2018}. To exploit this difference, we re-estimate the specification separately for the two subsamples. The gap in statin volume and general-practitioner visits is concentrated in the heart-attack subsample, where it is large and precise, while the stroke subsample shows a much smaller and less precise gap (Appendix~\ref{sec:individual_treatment}). This pattern is what a delivery-side reading predicts, since lipid-lowering is the centrepiece of heart-attack follow-up. It is harder to reconcile with explanations based on patient demand or general health-seeking, which would predict a similar gap after strokes \citep{Peters2018sex,nanna2019sex,Vaccarino2011}.

\section{Does the gap improve men's outcomes?} \label{sec:outcomes_men}

Men receive more secondary-prevention pharmacotherapy and make more primary-care contacts than women after the shock. We now ask whether this additional care translates into better outcomes. We turn to the downstream margins we observe directly: long-run mortality, the accumulation of new comorbidities, and post-shock labour-market trajectories.

\subsection{Long-run mortality}\label{sec:longrun}

Mortality is the welfare endpoint most often invoked in the literature on sex differences in cardiovascular care, so we examine it first. We use a twenty-year follow-up window for the cohorts where this is feasible. Table~\ref{tab:mortality_20yr} reports the dynamic average treatment effect on the treated for annual post-shock mortality, separately by gender, estimated with the \citet{callaway2021difference} doubly robust estimator on the maximally feasible cohort at each horizon; the five-year dynamic profile is shown in Figure~\ref{fig:mortality_ATT}, and the full 0--19 year table is in Appendix~\ref{app:mortality_20yr}.

\begin{table}[htbp]
\centering
\caption{Long-run mortality: dynamic ATT over a 20-year follow-up window.}
\label{tab:mortality_20yr}
\footnotesize
\begin{tabular}{ccccccc}
\toprule
\textbf{Years post-shock} & \multicolumn{3}{c}{\textbf{Women}} & \multicolumn{3}{c}{\textbf{Men}} \\
\cmidrule(lr){2-4} \cmidrule(lr){5-7}
& ATT & 95\% CI & $N$ & ATT & 95\% CI & $N$ \\
\midrule
1 & 0.04 & [0.03, 0.04] & 13{,}172 & 0.04 & [0.04, 0.04] & 25{,}621 \\
5 & 0.03 & [0.02, 0.03] & 8{,}752 & 0.03 & [0.03, 0.03] & 17{,}268 \\
10 & 0.04 & [0.03, 0.04] & 4{,}860 & 0.04 & [0.03, 0.04] & 9{,}475 \\
15 & 0.04 & [0.03, 0.06] & 1{,}990 & 0.05 & [0.04, 0.06] & 3{,}988 \\
19 & 0.04 & [0.02, 0.06] & 503    & 0.05 & [0.04, 0.07] & 979 \\
\bottomrule
\end{tabular}
\\[0.5em]
\begin{minipage}{0.95\textwidth}
\footnotesize \textit{Notes:} Selected event times from the full 0--19 year ATT (full table in Appendix~\ref{app:mortality_20yr}). ATT on the annual probability of death, estimated using the \cite{callaway2021difference} doubly robust estimator with the maximally feasible cohort selection at each horizon; estimates at short horizons differ slightly from Table~\ref{tab:mortality} because of the different cohort selection. Confidence intervals at the 95\% level via the multiplier bootstrap with individual-level clustering. The sample sizes shrink with event time because only cohorts treated sufficiently early in the panel contribute to estimates at long horizons.
\end{minipage}
\end{table}

At every horizon from one to nineteen years post-shock, we find no evidence of a gender difference in the shock's mortality effect. The point estimates differ by at most two percentage points, the confidence intervals overlap throughout, and the time profile of the effect is similar across genders. Men's larger post-shock increase in pharmacotherapy is therefore not associated with lower long-run mortality.

Mortality is a binary and coarse outcome. It aggregates the cardiovascular cause of death, where a gap in care would most plausibly register, with competing-risk causes, where it would not, and it captures neither recurrent cardiovascular events and hospitalisations nor functional status, quality of life and labour-market participation. This is why we also examine comorbidity accumulation and labour-market outcomes in Section~\ref{sec:other_margins}. Women's baseline cardiovascular mortality exceeds men's in observational data \citep{Vaccarino2011}, so a constant gap in care can coexist with equal observed all-cause mortality. We read the mortality null as evidence on one margin where the welfare consequences of the gap are not visible, rather than as evidence that the gap has no welfare consequences.

\subsection{Morbidity accumulation and labour-market outcomes}\label{sec:other_margins}

We next ask whether the additional care men receive shows benefits on two further sets of downstream outcomes: the accumulation of new comorbidities and post-shock labour-market trajectories. We estimate both with the gender-specific \citet{callaway2021difference} doubly robust estimator used throughout. Table~\ref{tab:other_margins} reports the aggregate effects, and Appendix~\ref{app:other_margins} reports the underlying dynamic estimates.

\begin{table}[htbp]
\centering
\caption{Post-shock effects on comorbidity accumulation and labour-market outcomes.}
\label{tab:other_margins}
\footnotesize
\begin{tabular}{lcccccccc}
  \hline\hline
& ATT & CI lower & CI upper & Std.\ err. & No.\ obs. & $p$-value & Baseline & Rel. \\
&  &  &  &  &  &  & mean & \% change \\
  \hline
\multicolumn{9}{l}{\textbf{Cardiac comorbidities}} \\ \hline
Women & 0.07 & 0.06 & 0.08 & $<$0.01 & 10{,}907 & $<0.01$ & 0.03 & 227\% \\
Men   & 0.09 & 0.08 & 0.09 & $<$0.01 & 23{,}079 & $<0.01$ & 0.03 & 292\% \\ \hline
\multicolumn{9}{l}{\textbf{Non-cardiac comorbidities}} \\ \hline
Women & 0.11 & 0.10 & 0.12 & $<$0.01 & 10{,}907 & $<0.01$ & 0.13 & 85\% \\
Men   & 0.11 & 0.10 & 0.11 & $<$0.01 & 23{,}079 & $<0.01$ & 0.09 & 122\% \\ \hline
\multicolumn{9}{l}{\textbf{Work absence}} \\ \hline
Women & 39.52  & 34.57 & 44.46 & 2.69 & 689   & $<0.01$ & 25.63 & 154\% \\
Men   & 36.38 & 31.35 & 41.41 & 2.56 & 821   & $<0.01$ & 16.97  & 214\% \\ 
\hline\hline
\end{tabular}
\\[0.5em]
\begin{minipage}{0.95\textwidth}
\footnotesize \textit{Notes:} Gender-specific ATTs from the \cite{callaway2021difference} doubly robust estimator, averaged over the post-shock event years and cohort-weighted as in Table~\ref{tab:aggregate_att}. Confidence intervals at the 95\% level via the multiplier bootstrap with individual-level clustering; standard errors reported as $<$0.01 are non-zero but round to zero at two decimals. The comorbidity outcomes (cumulative first-occurrence counts of distinct Elixhauser conditions, defined in Section~\ref{sec:outcomes}) are estimated on 23{,}079 men and 10{,}907 women; work absence days are measured for the working-age, labour-market-attached subsample, which is smaller still and differs in size across genders. 
The dynamic ATTs underlying these aggregates are in Appendix~\ref{app:other_margins}.
\end{minipage}
\end{table}

We consider comorbidities first. Treated individuals accumulate more cardiac comorbidities after the shock than their not-yet-treated counterparts, and the increase is, if anything, larger for men than for women, at about 0.09 against 0.07. This male excess should be interpreted with caution, because men also make more frequent primary-care contacts after the shock (Section~\ref{sec:results}) and therefore present more occasions on which a new diagnosis can be recorded. The non-cardiac comorbidity index helps separate a genuine difference in morbidity from this difference in diagnostic contact, since it aggregates conditions outside the cardiovascular pathway. On this placebo outcome the two genders are statistically indistinguishable, at about 0.11 each, which indicates that the cardiac asymmetry is specific to the cardiovascular pathway. Women's smaller post-shock increase in secondary-prevention pharmacotherapy is not accompanied by a larger accumulation of cardiac comorbidities, where a gap in care would be expected to have an effect on morbidity. However, a diagnosis-count index cannot separate true morbidity from detected morbidity, and if providers attend less closely to women's cardiovascular disease, the same behaviour would leave some of women's new cardiac conditions unrecorded, so the index may understate women's morbidity rather than overstate it. We therefore read the cardiac result as no detectable excess morbidity among women, not as proof of equal underlying morbidity.

We then turn to labour-market outcomes. Work absence days rise sharply around the shock, and by similar amounts for both genders: about 36 days for men and 40 for women. The larger relative increase for men reflects only their lower pre-shock baseline, about 31 days against 35 for women, exactly as for primary-care visits in Section~\ref{sec:results}. 

The labour-market analysis is based on a substantially smaller sample of working-age individuals attached to the labour force, with unequal numbers of men and women, making these estimates less precise than the main results. In addition, disability-pension receipt and labour-force exit are absorbing transitions with virtually no pre-shock variation and therefore cannot be estimated within the event-study framework.

The pattern across these margins is consistent. Men's care and spending rise by more, yet on every outcome we observe they fare no better than women.

\section{Conclusion} \label{sec:conclusion}

Using Danish register data on first-time non-fatal heart attacks and strokes among 50-to-70-year-olds between 1995 and 2018, we provide causal evidence on how men's and women's care responds to an acute cardiovascular shock. We document a large gender gap in post-shock secondary-prevention care: men's secondary-prevention pharmacotherapy and primary-care contacts rise substantially more than women's after the event. The gap in inpatient days is small in absolute terms, while the gap on the discretionary outpatient and prescribing margins is large.

We find that this gap is not explained by the standard alternatives. It does not reflect differences in severity, since it survives within both types of cardiovascular shock and across severity strata. It does not reflect survival selection. Nor is it explained by compositional differences on observables. It does not reflect fill behaviour: the shortfall comes from lower prescribed intensity (lower-dose prescriptions per pickup, less frequent switching to the highest-intensity statin molecule) rather than from the number of fills, which rises almost identically for the two genders. It does not reflect a mechanical ceiling on women's already higher primary-care use, an explanation that cannot account for the prescribing gap. It does not reflect weaker patient demand. What remains is consistent with how care is delivered: the gap rises across drug classes in step with the latitude a prescriber has to exercise judgement, from no statistically significant gap for antiplatelets to the widest gap for ACE inhibitors, consistent with a supply-side channel in the prescribing decision. The gap is also concentrated after heart attacks rather than strokes, the pathway where lipid-lowering therapy is central.

We then ask whether the additional care men receive improves their outcomes, and we find no measurable advantage. We detect no gender difference in the shock's mortality effect at any horizon out to twenty years. The labour-market outcomes we measure show no gap to women's disadvantage, with post-shock work absence days rising almost identically for both genders. Women accumulate no more new comorbidities than men. The extra prescribing men receive therefore leaves no detectable footprint on the survival, labour-market, or morbidity outcomes we observe.

These results are particularly striking given women's general survival advantage. Women fare at least as well as men on every outcome we observe while their secondary-prevention care rises by substantially less, so they might have fared even better had their treatment matched men's.

There are a few questions the data do not allow us to address. We do not observe the treating physician, so the prescribing-intensity finding cannot distinguish gender concordance from practice style; linking patients to prescribers is the natural next step. Finer severity measures such as infarct size and troponin are unobserved, though the gap is present within both the STEMI and the NSTEMI stratum. Part of the visit gap could reflect women's higher baseline primary-care use, though such a ceiling cannot explain the statin gap. The four drug classes we cover are the core of post-shock pharmacotherapy. Finally, we cannot study disability-pension receipt and labour-force exit with our difference-in-differences design: both are essentially absent before the shock, so there is no pre-treatment outcome to compare.

We provide a quantitative anchor for the male-female health-survival paradox: women's longer expected lifespans coexist with a materially smaller increase in post-shock secondary-prevention care, in a way that cannot be reduced to access constraints, severity, selection, or composition. The gap persists in a country with universal, free healthcare and low gender inequality, so structural barriers are not its primary driver. Our estimates are for that reason plausibly conservative relative to the magnitudes that less equitable systems would likely produce. If the gap reflects prescriber behaviour, as our evidence suggests, and if women's marginal returns to secondary-prevention care are positive, then closing it will require acting on prescribing intensity and follow-up care rather than on access alone.

\newpage

\justify
\bibliographystyle{chicago}
\bibliography{fullref}

@Preamble{ " \newcommand{\noop}[1]{} " }

@article{chandra2007productivity,
  title={{Productivity Spillovers in Health Care: Evidence From the Treatment of Heart Attacks}},
  author={Chandra, Amitabh and Staiger, Douglas O},
  journal={Journal of Political Economy},
  volume={115},
  number={1},
  pages={103--140},
  year={2007},
  publisher={The University of Chicago Press}
}

@article{doyle2011returns,
  title={{Returns to Local-Area Health Care Spending: Evidence From Health Shocks to Patients Far From Home}},
  author={Doyle, Joseph J},
  journal={American Economic Journal: Applied Economics},
  volume={3},
  number={3},
  pages={221--43},
  year={2011}
}

@article{grossman,
  title={{On the Concept of Health Capital and the Demand for Health}},
  author={Grossman, Michael},
  journal={Journal of Political Economy},
  volume={80},
  number={2},
  pages={223--255},
  year={1972},
  publisher={The University of Chicago Press}
}

@article{nielsen,
Author = {Fadlon, Itzik and Nielsen, Torben Heien},
Title = {{Family Health Behaviors}},
Journal = {American Economic Review},
Volume = {109},
Number = {9},
Year = {2019},
Month = {September},
Pages = {3162-91},
}

@incollection{grossman2000human,
  title={{The Human Capital Model}},
  author={Grossman, Michael},
  booktitle={Handbook of Health Economics},
  volume={1},
  pages={347--408},
  year={2000},
  publisher={Elsevier}
}

@article{schmidt2015danish,
  title={{The Danish National Patient Registry: A Review of Content, Data Quality, and Research Potential}},
  author={Schmidt, Morten and Schmidt, Sigrun Alba Johannesdottir and Sandegaard, Jakob Lynge and Ehrenstein, Vera and Pedersen, Lars and S{\o}rensen, Henrik Toft},
  journal={Clinical Epidemiology},
  volume={7},
  pages={449--490},
  year={2015},
  publisher={Dove Press},
  doi={10.2147/CLEP.S91125}
}

@article{ozcan2016danish,
  title={{The Danish Heart Registry}},
  author={{\"O}zcan, Cengiz and Juel, Knud and Lassen, Jens Flensted and von Kappelgaard, Lene Mia and Mortensen, Poul Erik and Gislason, Gunnar},
  journal={Clinical Epidemiology},
  volume={8},
  pages={503--508},
  year={2016},
  publisher={Dove Press},
  doi={10.2147/CLEP.S99475}
}

@article{oksuzyan2011sex,
  title={{Sex Differences in Medication and Primary Healthcare Use Before and After Spousal Bereavement at Older Ages in Denmark: Nationwide Register Study of Over 6000 Bereavements}},
  author={Oksuzyan, Anna and Jacobsen, Rune and Glaser, Karen and Tomassini, Cecilia and Vaupel, James W and Christensen, Kaare},
  journal={Journal of Aging Research},
  volume={2011},
  pages={678289},
  year={2011},
  publisher={Hindawi},
  doi={10.4061/2011/678289}
}

@article{nobili2011drug,
  title={{Drug Utilization and Polypharmacy in an Italian Elderly Population: The EPIFARM-Elderly Project}},
  author={Nobili, Alessandro and Franchi, Carlotta and Pasina, Luca and Tettamanti, Mauro and Baviera, Marta and Monesi, Lara and Roncaglioni, Carla and Riva, Emma and Lucca, Ugo and Bortolotti, Angela and others},
  journal={Pharmacoepidemiology and Drug Safety},
  volume={20},
  number={5},
  pages={488--496},
  year={2011},
  publisher={Wiley Online Library}
}

@article{case2005sex,
  title={{Sex Differences in Morbidity and Mortality}},
  author={Case, Anne and Paxson, Christina},
  journal={Demography},
  volume={42},
  number={2},
  pages={189--214},
  year={2005},
  publisher={Springer}
}

@article{oksuzyan2008men,
  title={{Men: Good Health and High Mortality. Sex Differences in Health and Aging}},
  author={Oksuzyan, Anna and Juel, Knud and Vaupel, James W and Christensen, Kaare},
  journal={Aging Clinical and Experimental Research},
  volume={20},
  number={2},
  pages={91--102},
  year={2008},
  publisher={Springer}
}

@article{juel2007men,
  title={{Are Men Seeking Medical Advice Too Late? Contacts to General Practitioners and Hospital Admissions in Denmark 2005}},
  author={Juel, Knud and Christensen, Kaare},
  journal={Journal of Public Health},
  volume={30},
  number={1},
  pages={111--113},
  year={2008},
  publisher={Oxford University Press},
  doi={10.1093/pubmed/fdm072}
}

@article{galdas2005men,
  title={{Men and Health Help-Seeking Behaviour: Literature Review}},
  author={Galdas, Paul M and Cheater, Francine and Marshall, Paul},
  journal={Journal of Advanced Nursing},
  volume={49},
  number={6},
  pages={616--623},
  year={2005},
  publisher={Wiley Online Library}
}

@article{blaakilde2012dode,
  title={{D{\o}de M{\ae}nd og Syge Kvinder. K{\o}n, Alder og Ulighed i Sundhed}},
  author={Blaakilde, Anne Leonora},
  journal={Kvinder, K{\o}n \& Forskning},
  number={4},
  year={2012}
}

@techreport{pedersen2014sundhed,
  title={{Sundhed og trivsel i et k{\o}nsperspektiv}},
  author={Pedersen, Pia Vivian and Johansen, Katrine Bindesb{\o}l Holm and Ekholm, Ola and Juel, Knud},
  year={2014},
  institution={Statens Institut for Folkesundhed, Syddansk Universitet},
  address={K{\o}benhavn},
  note={ISBN 978-87-7899-283-3}
}

@article{cutler2005explains,
  title={{What Explains Differences in Smoking, Drinking, and Other Health-Related Behaviors?}},
  author={Cutler, David M and Glaeser, Edward},
  journal={American Economic Review},
  volume={95},
  number={2},
  pages={238--242},
  year={2005}
}

@article{baji2018adaptation,
  title={{Adaptation or Recovery After Health Shocks? Evidence Using Subjective and Objective Health Measures}},
  author={Baji, Petra and B{\'\i}r{\'o}, Anik{\'o}},
  journal={Health Economics},
  volume={27},
  number={5},
  pages={850--864},
  year={2018},
  publisher={Wiley Online Library}
}

@article{legato2016consideration,
  title={{Consideration of Sex Differences in Medicine to Improve Health Care and Patient Outcomes}},
  author={Legato, Marianne J and Johnson, Paula A and Manson, JoAnn E},
  journal={JAMA},
  volume={316},
  number={18},
  pages={1865--1866},
  year={2016},
  publisher={American Medical Association}
}

@article{smith2009gender,
  title={{Gender Differences in the Colorado Stroke Registry}},
  author={Smith, Don B and Murphy, Paul and Santos, Patricia and Phillips, Merrilee and Wilde, Marsha},
  journal={Stroke},
  volume={40},
  number={4},
  pages={1078--1081},
  year={2009},
  publisher={Am Heart Assoc}
}

@article{goodman2021difference,
  title={{Difference-in-Differences with Variation in Treatment Timing}},
  author={Goodman-Bacon, Andrew},
  journal={Journal of Econometrics},
  volume={225},
  number={2},
  pages={254--277},
  year={2021},
  publisher={Elsevier}
}

@article{de2020two,
  title={{Two-Way Fixed Effects Estimators with Heterogeneous Treatment Effects}},
  author={de Chaisemartin, Cl{\'e}ment and D’Haultf{\oe}uille, Xavier},
  journal={American Economic Review},
  volume={110},
  number={9},
  pages={2964--2996},
  year={2020},
  publisher={American Economic Association},
  doi={10.1257/aer.20181169}
}

@article{callaway2021difference,
  title={{Difference-in-Differences with Multiple Time Periods}},
  author={Callaway, Brantly and Sant’Anna, Pedro H. C.},
  journal={Journal of Econometrics},
  volume={225},
  number={2},
  pages={200--230},
  year={2021},
  publisher={Elsevier},
  doi={10.1016/j.jeconom.2020.12.001}
}

@article{sant2020doubly,
  title={{Doubly Robust Difference-in-Differences Estimators}},
  author={Sant’Anna, Pedro H. C. and Zhao, Jun},
  journal={Journal of Econometrics},
  volume={219},
  number={1},
  pages={101--122},
  year={2020},
  publisher={Elsevier},
  doi={10.1016/j.jeconom.2020.06.003}
}

@article{kiss2024sex,
  title={{Sex Differences in the Intensity of Statin Prescriptions at Initiation in a Primary Care Setting}},
  author={Kiss, Pauline AJ and Uijl, Alicia and de Boer, Annemarijn R and Duk, Tessa CX and Grobbee, Diederick E and Hollander, Monika and Smits, Elisabeth and Sturkenboom, Miriam CJM and Peters, Sanne AE},
  journal={Heart},
  volume={110},
  number={15},
  pages={981--987},
  year={2024},
  publisher={BMJ Publishing Group Ltd and British Cardiovascular Society}
}

@article{nanna2019sex,
  title={{Sex Differences in the Use of Statins in Community Practice: Patient and Provider Assessment of Lipid Management Registry}},
  author={Nanna, Michael G and Wang, Tracy Y and Xiang, Qun and Goldberg, Anne C and Robinson, Jennifer G and Roger, Veronique L and Virani, Salim S and Wilson, Peter WF and Louie, Michael J and Koren, Andrew and others},
  journal={Circulation: Cardiovascular Quality and Outcomes},
  volume={12},
  number={8},
  pages={e005562},
  year={2019},
  publisher={Am Heart Assoc}
}

@article{sibbritt2024demographic,
  title={{Demographic Factors Effect Stroke-Related Healthcare Utilisation Among Australian Stroke Survivors}},
  author={Sibbritt, D and Bayes, J and Peng, W and Adams, J},
  journal={Scientific Reports},
  volume={14},
  number={1},
  pages={21241},
  year={2024},
  publisher={Nature Publishing Group UK London},
  doi={10.1038/s41598-024-72092-w}
}

@article{borusyak2024revisiting,
  title={{Revisiting Event-Study Designs: Robust and Efficient Estimation}},
  author={Borusyak, Kirill and Jaravel, Xavier and Spiess, Jann},
  journal={Review of Economic Studies},
  volume={91},
  number={6},
  pages={3253--3285},
  year={2024},
  publisher={Oxford University Press UK}
}

@article{sun2021estimating,
  title={{Estimating Dynamic Treatment Effects in Event Studies With Heterogeneous Treatment Effects}},
  author={Sun, Liyang and Abraham, Sarah},
  journal={Journal of Econometrics},
  volume={225},
  number={2},
  pages={175--199},
  year={2021},
  publisher={Elsevier}
}

@article{fadlon2021family,
  title={{Family Labor Supply Responses to Severe Health Shocks: Evidence From Danish Administrative Records}},
  author={Fadlon, Itzik and Nielsen, Torben Heien},
  journal={American Economic Journal: Applied Economics},
  volume={13},
  number={3},
  pages={1--30},
  year={2021},
  publisher={American Economic Association 2014 Broadway, Suite 305, Nashville, TN 37203-2425}
}

@article{chandra2020identifying,
  title={{Identifying Sources of Inefficiency in Healthcare}},
  author={Chandra, Amitabh and Staiger, Douglas O},
  journal={The Quarterly Journal of Economics},
  volume={135},
  number={2},
  pages={785--843},
  year={2020},
  publisher={Oxford University Press}
}

@article{schulman1999effect,
  title={{The Effect of Race and Sex on Physicians’ Recommendations for Cardiac Catheterization}},
  author={Schulman, Kevin A and Berlin, Jesse A and Harless, William and Kerner, Jon F and Sistrunk, Shyrl and Gersh, Bernard J and Dub{\'e}, Ross and Taleghani, Christopher K and Burke, Jennifer E and Williams, Sankey and others},
  journal={New England Journal of Medicine},
  volume={340},
  number={8},
  pages={618--626},
  year={1999},
  publisher={Massachusetts Medical Society},
  doi={10.1056/NEJM199902253400806}
}

@article{cabral2024gender,
  title={{Gender Differences in Medical Evaluations: Evidence From Randomly Assigned Doctors}},
  author={Cabral, Marika and Dillender, Marcus},
  journal={American Economic Review},
  volume={114},
  number={2},
  pages={462--499},
  year={2024},
  publisher={American Economic Association 2014 Broadway, Suite 305, Nashville, TN 37203}
}

@techreport{conti2025menopause,
  title={{The Menopause ``Penalty"}},
  author={Conti, Gabriella and Ginja, Rita and Persson, Petra and Willage, Barton},
  year={2025},
  month={March},
  institution={National Bureau of Economic Research},
  type={NBER Working Paper},
  number={33621},
  doi={10.3386/w33621}
}

@misc{hirsh2019guidelines,
  author       = {De Jesus, Sasha and Myerson, Merle and Gianos, Eugenia},
  title        = {{Guidelines for Lipid-Lowering in Women -- What Have We Learned?}},
  howpublished = {LipidSpin, vol.~17, no.~4, pp.~24--28, National Lipid Association},
  year         = {2019}
}

@article{witting2022review,
  title={{Review of Lipid-Lowering Therapy in Women from Reproductive to Postmenopausal Years}},
  author={Witting, Celeste and Devareddy, Ankita and Rodriguez, Fatima},
  journal={Reviews in Cardiovascular Medicine},
  volume={23},
  number={5},
  pages={183},
  year={2022}
}

@article{bockerman2025family,
  title={{A Family Affair? Long-Term Economic and Mental Health Effects of Spousal Cancer}},
  author={B{\"o}ckerman, Petri and Kortelainen, Mika and Salokangas, Henri and Vaalavuo, Maria},
  journal={Journal of Population Economics},
  volume={38},
  number={1},
  pages={19},
  year={2025},
  publisher={Springer},
  doi={10.1007/s00148-025-01070-x}
}

@article{Li2020,
  author    = {Li, Linxin and Scott, Catherine A. and Rothwell, Peter M.},
  title     = {{Trends in Stroke Incidence in High-Income Countries in the 21st Century: Population-Based Study and Systematic Review}},
  journal   = {Stroke},
  year      = {2020},
  volume    = {51},
  number    = {5},
  pages     = {1372--1380},
  doi       = {10.1161/STROKEAHA.119.028484}
}

@article{Shah2021,
  author    = {Timmis, Adam and Vardas, Panos and Townsend, Nick and Torbica, Aleksandra and Katus, Hugo and De Smedt, Delphine and others},
  title     = {{European Society of Cardiology: Cardiovascular Disease Statistics 2021}},
  journal   = {European Heart Journal},
  year      = {2022},
  volume    = {43},
  number    = {8},
  pages     = {716--799},
  doi       = {10.1093/eurheartj/ehab892}
}

@article{Canto2000,
  author    = {Canto, Jorge G. and Shlipak, Michael G. and Rogers, William J. and Malmgren, Judith A. and Frederick, Peter D. and Lambrew, Christopher T. and Ornato, Joseph P. and Barron, Hilton V. and Kiefe, Catarina I.},
  title     = {{Prevalence, Clinical Characteristics, and Mortality Among Patients With Myocardial Infarction Presenting Without Chest Pain}},
  journal   = {JAMA},
  year      = {2000},
  volume    = {283},
  number    = {24},
  pages     = {3223--3229},
  doi       = {10.1001/jama.283.24.3223}
}

@article{Schulte2023,
  author    = {Schulte, K. J. and Mayrovitz, Harvey N.},
  title     = {{Myocardial Infarction Signs and Symptoms: Females vs. Males}},
  journal   = {Cureus},
  year      = {2023},
  volume    = {15},
  number    = {4},
  pages     = {e37522},
  doi       = {10.7759/cureus.37522}
}

@article{Joseph2021,
  author    = {Joseph, Neethu Maria and Ramamoorthy, Lakshmi and Satheesh, Santhosh},
  title     = {{Atypical Manifestations of Women Presenting with Myocardial Infarction at Tertiary Health Care Center: An Analytical Study}},
  journal   = {Journal of Mid-life Health},
  year      = {2021},
  volume    = {12},
  number    = {3},
  pages     = {219--224},
  doi       = {10.4103/jmh.JMH_20_20}
}

@article{Cardeillac2022,
  title={{Symptoms of Infarction in Women: Is There a Real Difference Compared to Men? A Systematic Review of the Literature with Meta-Analysis}},
  author={Cardeillac, Martin and Lefebvre, Fran{\c{c}}ois and Baicry, Florent and Le Borgne, Pierrick and Gil-Jardin{\'e}, C{\'e}dric and Cipolat, Lauriane and Peschanski, Nicolas and Abensur Vuillaume, Laure},
  journal={Journal of Clinical Medicine},
  volume={11},
  number={5},
  pages={1319},
  year={2022},
  publisher={MDPI}
}

@article{StuartShor2009,
  author    = {Stuart-Shor, Eileen M. and Wellenius, Gregory A. and DelloIacono, Donna M. and Mittleman, Murray A.},
  title     = {{Gender Differences in Presenting and Prodromal Stroke Symptoms}},
  journal   = {Stroke},
  year      = {2009},
  volume    = {40},
  number    = {4},
  pages     = {1121--1126},
  doi       = {10.1161/STROKEAHA.108.543371}
}

@article{Ali2021,
  author    = {Ali, Mariam and van Os, Hendrikus J. A. and van der Weerd, Nelleke and Schoones, Jan W. and Heymans, Martijn W. and Kruyt, Nyika D. and Visser, Marieke C. and Wermer, Marieke J. H.},
  title     = {{Sex Differences in Presentation of Stroke: A Systematic Review and Meta-Analysis}},
  journal   = {Stroke},
  year      = {2022},
  volume    = {53},
  number    = {2},
  pages     = {345--354},
  doi       = {10.1161/STROKEAHA.120.034040}
}

@article{mangione2022statin,
  title={{Statin Use for the Primary Prevention of Cardiovascular Disease in Adults: US Preventive Services Task Force Recommendation Statement}},
  author={Mangione, Carol M and Barry, Michael J and Nicholson, Wanda K and Cabana, Michael and Chelmow, David and Coker, Tumaini Rucker and Davis, Esa M and Donahue, Katrina E and Ja{\'e}n, Carlos Roberto and Kubik, Martha and others},
  journal={JAMA},
  volume={328},
  number={8},
  pages={746--753},
  year={2022},
  publisher={American Medical Association}
}

@article{Peters2018sex,
  title   = {{Sex Differences in High-Intensity Statin Use Following Myocardial Infarction in the United States}},
  author  = {Peters, Sanne A. E. and Colantonio, Lisandro D. and Zhao, Hong and Bittner, Vera and Dai, Yuling and Farkouh, Michael E. and Monda, Keri L. and Safford, Monika M. and Muntner, Paul and Woodward, Mark},
  journal = {Journal of the American College of Cardiology},
  volume  = {71},
  number  = {16},
  pages   = {1729--1737},
  year    = {2018},
  publisher = {American College of Cardiology Foundation},
  doi     = {10.1016/j.jacc.2018.02.032}
}

@article{greenwood2018patient,
  title   = {{Patient--Physician Gender Concordance and Increased Mortality among Female Heart Attack Patients}},
  author  = {Greenwood, Brad N. and Carnahan, Seth and Huang, Laura},
  journal = {Proceedings of the National Academy of Sciences},
  volume  = {115},
  number  = {34},
  pages   = {8569--8574},
  year    = {2018},
  publisher = {National Academy of Sciences},
  doi     = {10.1073/pnas.1800097115}
}

@article{currie2016provider,
  title   = {{Provider Practice Style and Patient Health Outcomes: The Case of Heart Attacks}},
  author  = {Currie, Janet and MacLeod, W. Bentley and Van Parys, Jessica},
  journal = {Journal of Health Economics},
  volume  = {47},
  pages   = {64--80},
  year    = {2016},
  publisher = {Elsevier},
  doi     = {10.1016/j.jhealeco.2016.01.013}
}

@article{finkelstein2016sources,
  title   = {Sources of Geographic Variation in Health Care: Evidence from Patient Migration},
  author  = {Finkelstein, Amy and Gentzkow, Matthew and Williams, Heidi},
  journal = {Quarterly Journal of Economics},
  volume  = {131},
  number  = {4},
  pages   = {1681--1726},
  year    = {2016},
  publisher = {Oxford University Press},
  doi     = {10.1093/qje/qjw023}
}

@article{Heidenreich2011,
  title   = {{Forecasting the Future of Cardiovascular Disease in the United States: A Policy Statement from the American Heart Association}},
  author  = {Heidenreich, Paul A. and Trogdon, Justin G. and Khavjou, Olga A. and Butler, Javed and Dracup, Kathleen and Ezekowitz, Michael D. and Finkelstein, Eric Andrew and Hong, Yuling and Johnston, S. Claiborne and Khera, Amit and Lloyd-Jones, Donald M. and Nelson, Sue A. and Nichol, Graham and Orenstein, Diane and Wilson, Peter W. F. and Woo, Y. Joseph},
  journal = {Circulation},
  volume  = {123},
  number  = {8},
  pages   = {933--944},
  year    = {2011},
  publisher = {American Heart Association},
  doi     = {10.1161/CIR.0b013e31820a55f5}
}

@article{Roth2018,
  title   = {{Global, Regional, and National Age-Sex-Specific Mortality for 282 Causes of Death in 195 Countries and Territories, 1980--2017: A Systematic Analysis for the Global Burden of Disease Study 2017}},
  author  = {Roth, Gregory A. and Abate, Degu and Abate, Kalkidan Hassen and Abay, Solomon M. and Abbafati, Cristiana and Abbasi, Nooshin and Abbastabar, Hedayat and Abd-Allah, Foad and Abdela, Jemal and Abdelalim, Ahmed and others},
  journal = {The Lancet},
  volume  = {392},
  number  = {10159},
  pages   = {1736--1788},
  year    = {2018},
  publisher = {Elsevier},
  doi     = {10.1016/S0140-6736(18)32203-7}
}

@article{Vaccarino2011,
  title   = {{Ischaemic Heart Disease in Women: Are There Sex Differences in Pathophysiology and Risk Factors? Position Paper from the Working Group on Coronary Pathophysiology and Microcirculation of the European Society of Cardiology}},
  author  = {Vaccarino, Viola and Badimon, Lina and Corti, Roberto and de Wit, Cor and Dorobantu, Maria and Hall, Alistair and Koller, Akos and Marzilli, Mario and Pries, Axel and Bugiardini, Raffaele},
  journal = {Cardiovascular Research},
  volume  = {90},
  number  = {1},
  pages   = {9--17},
  year    = {2011},
  publisher = {Oxford University Press},
  doi     = {10.1093/cvr/cvq394}
}

@article{croson2009gender,
  title   = {{Gender Differences in Preferences}},
  author  = {Croson, Rachel and Gneezy, Uri},
  journal = {Journal of Economic Literature},
  volume  = {47},
  number  = {2},
  pages   = {448--474},
  year    = {2009},
  publisher = {American Economic Association},
  doi     = {10.1257/jel.47.2.448}
}

@article{hanaoka2018risk,
  title   = {{Do Risk Preferences Change? Evidence from the Great East Japan Earthquake}},
  author  = {Hanaoka, Chie and Shigeoka, Hitoshi and Watanabe, Yasutora},
  journal = {American Economic Journal: Applied Economics},
  volume  = {10},
  number  = {2},
  pages   = {298--330},
  year    = {2018},
  publisher = {American Economic Association},
  doi     = {10.1257/app.20170048}
}

@article{powell1997gender,
  title   = {{Gender Differences in Risk Behaviour in Financial Decision-Making: An Experimental Analysis}},
  author  = {Powell, Melanie and Ansic, David},
  journal = {Journal of Economic Psychology},
  volume  = {18},
  number  = {6},
  pages   = {605--628},
  year    = {1997},
  publisher = {Elsevier},
  doi     = {10.1016/S0167-4870(97)00026-3}
}

@article{byrne2023esc,
  title   = {{2023 ESC Guidelines for the Management of Acute Coronary Syndromes}},
  author  = {Byrne, Robert A. and others},
  journal = {European Heart Journal},
  volume  = {44},
  number  = {38},
  pages   = {3720--3826},
  year    = {2023},
  doi     = {10.1093/eurheartj/ehad191}
}

@article{rao2025acc,
  title   = {{2025 ACC/AHA/ACEP/NAEMSP/SCAI Guideline for the Management of Patients with Acute Coronary Syndromes}},
  author  = {Rao, Sunil V. and others},
  journal = {Circulation},
  volume  = {151},
  number  = {13},
  pages   = {e771--e862},
  year    = {2025},
  doi     = {10.1161/CIR.0000000000001309}
}

@article{mach2020esc,
  title   = {{2019 ESC/EAS Guidelines for the Management of Dyslipidaemias: Lipid Modification to Reduce Cardiovascular Risk}},
  author  = {Mach, Fran{\c c}ois and others},
  journal = {European Heart Journal},
  volume  = {41},
  number  = {1},
  pages   = {111--188},
  year    = {2020},
  doi     = {10.1093/eurheartj/ehz455}
}

@article{grundy2019aha,
  title   = {{2018 AHA/ACC Multisociety Guideline on the Management of Blood Cholesterol}},
  author  = {Grundy, Scott M. and others},
  journal = {Circulation},
  volume  = {139},
  number  = {25},
  pages   = {e1082--e1143},
  year    = {2019},
  doi     = {10.1161/CIR.0000000000000625}
}

@article{kleindorfer2021guideline,
  title   = {{2021 Guideline for the Prevention of Stroke in Patients with Stroke and Transient Ischemic Attack}},
  author  = {Kleindorfer, Dawn O. and others},
  journal = {Stroke},
  volume  = {52},
  number  = {7},
  pages   = {e364--e467},
  year    = {2021},
  doi     = {10.1161/STR.0000000000000375}
}

@article{dawson2022european,
  title   = {{European Stroke Organisation (ESO) Guideline on Pharmacological Interventions for Long-Term Secondary Prevention after Ischaemic Stroke or Transient Ischaemic Attack}},
  author  = {Dawson, Jesse and others},
  journal = {European Stroke Journal},
  volume  = {7},
  number  = {3},
  pages   = {I--XLI},
  year    = {2022},
  doi     = {10.1177/23969873221100032}
}

@article{yndigegn2024beta,
  title   = {{Beta-Blockers after Myocardial Infarction and Preserved Ejection Fraction}},
  author  = {Yndigegn, Troels and others},
  journal = {New England Journal of Medicine},
  volume  = {390},
  number  = {15},
  pages   = {1372--1381},
  year    = {2024},
  doi     = {10.1056/NEJMoa2401479}
}

@article{ibanez2025beta,
  title   = {{Beta-Blockers after Myocardial Infarction without Reduced Ejection Fraction}},
  author  = {Ibanez, Borja and others},
  journal = {New England Journal of Medicine},
  volume  = {393},
  number  = {19},
  pages   = {1889--1900},
  year    = {2025},
  doi     = {10.1056/NEJMoa2504735}
}

@article{munkhaugen2025beta,
  title   = {{Beta-Blockers after Myocardial Infarction in Patients without Heart Failure}},
  author  = {Munkhaugen, John and others},
  journal = {New England Journal of Medicine},
  volume  = {393},
  number  = {19},
  pages   = {1901--1911},
  year    = {2025},
  doi     = {10.1056/NEJMoa2505985}
}

@article{kristensen2026beta,
  title   = {{Beta-Blockers after Myocardial Infarction with Normal Ejection Fraction}},
  author  = {Kristensen, Anna Meta Dyrvig and others},
  journal = {New England Journal of Medicine},
  volume  = {394},
  number  = {6},
  pages   = {540--550},
  year    = {2026},
  doi     = {10.1056/NEJMoa2512686}
}

@article{rossello2025beta,
  title   = {{Beta Blockers after Myocardial Infarction with Mildly Reduced Ejection Fraction: An Individual Patient Data Meta-Analysis of Randomised Controlled Trials}},
  author  = {Rossello, Xavier and others},
  journal = {The Lancet},
  volume  = {406},
  number  = {10508},
  pages   = {1128--1137},
  year    = {2025},
  doi     = {10.1016/S0140-6736(25)01592-2}
}

\newpage
\appendix

\part*{Appendix}

\section{Trend plots} \label{sec:trendplots}

Because every individual in the sample is eventually treated, the control series in these plots is constructed from future-treated cohorts: individuals who remain untreated at the time of a given cohort's shock but are treated within the subsequent five years. Each control individual is assigned a ``placebo'' event date coinciding with the focal cohort's actual treatment year, and outcomes are extracted over a five-year window around this placebo date. The construction closely resembles that of \cite{nielsen}, who similarly use future-treated cohorts as comparison groups; whereas they require control individuals to be treated exactly five years after the focal cohort, we adopt a more inclusive criterion that incorporates individuals treated at any point during the subsequent five-year window, which increases statistical power while preserving temporal comparability.

\begin{figure}[H]
    \centering

    \begin{minipage}{0.48\textwidth}
        \centering
        \includegraphics[width=\textwidth]{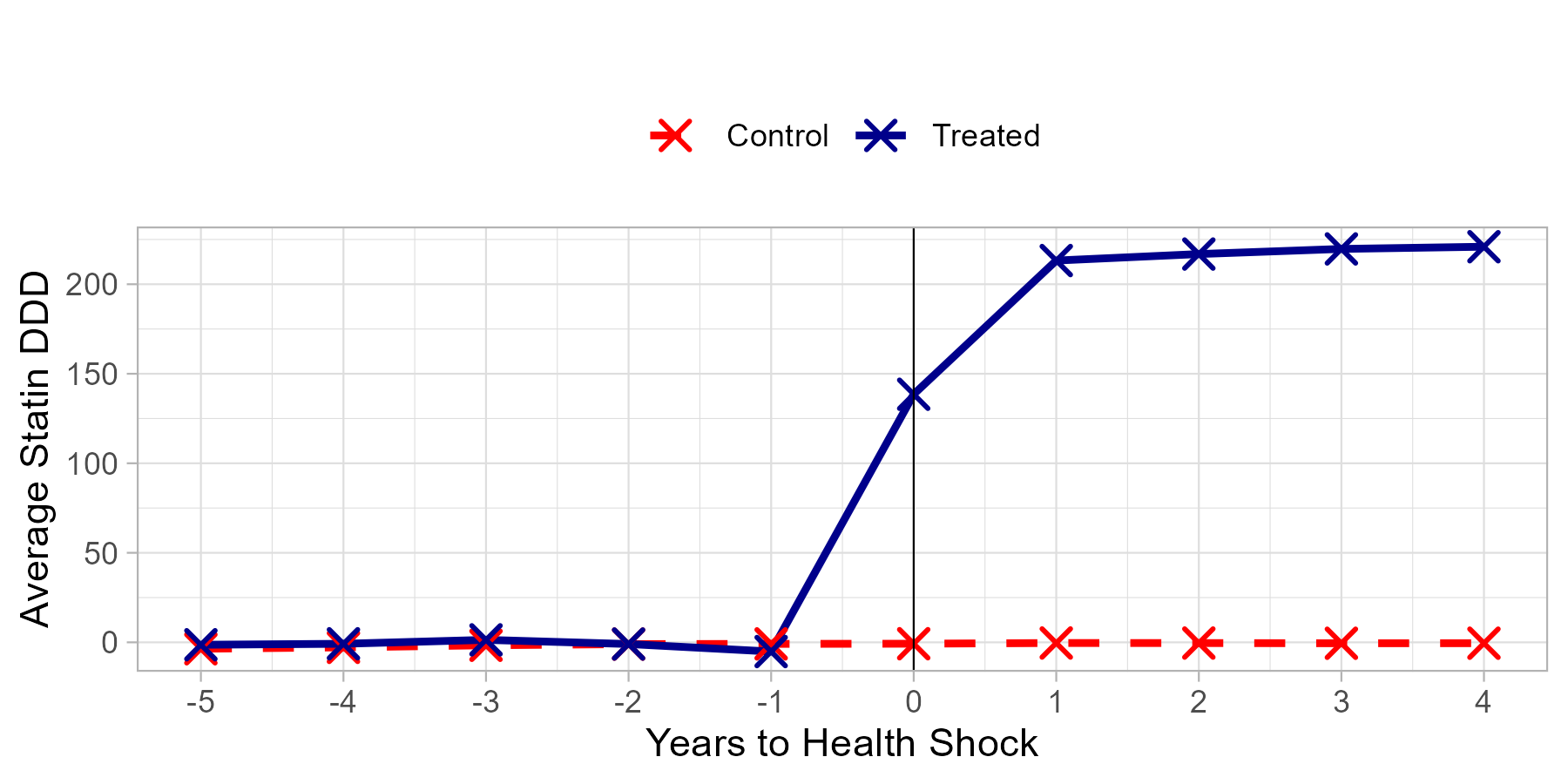}
        \captionof{figure}{Statin DDD, women.}
    \end{minipage}%
    \hfill
    \begin{minipage}{0.48\textwidth}
        \centering
        \includegraphics[width=\textwidth]{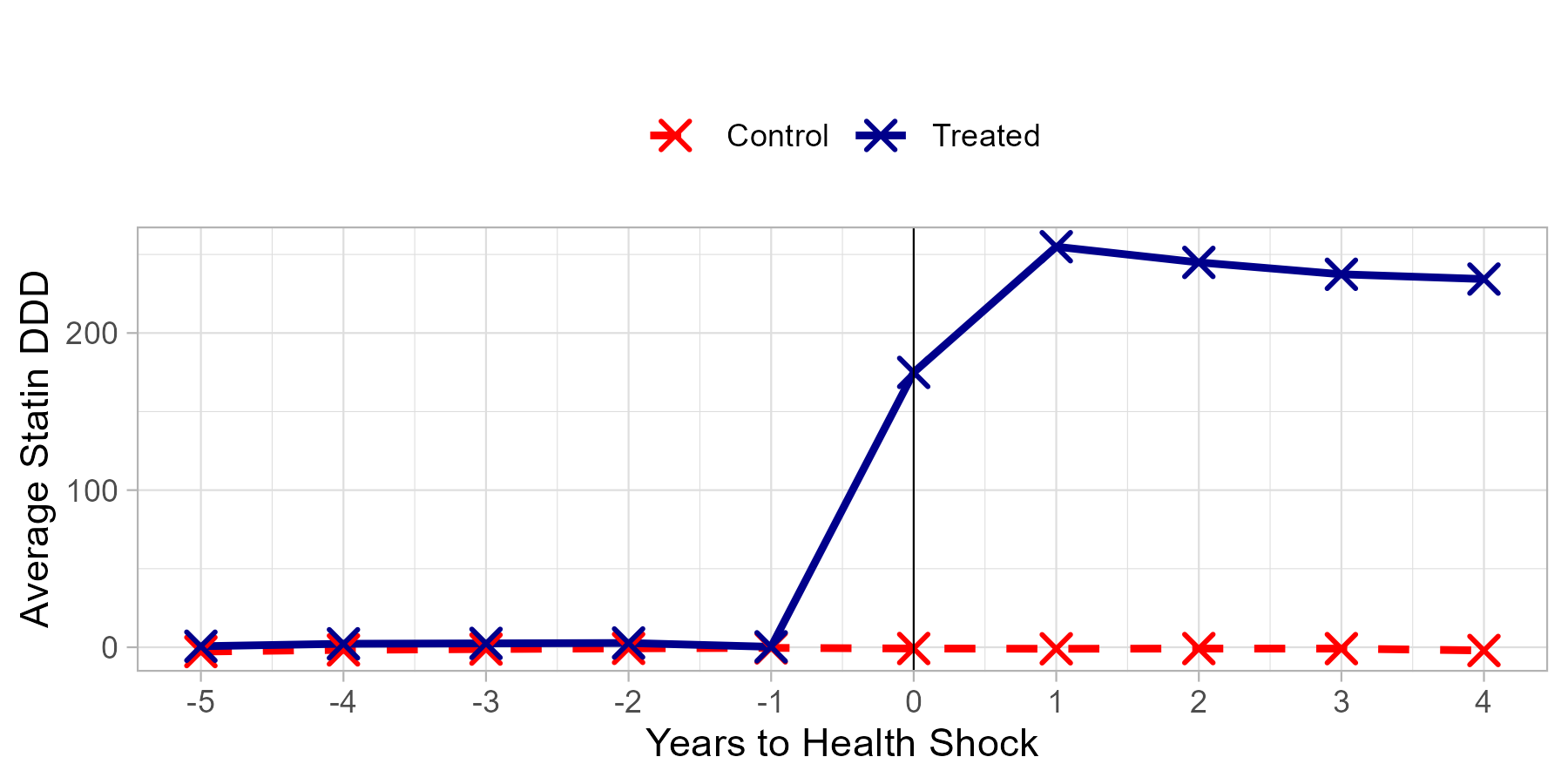}
        \captionof{figure}{Statin DDD, men.}
    \end{minipage}

    \vspace{0.5cm}

    \begin{minipage}{0.48\textwidth}
        \centering
        \includegraphics[width=\textwidth]{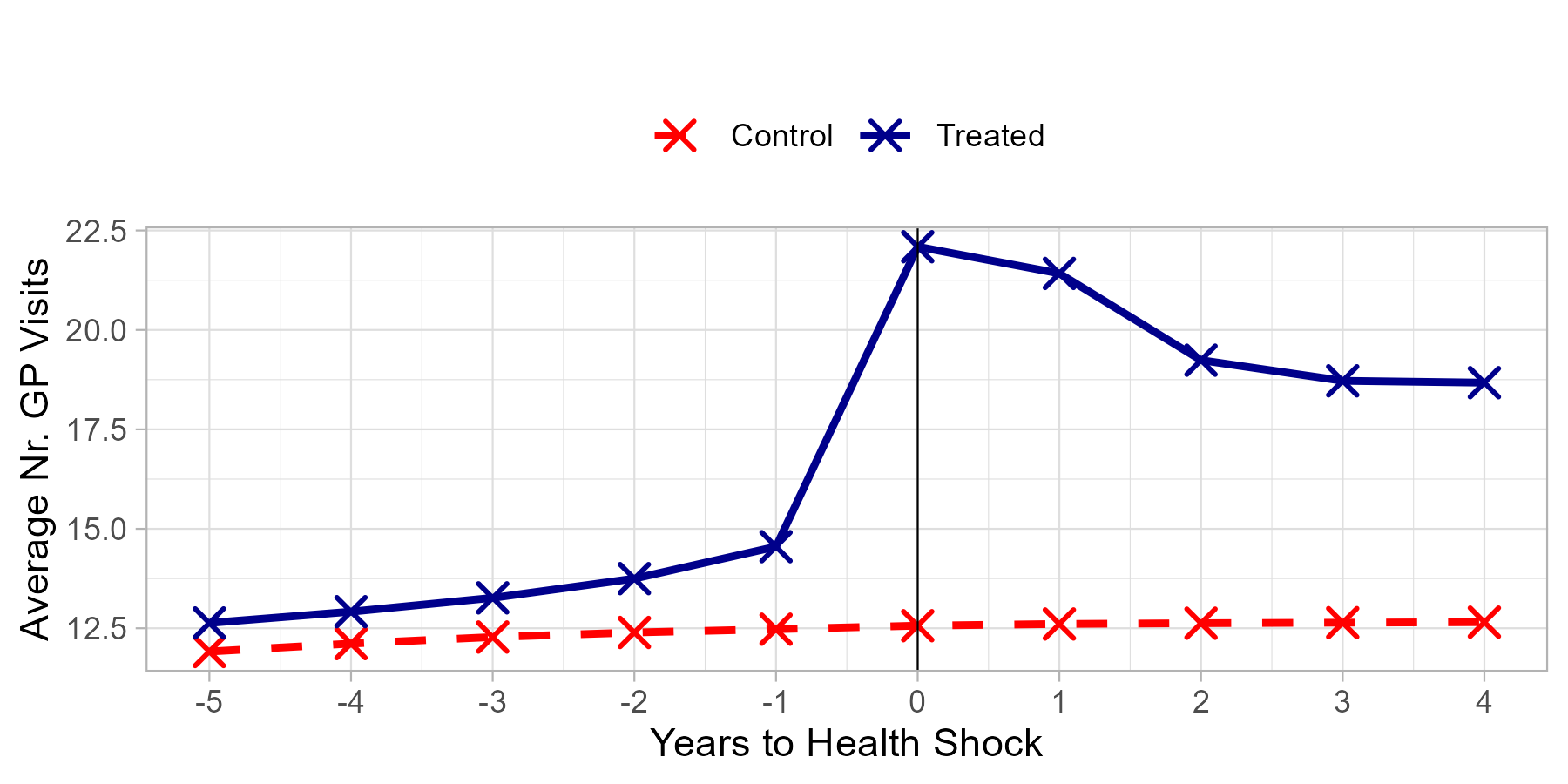}
        \captionof{figure}{GP visits, women.}
    \end{minipage}%
    \hfill
    \begin{minipage}{0.48\textwidth}
        \centering
        \includegraphics[width=\textwidth]{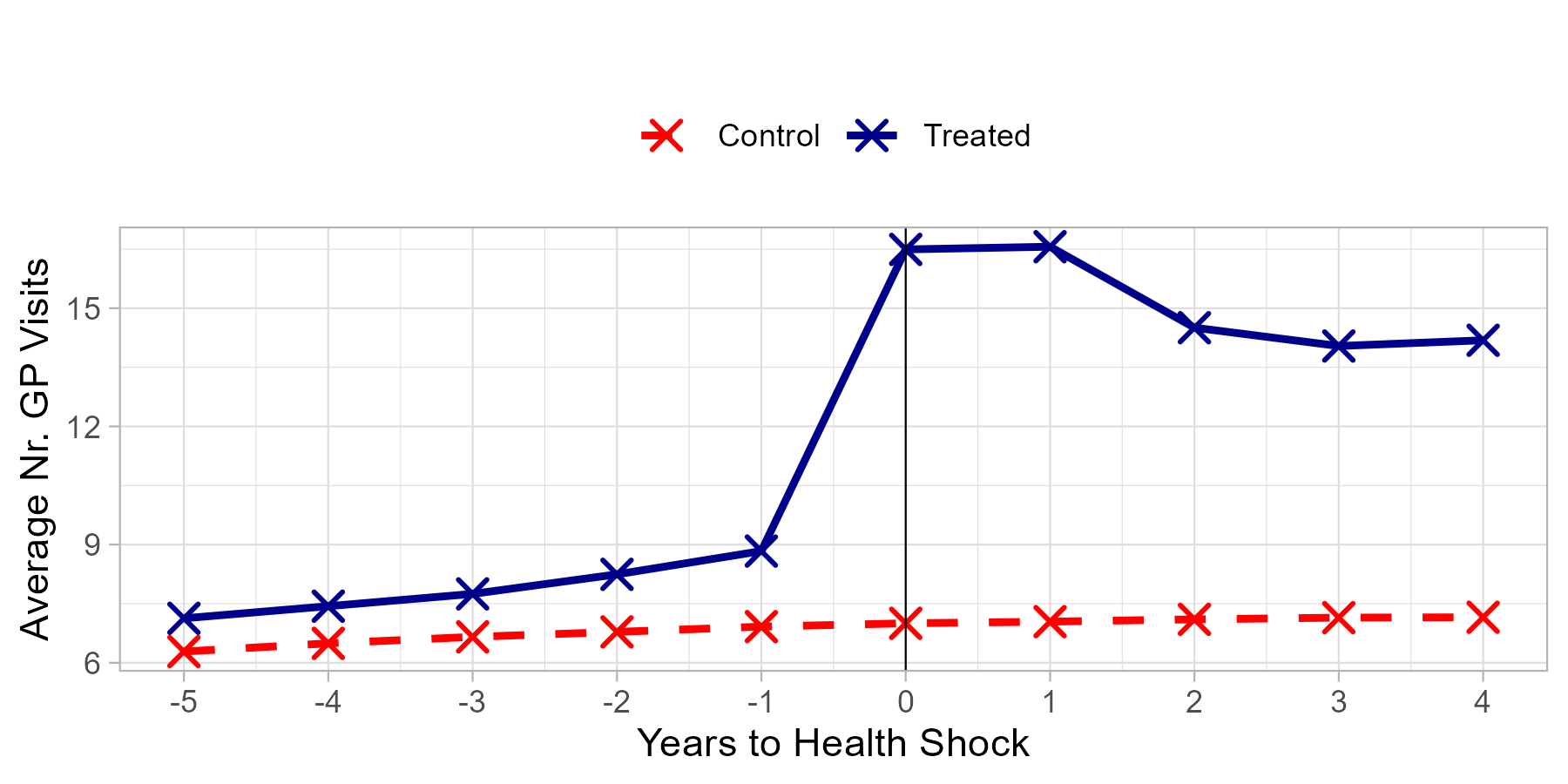}
        \captionof{figure}{GP visits, men.}
    \end{minipage}

    \vspace{0.5cm}

    \begin{minipage}{0.48\textwidth}
        \centering
        \includegraphics[width=\textwidth]{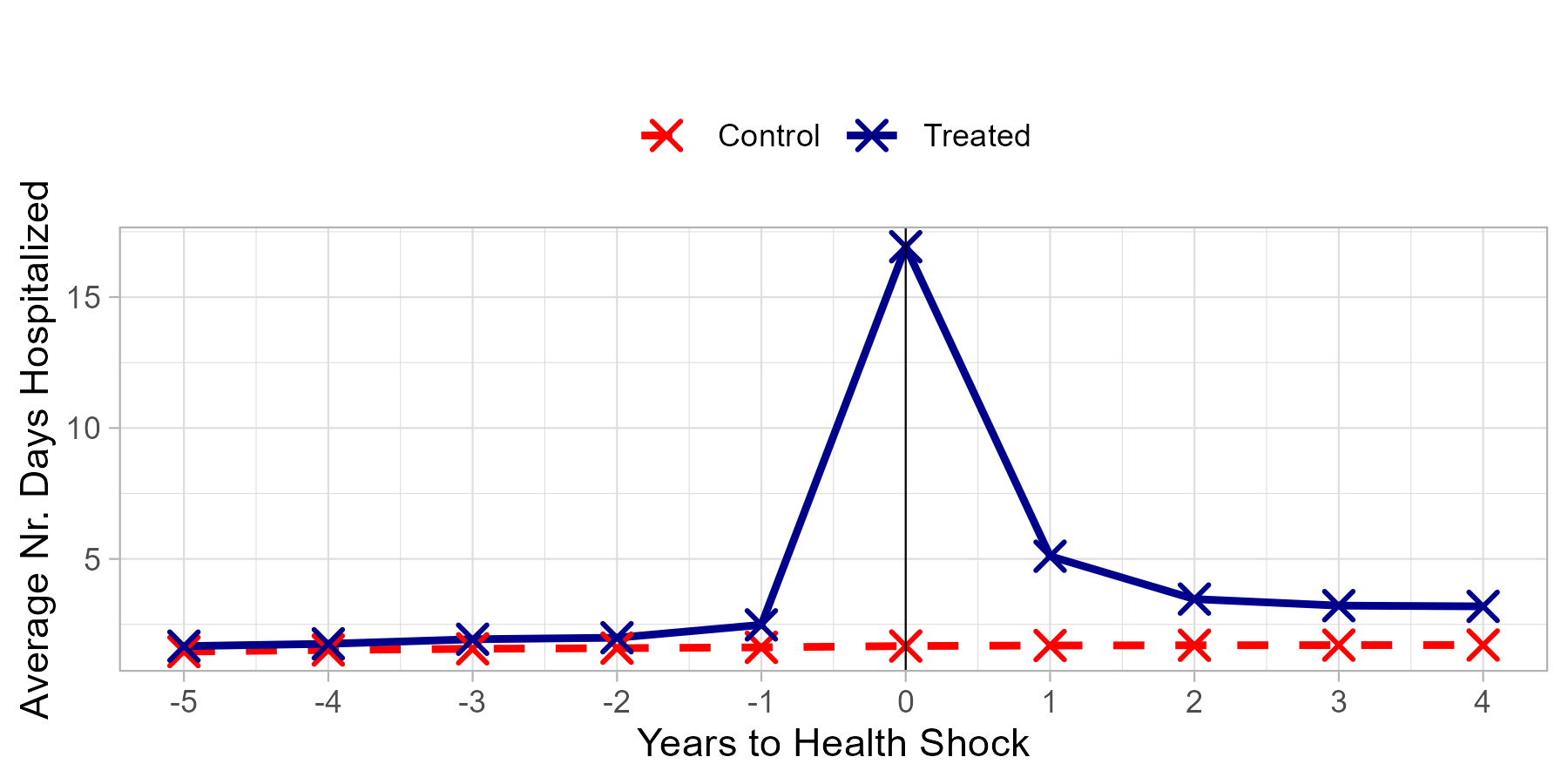}
        \captionof{figure}{Days hospitalised, women.}
    \end{minipage}%
    \hfill
    \begin{minipage}{0.48\textwidth}
        \centering
        \includegraphics[width=\textwidth]{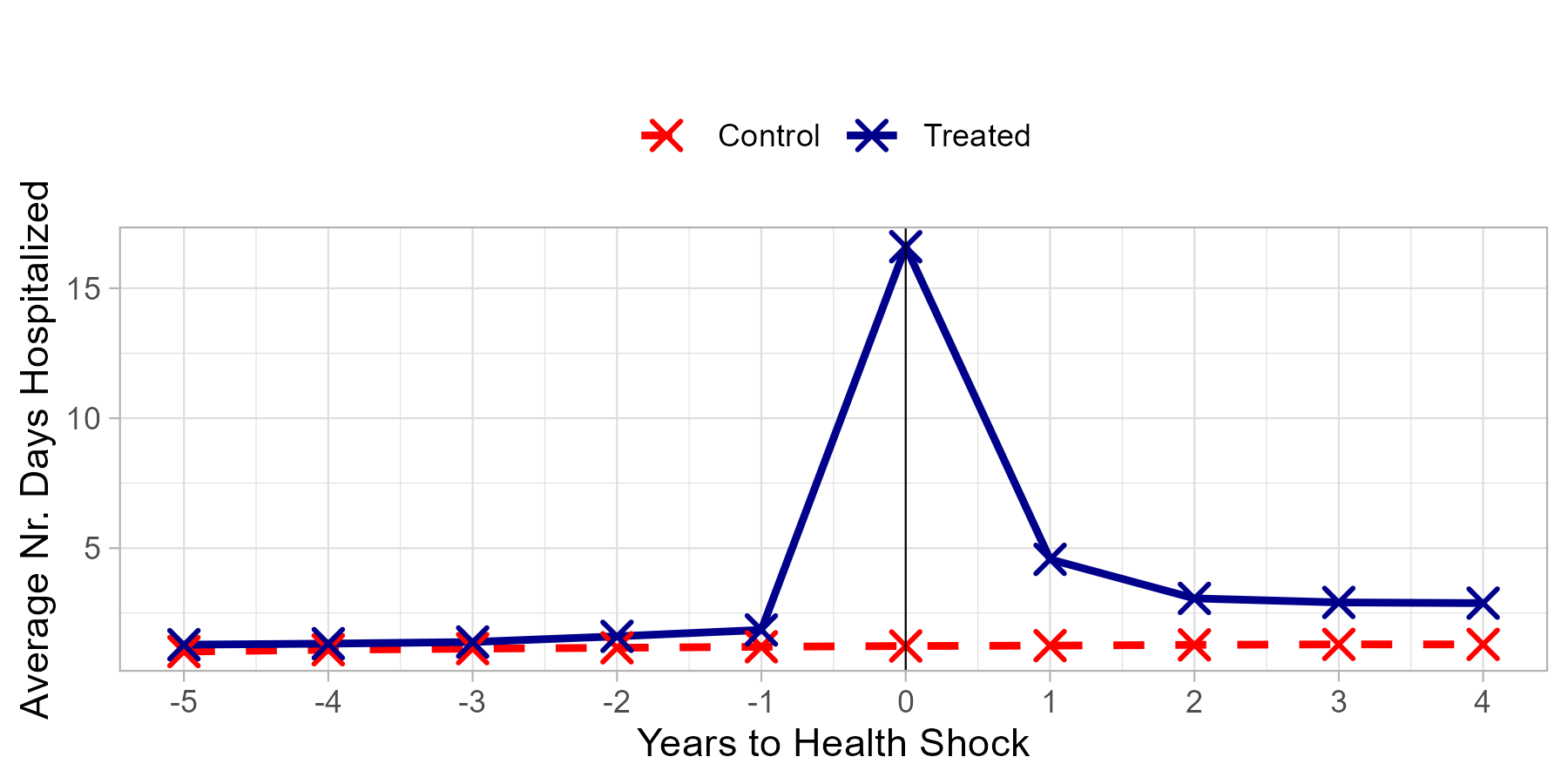}
        \captionof{figure}{Days hospitalised, men.}
    \end{minipage}

    \caption{Parallel trends for treated and placebo control groups, by gender, with year fixed effects partialled out. Control individuals are not yet treated at the time of the focal cohort's shock but are treated within the subsequent five-year window; each is assigned a placebo event date coinciding with the focal cohort's treatment year.}
\end{figure}

\clearpage

\section{Tables corresponding to main results} \label{sec:tables_main}

\begin{table}[ht]
\centering
\caption{Statin DDD: dynamic ATT.}
\label{tab:Statinmain}
\begin{tabular}{m{2em}cccccc}
  \hline\hline
    \multicolumn{1}{p{2cm}}{\textbf{Event Time}} & \textbf{Avg ATT} & \textbf{CI-Lower}  & \multicolumn{1}{p{5em}}{\textbf{CI-Upper}}  & \multicolumn{1}{p{4em}}{\textbf{No.\ unique treated}} & \textbf{$p$-value} \\
  \hline
  \multicolumn{2}{l}{\textbf{Panel A: Women}} & & & & & \\
  \hline
$-4$ & $-0.37$ & $-3.28$ & $2.53$  & 12{,}927 & 0.75 \\
$-3$ & $1.96$  & $-1.72$ & $5.65$  & 13{,}616 & 0.16 \\
$-2$ & $-1.46$ & $-4.13$ & $1.20$  & 14{,}313 & 0.13 \\
$-1$ & $1.59$  & $-1.34$ & $4.53$  & 14{,}342 & 0.15 \\
$0$  & $151.76$ & $145.55$ & $157.96$  & 15{,}140 & $<0.01$ \\
$1$  & $234.14$ & $223.37$ & $244.92$  & 14{,}363 & $<0.01$ \\
$2$  & $240.36$ & $226.21$ & $254.51$  & 13{,}046 & $<0.01$ \\
$3$  & $243.78$ & $228.59$ & $258.97$  & 11{,}934 & $<0.01$ \\
$4$  & $242.62$ & $225.82$ & $259.41$  & 10{,}918 & $<0.01$ \\
  \hline
  \multicolumn{2}{l}{\textbf{Panel B: Men}} & & & & & \\
  \hline
$-4$ & $0.20$  & $-1.32$ & $1.71$  & 24{,}934 & 0.74 \\
$-3$ & $-1.28$ & $-2.71$ & $0.16$  & 26{,}468 & 0.02 \\
$-2$ & $0.24$  & $-1.33$ & $1.80$  & 28{,}110 & 0.70 \\
$-1$ & $2.94$  & $1.26$  & $4.62$  & 28{,}307 & $<0.01$ \\
$0$  & $185.29$ & $181.01$ & $189.56$ & 30{,}103 & $<0.01$ \\
$1$  & $271.23$ & $264.23$ & $278.23$  & 28{,}547 & $<0.01$ \\
$2$  & $261.61$ & $253.39$ & $269.84$  & 25{,}945 & $<0.01$ \\
$3$  & $254.04$ & $244.75$ & $263.34$  & 23{,}790 & $<0.01$ \\
$4$  & $250.89$ & $240.41$ & $261.38$  & 21{,}944 & $<0.01$ \\ \hline\hline
\end{tabular}
\\[0.5em]
\begin{minipage}{0.95\textwidth}
\footnotesize \textit{Notes:} Dynamic ATT for annual statin DDD, by event time, separately by gender. Estimator: \cite{callaway2021difference} doubly robust with a five-year non-shifting comparison window, conditioning on age, log income, region, household type and education. Confidence intervals at the 95\% level via the multiplier bootstrap with 1{,}000 replications and individual-level clustering.
\end{minipage}
\end{table}

\begin{table}[ht]
\centering
\caption{GP visits: dynamic ATT.}
\label{tab:GPmain}
\begin{tabular}{m{2em}cccccc}
  \hline\hline
    \multicolumn{1}{p{2cm}}{\textbf{Event Time}} & \textbf{Avg ATT} & \textbf{CI-Lower}  & \multicolumn{1}{p{5em}}{\textbf{CI-Upper}}  & \multicolumn{1}{p{4em}}{\textbf{No.\ unique treated}} & \textbf{$p$-value} \\
  \hline
  \multicolumn{2}{l}{\textbf{Panel A: Women}} & & & & & \\
  \hline
$-4$ & $-0.18$ & $-0.44$ & $0.08$  & 12{,}927 & 0.08 \\
$-3$ & $-0.11$ & $-0.39$ & $0.17$  & 13{,}616 & 0.29 \\
$-2$ & $0.08$  & $-0.23$ & $0.39$  & 14{,}313 & 0.51 \\
$-1$ & $0.56$  & $0.28$  & $0.85$  & 14{,}342 & $<0.01$ \\
$0$  & $7.20$  & $6.80$  & $7.59$  & 15{,}140 & $<0.01$ \\
$1$  & $6.40$  & $5.95$  & $6.86$  & 14{,}363 & $<0.01$ \\
$2$  & $3.98$  & $3.51$  & $4.45$  & 13{,}046 & $<0.01$ \\
$3$  & $2.94$  & $2.44$  & $3.44$  & 11{,}934 & $<0.01$ \\
$4$  & $2.22$  & $1.73$  & $2.70$  & 10{,}918 & $<0.01$ \\
  \hline
  \multicolumn{2}{l}{\textbf{Panel B: Men}} & & & & & \\
\hline
$-4$ & $-0.06$ & $-0.22$ & $0.10$  & 24{,}934 & 0.31 \\
$-3$ & $-0.06$ & $-0.22$ & $0.10$  & 26{,}468 & 0.34 \\
$-2$ & $0.11$  & $-0.05$ & $0.27$  & 28{,}110 & 0.05 \\
$-1$ & $0.23$  & $0.06$  & $0.40$  & 28{,}307 & $<0.01$ \\
$0$  & $7.54$  & $7.32$  & $7.75$  & 30{,}103 & $<0.01$ \\
$1$  & $7.45$  & $7.19$  & $7.71$  & 28{,}547 & $<0.01$ \\
$2$  & $5.12$  & $4.85$  & $5.38$  & 25{,}945 & $<0.01$ \\
$3$  & $4.28$  & $4.01$  & $4.56$  & 23{,}790 & $<0.01$ \\
$4$  & $3.96$  & $3.66$  & $4.26$  & 21{,}944 & $<0.01$ \\ \hline\hline
\end{tabular}
\\[0.5em]
\begin{minipage}{0.95\textwidth}
\footnotesize \textit{Notes:} As in Table~\ref{tab:Statinmain}; outcome is the annual count of GP visits (all visits, not only those directly related to the cardiovascular event).
\end{minipage}
\end{table}

\begin{table}[ht]
\centering
\caption{Days hospitalised: dynamic ATT.}
\label{tab:hospmain}
\begin{tabular}{m{2em}cccccc}
  \hline\hline
    \multicolumn{1}{p{2cm}}{\textbf{Event Time}} & \textbf{Avg ATT} & \textbf{CI-Lower}  & \multicolumn{1}{p{5em}}{\textbf{CI-Upper}}  & \multicolumn{1}{p{4em}}{\textbf{No.\ unique treated}} & \textbf{$p$-value} \\
  \hline
  \multicolumn{2}{l}{\textbf{Panel A: Women}} & & & & & \\
  \hline
$-4$ & $-0.09$ & $-0.29$ & $0.11$  & 12{,}927 & 0.19 \\
$-3$ & $0.01$  & $-0.21$ & $0.22$  & 13{,}616 & 0.92 \\
$-2$ & $-0.11$ & $-0.32$ & $0.09$  & 14{,}313 & 0.11 \\
$-1$ & $0.45$  & $0.20$  & $0.69$  & 14{,}342 & $<0.01$ \\
$0$  & $14.51$ & $13.97$ & $15.05$ & 15{,}140 & $<0.01$ \\
$1$  & $2.53$  & $2.14$  & $2.91$  & 14{,}363 & $<0.01$ \\
$2$  & $0.82$  & $0.49$  & $1.14$  & 13{,}046 & $<0.01$ \\
$3$  & $0.46$  & $0.12$  & $0.80$  & 11{,}934 & $<0.01$ \\
$4$  & $0.06$  & $-0.28$ & $0.41$  & 10{,}918 & 0.61 \\
  \hline
  \multicolumn{2}{l}{\textbf{Panel B: Men}} & & & & & \\
\hline
$-4$ & $-0.08$ & $-0.22$ & $0.06$  & 24{,}934 & 0.10 \\
$-3$ & $-0.08$ & $-0.21$ & $0.05$  & 26{,}468 & 0.07 \\
$-2$ & $0.10$  & $-0.04$ & $0.24$  & 28{,}110 & 0.04 \\
$-1$ & $0.17$  & $0.01$  & $0.32$  & 28{,}307 & $<0.01$ \\
$0$  & $14.90$ & $14.53$ & $15.26$ & 30{,}103 & $<0.01$ \\
$1$  & $2.76$  & $2.52$  & $3.01$  & 28{,}547 & $<0.01$ \\
$2$  & $1.24$  & $1.02$  & $1.47$  & 25{,}945 & $<0.01$ \\
$3$  & $1.05$  & $0.82$  & $1.28$  & 23{,}790 & $<0.01$ \\
$4$  & $0.84$  & $0.62$  & $1.06$  & 21{,}944 & $<0.01$ \\
\hline\hline
\end{tabular}
\\[0.5em]
\begin{minipage}{0.95\textwidth}
\footnotesize \textit{Notes:} As in Table~\ref{tab:Statinmain}; outcome is the annual count of inpatient days.
\end{minipage}
\end{table}

\FloatBarrier
\clearpage

\section{Tables and figures for robustness checks}

\subsection{Matched sample}\label{sec:matched_sample}
The matched sample is constructed by nearest-neighbour matching on the propensity score from a logit of female on $X$, evaluated at $t=g-1$.

\begin{figure}[H]
    \centering
    \includegraphics[width=0.85\linewidth]{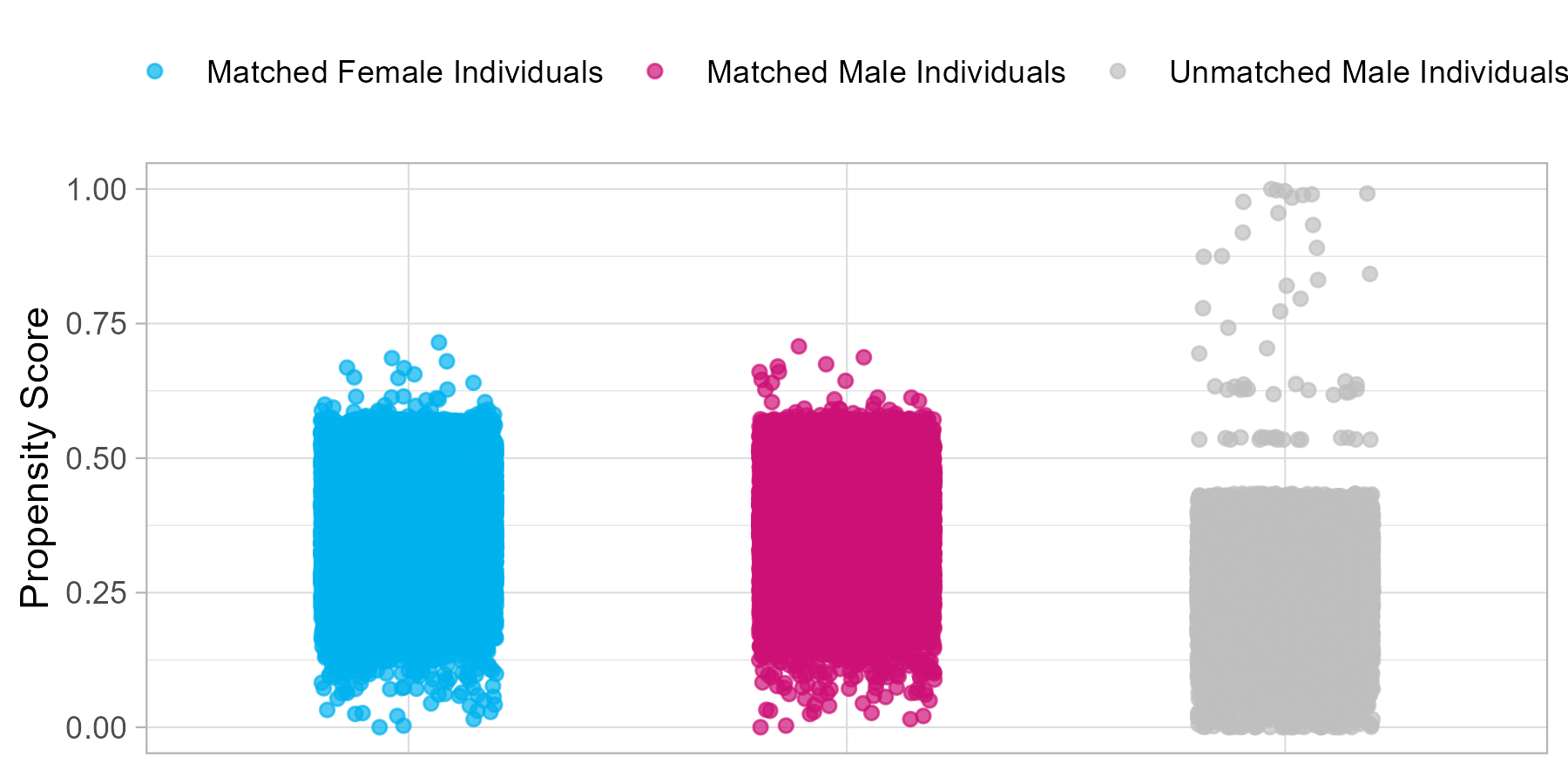}
    \caption{Propensity-score distribution before and after matching, illustrating overlap between men and women.}
    \begin{minipage}{0.9\textwidth}
    \footnotesize \textit{Notes:} Estimated propensity scores from the logit of female on age, region, education, household type and equivalised income, evaluated at $t=g-1$, shown before and after nearest-neighbour matching. Table~\ref{tab:match_balance} reports the corresponding balance statistics.
    \end{minipage}
\end{figure}

\begin{figure}[H]
    \centering
    \includegraphics[width=0.95\linewidth]{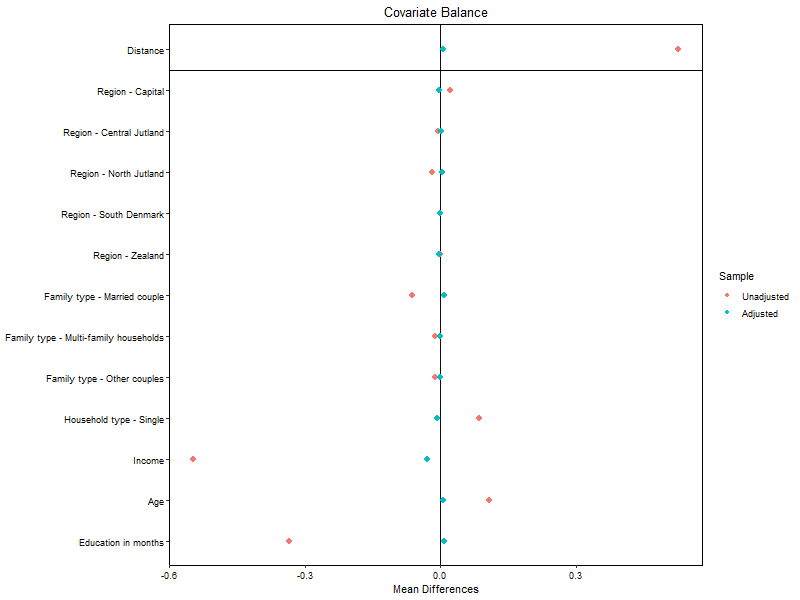}
    \caption{Standardised mean differences of baseline covariates before and after propensity-score matching, to assess covariate balance between men and women.}
    \begin{minipage}{0.9\textwidth}
    \footnotesize \textit{Notes:} Standardised mean differences for each covariate in the propensity-score model, before and after nearest-neighbour matching of women to men. All post-matching differences are no larger than 0.03 in absolute value (Table~\ref{tab:match_balance}).
    \end{minipage}
\end{figure}

\begin{table}[H]
\centering
\caption{Covariate balance in the matched sample.}
\label{tab:match_balance}
\footnotesize
\begin{tabular}{llllllll}
  \hline\hline
     & \multicolumn{1}{p{3em}}{\textbf{Means Men}} & \multicolumn{1}{p{3em}}{\textbf{Means Women}}  & \multicolumn{1}{p{3em}}{\textbf{Std Mean Diff}} & \multicolumn{1}{p{3em}}{\textbf{Var. Ratio}} & \multicolumn{1}{p{3em}}{\textbf{eCDF Mean}} & \multicolumn{1}{p{3em}}{\textbf{eCDF Max}} & \multicolumn{1}{p{4em}}{\textbf{Std. Pair Dist.}} \\
  \hline
  Distance & 0.37 & 0.37 & 0.01 & 1.02 & 0.00 & 0.01 & 0.01 \\
  Region Capital & 0.28 & 0.29 & -0.01 &  & 0.00 & 0.00 & 0.86 \\
  Region Central Jutland & 0.21 & 0.21 & 0.01 &  & 0.00 & 0.00 & 0.81 \\
  Region North Jutland & 0.13 & 0.12 & 0.01 &  & 0.00 & 0.00 & 0.64 \\
  Region South Denmark & 0.22 & 0.22 & -0.00 &  & 0.00 & 0.00 & 0.82 \\
  Region Zealand & 0.16 & 0.16 & -0.01 &  & 0.00 & 0.00 & 0.72 \\
    Married couple & 0.53 & 0.52 & 0.02 &  & 0.01 & 0.01 & 0.88 \\
   Multi-family households & 0.09 & 0.10 & -0.00 &  & 0.00 & 0.00 & 0.58 \\
   Other couples & 0.07 & 0.07 & 0.00 &  & 0.00 & 0.00 & 0.12 \\
   Single & 0.31 & 0.31 & -0.02 &  & 0.01 & 0.01 & 0.70 \\
  Income (DKK) & 176338 & 179615 & -0.03 & 0.98 & 0.01 & 0.04 & 0.59 \\
  Age & 59.94 & 59.90 & 0.01 & 1.01 & 0.00 & 0.01 & 1.11 \\
  Education in months & 138.11 & 137.75 & 0.01 & 0.97 & 0.04 & 0.14 & 0.70 \\
\hline\hline
\end{tabular}
\\[0.5em]
\begin{minipage}{0.95\textwidth}
\footnotesize \textit{Notes:} Post-matching balance for the covariates in the propensity-score model (logit of female on age, region, education, household type and equivalised income, evaluated at $t=g-1$), after nearest-neighbour matching of women to men. ``Distance'' is the estimated propensity score. Income is equivalised disposable household income in DKK; education is in months. Standardised mean differences are no larger than 0.03 in absolute value for every covariate; see also the love plot above.
\end{minipage}
\end{table}

\begin{figure}[H]
    \centering
    \begin{subfigure}{0.65\textwidth}
        \centering
        \includegraphics[width=\textwidth]{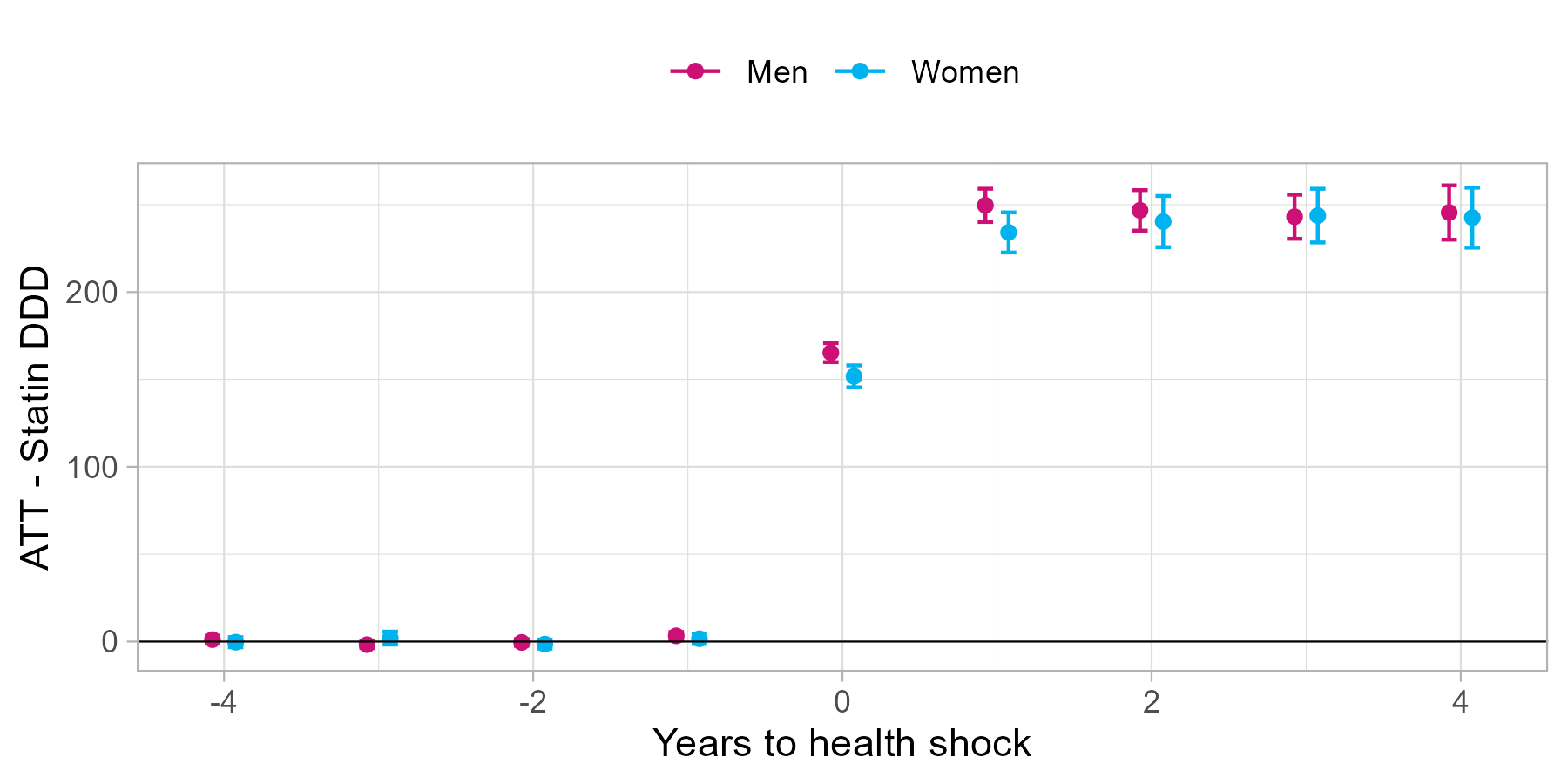}
        \caption{Statin DDD, matched sample.}
    \end{subfigure}

    \vspace{0.4cm}

    \begin{subfigure}{0.65\textwidth}
        \centering
        \includegraphics[width=\textwidth]{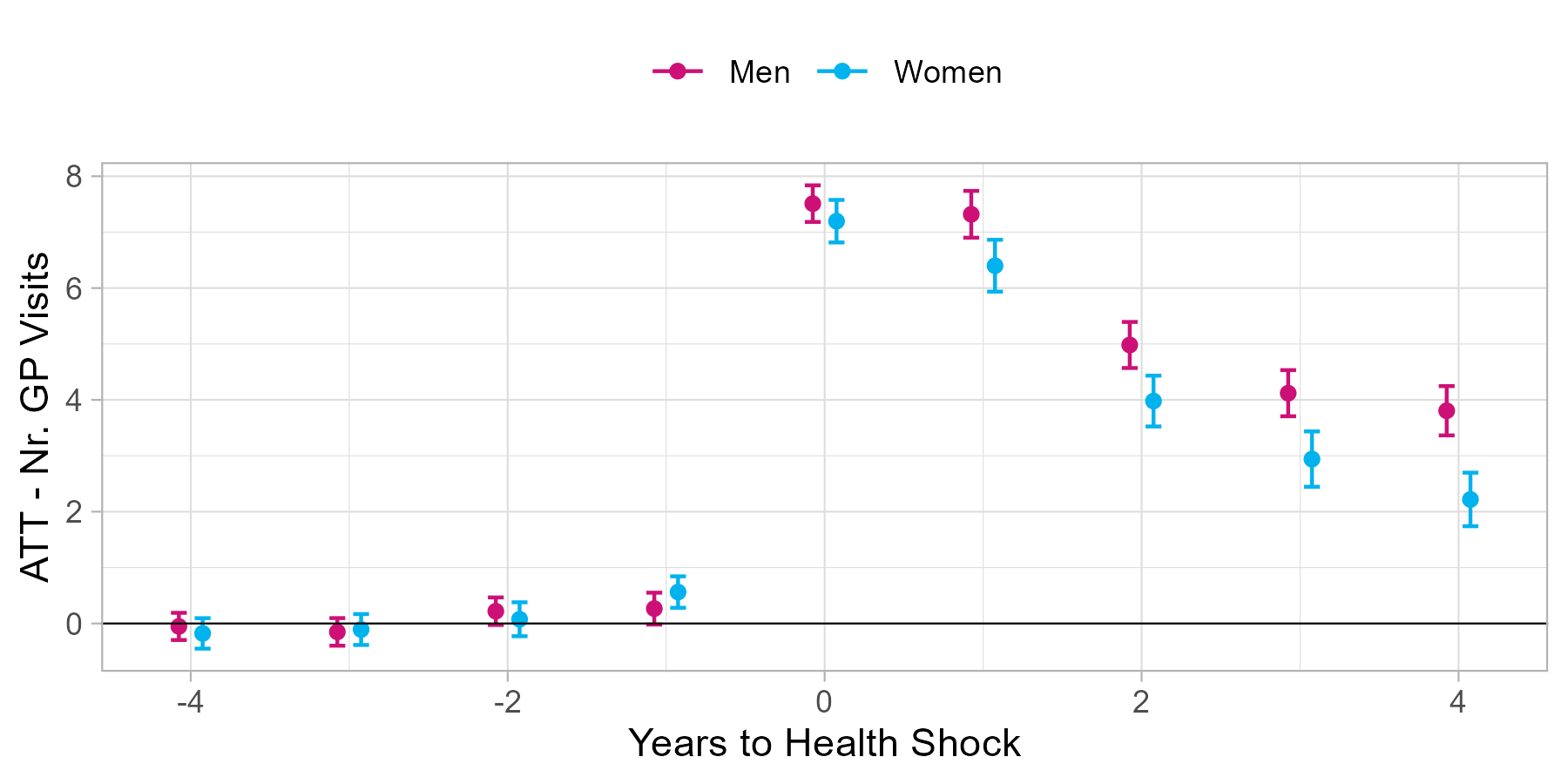}
        \caption{GP visits, matched sample.}
    \end{subfigure}

    \vspace{0.4cm}

    \begin{subfigure}{0.65\textwidth}
        \centering
        \includegraphics[width=\textwidth]{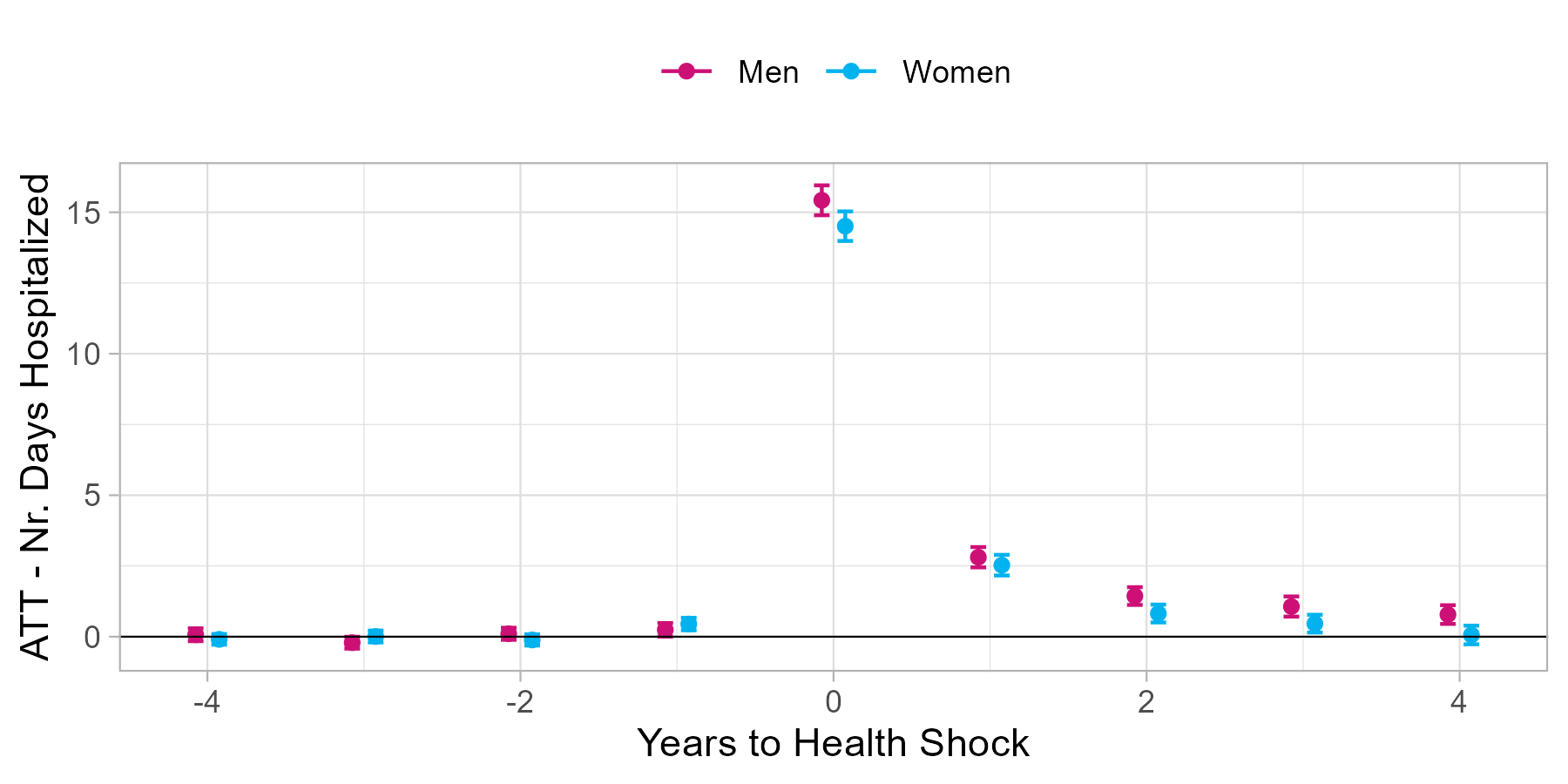}
        \caption{Days hospitalised, matched sample.}
    \end{subfigure}

    \caption{Matched-sample dynamic ATT estimates.}
    \label{fig:matched_results}
    \begin{minipage}{0.9\textwidth}
    \footnotesize \textit{Notes:} Estimator and inference as in Figure~\ref{fig:dynamic_att_full} (simultaneous 95\% multiplier-bootstrap confidence bands, individual-level clustering), on the propensity-score-matched sample of Table~\ref{tab:match_balance}.
    \end{minipage}
\end{figure}

\FloatBarrier
\clearpage
\subsection{Triple-difference}\label{sec:tdid}

\begin{figure}[htbp]
    \centering
    \begin{subfigure}{0.65\textwidth}
        \centering
        \includegraphics[width=\textwidth]{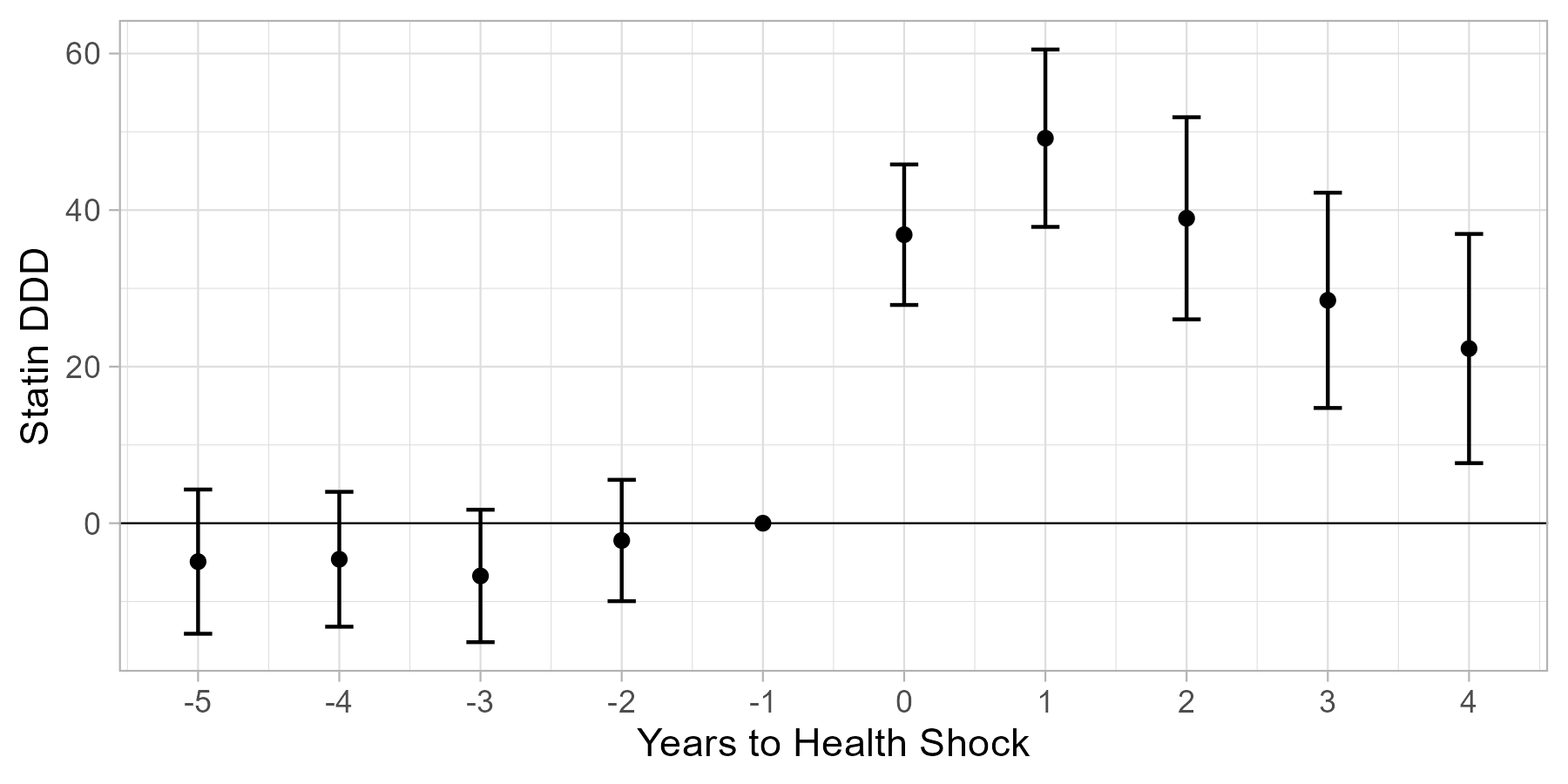}
        \caption{Statin DDD.}
    \end{subfigure}

    \vspace{0.4cm}

    \begin{subfigure}{0.65\textwidth}
        \centering
        \includegraphics[width=\textwidth]{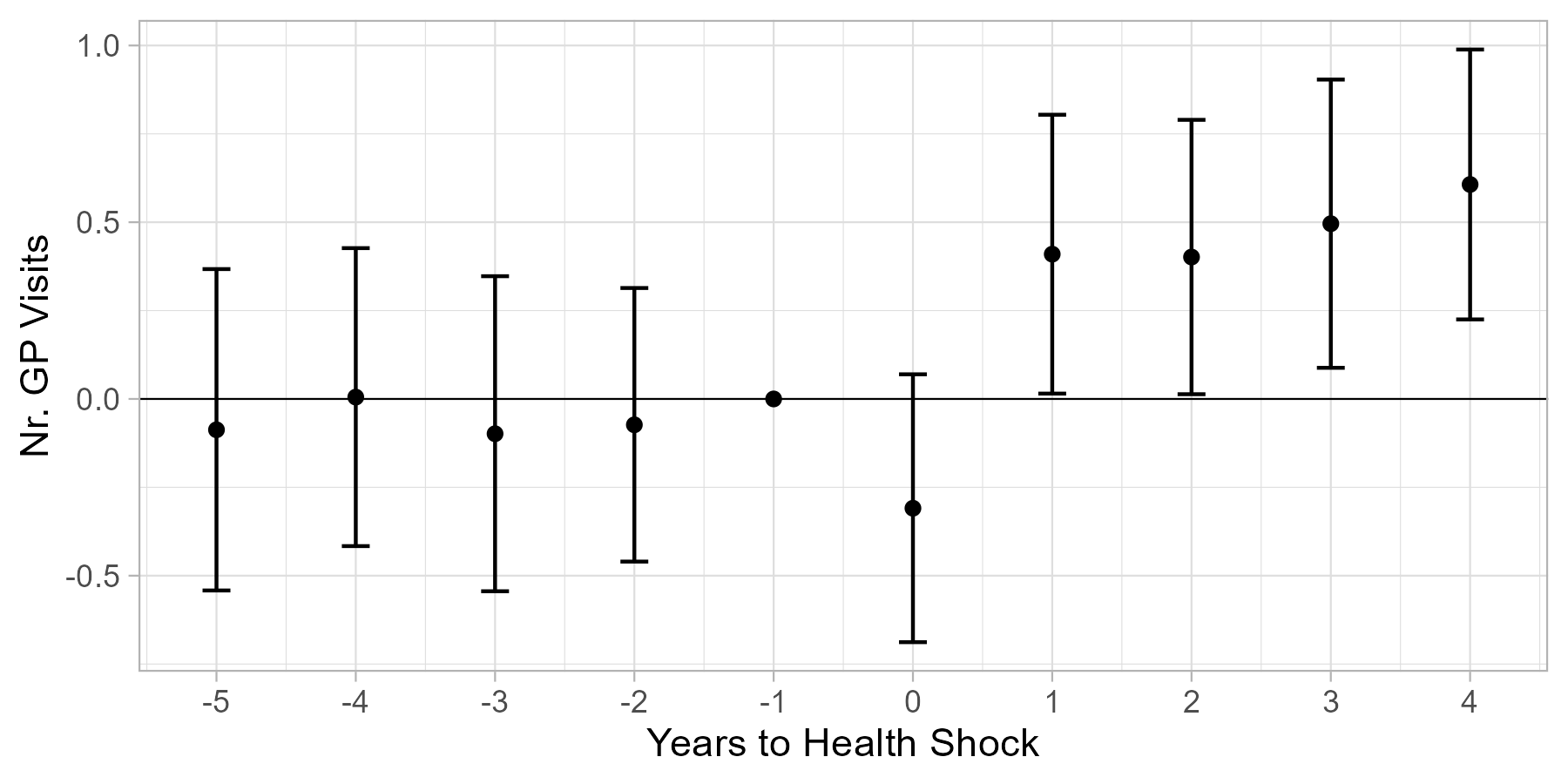}
        \caption{GP visits.}
    \end{subfigure}

    \vspace{0.4cm}

    \begin{subfigure}{0.65\textwidth}
        \centering
        \includegraphics[width=\textwidth]{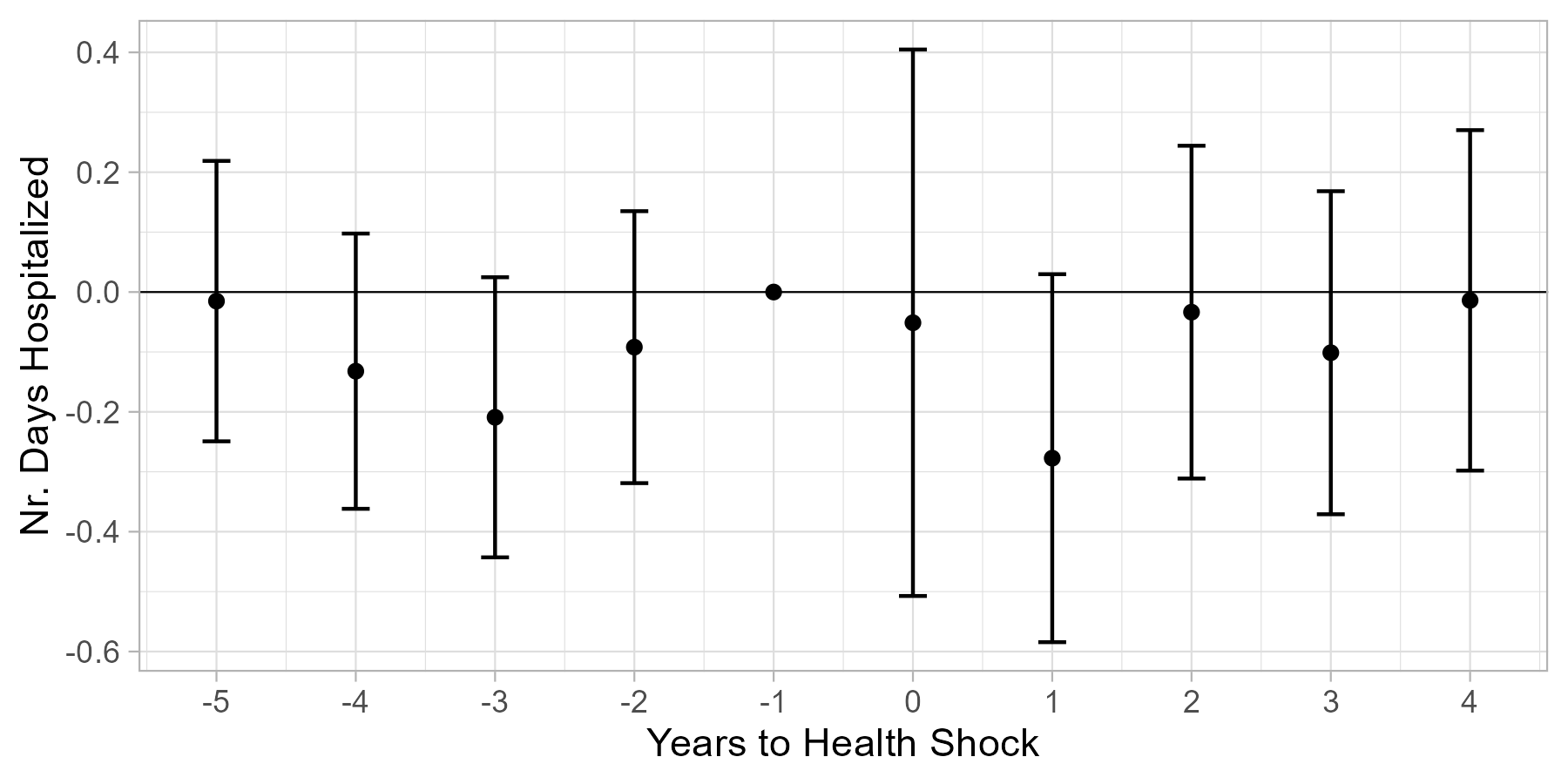}
        \caption{Days hospitalised.}
    \end{subfigure}
    \caption{Triple-difference estimates of the gender contrast.}
    \label{fig:triple_did}
    \begin{minipage}{0.9\textwidth}
    \footnotesize \textit{Notes:} $\widehat{\beta}^{\text{DDD}}_{\tau}$ from Equation~\eqref{eq:tripledid}, estimated using the \cite{nielsen} comparison-group construction (individuals treated within five years of the focal cohort). The plotted coefficient is the time-varying interaction female $\times$ event-time $\times$ treated. Shaded areas denote 95\% confidence intervals.
    \end{minipage}
\end{figure}

\subsection{Alternative specification without covariates} \label{app:unconditional}

\begin{figure}[H]
    \centering
    \begin{subfigure}{0.65\textwidth}
        \centering
        \includegraphics[width=\textwidth]{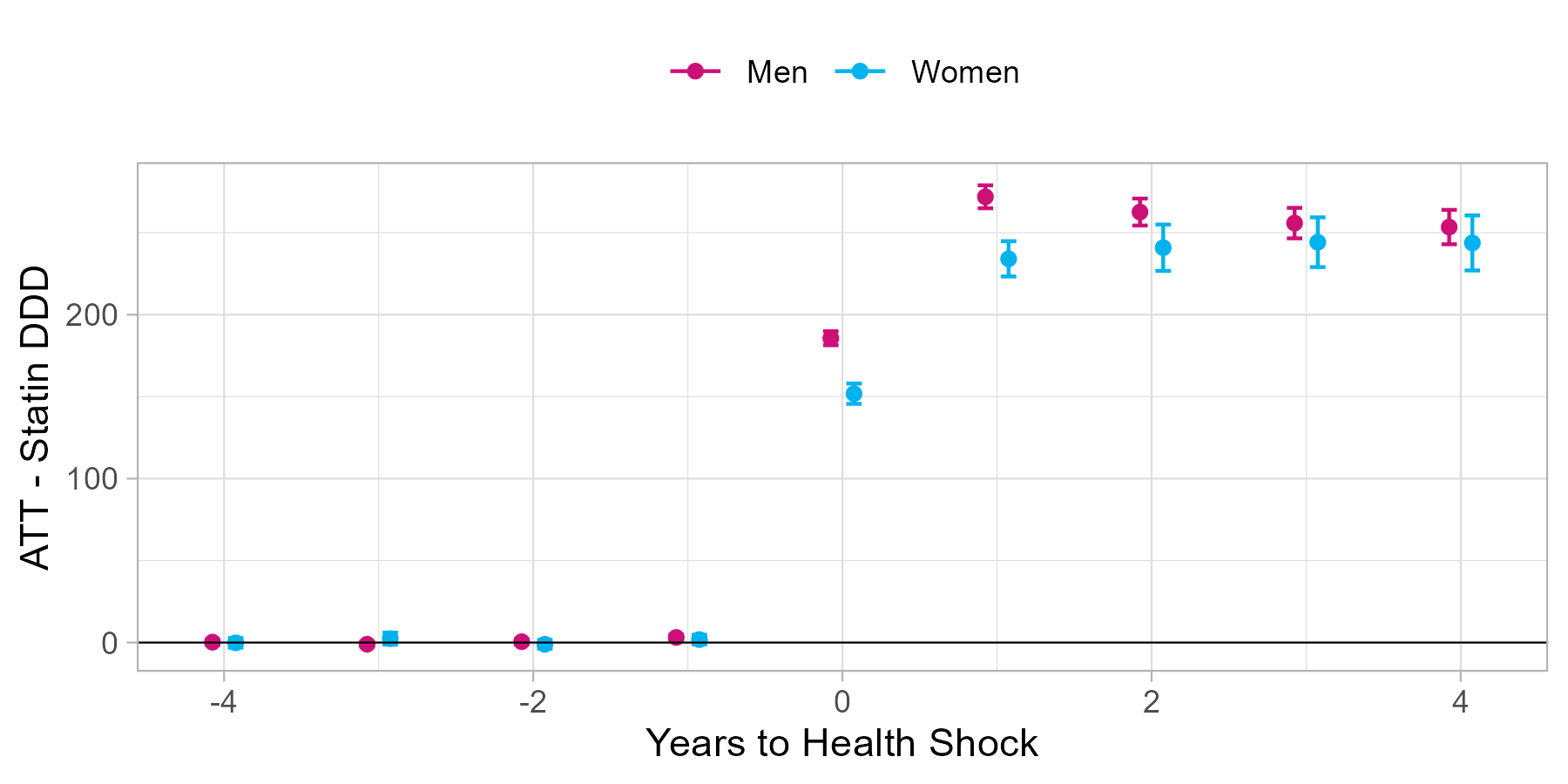}
        \caption{Statin DDD.}
    \end{subfigure}
    \vspace{0.4cm}
    \begin{subfigure}{0.65\textwidth}
        \centering
        \includegraphics[width=\textwidth]{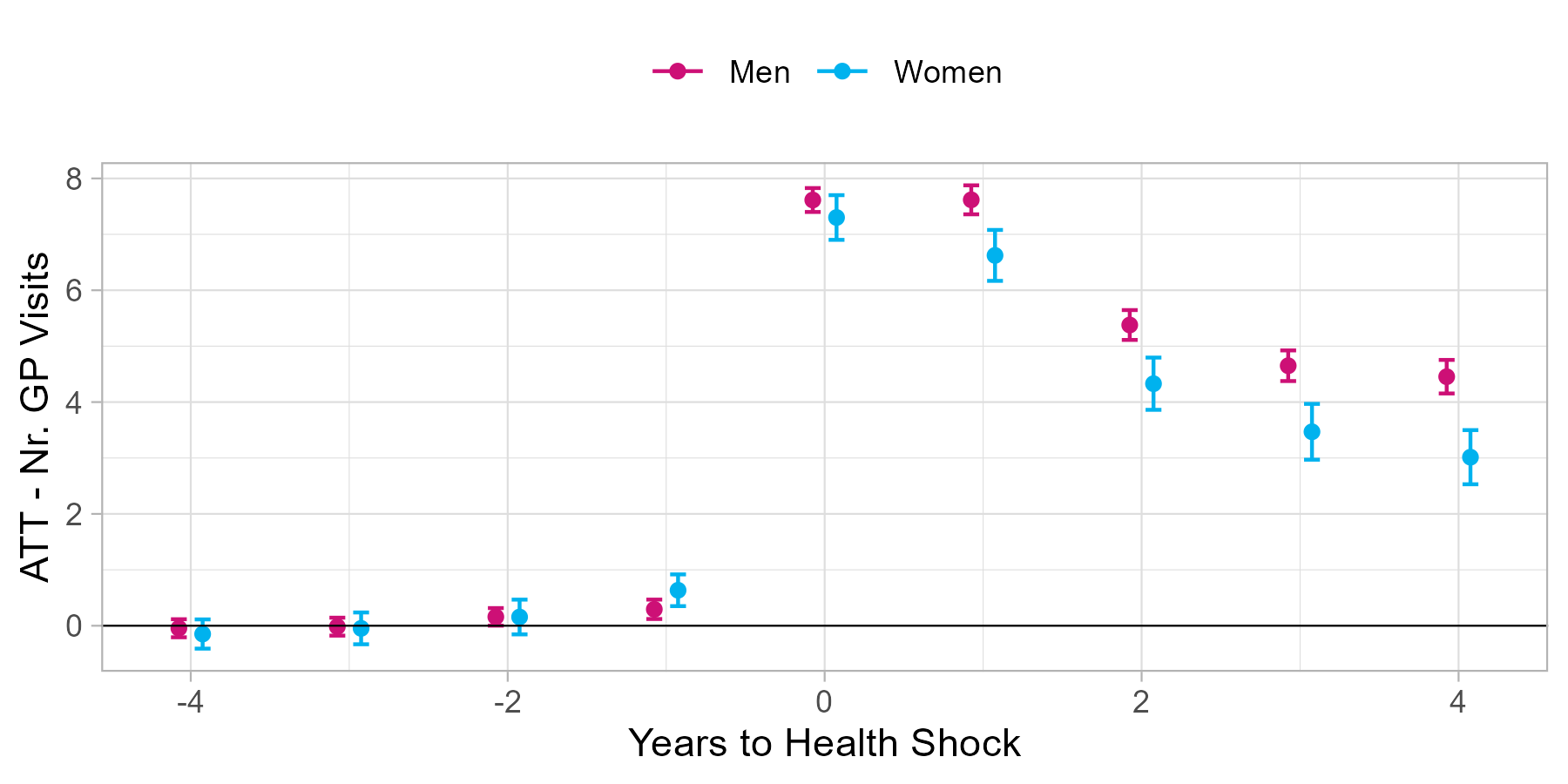}
        \caption{GP visits.}
    \end{subfigure}
    \vspace{0.4cm}
    \begin{subfigure}{0.65\textwidth}
        \centering
        \includegraphics[width=\textwidth]{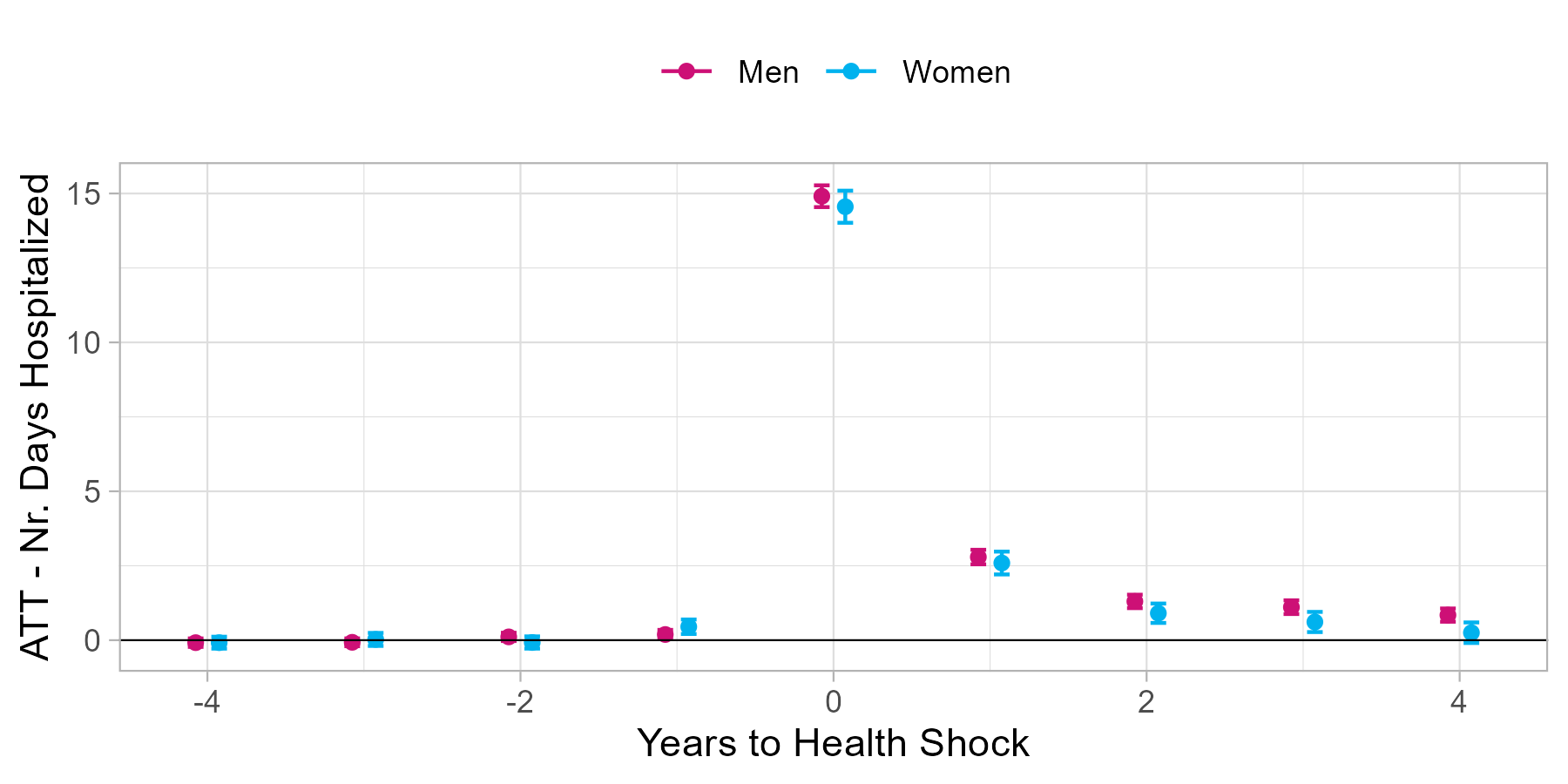}
        \caption{Days hospitalised.}
    \end{subfigure}
    \caption{Dynamic ATT estimates under unconditional parallel trends (no covariates).}
    \label{fig:unconditional}
    \begin{minipage}{0.9\textwidth}
    \footnotesize \textit{Notes:} Estimator and inference as in Figure~\ref{fig:dynamic_att_full}, dropping all covariates and relying on unconditional parallel trends.
    \end{minipage}
\end{figure}

\FloatBarrier
\clearpage

\subsection{Two-way fixed effects} \label{sec:twfe_estimates_main}
\begin{figure}[H]
    \centering
    \begin{subfigure}{0.65\textwidth}
        \centering
        \includegraphics[width=\textwidth]{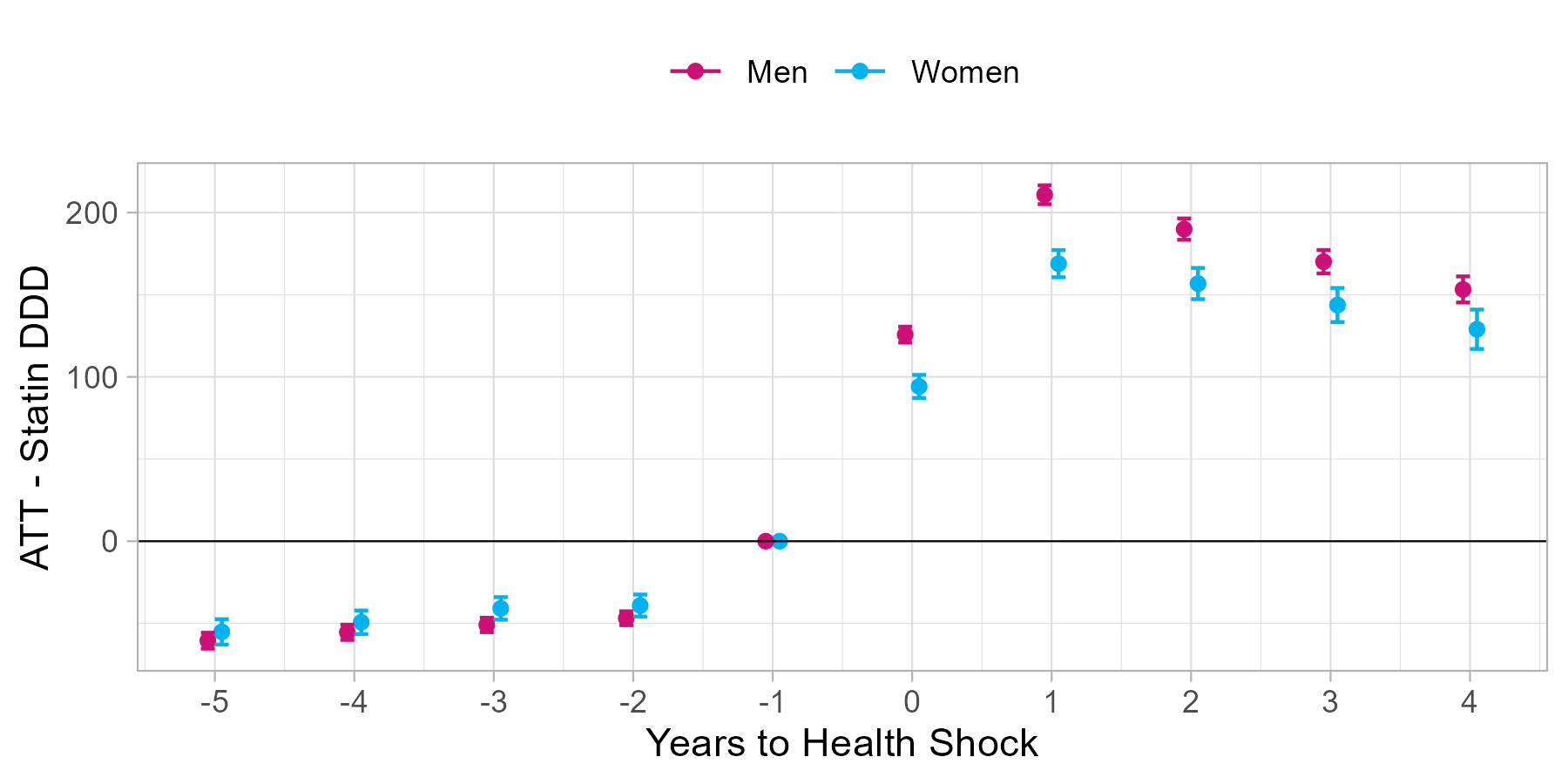}
        \caption{Statin DDD.}
        \label{fig:dynamic_twfe_statin_ddd_unbalanced_fadlon_nielsen}
    \end{subfigure}
    \vspace{0.4cm}
    \begin{subfigure}{0.65\textwidth}
        \centering
        \includegraphics[width=\textwidth]{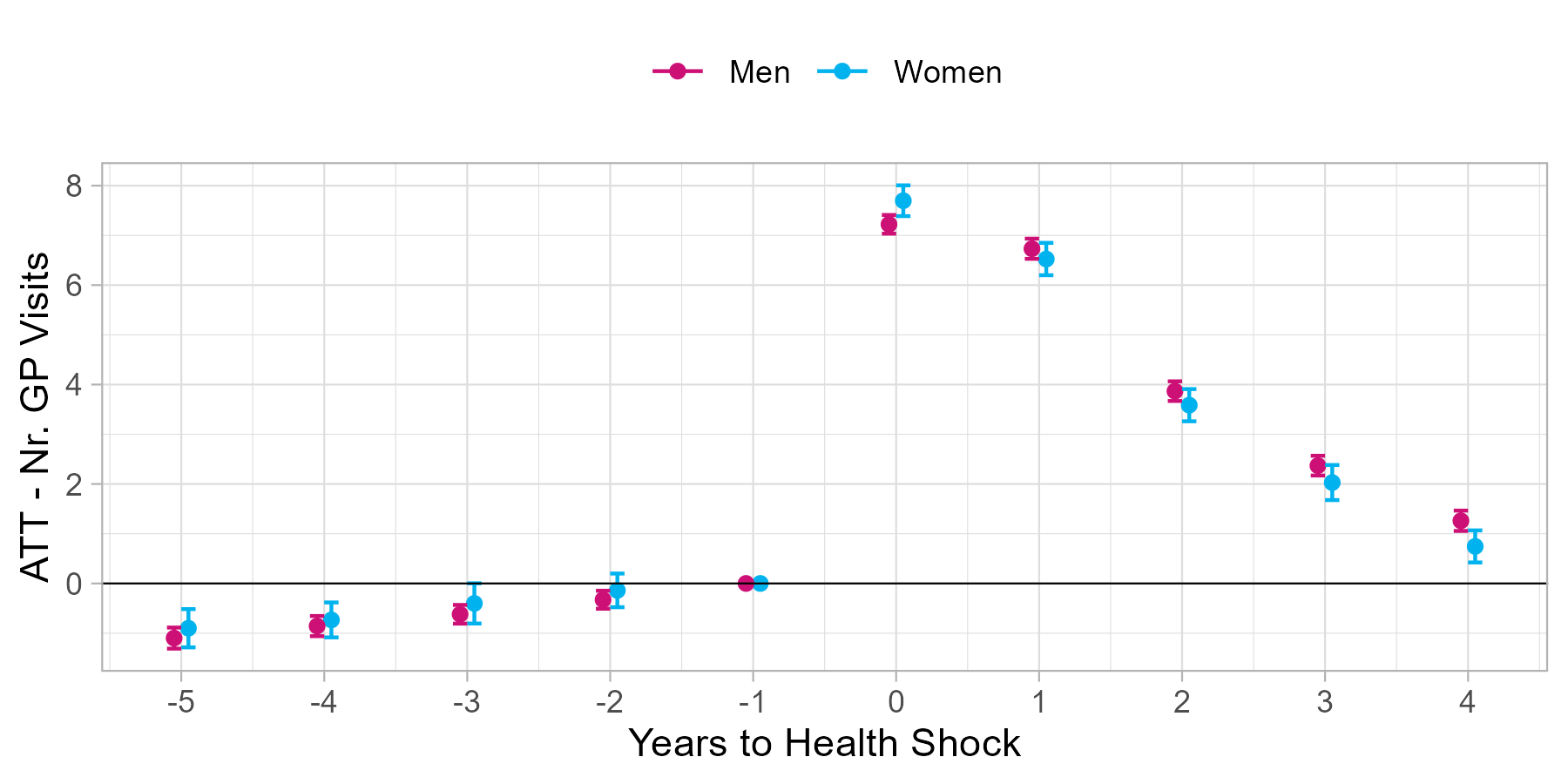}
        \caption{GP visits.}
        \label{fig:dynamic_twfe_nr_gp_visits_unbalanced_fadlon_nielsen}
    \end{subfigure}
    \vspace{0.4cm}
    \begin{subfigure}{0.65\textwidth}
        \centering
        \includegraphics[width=\textwidth]{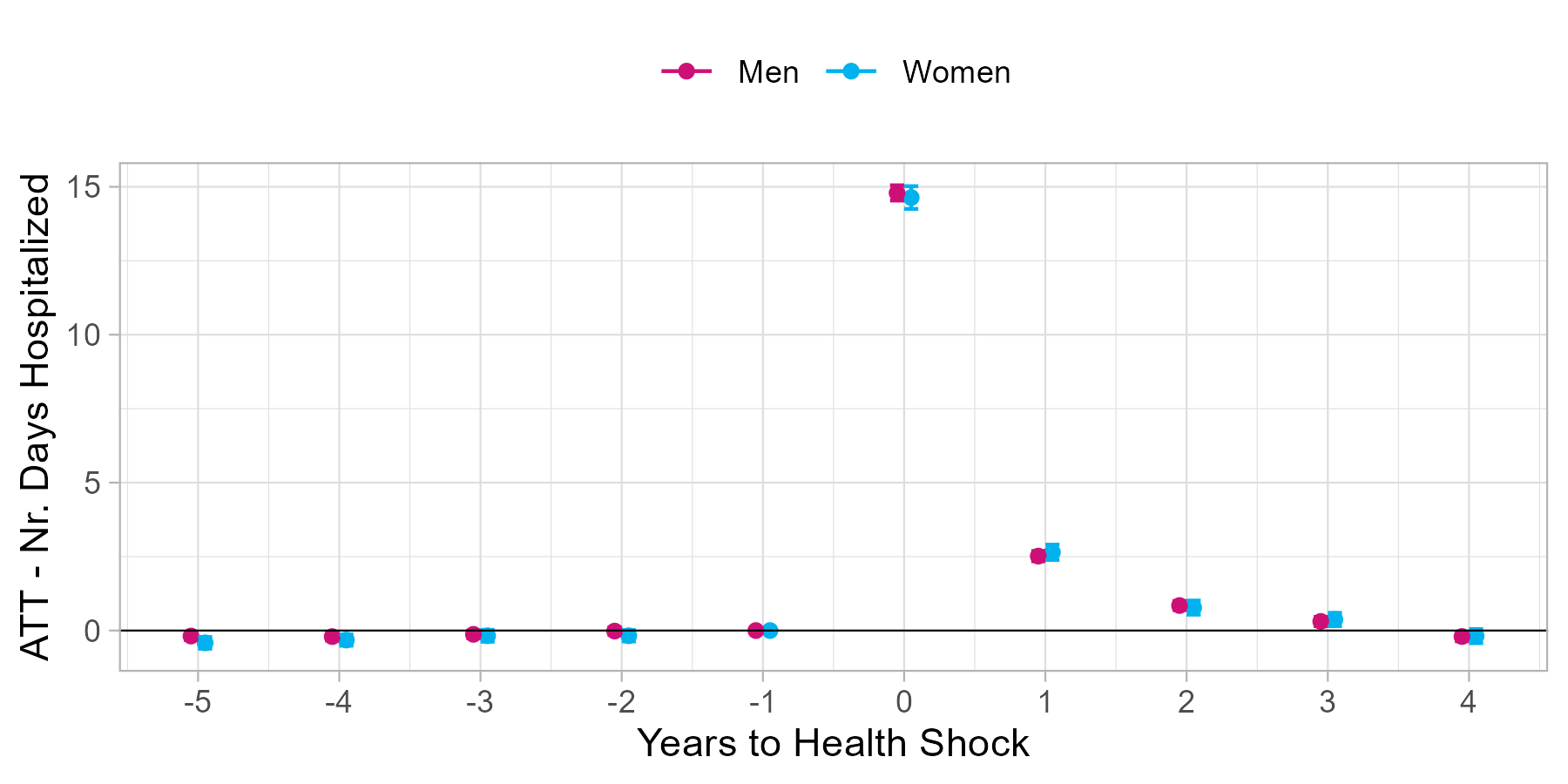}
        \caption{Days hospitalised.}
    \end{subfigure}
    \caption{Dynamic TWFE estimates as in \cite{nielsen}.}
    \label{fig:twfe_estimates_main}
    \begin{minipage}{0.9\textwidth}
    \footnotesize \textit{Notes:} Dynamic two-way fixed-effects specification following \cite{nielsen}, with the control group defined as individuals treated exactly five years after the focal cohort. Shaded areas denote 95\% confidence intervals based on standard errors clustered at the individual level.
    \end{minipage}
\end{figure}

\FloatBarrier
\clearpage

\subsection{Three- and seven-year comparison windows} \label{sec:estimates_7year_window}

\begin{figure}[H]
    \centering
    \begin{subfigure}{0.75\textwidth}
        \centering
        \includegraphics[width=\textwidth]{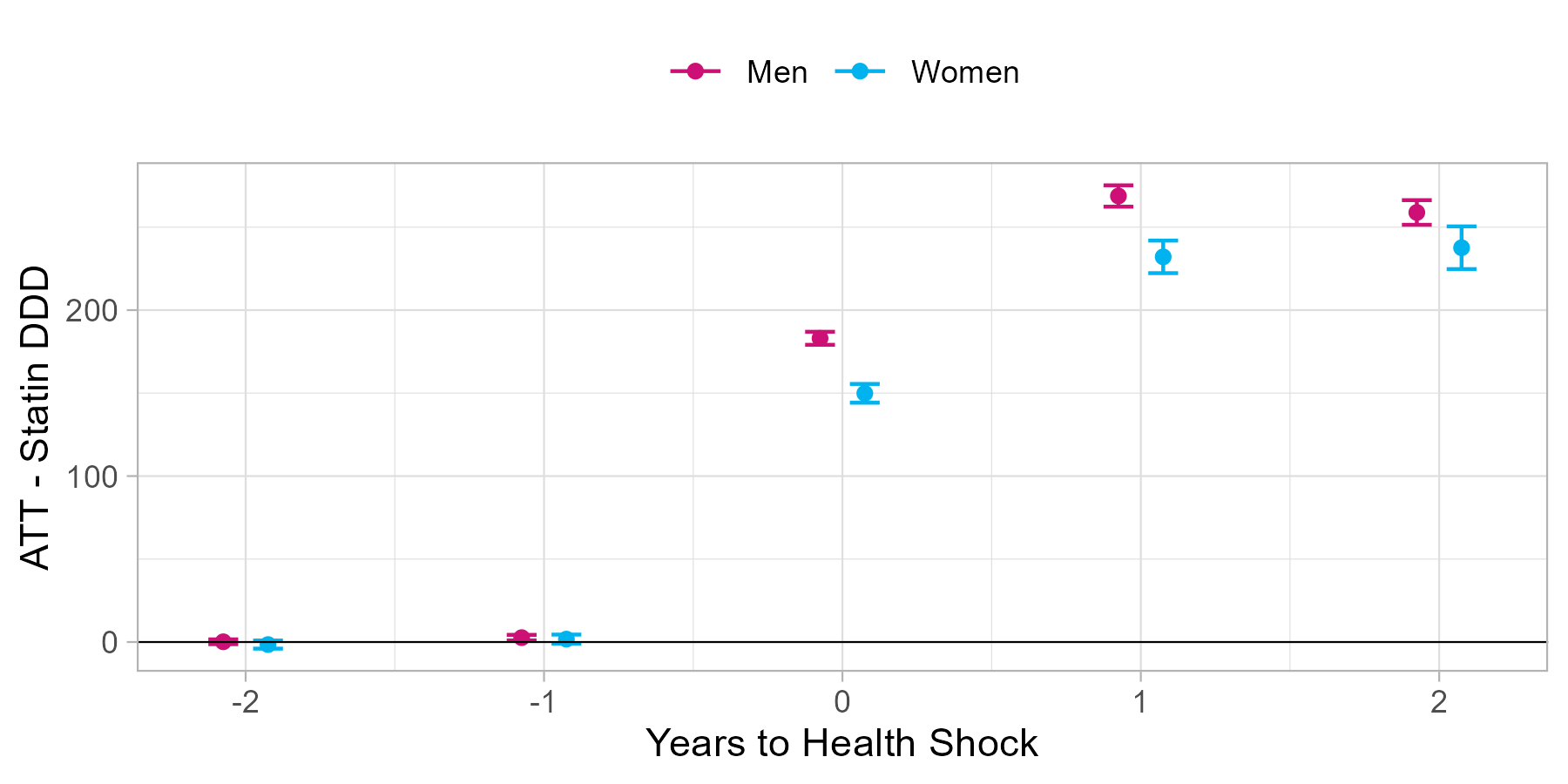}
        \caption{Statin DDD, 3-year window.}
    \end{subfigure}
    \vspace{0.4cm}
    \begin{subfigure}{0.75\textwidth}
        \centering
        \includegraphics[width=\textwidth]{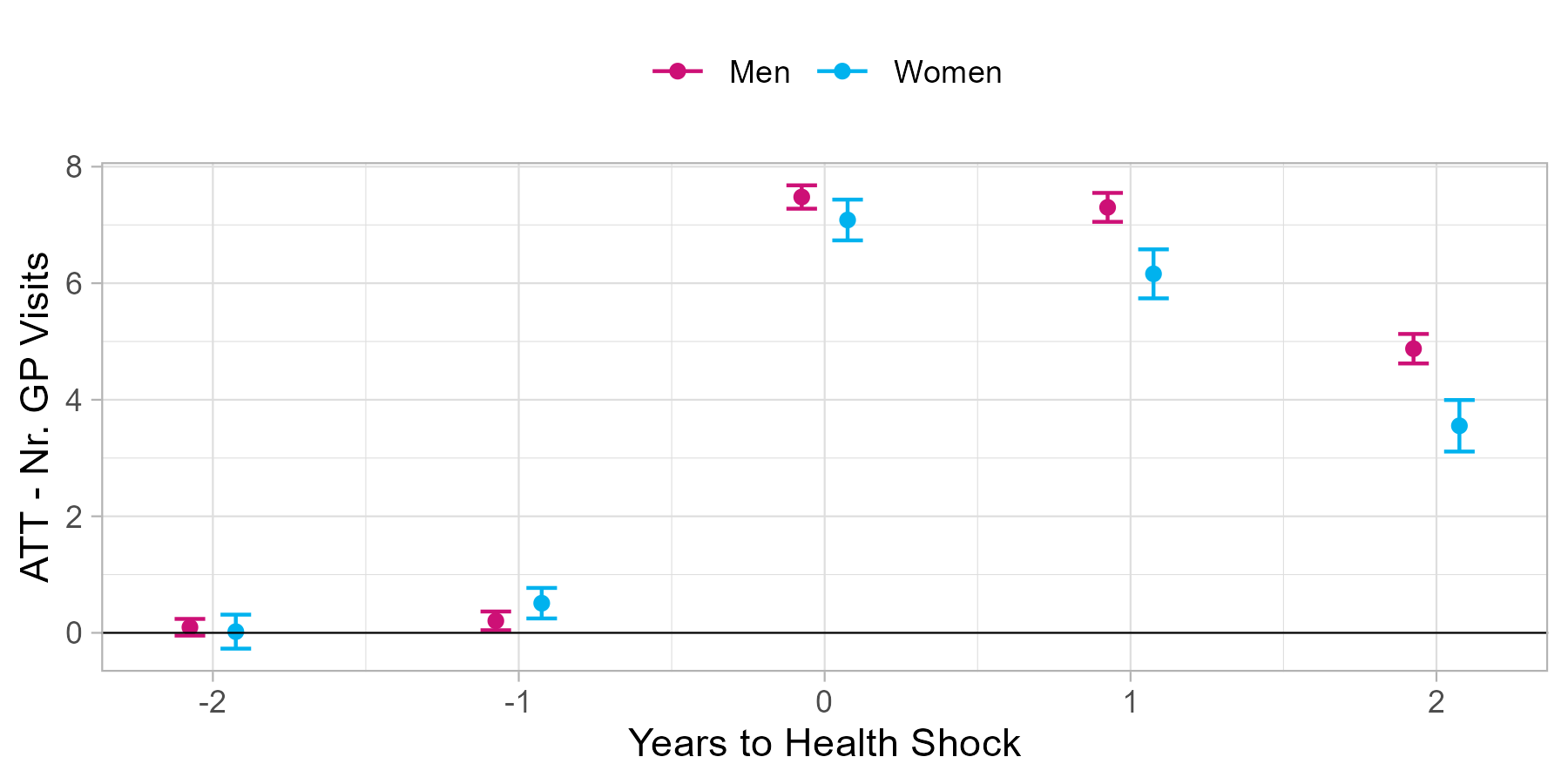}
        \caption{GP visits, 3-year window.}
    \end{subfigure}
    \vspace{0.4cm}
    \begin{subfigure}{0.75\textwidth}
        \centering
        \includegraphics[width=\textwidth]{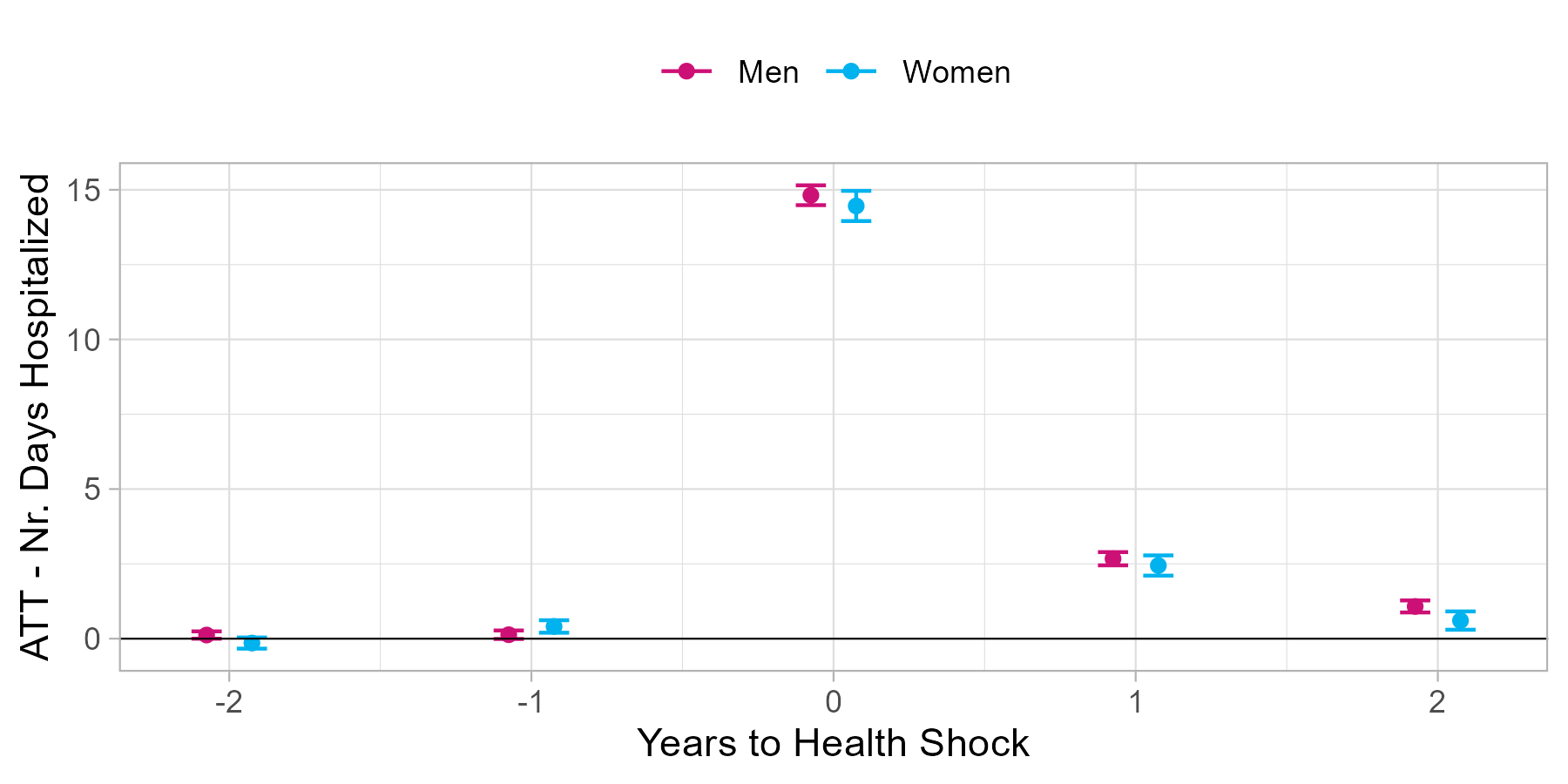}
        \caption{Days hospitalised, 3-year window.}
    \end{subfigure}
    \caption{Dynamic ATT estimates: 3-year non-shifting window.}
    \label{fig:3year_window}
    \begin{minipage}{0.9\textwidth}
    \footnotesize \textit{Notes:} Estimator and inference as in Figure~\ref{fig:dynamic_att_full}, with a three-year non-shifting comparison window.
    \end{minipage}
\end{figure}

\begin{figure}[H]
    \centering
    \begin{subfigure}{0.75\textwidth}
        \centering
        \includegraphics[width=\textwidth]{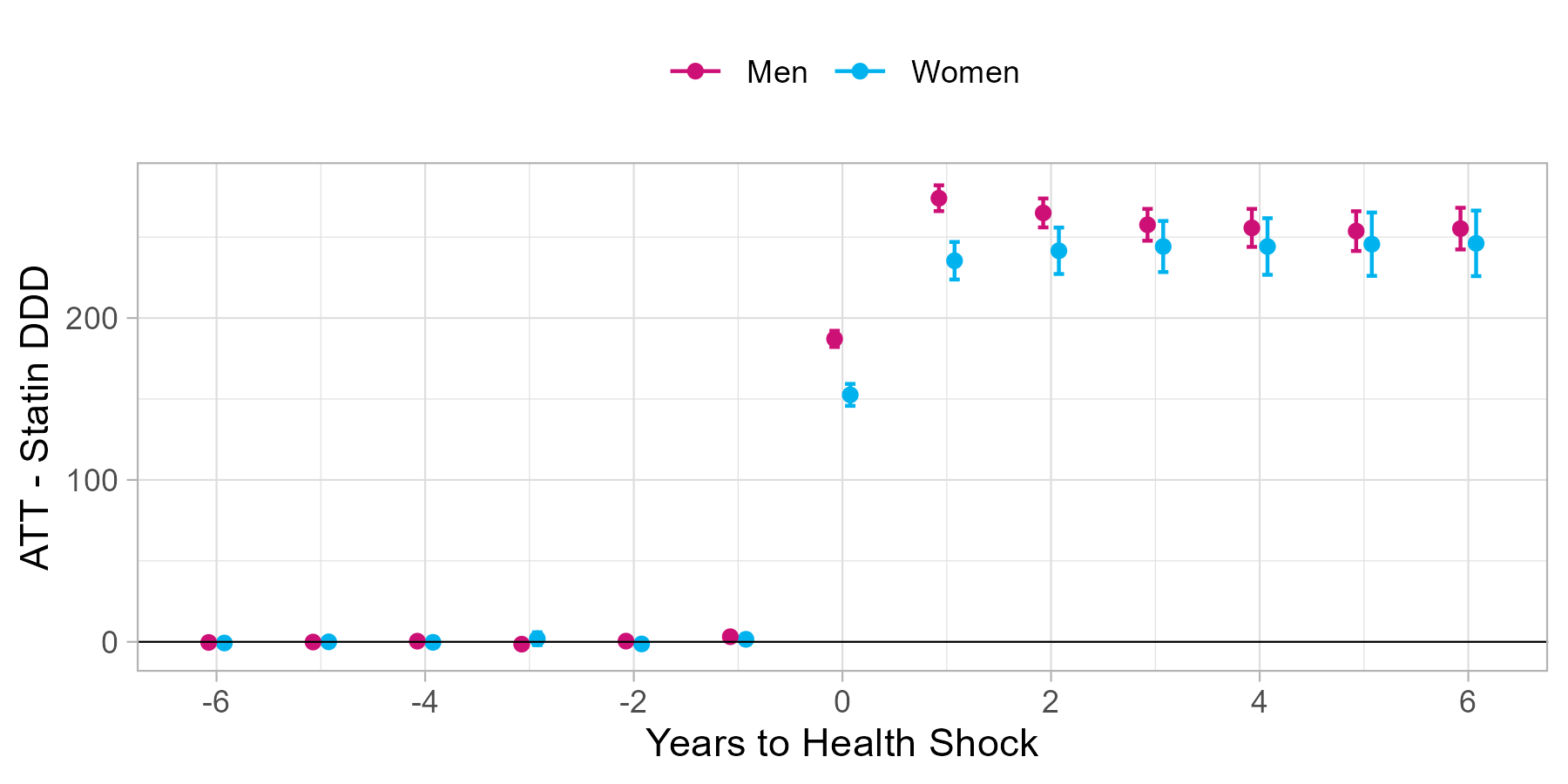}
        \caption{Statin DDD, 7-year window.}
    \end{subfigure}
    \vspace{0.4cm}
    \begin{subfigure}{0.75\textwidth}
        \centering
        \includegraphics[width=\textwidth]{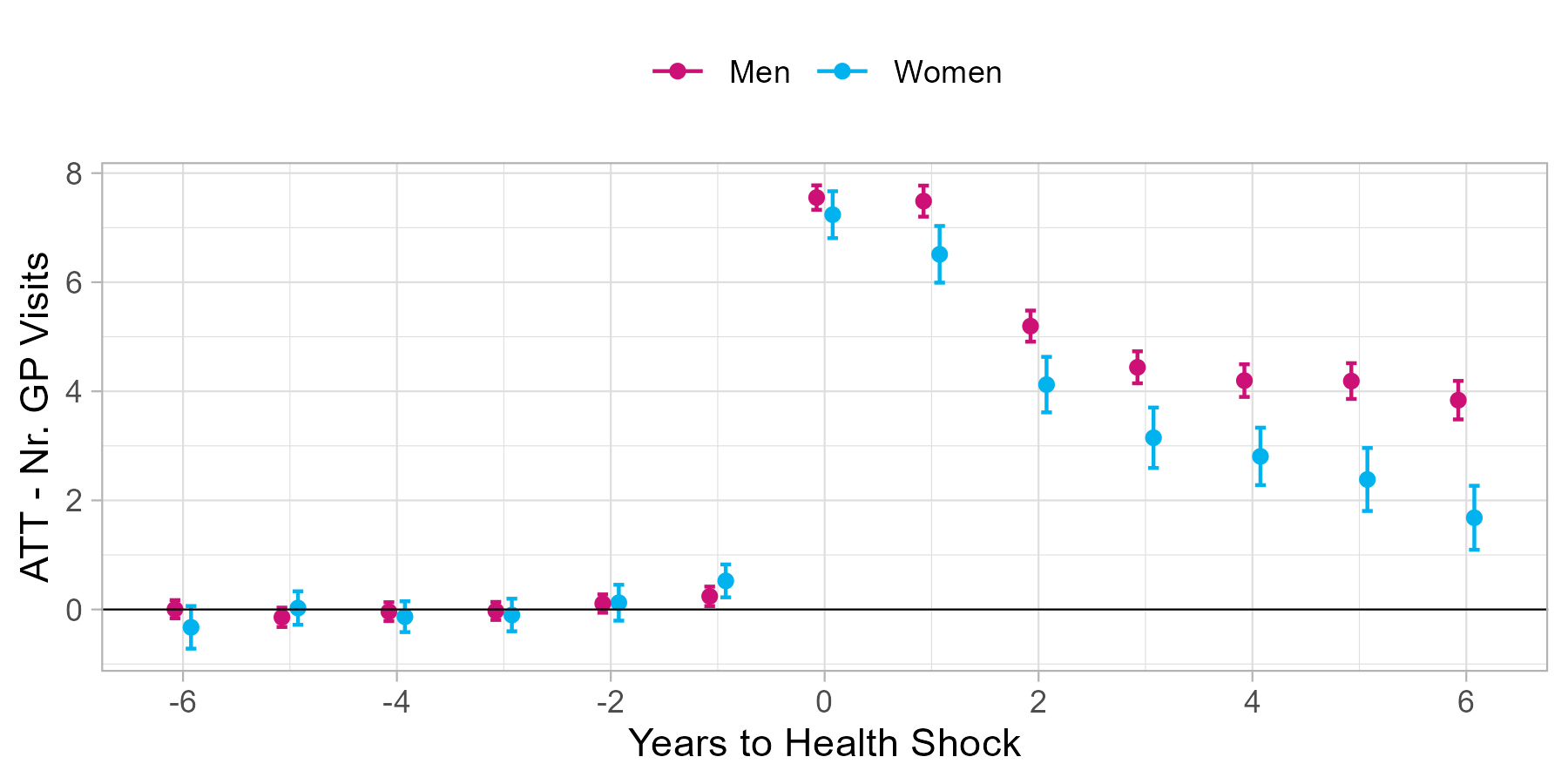}
        \caption{GP visits, 7-year window.}
    \end{subfigure}
    \vspace{0.4cm}
    \begin{subfigure}{0.75\textwidth}
        \centering
        \includegraphics[width=\textwidth]{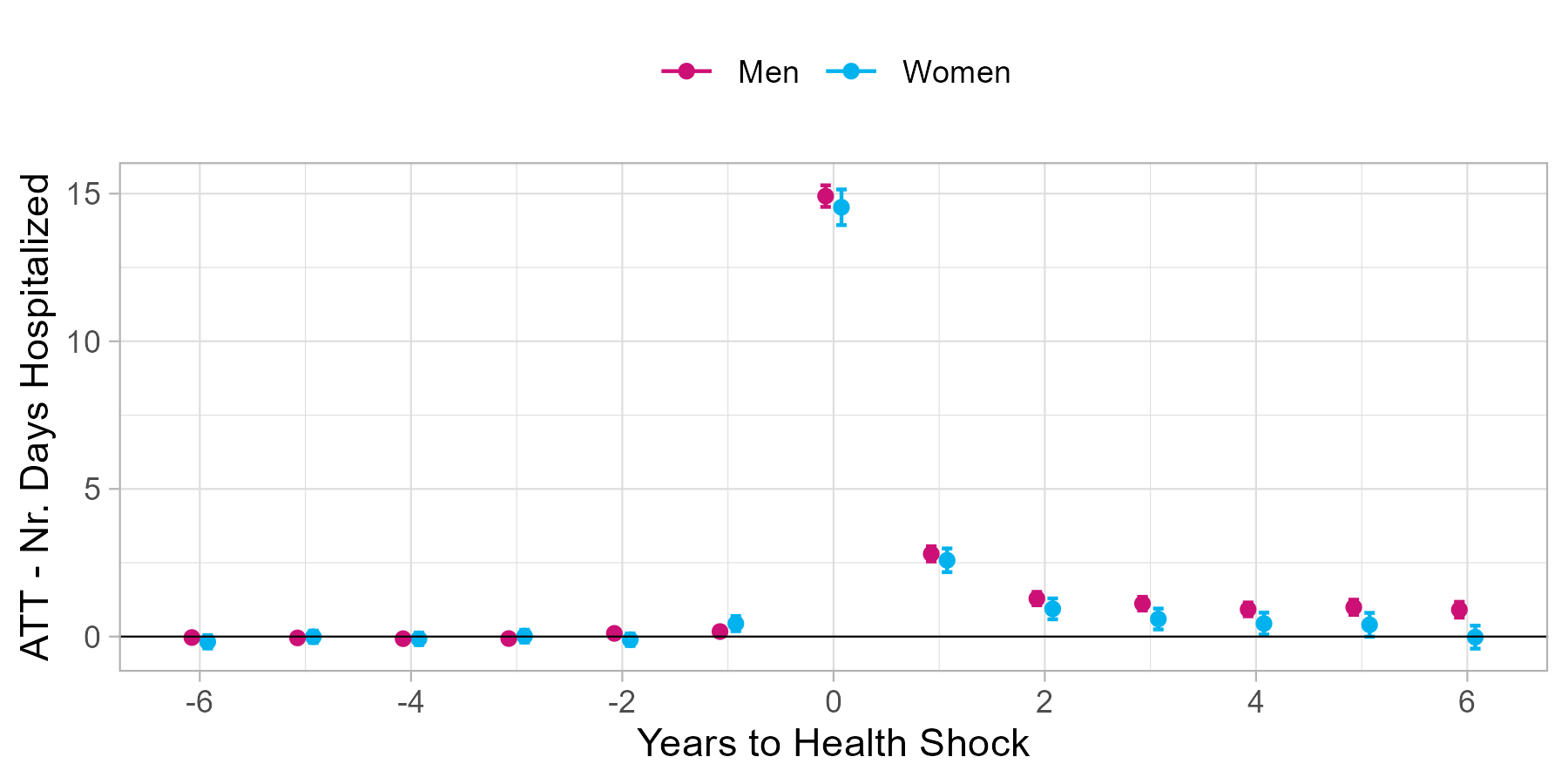}
        \caption{Days hospitalised, 7-year window.}
    \end{subfigure}
    \caption{Dynamic ATT estimates: 7-year non-shifting window.}
    \label{fig:7year_window}
    \begin{minipage}{0.9\textwidth}
    \footnotesize \textit{Notes:} Estimator and inference as in Figure~\ref{fig:dynamic_att_full}, with a seven-year non-shifting comparison window.
    \end{minipage}
\end{figure}

\section{Additional tables and figures} \label{mechanisms}
\subsection{Severity of the shock (STEMI vs.\ NSTEMI)} \label{sec:intensity}
\begin{figure}[H]
    \centering
    \begin{subfigure}{0.86\textwidth}
        \centering
        \includegraphics[width=\textwidth]{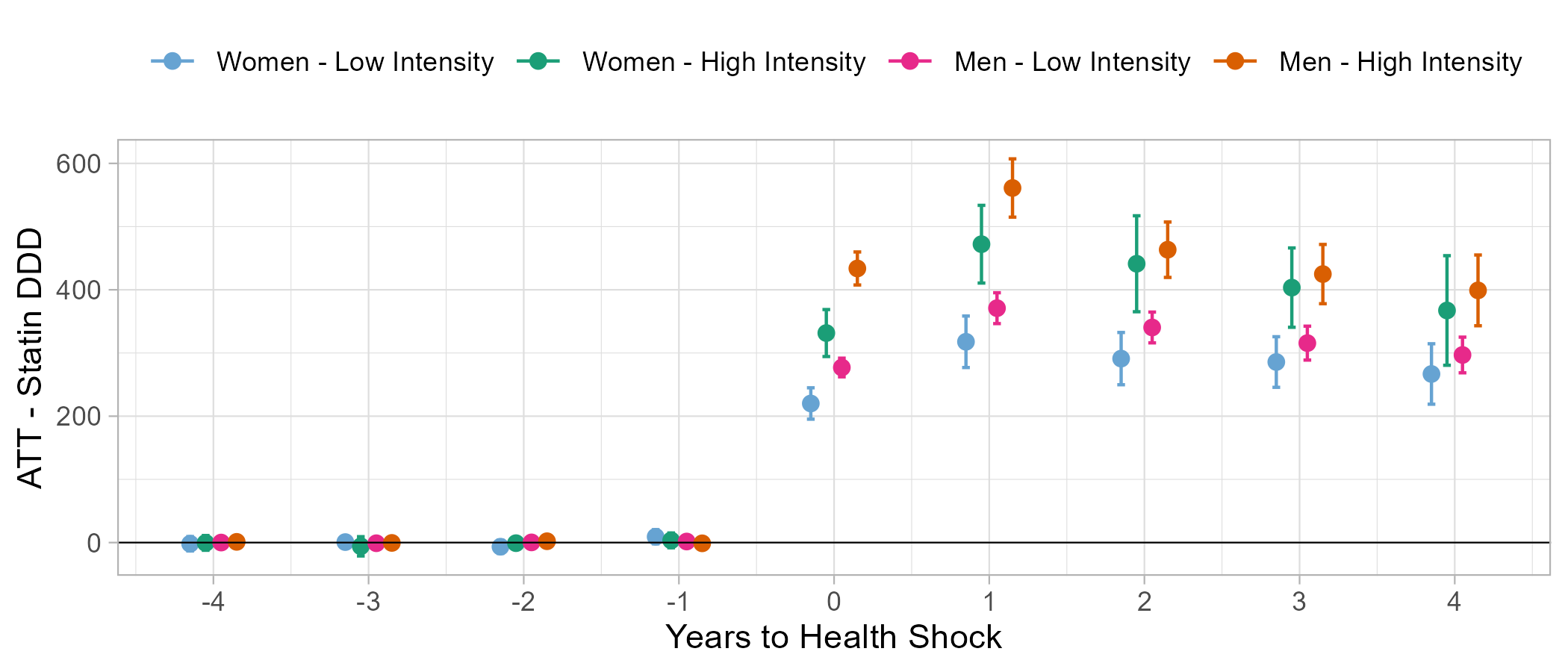}
        \caption{Statin DDD.}
    \end{subfigure}
    \vspace{0.15cm}
    \begin{subfigure}{0.86\textwidth}
        \centering
        \includegraphics[width=\textwidth]{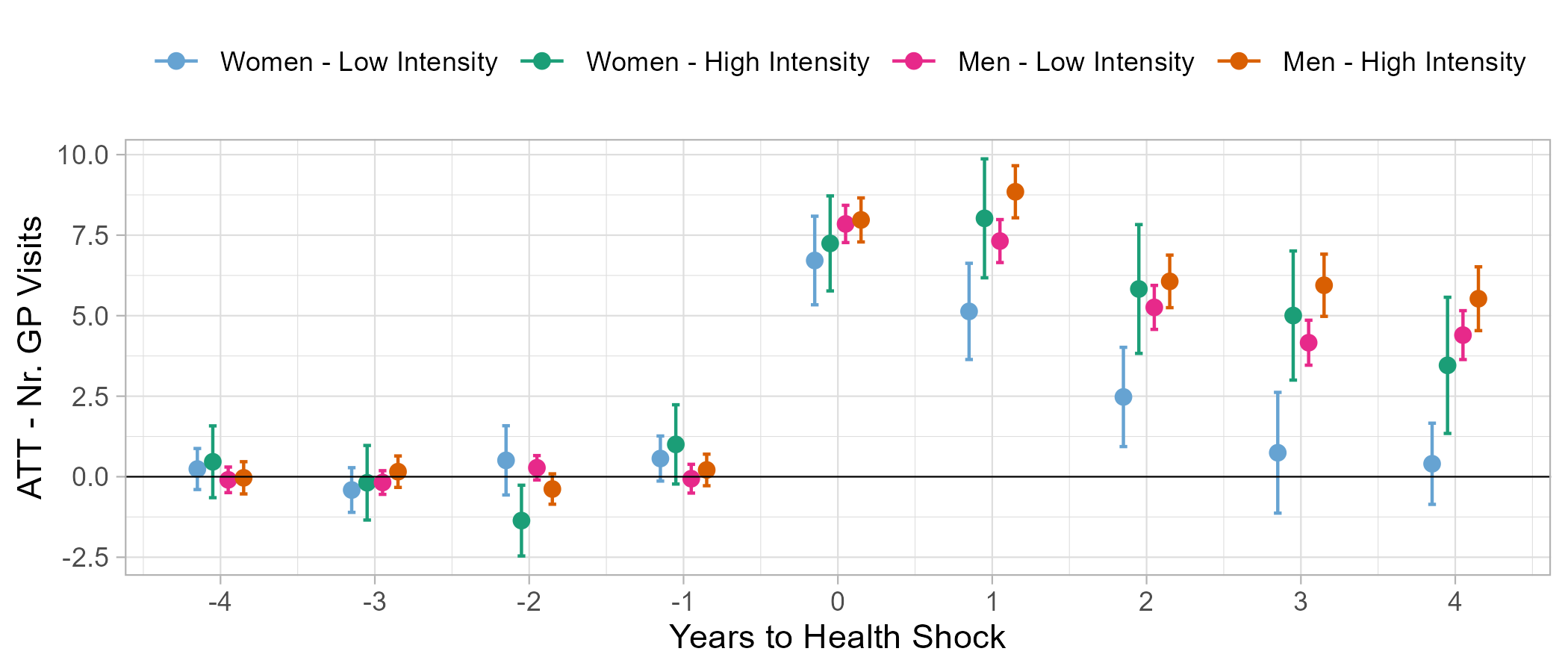}
        \caption{GP visits.}
    \end{subfigure}
    \vspace{0.15cm}
    \begin{subfigure}{0.86\textwidth}
        \centering
        \includegraphics[width=\textwidth]{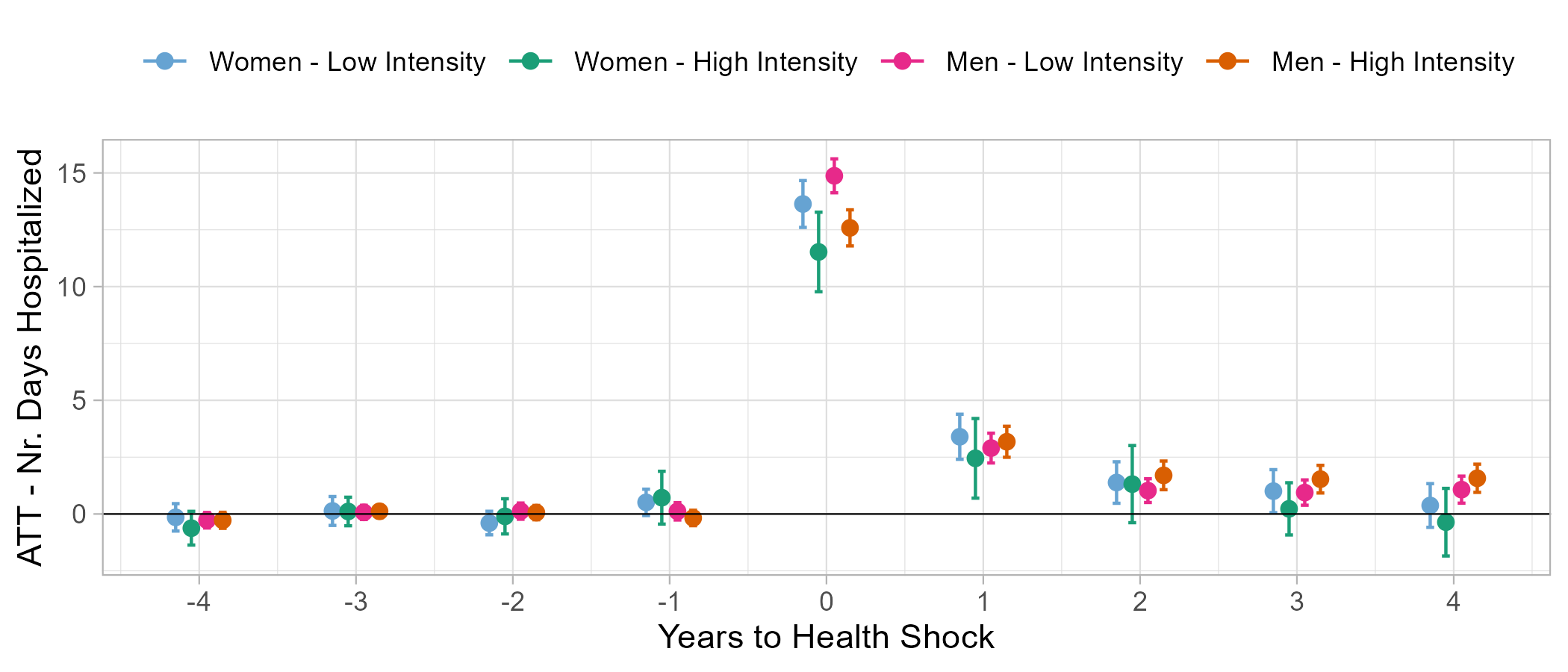}
        \caption{Days hospitalised.}
    \end{subfigure}
    \caption{Dynamic ATT by shock severity (STEMI vs.\ NSTEMI), heart-attack subsample.}
    \label{fig:subgrouping_intensity}
    \begin{minipage}{0.9\textwidth}
    \footnotesize \textit{Notes:} Estimator and inference as in Figure~\ref{fig:dynamic_att_full}, estimated separately by gender and by infarct type (STEMI: ICD-10 I21.0--I21.3; NSTEMI: I21.4) within the heart-attack subsample.
    \end{minipage}
\end{figure}

\FloatBarrier

\subsection{Balanced panel}
\label{app:twfe_balanced}

\begin{figure}[H]
    \centering
    \begin{minipage}{0.65\textwidth}
        \centering
        \includegraphics[width=\textwidth]{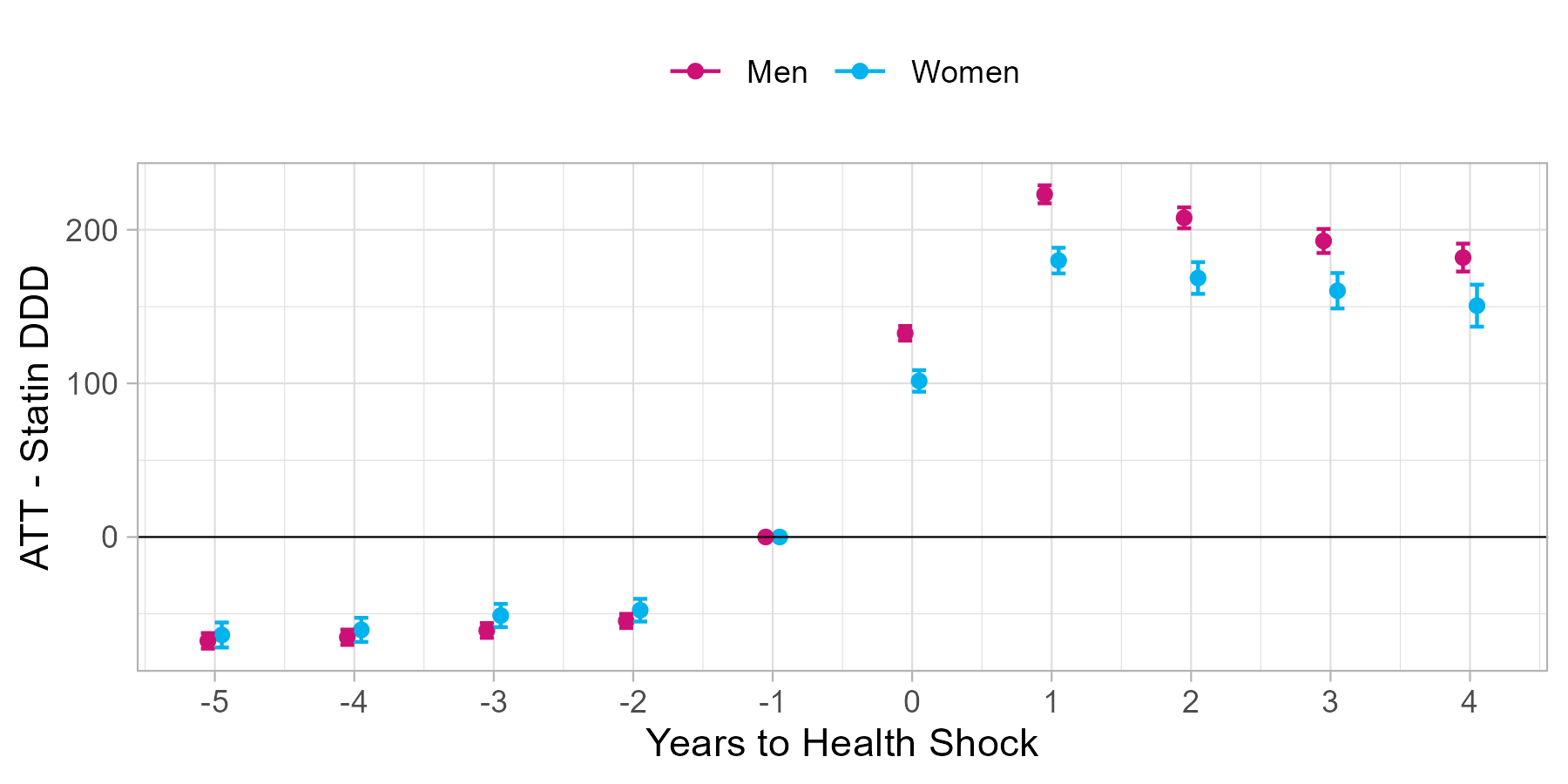}
        \captionof{figure}{Statin DDD, balanced panel.}
    \end{minipage}
    \vspace{0.4cm}
    \begin{minipage}{0.65\textwidth}
        \centering
        \includegraphics[width=\textwidth]{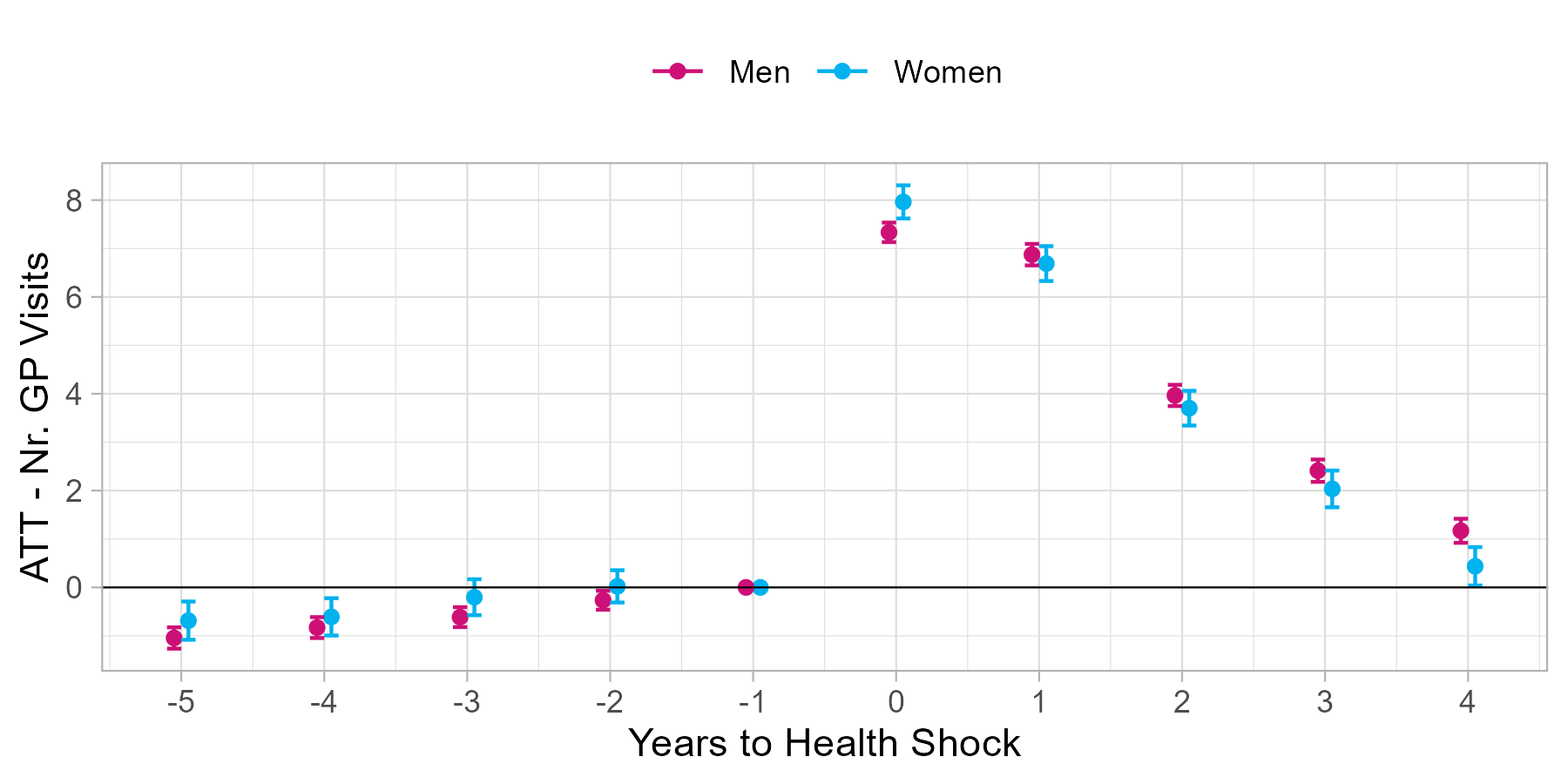}
        \captionof{figure}{GP visits, balanced panel.}
    \end{minipage}
    \vspace{0.4cm}
    \begin{minipage}{0.65\textwidth}
        \centering
        \includegraphics[width=\textwidth]{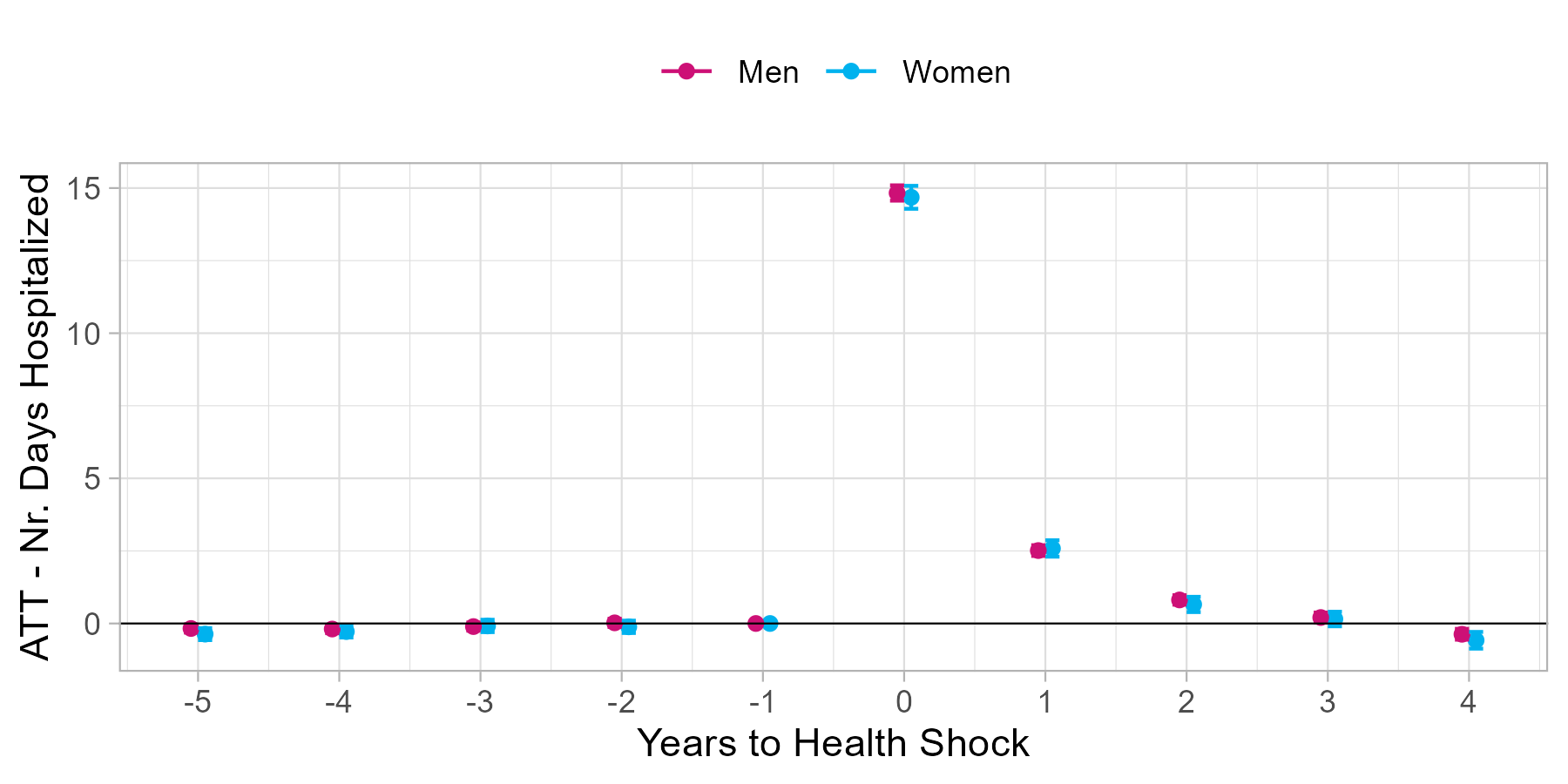}
        \captionof{figure}{Days hospitalised, balanced panel.}
    \end{minipage}
    \caption{Dynamic TWFE estimates: balanced panel restricted to individuals observed over all five pre- and five post-shock years.}
    \begin{minipage}{0.9\textwidth}
    \footnotesize \textit{Notes:} Specification and inference as in Figure~\ref{fig:twfe_estimates_main}, restricted to individuals observed over all five pre- and five post-shock years (74\% of men and 77\% of women retained).
    \end{minipage}
\end{figure}

\subsection{Heart attacks vs.\ strokes}
\label{sec:individual_treatment}

\begin{figure}[H]
    \centering
    \begin{subfigure}{0.95\textwidth}
        \centering
        \includegraphics[width=\textwidth]{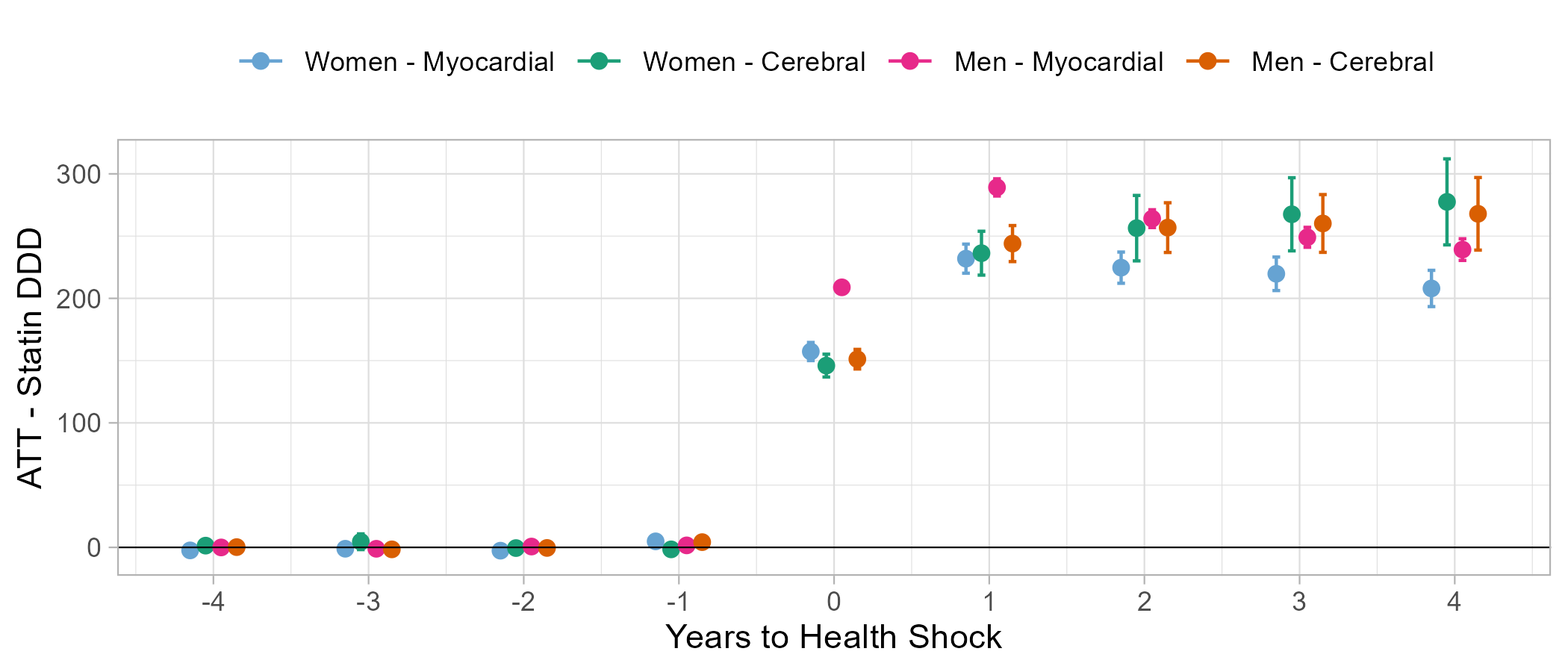}
        \caption{Statin DDD.}
    \end{subfigure}
    \vspace{0.4cm}
    \begin{subfigure}{0.95\textwidth}
        \centering
        \includegraphics[width=\textwidth]{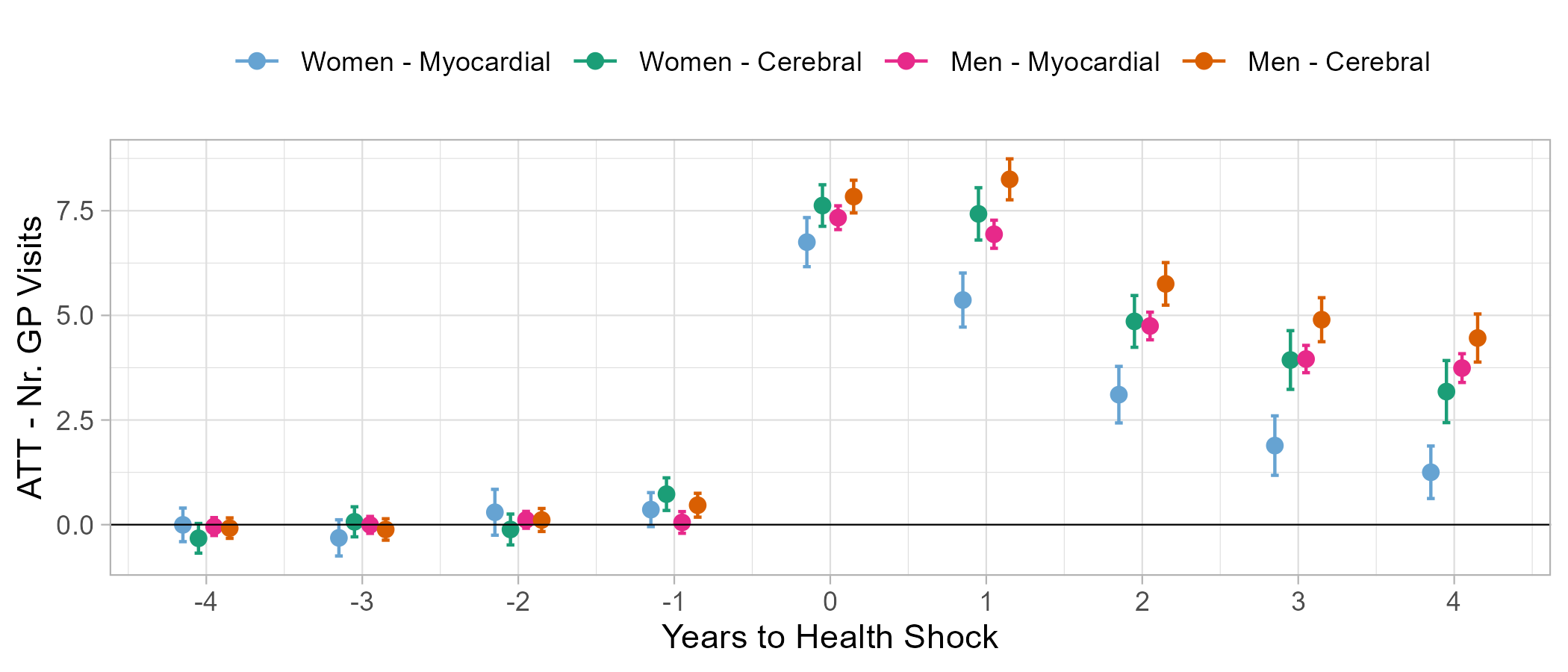}
        \caption{GP visits.}
    \end{subfigure}
    \vspace{0.4cm}
    \begin{subfigure}{0.95\textwidth}
        \centering
        \includegraphics[width=\textwidth]{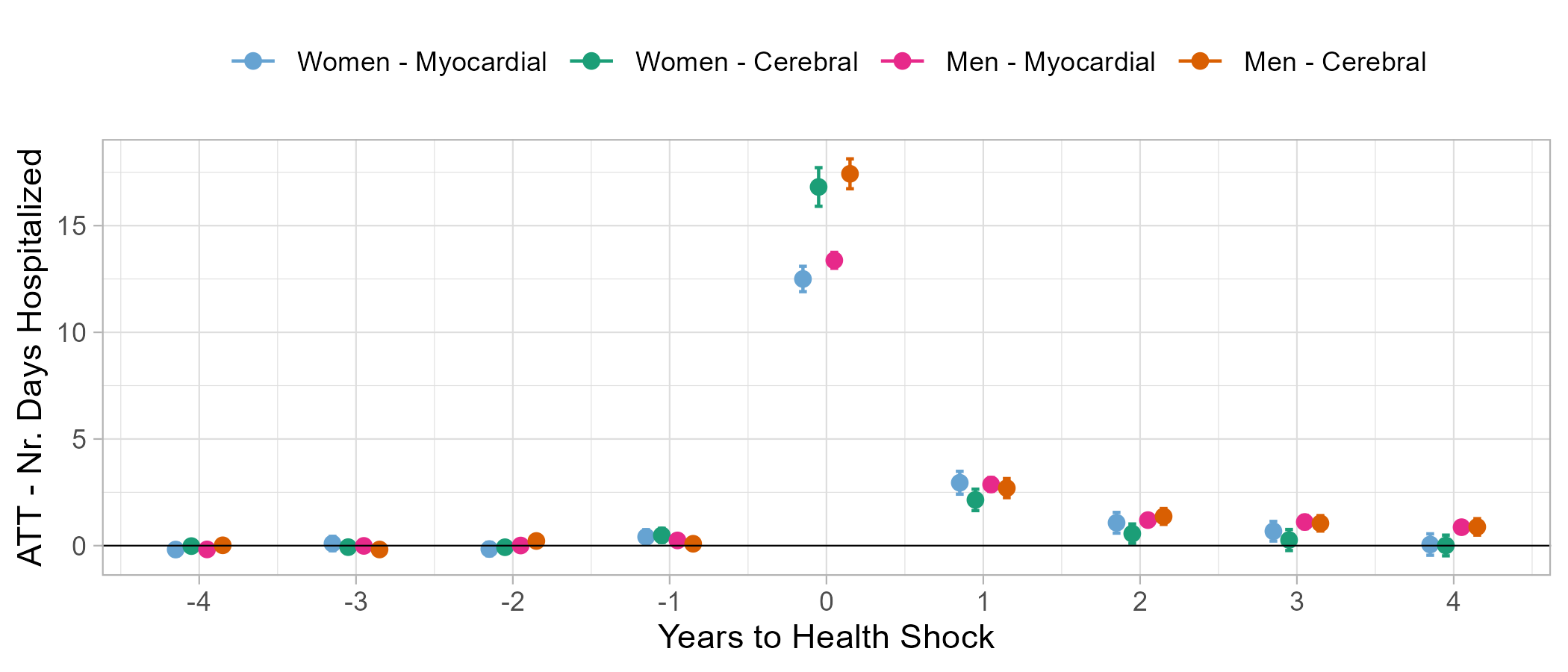}
        \caption{Days hospitalised.}
    \end{subfigure}
    \caption{Dynamic ATT by shock type (heart attack vs.\ stroke).}
    \label{fig:subgrouping_treatment}
    \begin{minipage}{0.9\textwidth}
    \footnotesize \textit{Notes:} Estimator and inference as in Figure~\ref{fig:dynamic_att_full}, estimated separately by gender and by shock type (acute myocardial infarction vs.\ cerebral infarction).
    \end{minipage}
\end{figure}

\subsection{Heterogeneity by income}
\label{sec:income_heterogeneity}

\begin{figure}[H]
    \centering
    \begin{subfigure}{0.95\textwidth}
        \centering
        \includegraphics[width=\textwidth]{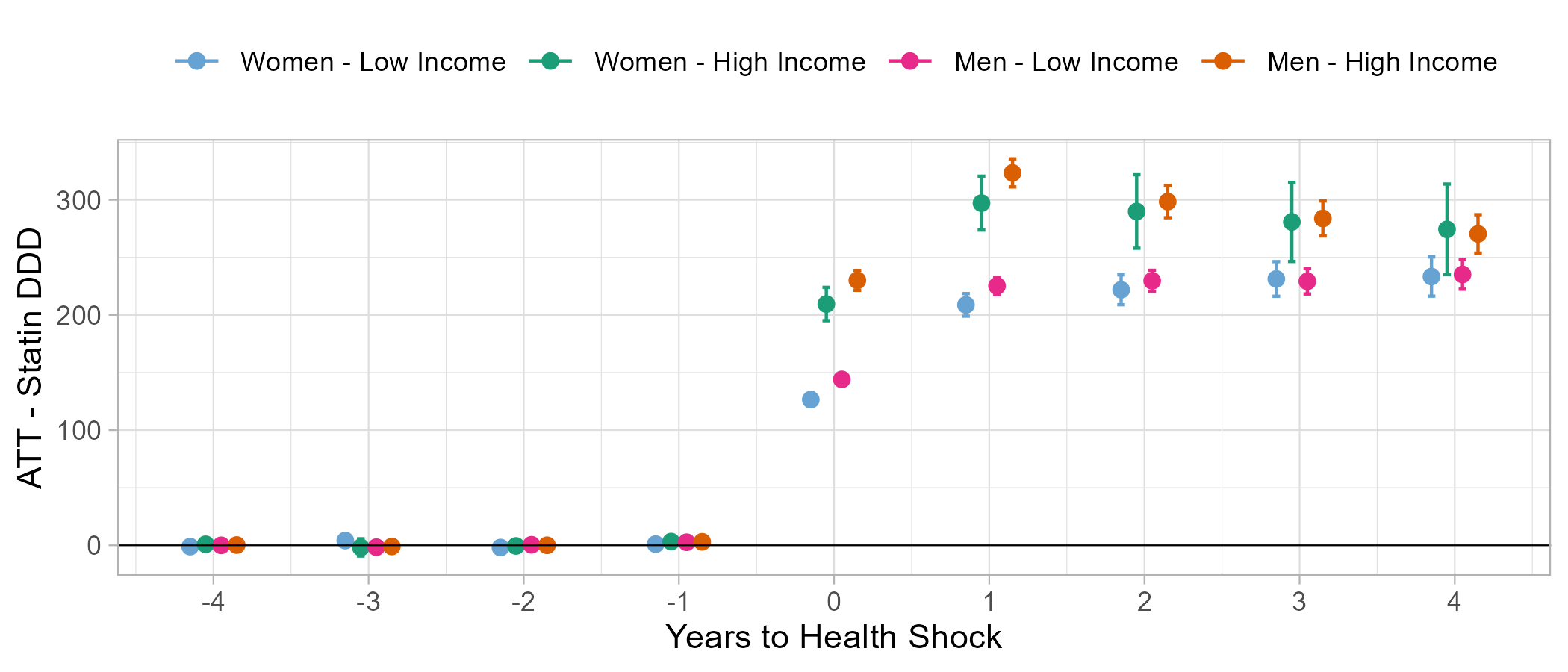}
        \caption{Statin DDD.}
    \end{subfigure}
    \vspace{0.4cm}
    \begin{subfigure}{0.95\textwidth}
        \centering
        \includegraphics[width=\textwidth]{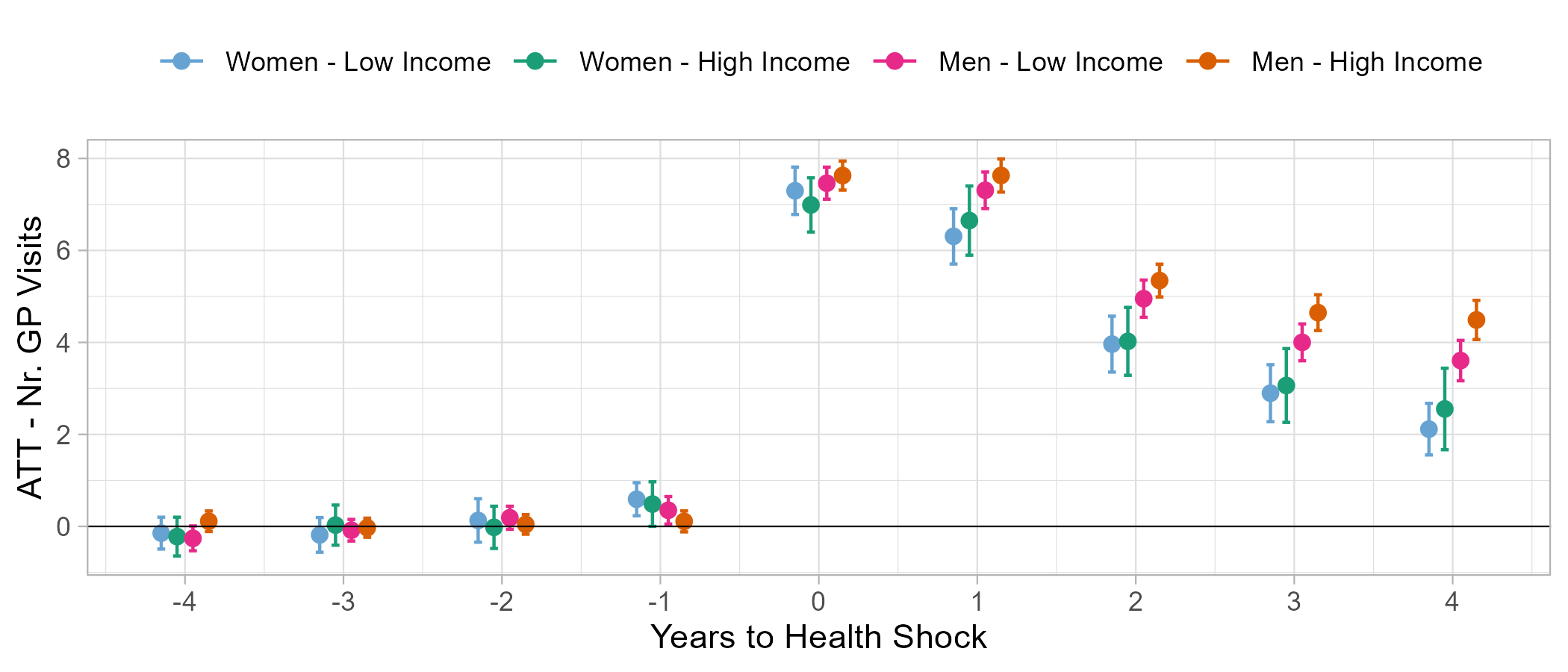}
        \caption{GP visits.}
    \end{subfigure}
    \vspace{0.4cm}
    \begin{subfigure}{0.95\textwidth}
        \centering
        \includegraphics[width=\textwidth]{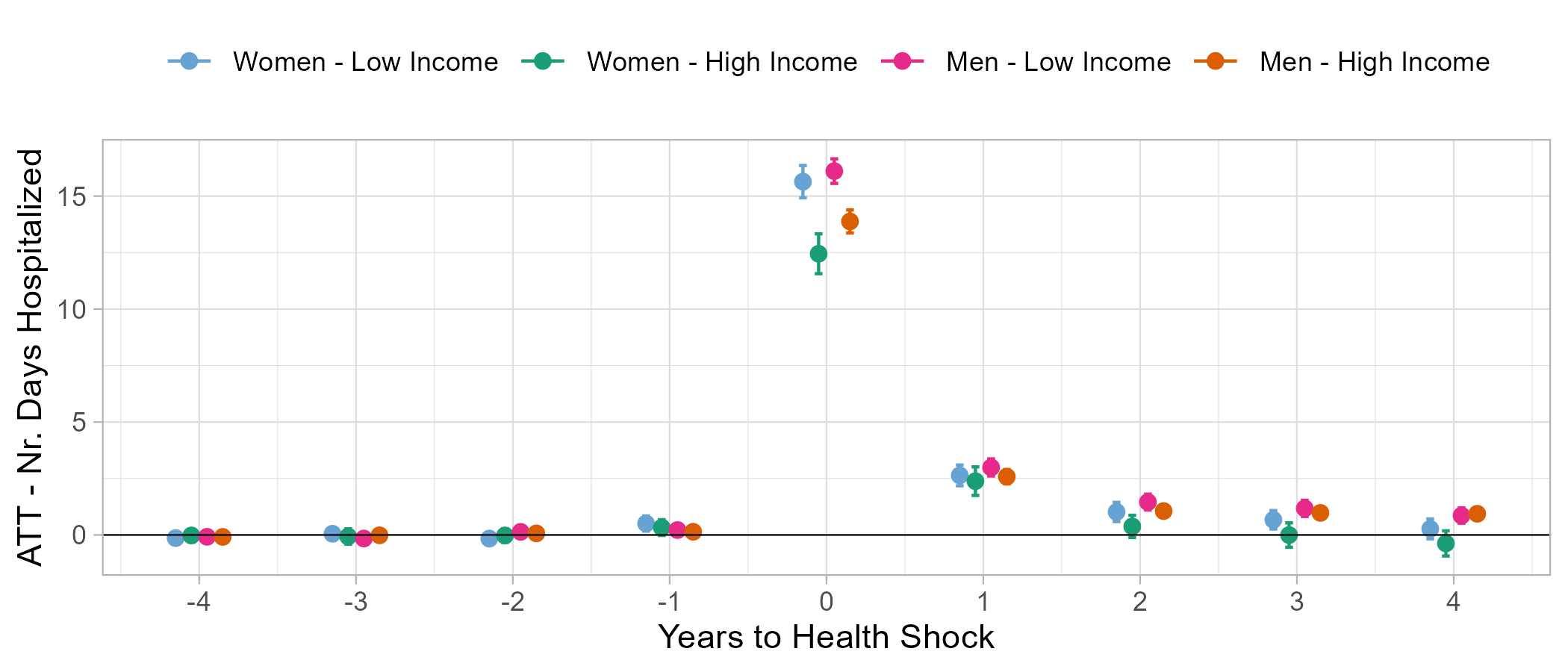}
        \caption{Days hospitalised.}
    \end{subfigure}
    \caption{Dynamic ATT by below- vs.\ above-median equivalised income.}
    \label{fig:subgrouping_income}
    \begin{minipage}{0.9\textwidth}
    \footnotesize \textit{Notes:} Estimator and inference as in Figure~\ref{fig:dynamic_att_full}, estimated separately by gender and by below- vs.\ above-median equivalised disposable household income at $t=g-1$.
    \end{minipage}
\end{figure}
\subsection{Heterogeneity by household type}
\label{sec:household_heterogeneity}

\begin{figure}[H]
    \centering
    \begin{subfigure}{0.95\textwidth}
        \centering
        \includegraphics[width=\textwidth]{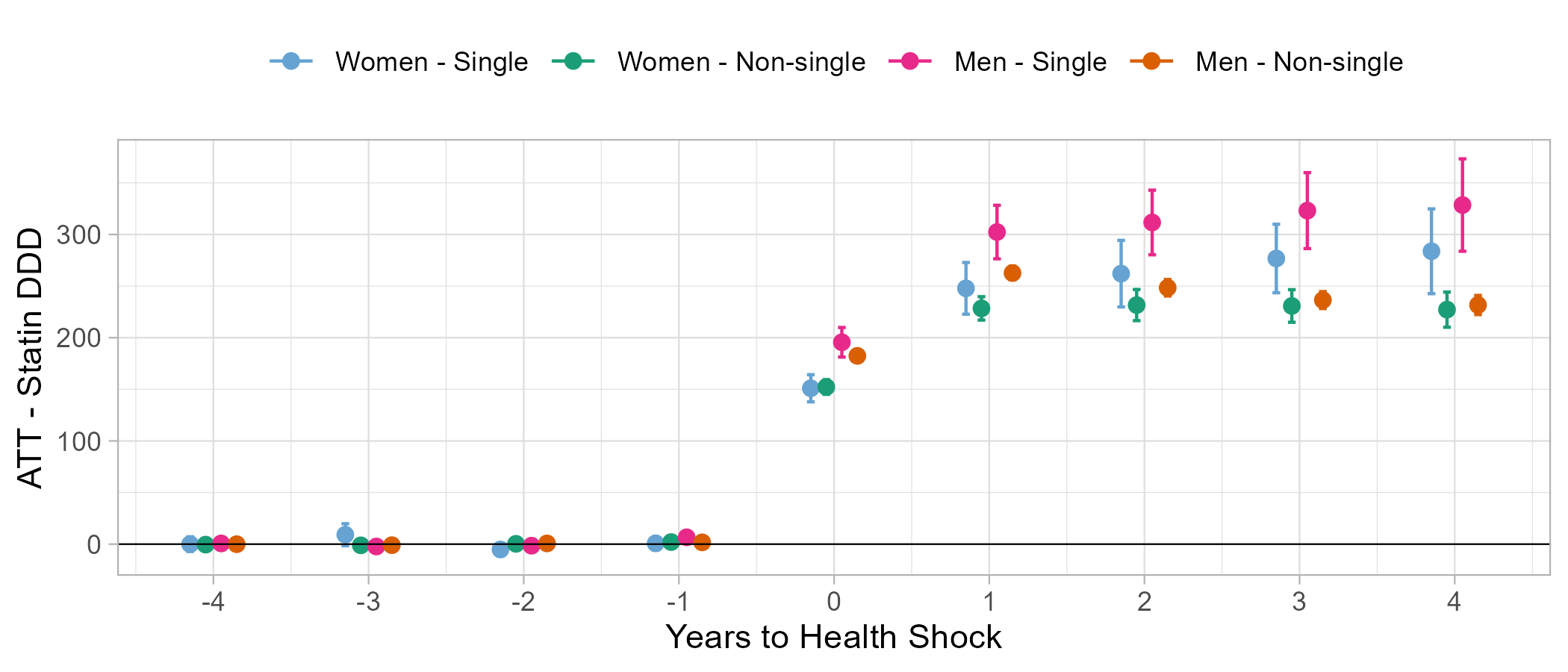}
        \caption{Statin DDD.}
    \end{subfigure}
    \vspace{0.4cm}
    \begin{subfigure}{0.95\textwidth}
        \centering
        \includegraphics[width=\textwidth]{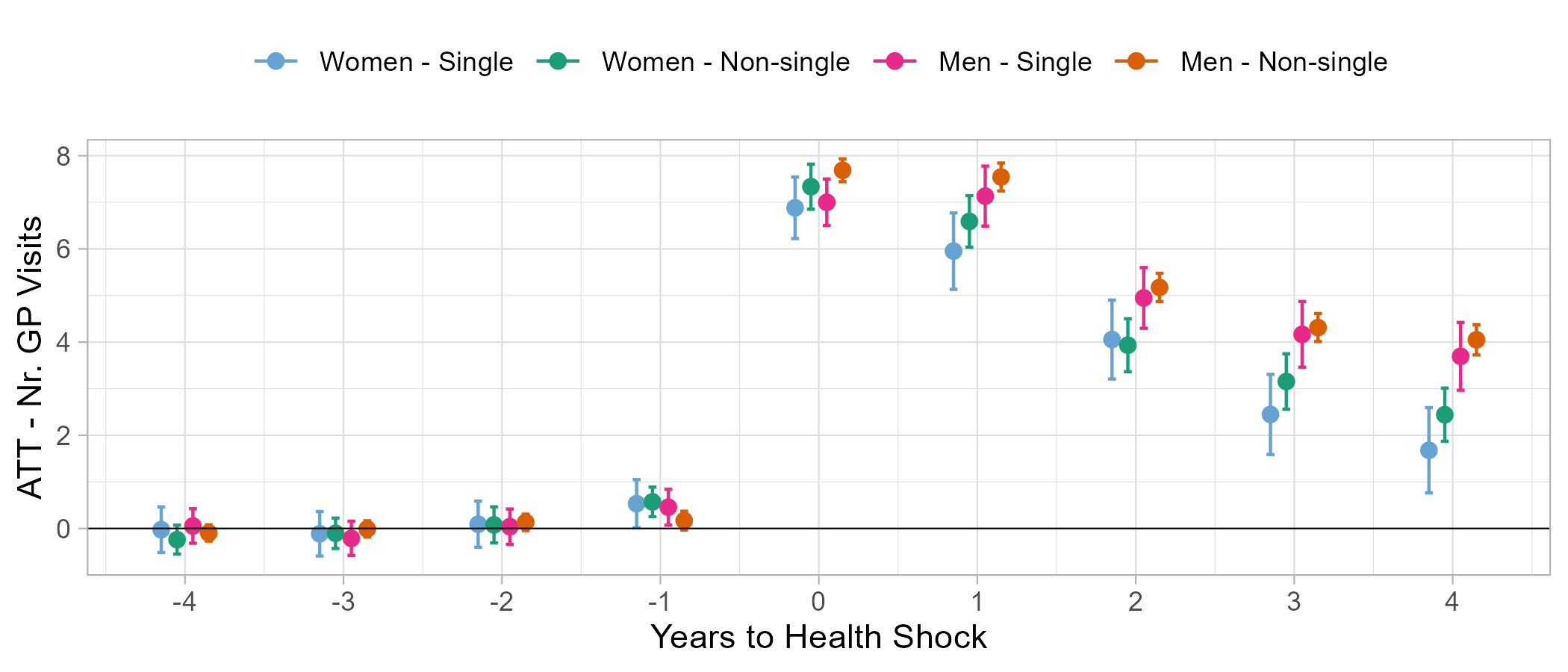}
        \caption{GP visits.}
    \end{subfigure}
    \vspace{0.4cm}
    \begin{subfigure}{0.95\textwidth}
        \centering
        \includegraphics[width=\textwidth]{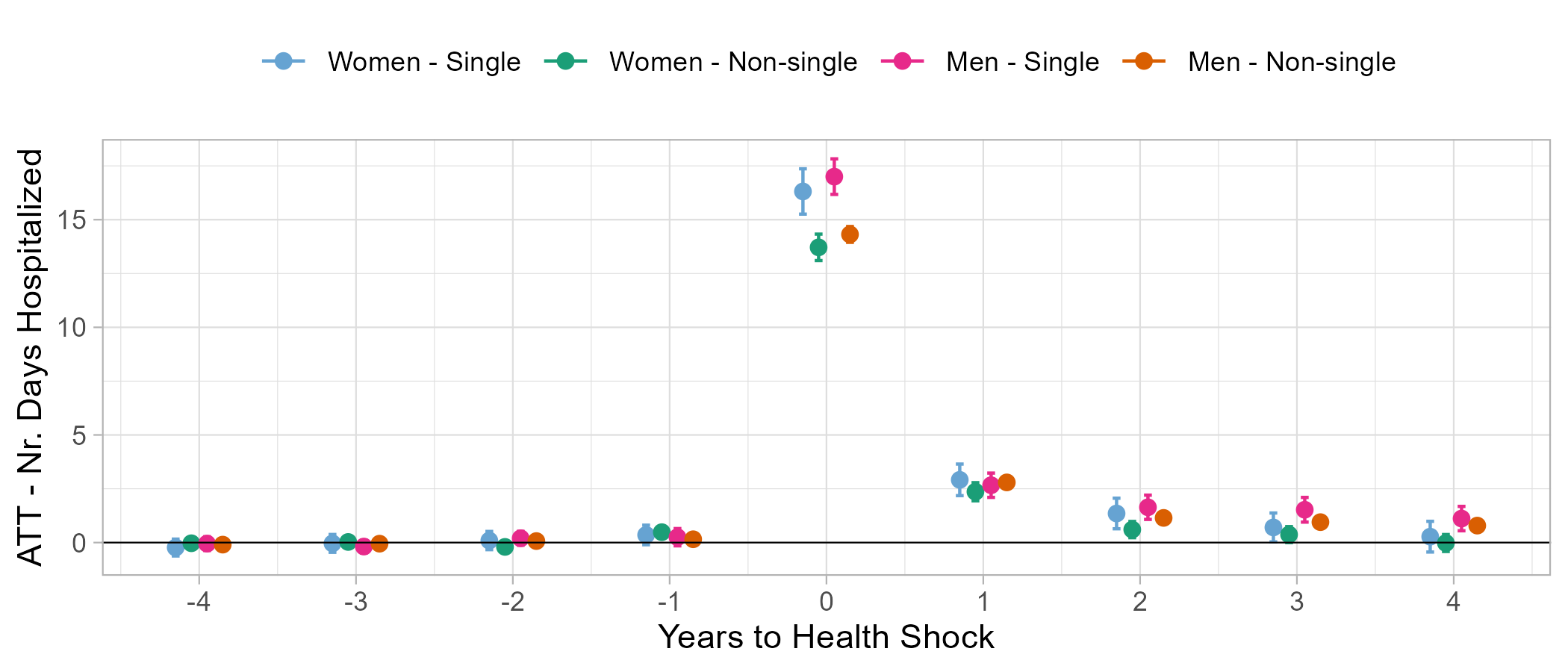}
        \caption{Days hospitalised.}
    \end{subfigure}
    \caption{Dynamic ATT by household type (single vs.\ non-single).}
    \label{fig:subgrouping_household}
    \begin{minipage}{0.9\textwidth}
    \footnotesize \textit{Notes:} Estimator and inference as in Figure~\ref{fig:dynamic_att_full}, estimated separately by gender and by household type (single vs.\ non-single) at $t=g-1$.
    \end{minipage}
\end{figure}

\FloatBarrier
\clearpage

\subsection{Statin pickups} \label{sec:statinpickup}
\begin{table}[ht]
\centering
\caption{Statin pickup count: dynamic ATT.}
\label{tab:s_pickup_overallATT}
\begin{tabular}{m{2em}ccccc}
  \hline\hline
    \multicolumn{1}{p{2cm}}{\textbf{Event Time}} & \textbf{Avg ATT} & \textbf{CI-Lower}  & \multicolumn{1}{p{5em}}{\textbf{CI-Upper}} & \multicolumn{1}{p{4em}}{\textbf{No.\ unique treated}} & \textbf{$p$-value}\\
  \hline
  \multicolumn{2}{l}{\textbf{Panel A: Women}} & &  & & \\
  \hline
$-4$ & $-0.01$ & $-0.03$ & $0.01$  & 12{,}927 & 0.29 \\
$-3$ & $-0.00$ & $-0.02$ & $0.02$  & 13{,}616 & 0.82 \\
$-2$ & $-0.01$ & $-0.04$ & $0.01$  & 14{,}313 & 0.08 \\
$-1$ & $0.04$  & $0.02$  & $0.06$  & 14{,}342 & $<0.01$ \\
$0$  & $1.26$  & $1.21$  & $1.30$  & 15{,}140 & $<0.01$ \\
$1$  & $1.99$  & $1.91$  & $2.06$  & 14{,}363 & $<0.01$ \\
$2$  & $2.03$  & $1.94$  & $2.12$  & 13{,}046 & $<0.01$ \\
$3$  & $2.06$  & $1.96$  & $2.16$  & 11{,}934 & $<0.01$ \\
$4$  & $2.02$  & $1.92$  & $2.13$  & 10{,}918 & $<0.01$ \\
  \hline
  \multicolumn{2}{l}{\textbf{Panel B: Men}} & & &  & \\
  \hline
$-4$ & $-0.00$ & $-0.02$ & $0.01$  & 24{,}934 & 0.43 \\
$-3$ & $-0.01$ & $-0.02$ & $0.01$  & 26{,}468 & 0.16 \\
$-2$ & $0.00$  & $-0.01$ & $0.02$  & 28{,}110 & 0.84 \\
$-1$ & $0.02$  & $0.00$  & $0.03$  & 28{,}307 & $<0.01$ \\
$0$  & $1.37$  & $1.34$  & $1.40$  & 30{,}103 & $<0.01$ \\
$1$  & $2.10$  & $2.05$  & $2.15$  & 28{,}547 & $<0.01$ \\
$2$  & $2.06$  & $2.01$  & $2.12$  & 25{,}945 & $<0.01$ \\
$3$  & $2.04$  & $1.98$  & $2.10$  & 23{,}790 & $<0.01$ \\
$4$  & $2.01$  & $1.95$  & $2.08$  & 21{,}944 & $<0.01$ \\
\hline\hline
\end{tabular}
\\[0.5em]
\begin{minipage}{0.95\textwidth}
\footnotesize \textit{Notes:} Annual count of separate statin pickups at a Danish pharmacy. Estimator and inference as in Table~\ref{tab:Statinmain}.
\end{minipage}
\end{table}

\clearpage

\subsection{Healthcare expenditure} \label{app:healthcare_expenditure}

\begin{table}[ht]
\centering
\caption{Statin out-of-pocket expenditure: dynamic ATT.}
\label{tab:ATT_expenditure_statins}
\begin{tabular}{m{2em}ccccc}
  \hline\hline
  \multicolumn{1}{p{2cm}}{\textbf{Event Time}} & \textbf{Avg ATT (DKK)} & \textbf{CI-Lower}  & \multicolumn{1}{p{5em}}{\textbf{CI-Upper}}  & \multicolumn{1}{p{4em}}{\textbf{No.\ unique treated}} & \textbf{$p$-value}\\
  \hline
  \multicolumn{2}{l}{\textbf{Panel A: Women}} & & & & \\
  \hline
$-4$ & $0.17$  & $-3.72$ & $4.06$  & 12{,}927 & 0.91 \\
$-3$ & $0.78$  & $-3.83$ & $5.38$  & 13{,}616 & 0.63 \\
$-2$ & $-3.54$ & $-7.84$ & $0.77$  & 14{,}313 & 0.02 \\
$-1$ & $2.68$  & $-1.82$ & $7.17$  & 14{,}342 & 0.10 \\
$0$  & $147.61$ & $138.05$ & $157.18$  & 15{,}140 & $<0.01$ \\
$1$  & $218.18$ & $205.16$ & $231.20$  & 14{,}363 & $<0.01$ \\
$2$  & $208.16$ & $194.65$ & $221.66$  & 13{,}046 & $<0.01$ \\
$3$  & $193.05$ & $179.33$ & $206.77$  & 11{,}934 & $<0.01$ \\
$4$  & $178.60$ & $164.04$ & $193.16$  & 10{,}918 & $<0.01$ \\
  \hline
  \multicolumn{2}{l}{\textbf{Panel B: Men}} & & & & \\
  \hline
$-4$ & $0.92$  & $-2.54$ & $4.38$  & 24{,}934 & 0.47 \\
$-3$ & $-1.61$ & $-4.99$ & $1.77$  & 26{,}468 & 0.19 \\
$-2$ & $-1.52$ & $-4.89$ & $1.85$  & 28{,}110 & 0.21 \\
$-1$ & $2.79$  & $-0.43$ & $6.01$  & 28{,}307 & 0.02 \\
$0$  & $184.01$ & $176.28$ & $191.74$  & 30{,}103 & $<0.01$ \\
$1$  & $267.82$ & $257.63$ & $278.02$  & 28{,}547 & $<0.01$ \\
$2$  & $250.34$ & $239.82$ & $260.87$  & 25{,}945 & $<0.01$ \\
$3$  & $227.56$ & $216.81$ & $238.31$  & 23{,}790 & $<0.01$ \\
$4$  & $210.80$ & $199.29$ & $222.31$  & 21{,}944 & $<0.01$ \\
\hline\hline
\end{tabular}
\\[0.5em]
\begin{minipage}{0.95\textwidth}
\footnotesize \textit{Notes:} Annual patient out-of-pocket statin expenditure in 2015 DKK (1 EUR $\approx$ 7.46 DKK). Estimator and inference as in Table~\ref{tab:Statinmain}.
\end{minipage}
\end{table}

\begin{table}[ht]
\centering
\caption{GP visit fees (paid by public insurance): dynamic ATT.}
\label{tab:ATT_expenditure_gp_visits}
\begin{tabular}{m{2em}ccccc}
  \hline\hline
  \multicolumn{1}{p{2cm}}{\textbf{Event Time}} & \textbf{Avg ATT (DKK)} & \textbf{CI-Lower}  & \multicolumn{1}{p{5em}}{\textbf{CI-Upper}} & \multicolumn{1}{p{4em}}{\textbf{No.\ unique treated}} & \textbf{$p$-value}\\
  \hline
  \multicolumn{2}{l}{\textbf{Panel A: Women}} & & & & \\
  \hline
$-4$ & $143.50$ & $-1{,}285.31$ & $1{,}572.32$  & 12{,}927 & 0.78 \\
$-3$ & $94.24$  & $-1{,}119.01$ & $1{,}307.48$  & 13{,}616 & 0.84 \\
$-2$ & $-448.47$ & $-1{,}473.15$ & $576.21$  & 14{,}313 & 0.23 \\
$-1$ & $608.61$ & $-399.79$ & $1{,}617.01$  & 14{,}342 & 0.10 \\
$0$  & $3{,}997.68$ & $2{,}892.60$ & $5{,}102.75$  & 15{,}140 & $<0.01$ \\
$1$  & $3{,}790.05$ & $2{,}291.64$ & $5{,}288.47$  & 14{,}363 & $<0.01$ \\
$2$  & $1{,}386.85$ & $223.43$ & $2{,}550.28$  & 13{,}046 & $<0.01$ \\
$3$  & $1{,}021.10$ & $-207.49$ & $2{,}249.68$  & 11{,}934 & 0.02 \\
$4$  & $683.02$ & $-879.70$ & $2{,}245.74$  & 10{,}918 & 0.21 \\
  \hline
  \multicolumn{2}{l}{\textbf{Panel B: Men}} & & & & \\
  \hline
$-4$ & $-294.08$ & $-994.09$ & $405.94$  & 24{,}934 & 0.25 \\
$-3$ & $-340.93$ & $-947.54$ & $265.68$  & 26{,}468 & 0.12 \\
$-2$ & $342.32$  & $-268.35$ & $953.00$  & 28{,}110 & 0.13 \\
$-1$ & $13.76$   & $-605.79$ & $633.32$  & 28{,}307 & 0.96 \\
$0$  & $4{,}353.89$ & $3{,}751.14$ & $4{,}956.64$  & 30{,}103 & $<0.01$ \\
$1$  & $4{,}286.78$ & $3{,}597.15$ & $4{,}976.42$  & 28{,}547 & $<0.01$ \\
$2$  & $3{,}845.77$ & $2{,}977.85$ & $4{,}713.69$  & 25{,}945 & $<0.01$ \\
$3$  & $2{,}858.35$ & $2{,}082.19$ & $3{,}634.51$  & 23{,}790 & $<0.01$ \\
$4$  & $2{,}579.34$ & $1{,}424.38$ & $3{,}734.30$  & 21{,}944 & $<0.01$ \\
  \hline\hline
\end{tabular}
\\[0.5em]
\begin{minipage}{0.95\textwidth}
\footnotesize \textit{Notes:} Annual gross fee paid by the public insurance system to primary-care providers, in 2015 DKK (1 EUR $\approx$ 7.46 DKK). Excludes patient co-payments. Estimator and inference as in Table~\ref{tab:Statinmain}.
\end{minipage}
\end{table}

\clearpage

\subsection{Long-run mortality: 20-year window}\label{app:mortality_20yr}

Table~\ref{tab:mortality_20yr_full} reports the full dynamic ATT on annual mortality using a 20-year follow-up window. Sample sizes shrink with event time because only cohorts treated sufficiently early in the panel contribute to estimates at long horizons.

\begin{table}[ht]
\centering
\caption{Dynamic ATT on annual mortality, 20-year window.}
\label{tab:mortality_20yr_full}
\footnotesize
\begin{tabular}{ccccc|ccc}
\toprule
& \multicolumn{3}{c}{\textbf{Women}} & & \multicolumn{3}{c}{\textbf{Men}} \\
\cmidrule(lr){2-4} \cmidrule(lr){6-8}
\textbf{Year} & ATT & 95\% CI & $N$ & & ATT & 95\% CI & $N$ \\
\midrule
0  & 0.00 & [0.00, 0.00] & 13{,}949 & & 0.00 & [0.00, 0.00] & 27{,}175 \\
1  & 0.04 & [0.03, 0.04] & 13{,}172 & & 0.04 & [0.04, 0.04] & 25{,}621 \\
2  & 0.03 & [0.03, 0.03] & 11{,}855 & & 0.03 & [0.03, 0.03] & 23{,}021 \\
3  & 0.03 & [0.02, 0.03] & 10{,}745 & & 0.03 & [0.03, 0.03] & 20{,}869 \\
4  & 0.03 & [0.02, 0.03] & 9{,}730  & & 0.03 & [0.02, 0.03] & 19{,}025 \\
5  & 0.03 & [0.02, 0.03] & 8{,}752  & & 0.03 & [0.03, 0.03] & 17{,}268 \\
6  & 0.03 & [0.02, 0.04] & 7{,}886  & & 0.03 & [0.03, 0.04] & 15{,}570 \\
7  & 0.02 & [0.02, 0.03] & 7{,}014  & & 0.03 & [0.03, 0.04] & 13{,}962 \\
8  & 0.03 & [0.03, 0.04] & 6{,}223  & & 0.03 & [0.03, 0.04] & 12{,}370 \\
9  & 0.03 & [0.02, 0.04] & 5{,}505  & & 0.04 & [0.03, 0.04] & 10{,}910 \\
10 & 0.04 & [0.03, 0.04] & 4{,}860  & & 0.04 & [0.03, 0.04] & 9{,}475  \\
11 & 0.03 & [0.03, 0.04] & 4{,}191  & & 0.05 & [0.04, 0.05] & 8{,}196  \\
12 & 0.04 & [0.03, 0.05] & 3{,}607  & & 0.04 & [0.03, 0.05] & 7{,}014  \\
13 & 0.04 & [0.03, 0.05] & 2{,}992  & & 0.04 & [0.03, 0.05] & 5{,}927  \\
14 & 0.04 & [0.03, 0.05] & 2{,}490  & & 0.04 & [0.04, 0.05] & 4{,}909  \\
15 & 0.04 & [0.03, 0.06] & 1{,}990  & & 0.05 & [0.04, 0.06] & 3{,}988  \\
16 & 0.04 & [0.03, 0.05] & 1{,}516  & & 0.04 & [0.03, 0.05] & 3{,}050  \\
17 & 0.04 & [0.03, 0.06] & 1{,}105  & & 0.05 & [0.04, 0.06] & 2{,}274  \\
18 & 0.04 & [0.02, 0.06] & 792      & & 0.04 & [0.03, 0.05] & 1{,}586  \\
19 & 0.04 & [0.02, 0.06] & 503      & & 0.05 & [0.04, 0.07] & 979      \\
\bottomrule
\end{tabular}
\\[0.5em]
\begin{minipage}{0.95\textwidth}
\footnotesize \textit{Notes:} ATT on the annual probability of death, by event time, estimated using the \cite{callaway2021difference} doubly robust estimator with the maximally feasible cohort selection at each horizon. Confidence intervals at the 95\% level via the multiplier bootstrap with individual-level clustering. All $p$-values for non-zero post-event ATTs are below 0.01. The gender difference is statistically indistinguishable from zero at every horizon.
\end{minipage}
\end{table}

\FloatBarrier
\clearpage

\subsection{Cardiovascular drug battery: raw trends}\label{app:cardio_drugs}

\begin{figure}[H]
    \centering
    \begin{subfigure}{0.48\textwidth}
        \centering
        \includegraphics[width=\textwidth]{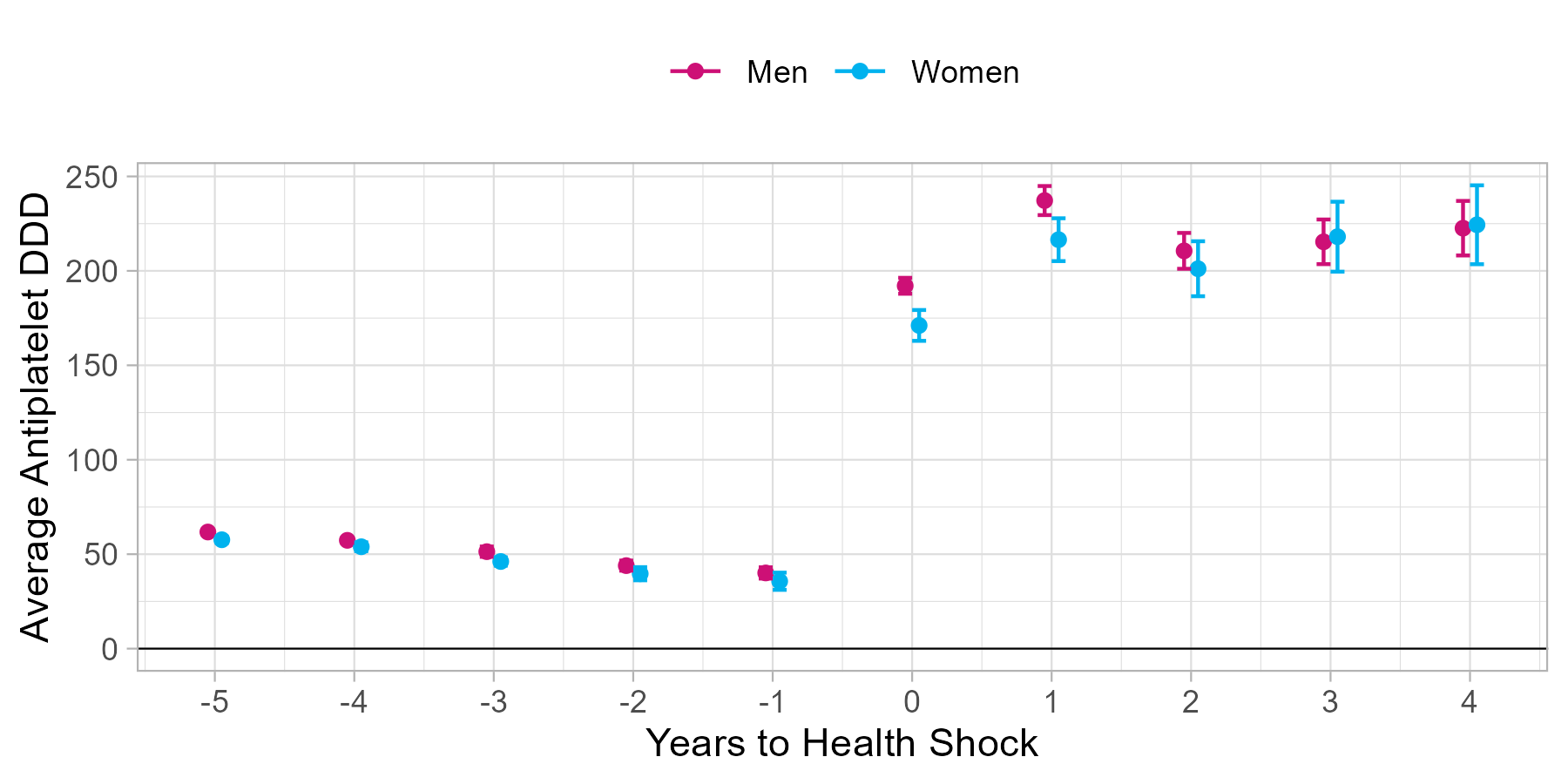}
        \caption{Antiplatelets (raw DDD trend).}
    \end{subfigure}
    \hfill
    \begin{subfigure}{0.48\textwidth}
        \centering
        \includegraphics[width=\textwidth]{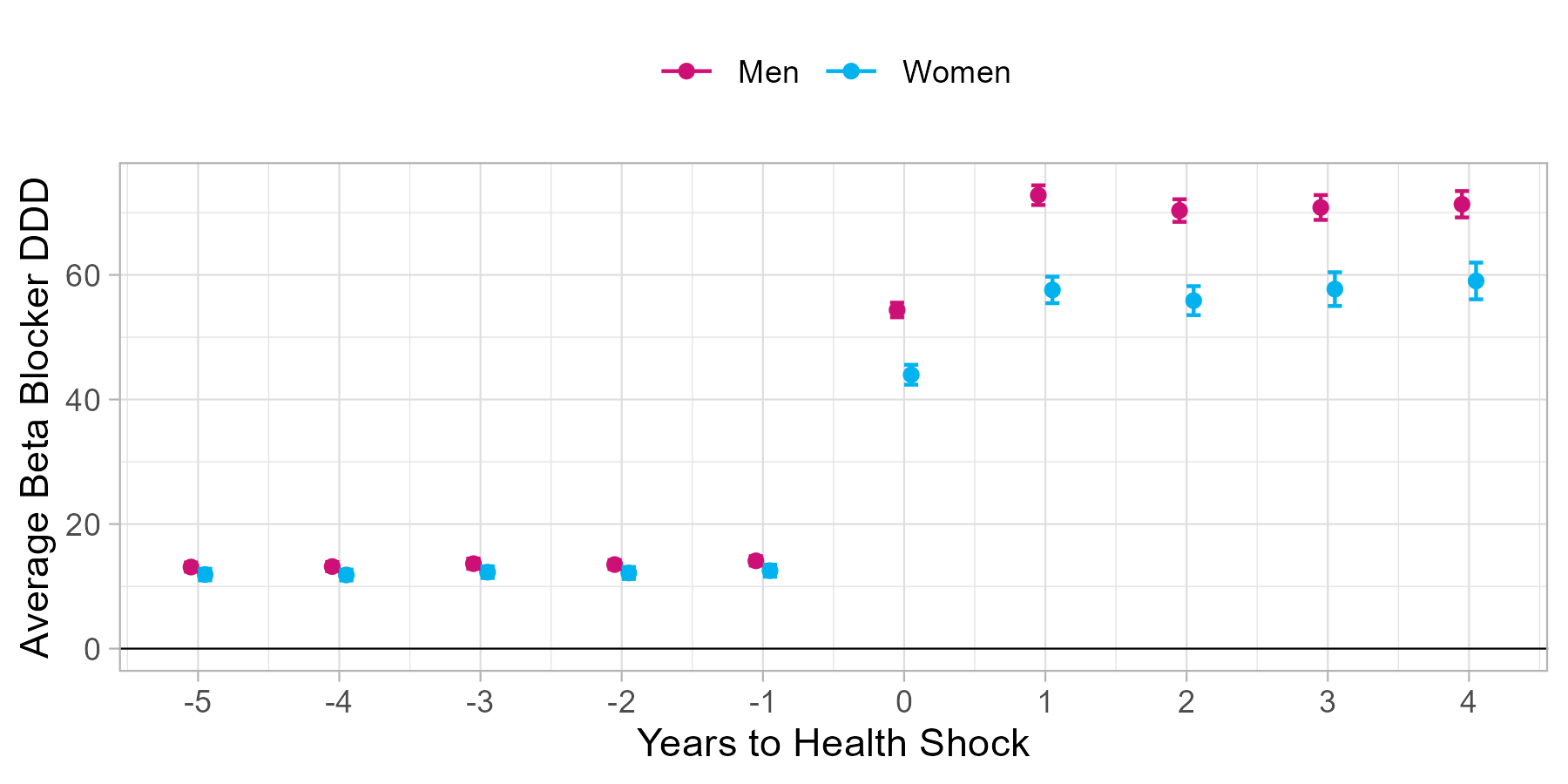}
        \caption{Beta blockers (raw DDD trend).}
    \end{subfigure}

    \vspace{0.3cm}

    \begin{subfigure}{0.48\textwidth}
        \centering
        \includegraphics[width=\textwidth]{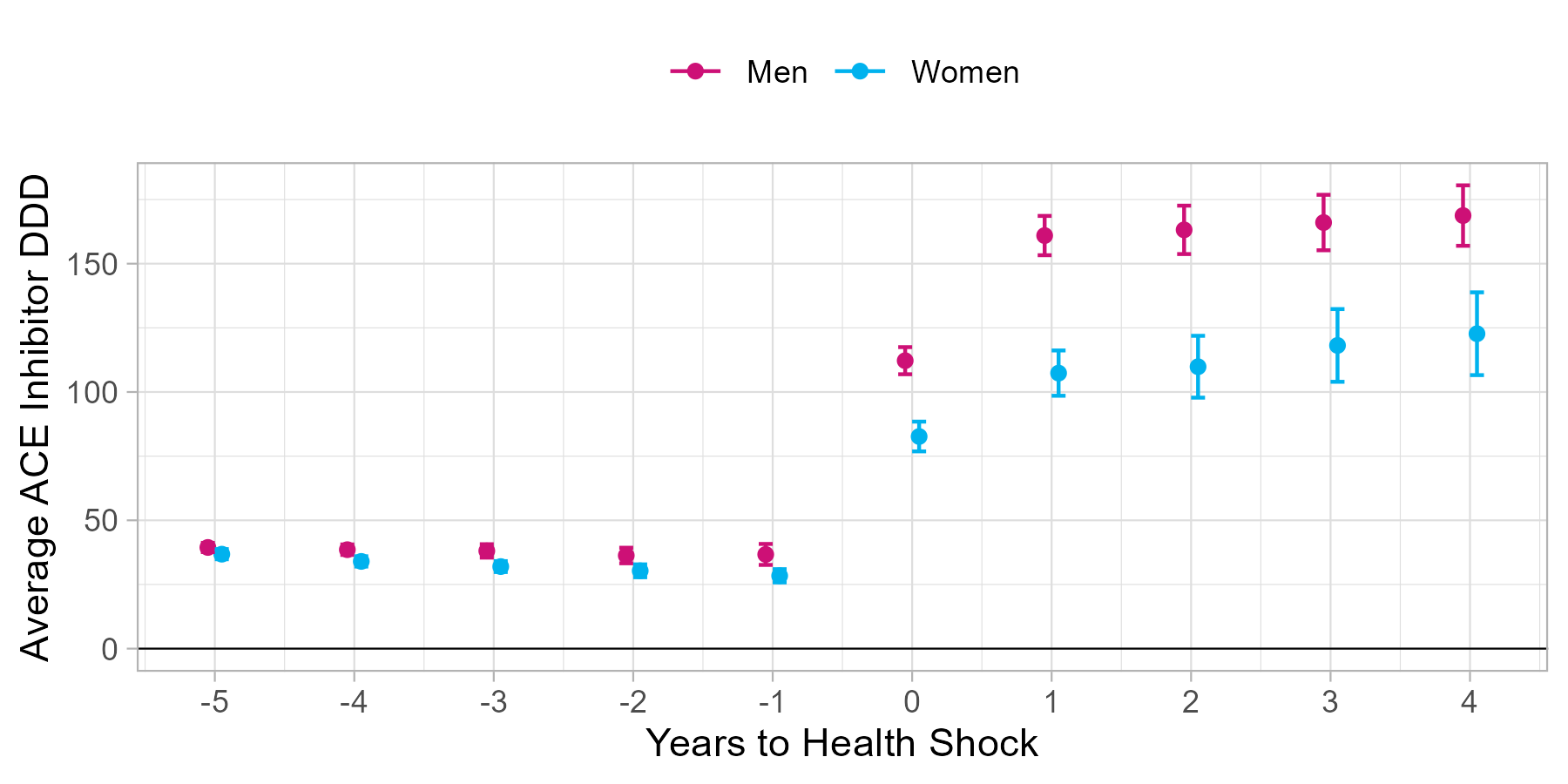}
        \caption{ACE inhibitors (raw DDD trend).}
    \end{subfigure}
    \hfill
    \begin{subfigure}{0.48\textwidth}
        \centering
        \includegraphics[width=\textwidth]{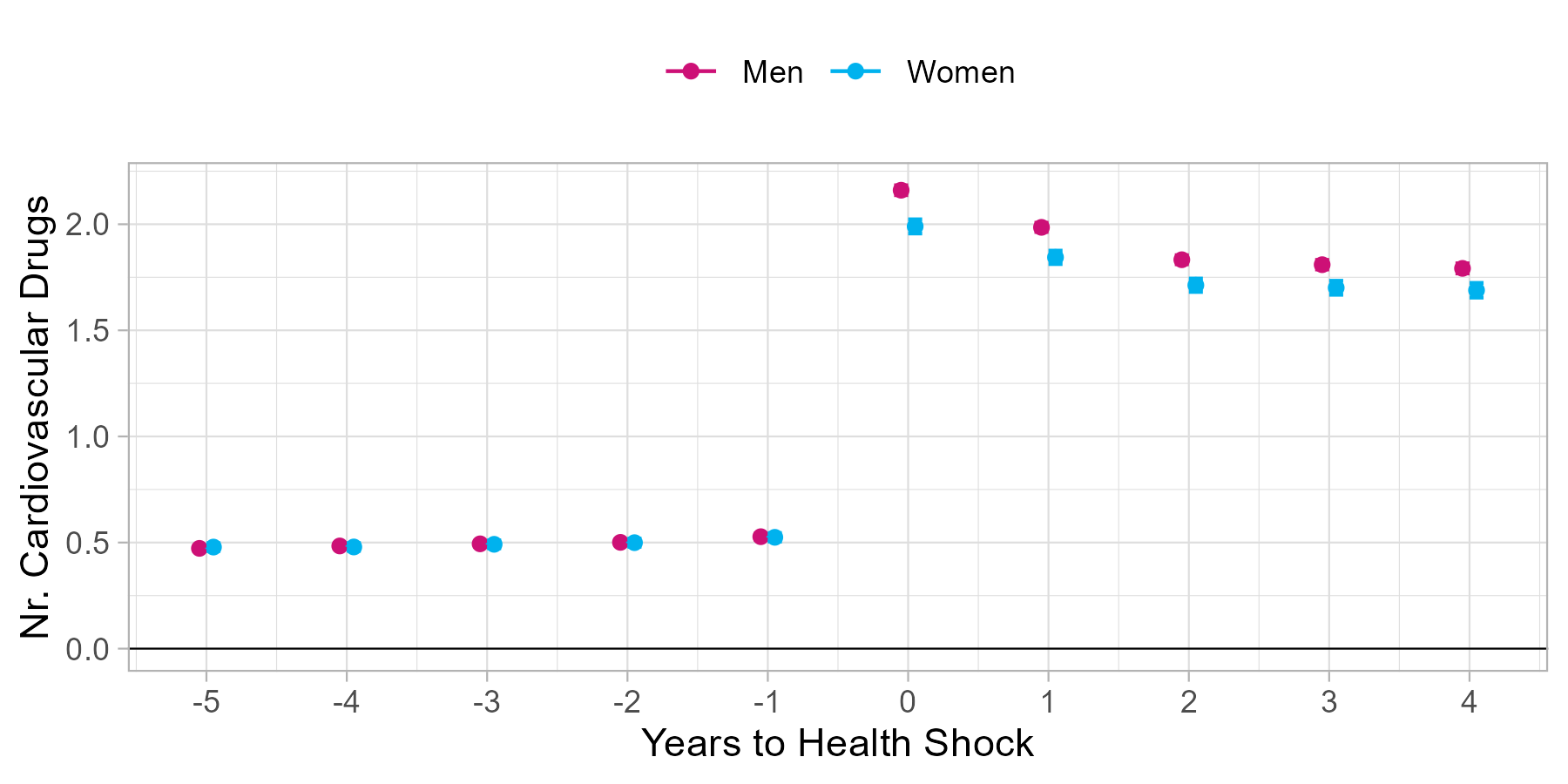}
        \caption{Number of distinct cardiovascular drugs (raw trend).}
    \end{subfigure}

    \caption{Raw cohort-mean trends in the cardiovascular drug battery, by gender and event time.}
    \label{fig:cardio_drug_trends}
    \begin{minipage}{0.9\textwidth}
    \footnotesize \textit{Notes:} Each panel reports the cohort-weighted mean of the indicated outcome by event time, separately for men and women, with year fixed effects partialled out. The corresponding dynamic ATT estimates are in Figure~\ref{fig:cardio_drug_battery}.
    \end{minipage}
\end{figure}

For completeness, Figure~\ref{fig:cardio_drug_trends} reports the raw cohort-mean trends for each of the four secondary-prevention drug classes documented in Section~\ref{sec:gradient}, separately by gender. The patterns parallel the dynamic ATTs in Figure~\ref{fig:cardio_drug_battery}.

\FloatBarrier
\clearpage

\subsection{Cohort-specific (group-specific) ATTs}
\label{sec:group_att}

\begin{figure}[H]
    \centering
    \begin{subfigure}{0.65\textwidth}
        \centering
        \includegraphics[width=\textwidth]{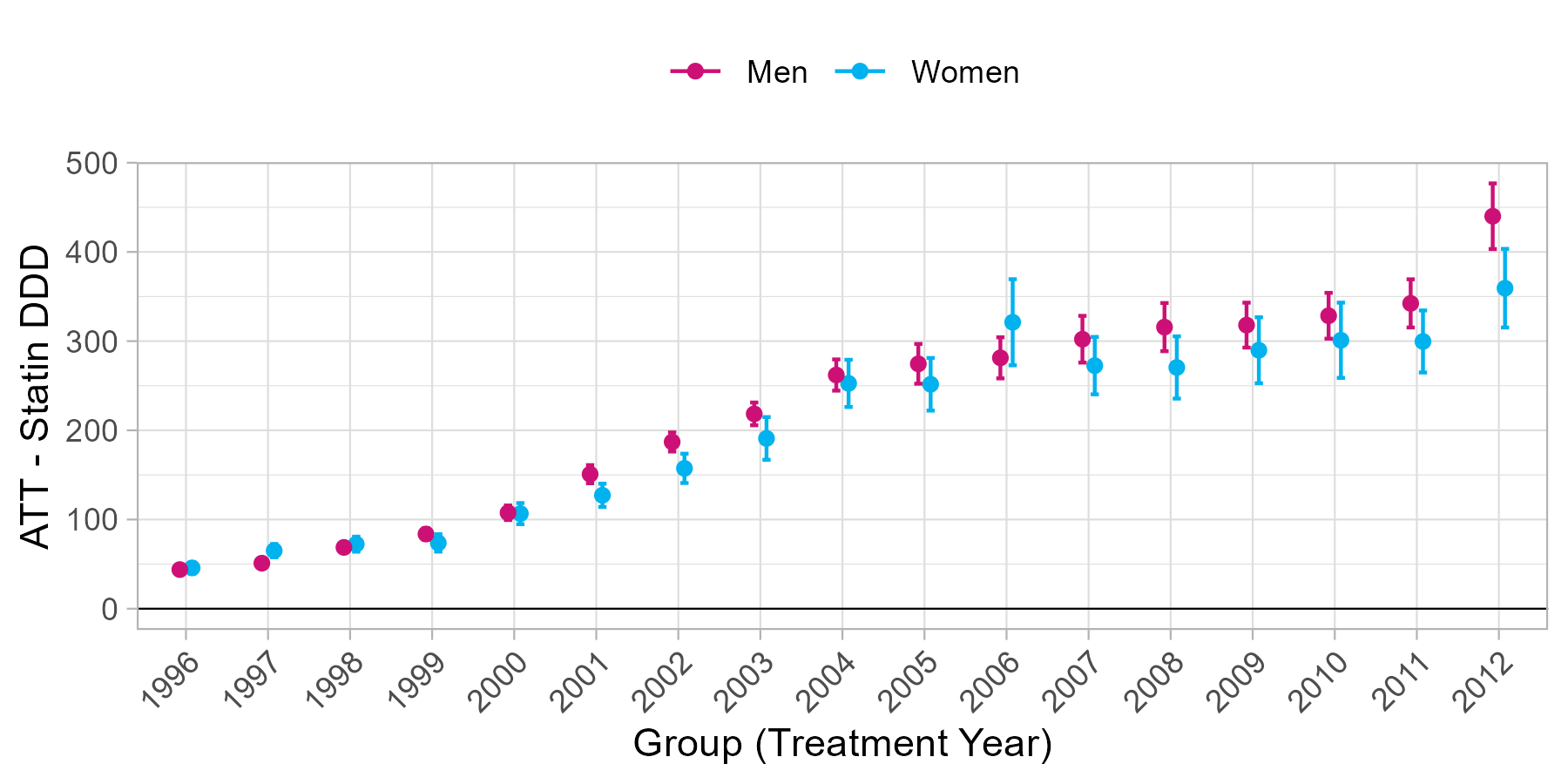}
        \caption{Statin DDD.}
    \end{subfigure}
    \vspace{0.4cm}
    \begin{subfigure}{0.65\textwidth}
        \centering
        \includegraphics[width=\textwidth]{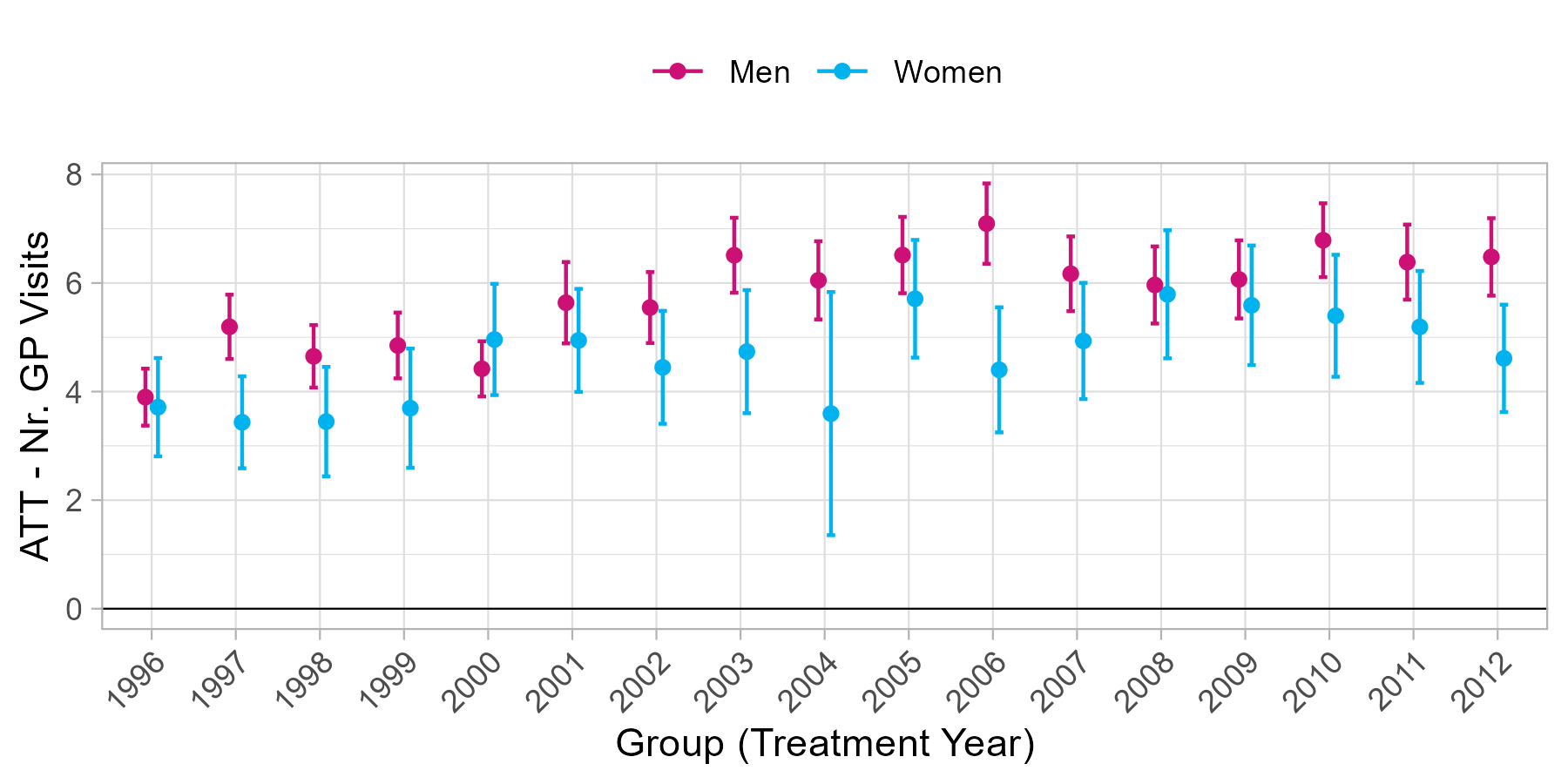}
        \caption{GP visits.}
    \end{subfigure}
    \vspace{0.4cm}
    \begin{subfigure}{0.65\textwidth}
        \centering
        \includegraphics[width=\textwidth]{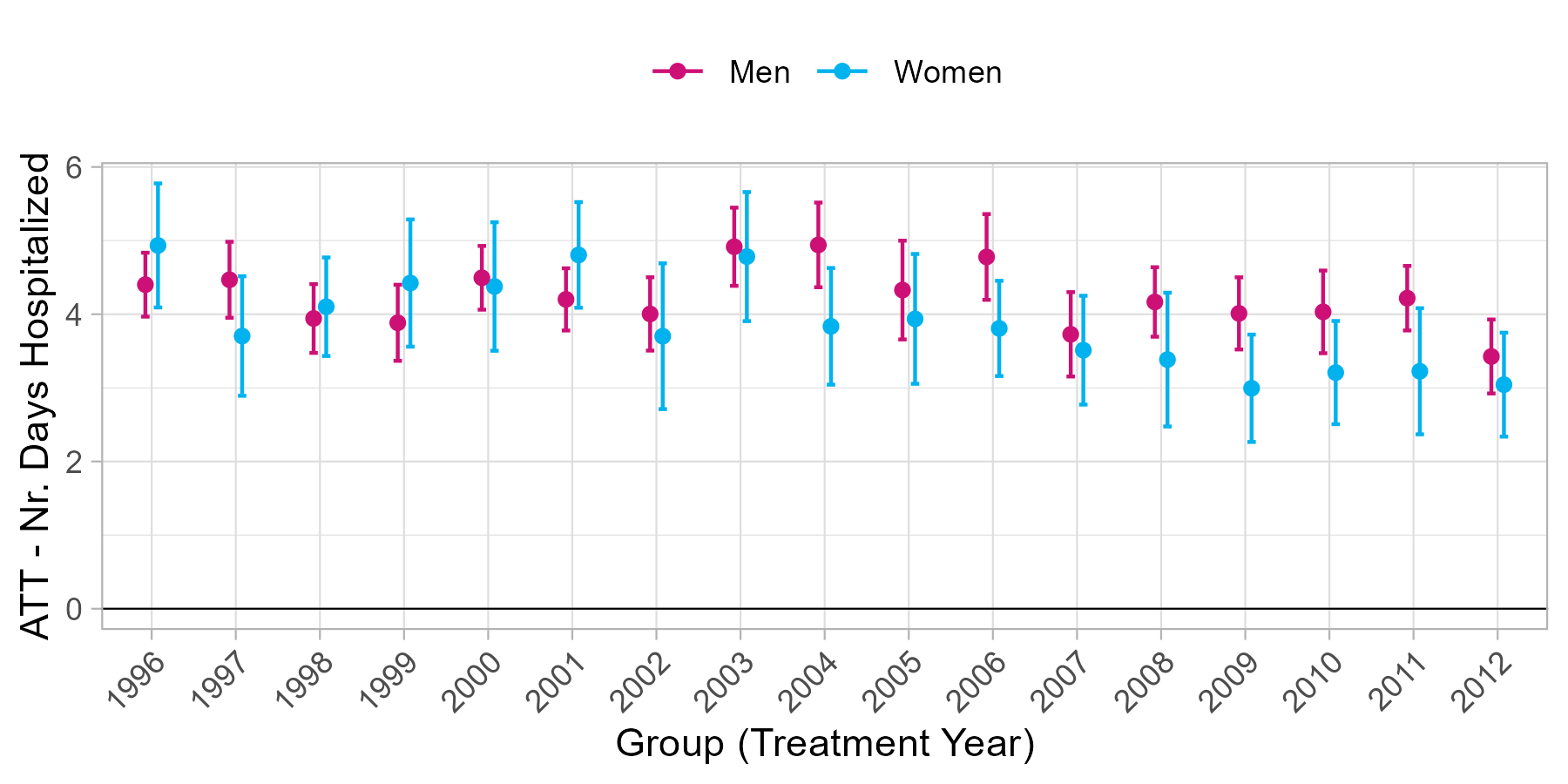}
        \caption{Days hospitalised.}
    \end{subfigure}
    \caption{Cohort-specific ATTs by treatment year.}
    \label{fig:group_att}
    \begin{minipage}{0.9\textwidth}
    \footnotesize \textit{Notes:} Cohort-specific (group) ATTs aggregated over post-shock event times, by treatment year and gender. Estimator and inference as in Figure~\ref{fig:dynamic_att_full}.
    \end{minipage}
\end{figure}

Figure~\ref{fig:group_att} reports cohort-specific (group) ATTs, aggregated over post-shock event times, separately by gender and by treatment year. Two features of the design shape these estimates at the panel boundaries: cohorts treated late in the panel contribute fewer post-shock event times, and the latest cohorts have fewer not-yet-treated individuals available as controls within the five-year window, so their estimates are correspondingly less precise. Across treatment years the gender pattern is stable, indicating that no single cohort drives the results.

\FloatBarrier

\FloatBarrier
\clearpage

\subsection{Morbidity and labour-market outcomes: dynamic ATTs}\label{app:other_margins}

\begin{figure}[H]
    \centering
    \begin{subfigure}{0.48\textwidth}
        \centering
        \includegraphics[width=\textwidth]{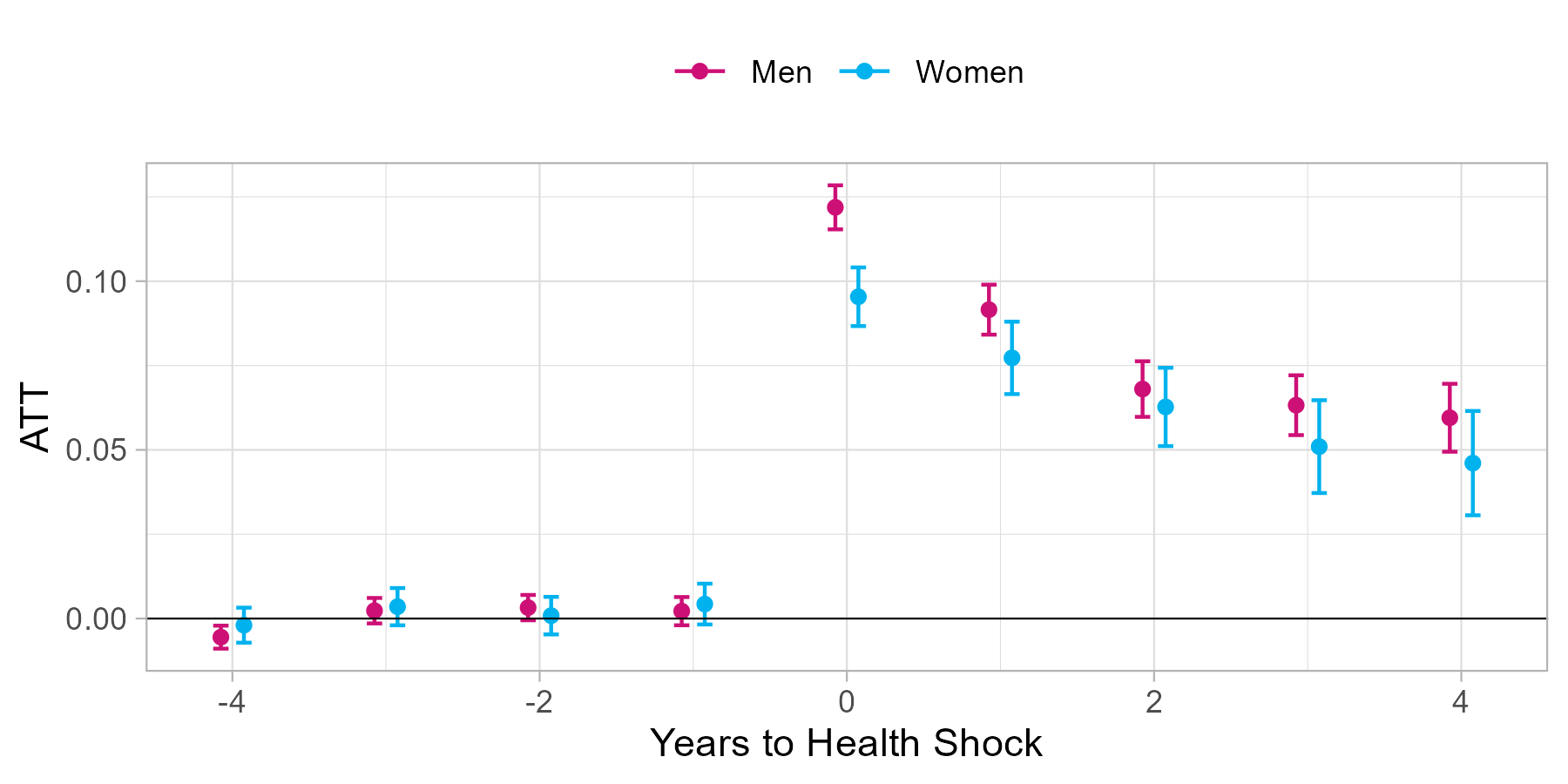}
        \caption{Cardiac comorbidities.}
    \end{subfigure}
    \hfill
    \begin{subfigure}{0.48\textwidth}
        \centering
        \includegraphics[width=\textwidth]{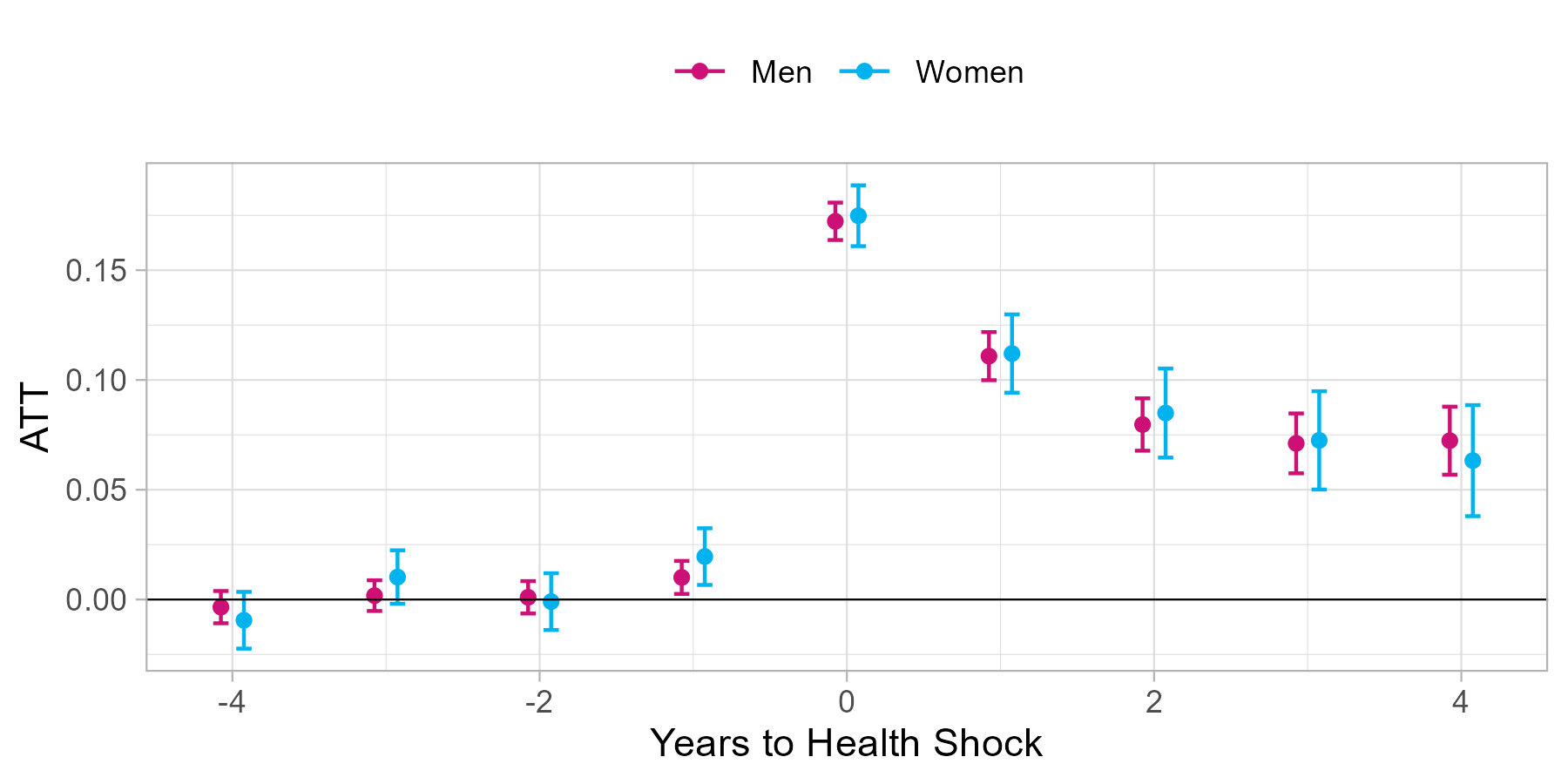}
        \caption{Non-cardiac comorbidities.}
    \end{subfigure}
    \caption{Dynamic ATT for comorbidity accumulation, by gender.}
    \label{fig:dynamic_other_morbidity}
    \begin{minipage}{0.9\textwidth}
    \footnotesize \textit{Notes:} Gender-specific dynamic ATTs from the \cite{callaway2021difference} doubly robust estimator with simultaneous 95\% multiplier-bootstrap confidence bands and individual-level clustering. Aggregate estimates in Table~\ref{tab:other_margins}.
    \end{minipage}
\end{figure}

\begin{figure}[H]
    \centering
    \includegraphics[width=0.7\textwidth]{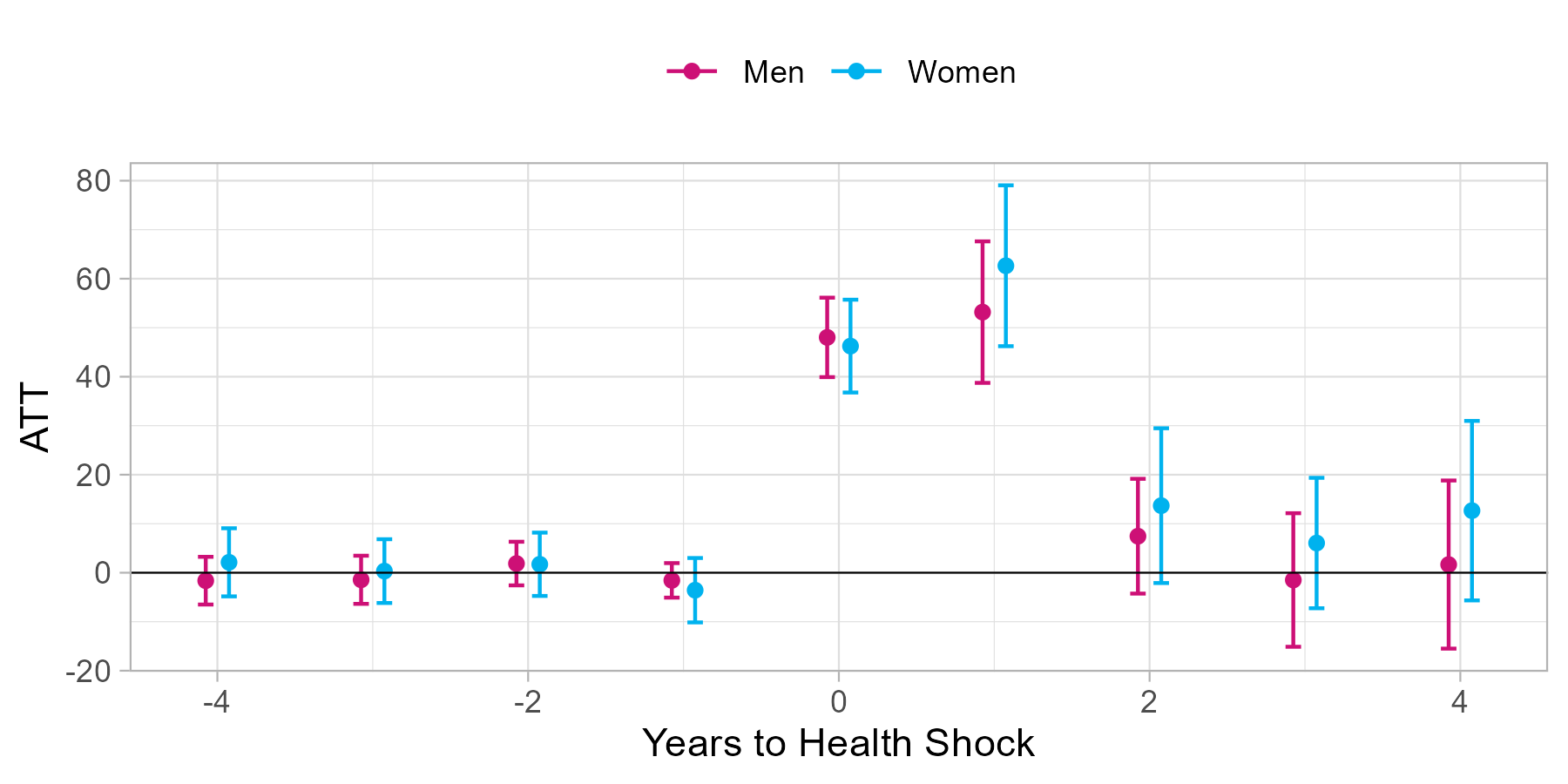}
    \caption{Dynamic ATT for work absence, by gender.}
    \label{fig:dynamic_other_labour}
    \begin{minipage}{0.9\textwidth}
    \footnotesize \textit{Notes:} Gender-specific dynamic ATTs from the \cite{callaway2021difference} doubly robust estimator with simultaneous 95\% multiplier-bootstrap confidence bands and individual-level clustering. The labour-market sample is restricted to working-age, labour-market-attached individuals (Table~\ref{tab:other_margins}). Aggregate estimates in Table~\ref{tab:other_margins}.
    \end{minipage}
\end{figure}

Figures~\ref{fig:dynamic_other_morbidity}--\ref{fig:dynamic_other_labour} report the gender-specific dynamic ATTs underlying the aggregate estimates in Table~\ref{tab:other_margins}.

\end{document}